\RequirePackage{fix-cm}
\documentclass[natbib,smallextended]{svjour3} 
\smartqed  
\usepackage{graphicx}

\usepackage{aas_macros}
\usepackage{color}
\usepackage[colorlinks=true,linkcolor=blue,citecolor=blue,urlcolor=blue]{hyperref}
\usepackage{txfonts}
\usepackage{lscape}
\usepackage{comment}
\usepackage{wrapfig}

\usepackage{amsmath}

\usepackage{xpatch}
\usepackage{hyperref}
\hypersetup{
	colorlinks=true,
	breaklinks=true,
	citecolor=blue,
	allcolors=blue,
	frenchlinks=true
}

\makeatletter
\xpatchcmd\NAT@citex
{%
	\@citea\NAT@hyper@{%
		\NAT@nmfmt{\NAT@nm}%
		\hyper@natlinkbreak{\NAT@aysep\NAT@spacechar}{\@citeb\@extra@b@citeb}%
		\NAT@date
	}%
}
{%
	\@citea
	\NAT@nmfmt{\NAT@nm}%
	\NAT@aysep\NAT@spacechar
	\NAT@hyper@{\NAT@date}%
}
{}{}
\xpatchcmd\NAT@citex
{%
	\@citea\NAT@hyper@{%
		\NAT@nmfmt{\NAT@nm}%
		\hyper@natlinkbreak{\NAT@spacechar\NAT@@open\if*#1*\else#1\NAT@spacechar\fi}%
		{\@citeb\@extra@b@citeb}%
		\NAT@date
	}%
}
{
	\@citea
	\NAT@nmfmt{\NAT@nm}%
	\NAT@spacechar\NAT@@open\if*#1*\else#1\NAT@spacechar\fi
	\NAT@hyper@{\NAT@date}%
}
{}{}
\makeatother

\newcommand{\mathbfit}[1]{\textbf{\textit{#1}}}

\newcommand{\mathbfss}[1]{\textbf{\textsf{#1}}}
\newcommand{\mat}{\ensuremath{\mathbfss}}
\newcommand{\bs}[1]{\boldsymbol{#1}}
\newcommand{\bcdot}{\bs{\cdot}}
\newcommand{\rmn}{\mathrm}
\newcommand{\CR}{\mathrm{cr}}
\newcommand{\eps}{\varepsilon}
\newcommand{\dd}{\mathrm{d}}
\newcommand{\bra}{\langle}
\newcommand{\ket}{\rangle}

\DeclareUnicodeCharacter{03B3}{$\gamma$}
\DeclareUnicodeCharacter{223C}{$\sim$}

\journalname{The Astronomy and Astrophysics Review}

\begin{document}

\title{Cosmic ray feedback in galaxies and galaxy clusters}
\subtitle{A pedagogical introduction and a topical review of the acceleration, transport, observables, and dynamical impact of cosmic rays}

\titlerunning{Cosmic ray feedback}

\author{Mateusz Ruszkowski${}^{1,3,*}$, Christoph Pfrommer${}^{2,*}$}
\authorrunning{M. Ruszkowski \& C. Pfrommer}

\institute{$^*$shared first authorship with the authors contributing equally to this work\\
$^1$Department of Astronomy, University of Michigan,
1085 S. University Ave., 323 West Hall, Ann Arbor, MI 48109-1107, USA\\
$^2$Leibniz Institute for Astrophysics Potsdam (AIP), 
An der Sternwarte 16, 14482 Potsdam, Germany\\
$^3$Max Planck Institute for Astrophysics
Karl-Schwarzschild-Str. 1
85748 Garching, Germany\\
\email{Mateusz Ruszkowski: mateuszr@umich.edu, Christoph Pfrommer: cpfrommer@aip.de}
}

\date{Received: date / Accepted: date}

\maketitle

\begin{abstract}
Understanding the physical mechanisms that control galaxy formation is a fundamental challenge in contemporary astrophysics. Recent advances in the field of astrophysical feedback strongly suggest that cosmic rays (CRs) may be crucially important for our understanding of cosmological galaxy formation and evolution. The appealing features of CRs are their relatively long cooling times and relatively strong dynamical coupling to the gas. In galaxies, CRs can be close to equipartition with the thermal, magnetic, and turbulent energy density in the interstellar medium, and can be dynamically very important in driving large-scale galactic winds. Similarly, CRs may provide a significant contribution to the pressure in the circumgalactic medium. In galaxy clusters, CRs may play a key role in addressing the classic cooling flow problem by facilitating efficient heating of the intracluster medium and preventing excessive star formation. Overall, the underlying physics of CR interactions with plasmas exhibit broad parallels across the entire range of scales characteristic of the interstellar, circumgalactic, and intracluster media. Here we present a review of the state-of-the-art of this field and provide a pedagogical introduction to cosmic ray plasma physics, including the physics of wave--particle interactions, acceleration processes, CR spatial and spectral transport, and important cooling processes. The field is ripe for discovery and will remain the subject of intense theoretical, computational, and observational research over the next decade with profound implications for the interpretation of the observations of stellar and supermassive black hole feedback spanning the entire width of the electromagnetic spectrum and multi-messenger data.  

\end{abstract}
\clearpage

\tableofcontents
\clearpage

\section{Introduction} \label{Introduction}
\subsection{Discovery of cosmic rays}
Cosmic rays (CRs) are relativistic charged particles that originate from outside the Solar System. The highest energy CRs are the fastest nuclei in the Universe, moving close to the speed of light. Most CRs are protons and alpha particles. Collisions between these primary CRs and atoms or ions in space produce secondary CRs. In particular, collisions of primary CRs with the atoms in Earth's atmosphere produce CR muons that can be directly detected at sea level (thousands of these particles will pass through your body in the few minutes it takes to read the introduction to this review). 
CRs were discovered over a century ago by an Austrian-American physicist Victor Hess in Bohemia, which was then part of the Austro-Hungarian Empire. In a series of high-altitude balloon flights, Victor Hess measured the rate of ionization as a function of height above ground using an electroscope. 
At the time these experiments were conducted it was generally believed that atmospheric ionization was due to radioactivity of the ground but results were inconclusive in part due to instrumentation defects. This made determining the source of the ionization a longstanding mystery ever since the discovery of radioactivity in 1896 by Henri Becquerel.
In 1912, Hess determined that the ionization rate increased with height above a certain altitude up to four times compared to the value measured at sea level, and this conclusion was unaffected even when the experiments were conducted during a partial solar eclipse \citep{hess_victor_uber_1912}.\footnote{Original publication in German: Hess \href{https://www.mpi-hd.mpg.de/hfm/HESS/public/HessArticle.pdf}{(1912)}; for English translation published in 2018 by Angelis and Schultz, see \citet{hess_observations_1912}.} These experiments firmly established that the source of ionization was extraterrestrial and that the Sun was not the main contributor to the ionizing flux.\footnote{It is worth noting that, in 1912 Domenico Pacini reported on the discovery that the ionizing rate decreased by $\sim$$20\%$ when measurements were made three meters below the surface of water, and concluded that ``a significant proportion of the pervasive radiation that is found in air had an origin that was independent of direct action of the active substances in the upper layers of the Earth’s surface'' \citep[][for English translation by De Angelis from 2017, see Pacini \href{https://arxiv.org/pdf/1002.1810.pdf}{1912}]{pacini_radiazione_1912}.}
The first evidence for cosmic ``rays'' being charged particles was provided by a Dutch physicist Jacob Clay, who observed a decrease in the ionizing flux closer to the equator during his sea voyages between Italy and Indonesia \citep{clay_penetrating_1927,clay_penetrating_1928}. This finding was later interpreted as the result of CRs deflection by the Earth's magnetic field, demonstrating that CRs are charged particles rather than ionizing radiation. In 1936, Victor Hess received a Nobel Prize for the discovery of CRs.\footnote{Victor F. Hess --  \href{https://www.nobelprize.org/prizes/physics/1936/hess/lecture/}{Nobel lecture}. Pacini passed away in 1934, two years before Victor Hess was awarded a Nobel prize for the discovery of CRs.} 
\subsection{Basic properties of cosmic rays}
\begin{figure}[tbp]
\begin{center}
\includegraphics[width=1.05\textwidth]{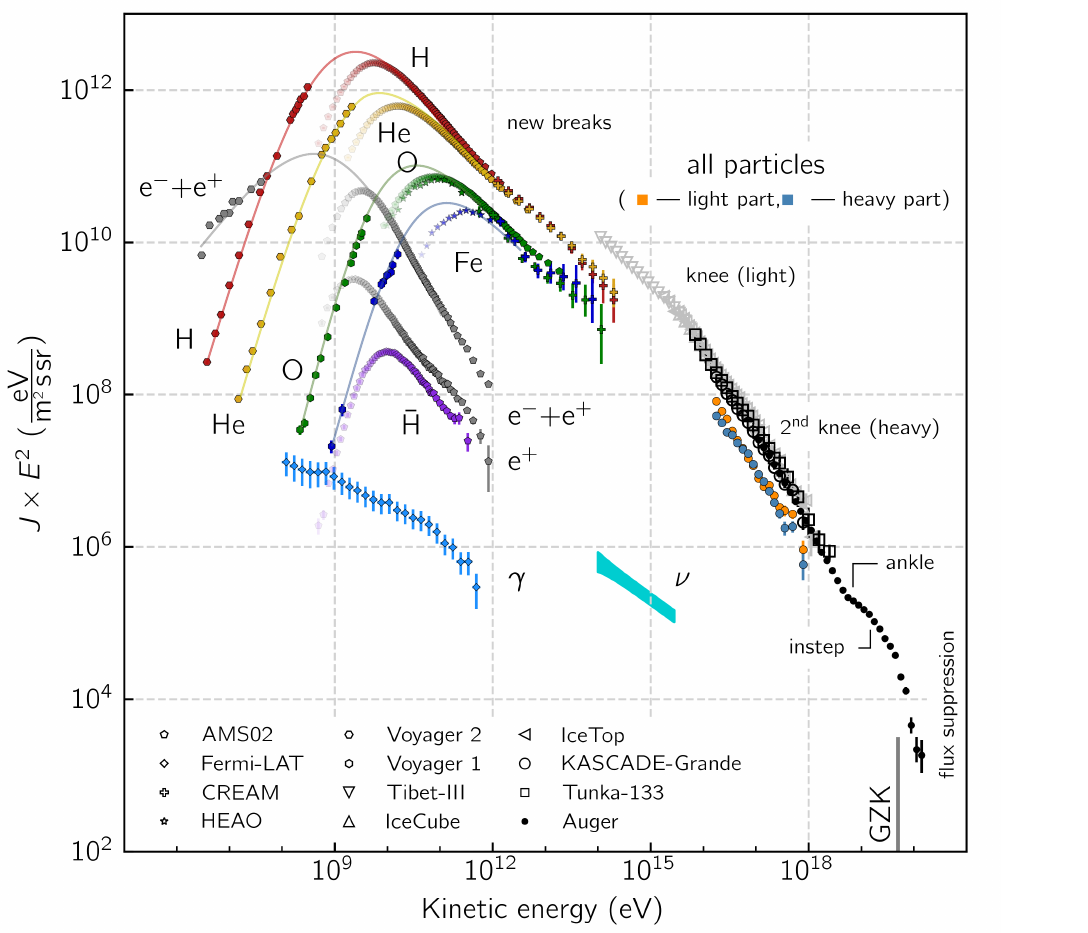}
\end{center}
\caption{Cosmic ray spectrum obtained using variety of instruments. Original figure adopted from \citet[][see references therein for data sources]{lenok_measurement_2022} and extended to include electron and positron data detected by \textit{Voyager 2} \citep{stone_cosmic_2019}. Note that \textit{Voyager 2} data has been measured outside the magnetopause and is thus  free from the effects of solar modulation (shown with a fading color gradient towards small energies below 10 GeV for protons and below higher energies for heavier ions). Thin colored lines connecting the \textit{Voyager} to Earth-bound low-energy CR data indicate the shape of the interstellar CR spectra in the Local Bubble. Notably the \textit{Voyager 2} data show an electron flux that dominates over that of protons at the lowest energies.}
\label{fig:fig1}
\end{figure}

A variety of modern instruments have been used to determine the spectrum of CRs across many orders of magnitude in particle energy. Figure~\ref{fig:fig1} combines the results of various measurements including data for the extragalactic $\gamma$-ray and neutrino backgrounds for comparison. The spectrum in this figure is multiplied by the square of particle energy in order to provide a proxy for the energy density per decade in particle energy.
As stated above, the energy density of CRs is dominated by contributions from hydrogen and helium, followed by that of heavier nuclei, electrons, positrons, and antiprotons. In particular, the electron-to-proton energy density ratio at energies of 10~GeV (which are large enough so that those CR fluxes are not affected by the magnetized solar wind) is about $K_\rmn{ep}\approx0.02$. Elemental abundances of CR nuclei closely track those seen in the Solar System, with the notable exception of lithium, beryllium, and boron (as well as the elements surrounding titanium, chromium, and manganese), which are more abundant in CRs. This deviation can be attributed to the spallation process in which CR carbon nuclei (for the light elements) and CR iron nuclei (for the heavy element groups) collide with interstellar medium (ISM) hydrogen atoms to form these more abundant elements.\\
\indent
The spectrum of CRs exhibits a number of important features. At proton energies well below several GeV (and for heavier nuclei below energies that are larger by a factor equal to the ion charge), it is a power-law characterized by a spectral index $\alpha\sim0$ ($\dd n/\dd E \propto E^{-\alpha}$). When observed close to Earth, the spectrum in this energy range is severely attenuated by \textit{solar modulation} -- a process in which CRs interacting with the turbulent and magnetized solar wind are effectively scattered and partially prevented from reaching Earth. The low-energy data shown in Fig.~\ref{fig:fig1} comes from the \textit{Voyager} probes after they had passed the heliopause, which is the boundary between the solar wind and the ISM, and is thus unaffected by the modulation. The maximum contribution to the CR energy density is provided by particles with energies of a few GeV. At larger particle energies, the spectrum declines as a power-law in energy, $\dd n/\dd E \propto E^{-2.7}$, where $n$ is the number density of particles of energy $E$. This power-law is commonly attributed to a combination of CR acceleration in supernova remnant (SNR) shock waves yielding a spectrum close to $\dd n/\dd E \propto E^{-2.2}$ and diffusive escape losses that steepen the spectral index of CRs remaining within the disk by $0.5$. Given the finite duration of the shock waves and the decreasing shock speed in the energy conserving Sedov--Taylor phase, the maximum energy to which CR protons can be accelerated via this process is limited to about a few $\times~10^{14}$~eV possibly reaching as far as the ``knee'' ($\sim3\times10^{15}$~eV) depending on the physics of magnetic field amplification, particle transport, and ambient ISM conditions. Past the knee, the composition of CRs includes substantial contributions from heavier ions and the all-particle spectrum steepens to a power law with spectral index $\alpha\approx3.1$ that extends up to the ``ankle'' ($\sim10^{18}$~eV), where it flattens again.\footnote{There have been more spectral features detected between the knee and ankle such as the second knee \citep{dembinski_data-driven_2017,schroder_high-energy_2019}, which we do not further discuss here.} \\
\indent
At the location of the ankle, the sources of CRs likely change from Galactic to extragalactic accelerators because, at this energy, the CR gyroradius is already comparable to the thickness of the Galactic disk \citep[e.g.,][]{kampert_cosmic_2007}. Collisions of CR protons with cosmic microwave background (CMB) photons can generate electron-positron pairs resulting in a pair production dip in the CR proton spectrum between $1\times10^{18}$ and $4\times10^{19}$~eV, which could be responsible for (some of) the observed spectral flattening \citep{berezinsky_astrophysical_2006}. Following the flattening, the spectrum exhibits a sharp cutoff at around $5\times10^{19}$~eV, the reason for which is still the subject of active research. Initially, this cutoff was thought to be due to the Greisen--Zatsepin--Kuzmin \citep[GZK,][]{greisen_end_1966,zatsepin_upper_1966,aloisio_propagation_2013} limit, which arises because of the collisions of CR protons with the CMB photons that lead to photo-production of pions, a process that drains energy from protons. Additional data from the Pierre Auger Collaboration \citep{aab_inferences_2017} established a change in trend from a proton-dominated CR composition at the ankle to an increasingly heavier composition with oxygen-group elements dominating the region at the cutoff \citep{aab_features_2020}. This casts doubt on the GZK interpretation and suggests that the maximum energies of heavier CR nuclei could be limited by photo-disintegration due to collisions with the CMB and extragalactic background light. Alternatively, the cutoff in the CR spectrum could signal a limiting voltage of the acceleration process because it is very challenging for astrophysical accelerators to reach the highest CR energies \citep{kotera_astrophysics_2011}. The largest detected CR energy is $\sim 3\times10^{20}$~eV, which coincidentally is the energy of a standard tennis ball travelling at a typical serve speed of 100 miles per hour (corresponding to $\sim160~\rmn{km}$ per hour), and exceeds the highest particle energy achievable at the world's largest particle accelerator, the Large Hadron Collider at CERN, by about $3\times10^3$ times.\footnote{Assuming proton-proton collisions, the center-of-mass energy $\sqrt{s}$ and the beam energy $E_\rmn{lab}$ in a fixed target experiment (measured in the laboratory frame) are related via $\sqrt{s}=(2 E_\rmn{lab} m_\rmn{p} c^2)^{1/2}$ so that the energy reachable by the Large Hadron Collider $\sqrt{s}\approx14$~TeV corresponds to a CR particle energy $E_\rmn{lab}\approx10^{17}$~eV, which falls short by a factor of $3\times10^3$ in comparison to the highest CR energies ever measured. Here, $m_\rmn{p}$ and $c$ denote the proton rest mass and light speed, respectively.}

\subsection{Role of cosmic rays in galaxy evolution}
\begin{figure}[tbp]
\begin{center}
\includegraphics[width=0.95\textwidth]{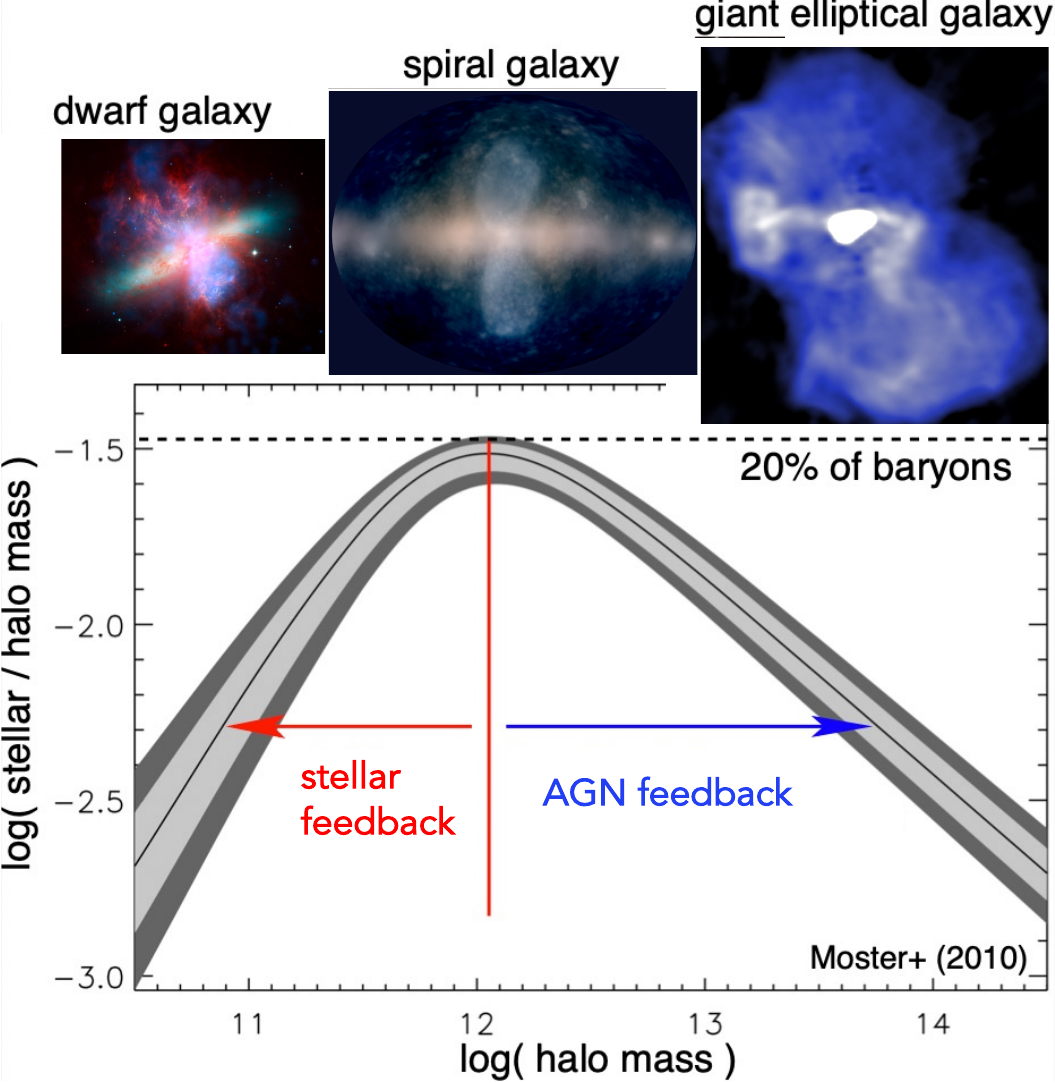}
\end{center}
\caption{\textit{Top:} From left to right the images show the starburst galaxy M82 (infrared in red: NASA/JPL-Caltech/Univ. of AZ/Engelbracht; optical in yellow: NASA/ESA/STScI/AURA/The Hubble Heritage Team; X-ray in blue: NASA/CXC/JHU/Strickland), \textit{Fermi Bubbles} in the Milky Way \citep[][reproduced with permission from A\&A]{platz_multi-component_2022}, and a LOFAR image of the giant elliptical galaxy M87 \citep[][reproduced with permission from A\&A]{de_gasperin_m_2012}. \textit{Bottom:} Stellar-to-halo mass ratio as a function of dark matter halo mass \citep{moster_constraints_2010}; reproduced with permission from ApJ.}
\label{fig:motivation}
\end{figure}

Switching the focus to a seemingly unrelated topic -- galaxy formation and evolution in a cosmologically expanding Universe
-- allows us to relate the underlying physics of galaxy formation to CR transport and to demonstrate that CRs could hold the key to solving some of the most important problems in this field. The standard cosmological cold dark matter model of the Universe with a cosmological constant $\Lambda$ predicts that baryons should follow dark matter potential wells and sink to their centers as a result of very efficient cooling. However, the baryon content of dark matter halos formed during the process of hierarchical structure formation falls significantly below the cosmological mean baryon density, which is known as the ``missing baryon problem'' \citep[e.g.,][]{bregman_search_2007,tumlinson_circumgalactic_2017}. In particular, the distribution of the ratios of stellar-to-dark matter halo mass as a function of halo mass \citep[][see bottom graph in Fig.~\ref{fig:motivation}]{moster_constraints_2010} reveals that even at halo masses around the Milky Way ($\sim10^{12}$~M$_{\odot}$), where star formation is most efficient, only $\sim20\,\%$ of the available baryons are converted into stars, and the star formation efficiency falls off rapidly toward both ends of the mass spectrum. These missing baryons either did not fall into dark matter potential wells (or remain undetected) or were expelled due to feedback processes, which can be ejective (when gas is expelled from galaxies before it can form stars) and/or preventative (when gas is heated and prevented from accreting onto galaxies). Star formation suppression is thought to be due to stellar and supermassive black hole feedback below and above this critical halo mass threshold, respectively \citep[e.g.,][]{somerville_physical_2015}.\\
\indent
Key challenges facing commonly employed thermal and radiation pressure--driven stellar feedback models are (i) the overcooling problem, where the thermal energy injected by supernovae (SNe) is quickly radiated away, thus diminishing the impact of SN explosions on regulating star formation (or, in case of high-resolution simulation models, 
causing the expelled gas to be recycled too fast and re-accreted in fountain flows), and (ii) poor coupling of stellar radiation to the gas and thus reduced momentum imparted to the gas due to photoionization and/or radiation pressure that 
open up low-density channels in the optically thick gas and dust enshrouding star forming environment, through which radiation can escape following the path of least resistance \citep{rosdahl_galaxies_2015}. Both of these issues severely reduce the efficiency of stellar feedback and galactic wind launching. However, as we will argue in this review, CRs accelerated at SNR shocks \citep{blandford_particle_1987} provide an efficient feedback mechanism that addresses both challenges because CR cooling times are typically much longer than those of the thermal gas and CRs are generally well coupled to the plasma via particle-wave interactions facilitated by CR driven plasma instabilities as well as direct inelastic collisions with the gas. These processes result in CRs imparting momentum to and heating of the plasmas, and can change in fundamental ways the properties and evolution of astrophysical objects. In fact, the CR energy density ($\sim1$~eV cm$^{-3}$) is close to equipartition with the thermal, magnetic, and turbulent energy densities in the ISM of our Milky Way \citep{boulares_galactic_1990}, suggesting that CRs are dynamically very important in regulating star formation.\\
\indent
Despite travelling at speeds close to the speed of light, $c$, arguments based on the production of secondary CRs via spallation reactions and the observed boron-to-carbon ratio reveal that CRs are confined to the Galactic disk over timescales ($\tau_{\rm diff}\sim3\times10^{7}$~yr) much longer than the light crossing time of the disk thickness ($\sim3\times10^{3}$~yr; assuming disk thickness $h_{\rm disk}\sim 1$~kpc). When the escape of CRs from the Galaxy over these long timescales is balanced by their acceleration in SNRs, CR pressure can build up to the observed dynamically important levels \citep{ginzburg_origin_1964}. Since a typical mean free path of these CRs is, on average, on the order of $\lambda_{\rm mfp,cr}\sim3h_{\rm disk}^{2}/(c\tau_{\rm diff})\sim$ 0.3 pc, i.e., much smaller than for photons, CRs may be well coupled to the ISM, and their diffusion from the disk can establish dynamically important pressure gradients that may drive strong galactic winds in late type galaxies, as first realized by \citet{ipavich_galactic_1975}. Transport of CRs into the circumgalactic medium (CGM) can affect its observational properties and provide non-thermal pressure support in the CGM, that could in turn affect thermal instability, precipitation, and thus feeding of star formation in the disk as we  will explain in this review. On even larger scales characteristic of active galactic nuclei (AGN) jet outflows in early type galaxies in galaxy groups and clusters, CRs accelerated in relativistic jets may play a key role in addressing the classic cooling flow (or cool core) problem by facilitating efficient heating of the intracluster medium (ICM) and preventing excessive star formation \citep[e.g.,][]{guo_feedback_2008}. Even when the CR pressure support is low in the CGM and ICM, CR heating of the gas can be competitive with radiative cooling, thus helping to prevent excessive mass accretion rates.\\
\indent
It is evident from the above considerations that CRs should be \textit{crucially} important for our understanding of cosmological galaxy formation and evolution. The last decade has seen tremendous progress in our understanding of the role of CRs in these feedback processes. On the theoretical side, our ability to model CRs has improved dramatically thanks to a significant and concerted effort in the community to model from first-principles the transport of CRs in the context of galactic wind launching and AGN feedback. One of the key challenges is to identify and fully explore the essential physics needed to realistically model CR feedback. This will likely be accomplished via a combination of research involving particle-in-cell (PIC) simulations of CR transport, one-dimensional models needed to interpret results from multi-dimensional and multi-physics simulations, and fully cosmological high-resolution simulations incorporating the physics-driven parameterizations of CR transport and momentum and energy deposition from stars and AGN.
On the observational side, new data from telescopes such as LOFAR, MeerKAT, Jansky VLA, \textit{Fermi}, H.E.S.S., MAGIC, VERITAS, HAWC, IceCube, and space missions such as \textit{Voyager}, AMS-02 and others have transformed our understanding of the role of CRs in feedback processes, and new and upcoming observatories such as Square-Kilometre Array (SKA; radio), James Webb Space Telescope (JWST; optical to the mid-infrared), and Cherenkov Telescope Array (CTA; $\gamma$ rays) will continue to drive progress in this very active field of research.
This conviction is shared by the Decadal Survey on Astronomy and Astrophysics 2020 panel which has voiced its appreciation of the importance of CR feedback in galaxy formation: ``The impact of CRs is one of the largest uncertainties in understanding feedback in galaxy formation. The primary uncertainty is how CRs are scattered by small-scale fluctuations in the magnetic field, which sets whether CRs can escape a region or whether their pressure builds up to the point where it can drive an outflow. [\ldots]
It is remarkable that tiny solar-system scale fluctuations in the galactic magnetic field are a key ingredient in understanding how galaxies drive winds on scales of tens of kiloparsecs, or that the large scale magnetic field properties or distant supernovae can affect the formation of pre-stellar cores'' 
\citep{decadal_survey_on_astronomy_and_astrophysics_2020_astro2020_pathways_2021}.\\
\indent
Perhaps the following closing statement from the Nobel lecture
by Cecil F. Powell in 1950 accurately captures not only the spirit of the heady years after the original CR discovery but also the current sentiment and excitement in the field of galaxy formation and CR feedback: ``It will indeed be of great interest if the contemporary studies of the primary radiation lead us [...] to the study of some of the most fundamental problems in the evolution of the cosmos.''\footnote{Closing sentence of the
\href{https://www.nobelprize.org/prizes/physics/1950/powell/lecture/}{Nobel lecture}
``The cosmic radiation'' by the British physicist and discoverer of the neutral pion -- Cecil F. Powell, December 11, 1950; The term ``primary radiation'' refers to primary CRs. Powell shared the Nobel prize with Carl D. Anderson -- an American physicist and discoverer of the positron and muon.}

\subsection{Scope and context of the review}
The overarching goal of this review is to present a comprehensive and pedagogical review of the field of the astrophysical CR feedback. Our approach encompasses an overview of the essential physical processes that play key roles in CR feedback followed by a systematic review of the role of CRs over a very wide range of physical scales emphasizing in the process the connections between CR physics operating on the various scales, and putting CR feedback processes in context of other stellar and supermassive black hole feedback mechanisms.\\
\indent
A number of other reviews discussing CRs have appeared recently. While these reviews mention CRs, our review occupies a distinct niche as it focuses almost exclusively on the in-depth discussion of the \textit{physics of CR feedback} across a wide range of physical scales -- from the scales of individual SNe to the scales of galaxy clusters. Here we list some the other reviews stating also their main focus and order them from small plasma kinetic to large cosmological scales, starting with
(i) \citet{zweibel_basis_2017}, a concise overview of CR physics with applications to feedback; 
(ii) \citet{caprioli_cosmic-ray_2016,pohl_pic_2020}, particle-in-cell simulations of CR acceleration;
(iii) \citet{marcowith_multi-scale_2020}, simulations of particle acceleration in astrophysical systems; 
(iv) \citet{padovani_impact_2020,gabici_low-energy_2022}, impact of low-energy CRs in the ISM; 
(v) \citet{grenier_nine_2015}, CR impact on the ISM, focusing on interstellar chemistry and CR propagation in molecular clouds; 
(vi) \citet{becker_tjus_closing_2020}, observational aspects of CRs and their secondaries; 
(vii) \citet{bykov_high-energy_2020}, high-energy particles and radiation in star-forming regions; 
(viii) \citet{veilleux_cool_2020}, in-depth review of the observational aspects of galactic outflows; 
(ix) \citet{amato_cosmic_2018}, CR transport in the Milky Way;
(x) \citet{recchia_cosmic_2020}, a concise overview of CR-driven winds including discussion of one-dimensional models;
(xi) \citet{yang_unveiling_2018}, origin of the \textit{Fermi Bubbles};
(xii) \citet{faucher-giguere_key_2023}, physical processes in the CGM; 
(xiii) \citet{hlavacek-larrondo_agn_2022}, AGN feedback in groups and clusters, 
(xiv) Yang \& Bourne (2023); macro- and micro-physics of AGN jet feedback in galaxy clusters,
(xv) \citet{kunz_plasma_2022}, plasma physics of the ICM; 
(xvi) \citet{vogelsberger_cosmological_2020}, short overview of recent advances in galaxy formation; 
(xvii) \citet{kotera_astrophysics_2011}, astrophysics of ultra-high energy CRs; and 
(xviii) \citet{hanasz_simulations_2021}, numerical methods for CR transport.

\subsection{Outline of the review}
We start the review with an in-depth discussion of the physics relevant to CR interactions with plasmas (Section~\ref{sec:physics}). In that section, we describe a number of processes that are essential for the modelling of the various astrophysical phenomena and the interpretation of multi-wavelength observations. These processes include particle-wave interactions and CR driven instabilities (Section~\ref{sec:CR-wave}), CR acceleration (Section~\ref{sec:acceleration_escape}), spatial and spectral CR transport (Section~\ref{sec:spatial_transport} and ~\ref{sec:spectral_transport}), and CR energy loss mechanisms (Section~\ref{sec:cooling_times}). In the process, we also present a brief overview of the approaches to study these processes (kinetic, hybrid, and fluid descriptions).\\
\indent 
Following this pedagogical physics introduction, we transition to the discussion in Section~\ref{Astrophysical_Systems} of the applications of these processes to astrophysical situations. In that section, we organize the discussion by relevant astrophysical scales, focusing first on CR impact on small scales before moving onto the discussion of the phenomena relevant to larger physical scales. We also separate the discussion of low-energy CRs (with energies $\lesssim$~GeV) that are critically important for ISM ionization, our understanding of non-thermal emission, and calibrating transport coefficients, but do not contribute significantly to gas pressure, from the discussion of CRs with energies above $\sim$~GeV that play a crucial role in dynamical feedback.
Specifically, we discuss the role of CR physics in the following astrophysical settings: low energy CR ionization in the ISM (Section~\ref{CR_ionization}), stellar feedback and CR-driven winds (starting with one-dimensional models and moving on to progressively more sophisticated descriptions of the dynamical interactions of CR with the ISM including the role of CRs in the dynamics of SNe, CR interactions with cold clouds and multiphase ISM, impact of CRs on star formation, and the physics of galactic wind launching; Section~\ref{galactic_winds}), impact of CRs in cosmological galaxy formation simulations (Section~\ref{Cosmological effects}), the role of CRs in thermal instability in the CGM and ICM (Section~\ref{titheorysection}), and the impact of CR feedback on the CGM and ICM in massive hot halos of the largest galaxies and galaxy clusters (Section~\ref{agntheorysection}).\\
\indent
As the next step, we discuss the observational signatures of CR feedback (Section~\ref{observational_signatures}), covering CR propagation in the Milky Way (Section~\ref{sec:CRprop}), CR aided outflows in the Milky Way (Section~\ref{sec:CR_driven_outflows}), non-thermal emission from galaxies (Section~\ref{Non-thermal emission from galaxies}), observational signatures of CR feedback in the CGM (Section~\ref{sec:CGM_feedback}) and ICM (Section~\ref{sec:AGN_feedback_clusters} and Section~\ref{content}), and current and future observatories (Section~\ref{sec:multi-messenger_observatories}). We conclude in Section~\ref{Open questions}, where we identify open questions and future directions.
While we discuss the GeV-TeV CR connection as a means of calibrating CR feedback, we refrain from discussing issues related to maximum energy of Galactic accelerators (e.g., hadronic PeVatrons) and do not cover the topic of ultra-high energy CRs.

\clearpage

\section{Physics}
\label{sec:physics}

Before we discuss the details and subtleties of CR feedback in galaxies and clusters, we introduce the basics of CR acceleration and transport. We put particular emphasis on the different physics models used to study CRs, from plasma-kinetic simulations to hydrodynamical models that characterize the CR population with a small number of thermodynamic variables. We start our discussion with CR interactions with electromagnetic waves in Section~\ref{sec:CR-wave} and explain the concept of CR pitch-angle scattering and CR driven instabilities. In Section~\ref{sec:acceleration_escape}, we review our knowledge on CR acceleration and escape from SNRs to the ISM. In particular, we provide an overview of the general picture of acceleration processes and discuss in detail diffusive shock acceleration of ions and electrons. We then transition to larger scales and discuss CR spatial transport in Section~\ref{sec:spatial_transport}, including a detailed explanation of one-moment and two-moment CR hydrodynamics. This is followed by a discussion of new results on the CR streaming instability, wave damping mechanisms and CR self-confinement as well as CR scattering with magneto-hydrodynamic (MHD) turbulence, which gives rise to external CR confinement by turbulence. In Section~\ref{sec:cooling_times}, we discuss radiative and non-radiative CR (ion and electron) interactions and their cooling times. Finally, we review the CR spectral transport in Section~\ref{sec:spectral_transport}. Specifically, we discuss how CR transport acquires a momentum dependence in the self-confinement and external-confinement picture of CR transport and provide an overview of the various numerical methods developed for evolving the CR momentum spectrum in space and time.

\subsection{Cosmic ray interactions with electromagnetic waves}
\label{sec:CR-wave}

In order to motivate the importance of collective interactions of CRs with plasma waves, we first provide some order of magnitude estimates of CR number densities. We then explain the various ways in which CRs interact with electromagnetic waves and end by introducing CR driven resonant plasma instabilities. 

\subsubsection{Estimates of cosmic ray number densities}
CRs are charged particles that form a collisionless species that is embedded in a magnetized background plasma, which is composed of various phases ranging from cold clouds (with temperatures $T\sim 10$~K) to ionized and neutral warm ($T\sim 10^4$~K) and hot phases ($T\sim 10^6$~K in galaxies and $T\sim 10^7$--$10^8$~K in galaxy clusters), which dominate by volume \citep{cox_three-phase_2005,draine_physics_2011}. In the midplane of the Milky Way, the energy densities of CRs, magnetic fields, turbulence, and the thermal population are in equipartition \citep{boulares_galactic_1990,cox_three-phase_2005,naab_theoretical_2017}, suggesting that these components are crucial for maintaining the energy equilibrium of the ISM. As discussed in Fig.~\ref{fig:fig1}, the Galactic CR population is dominated by particles with $\sim$~GeV energies while the warm ISM is dominated by thermal particles with energies around $\sim$~eV, which amounts to a ratio of $10^9$ in particle energy. Being in pressure equilibrium, we conclude that to order of magnitude, CRs must be extremely rare and only one in $\sim10^9$ ISM particles is a CR ion (i.e., mostly a CR proton), implying a CR number density of about $10^{-9}~\rmn{cm}^{-3}$ for a mean density of the warm ISM phase of $1~\rmn{cm}^{-3}$. The intracluster plasma is characterized by particle energies of $k_\rmn{B} T\sim1$--$10$~keV and densities $\sim10^{-3}~\rmn{cm}^{-3}$. The non-detection of $\gamma$ rays from clusters translates into an upper limit of the CR-to-thermal pressure ratio of $\lesssim10^{-2}$ in the bulk of the ICM \citep{ackermann_gev_2010,aleksic_magic_2010,aleksic_constraining_2012,arlen_constraints_2012,ackermann_search_2014} so that we obtain a CR-to-background density ratio of $\lesssim10^{-7}$, implying CR number densities in clusters of $\lesssim10^{-10}~\rmn{cm}^{-3}$. Because of the very low CR number density in the ISM and ICM, CRs almost never collide with each other. The interactions of CRs with background ions via Coulomb or hadronic particle collisions cool the CR population on too long time scales to establish a collisional equilibrium between CRs and thermal gas particles (see Section~\ref{sec:CR_ion_interactions}). Instead, CRs collectively interact with the background plasma predominantly via particle-wave scattering, which substantially reduces the effective mean free path of CRs \citep{wentzel_cosmic-ray_1974}.

\subsubsection{Cosmic ray-wave scattering and diffusion}\label{CR_scattering}
In the following we first explore the interaction of a single CR particle with an MHD wave and then discuss collective wave--particle interactions of CR populations. There are two types of CR interactions with MHD waves.\\
\indent
(i) A gyro-resonant interaction occurs when the Doppler-shifted rotation rate $\omega_\rmn{r}$ of a circularly polarized plasma wave is a multiple of the CR gyro frequency \citep{schlickeiser_cosmic_2002}, 
\begin{align}
\label{eq:resonance}
k_\parallel \mu \varv - \omega_\rmn{r} = \pm n \Omega,
\qquad
\mbox{where}
\qquad
\Omega=\frac{qB}{\gamma mc}
\end{align}
is the relativistic gyro frequency of a CR (ion or electron) population with characteristic Lorentz factor $\gamma=[1-(\varv/c)^2]^{-1/2}$, velocity $\varv$, particle mass $m$ and charge $q=Ze$ (where $Z$ is the charge number and $e$ is the elementary charge). Here, $c$ is the light speed, $k_\parallel$ is the component of the wave number parallel to the mean magnetic field, and $\mu=\cos\theta=\bs{p}\bcdot\bs{B}/(pB)$ is the cosine of the pitch angle between the magnetic field and momentum vectors, $\bs{B}$ and $\bs{p}$, while $B=|\bs{B}|$ and $p=|\bs{p}|$ are the magnitude of the magnetic field and the relativistic CR momentum, respectively, all measured in the frame that is comoving with the background plasma. For a positive sign of the wave vector, the sign $\pm$ denotes right and left circular polarization of the CR. Note that $n$ is a natural number and $n\geq1$ denotes the order of the resonant interaction. The case $n=1$ denotes the interaction with plasma waves propagating parallel to the magnetic field and $n>1$ accounts for the interaction with obliquely propagating waves, which can be seen by exploring the geometry of the wave--particle interaction.\\
\indent
To understand the physics of the resonant wave--particle interaction, we consider the motion of a particle in a uniform and constant magnetic field along which a small-amplitude shear Alfv\'en wave\footnote{A shear Alfv\'en wave (or Alfv\'en wave for short) is a type of plasma wave that arises due to oscillations of background ions in response to a restoring force provided by the tension of magnetic field lines. Evaluating the response of the plasma to such perturbations results in the dispersion relation for Alfv\'en waves, $\omega_\rmn{r}=k \varv_\rmn{a}$, where $\varv_\rmn{a}=B/\sqrt{4\pi\rho}$ is the Alfv\'en velocity in the Gaussian system of units. The ion mass density $\rho$ provides the inertia for the resulting electromagnetic wave and causes it to oscillate much slower than a light wave, provided the plasma is non-relativistic. As such, Alfv\'en waves are low-frequency (compared to the ion gyro frequency) oscillations and the perturbations of the magnetic field are transverse to the direction of propagation. The plasma $\beta$ parameter is the thermal-to-magnetic pressure ratio $\beta = P_\rmn{th}/P_B = 2 c_\rmn{s}/\varv_\rmn{a}$ and characterizes the magnetization of a warm plasma (and $c_\rmn{s}$ is the isothermal sound speed). In high-$\beta$ astrophysical plasmas, Alfv\'en waves propagate primarily alongside or opposite to the direction of the mean magnetic field with the Alfv\'en velocity because obliquely propagating Alfv\'en waves are quickly damped via the Landau damping process \citep{foote_hydromagnetic_1979} while this damping process is significantly reduced for low values of the plasma $\beta$; thereby enabling the anisotropic MHD cascade in the ISM.} travels. Because this represents a small perturbation to the constant magnetic field, the resulting particle motion is approximately described by a circular gyration in the plane perpendicular to the constant field, while the particle moves at uniform velocity along the field. Evaluating the two solutions of the resonance condition~\eqref{eq:resonance}, we find that for super-Alfv\'enic particle motions ($\mu\varv-\varv_\rmn{a}>0$), the magnetic field perturbation as seen by the particle has the same sense of polarization as the particle itself (i.e., the sense of rotation direction for a fixed coordinate along the field but for varying time). For sub-Alfv\'enic motions, the particle also probes the same wave polarization for each solution, but with an opposite polarization in comparison to the super-Alfv\'enic case. Hence, a resonant wave--particle interaction requires a Doppler-shifted Alfv\'en wave with the same rotation direction and frequency as that of the particle gyration frequency in its rest frame.\footnote{The process of pitch angle scattering of an electron interacting with a circularly polarized electromagnetic wave is beautifully visualized in \href{https://svs.gsfc.nasa.gov/4649}{videos} prepared by the NASA scientific visualization studio.}\\
\indent
(ii) The case $n=0$ in equation~\eqref{eq:resonance} (also called the ``Landau resonance'') describes to a non-resonant wave--particle interaction named ``transit time damping'' with $\omega_\rmn{r}=k_\parallel \mu\varv=k_\parallel \varv_\parallel$. This implies that the time it takes for a particle to traverse the confining region (i.e., the ``transit time'' of the particle), $\tau=\lambda_\parallel/\varv_\parallel = 2\pi/(k_\parallel \varv_\parallel)$ matches the wave period, $T=2\pi/\omega_\rmn{r}$. Physically, this means that an electron or proton is confined by a magnetic mirror force, which requires the presence of compressible electromagnetic perturbations. In MHD, those compressible perturbations are caused by fast and slow wave modes,\footnote{Fast and slow magnetosonic waves are compressible, longitudinal plasma waves for which the restoring force is provided by (thermal and magnetic) pressures. Fast magnetosonic waves propagate parallel and perpendicular to $\bs{B}$ and are equivalent to sound waves in high-$\beta$ plasmas.
Slow magnetosonic waves propagate predominantly parallel to $\bs{B}$ and are equivalent to compressible Alfv{\'e}n waves in high-$\beta$ plasmas.} with slow modes containing most of the compressible energy in subsonic turbulence. The particle surfs the wave (i.e., the particle experiences an accelerating electrostatic field of the wave) and gains energy because head-on interactions between particle and wave are more frequent than head-tail interactions.\\
\begin{figure}[tbp]
\begin{center}
\includegraphics[width=0.95\textwidth]{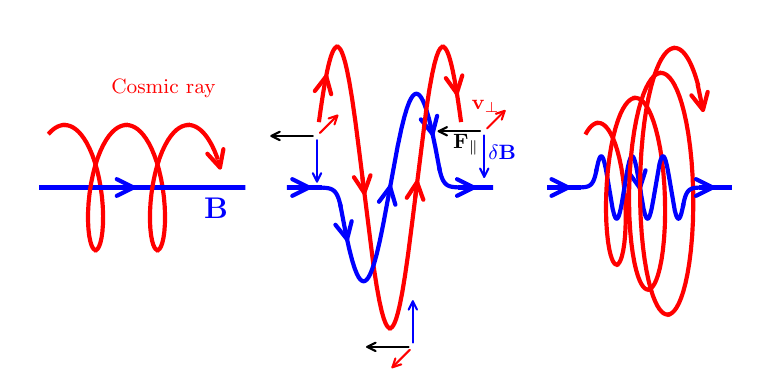}
\end{center}
\caption{The schematic drawing shows a CR proton (red) orbiting around a magnetic field line (left) and resonantly interacting with an Alfv\'en wave (center), i.e., when its wavelength equals the particle's gyroradius. In this drawing, we choose the phase of the Alfv\'en wave and gyrating CR such that the resulting Lorentz forces act opposite to the parallel velocity component of the proton along its entire orbit and decelerate it. Note that for a CR pitch angle $<90^\circ$ the magnetic vector of the circularly polarized Alfv\'en wave must be tilted such that the resulting Lorentz force decelerates the parallel velocity component. Since there are no electric fields in the reference frame of the moving wave, the proton energy (and total velocity $\varv^2=(\varv_\parallel - \varv_\rmn{a})^2+\varv_\perp^2$) is conserved and the perpendicular velocity component must increase, increasing the pitch angle between the momentum and magnetic field vectors. Switching the magnetic field perturbations ($\delta \bs{B}$) by $180^\circ$ would result in an accelerating Lorentz force along the CR orbit and a decreasing pitch angle. Hence, for a CR population with random phase shifts between the Alfv{\'e}n wave and CR gyro orbits, the CR protons would experience random gyro-averaged Lorentz forces with random net pitch angle changes. In consequence, proton scattering by an Alfv\'en wave packet (right) can be described by a diffusion process leading to an isotropization of the CRs in the reference frame of the wave for frequent CR-wave scatterings (see Fig.~\ref{fig:scattering}). In this case, CRs propagate on average at the Alfv\'en velocity, which in the ISM is about $10^4$ times smaller than the speed of light at which individual relativistic CR particles approximately travel (Jacob \& Pfrommer). }
\label{fig:Lorentz}
\end{figure}
\begin{figure}[tbp]
\begin{center}
\includegraphics[width=0.6\textwidth]{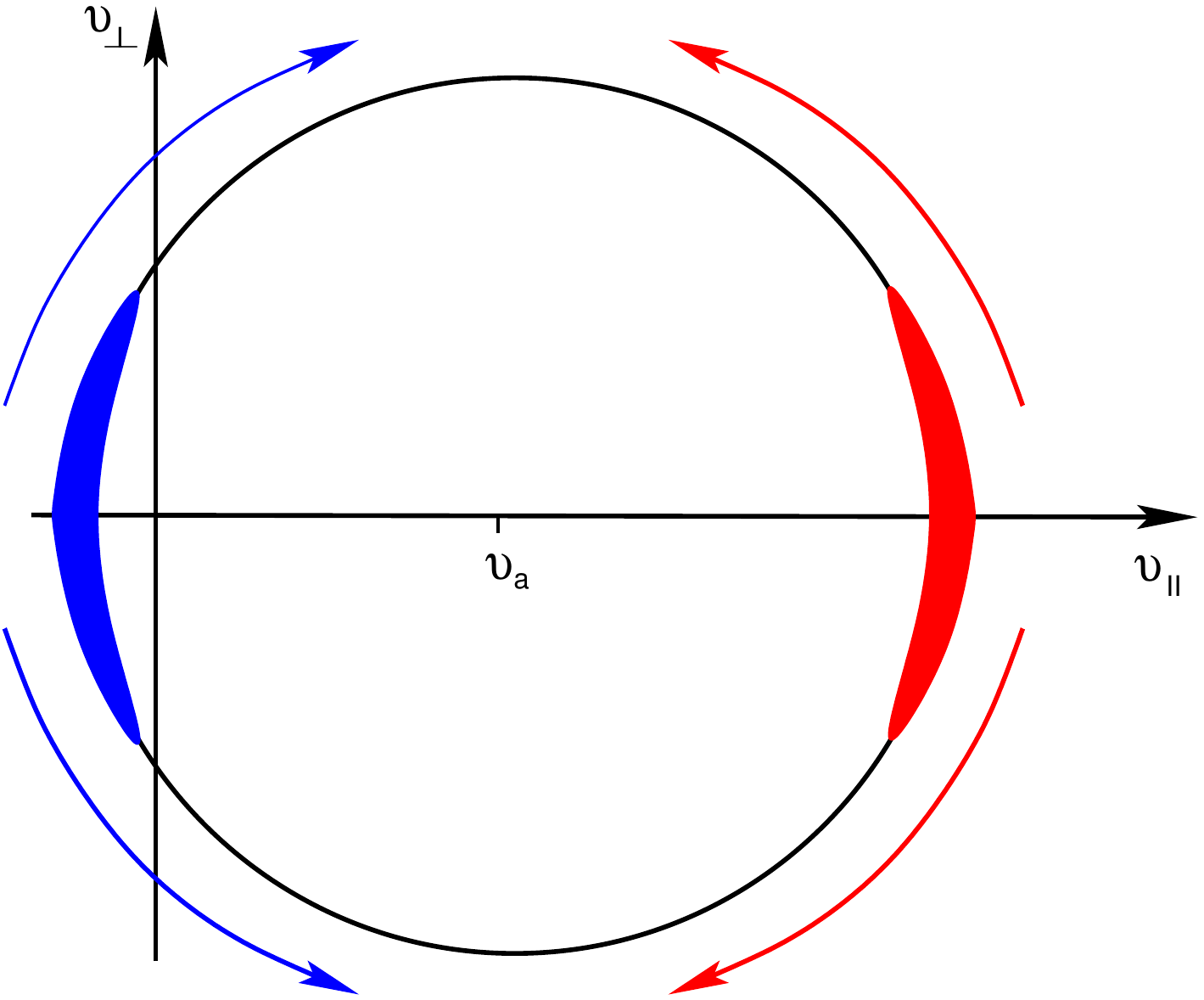}
\end{center}
\caption{A visualization of the CR scattering process with an Alfv\'en wave in CR velocity space, where velocities are measured in the frame that is comoving with the Alfv{\'e}n wave. An anisotropic CR distribution moving either rightward (red) or leftward (blue) has initial values of the cosine of the pitch angle $|\mu|=|\varv_\parallel/\varv|\lesssim 1$. Scattering off of an Alfv\'en wave leads to smaller (larger) values of $\varv_\parallel$ if initially $\varv_\parallel>\varv_\rmn{a}$ ($\varv_\parallel<\varv_\rmn{a}$ ) while conserving the particle energy in the rest frame of the Alfv\'en wave (see also Fig.~\ref{fig:Lorentz}). This process can be described as a diffusion process in $\mu$ along the equal-energy circle shown here so that the initially anisotropic distribution diffuses over time to assume a homogeneous distribution in $\mu$.}
\label{fig:scattering}
\end{figure}
\indent
Back to the resonant interaction of a CR particle with a shear Alfv\'en wave. As demonstrated in Fig.~\ref{fig:Lorentz}, upon resonantly interacting with an Alfv\'en wave, CRs change their pitch angle. This can be visualized by adopting the dispersion relation of Alfv\'en waves, $\omega_\rmn{r}=k \varv_\rmn{a}$ and evaluating the resonance condition~\eqref{eq:resonance} for wave propagation parallel to the magnetic field, yielding $\left(\varv_\parallel - \varv_\rmn{a} \right) k_\parallel = \pm \Omega$. This is a manifestation of gyroresonance and can be visualized as a circle in CR velocity space, centered on $\varv_\rmn{a}$. Shifting our frame of reference so that it is moving with the Alfv\'en speed has interesting consequences. In this frame, the Alfv\'en wave only retains a  magnetic field while the electric field vanishes. As a result, CRs can neither change their energy nor their total velocity, because this would require the presence of an electric field. In other words, $\varv^2 = (\varv_\parallel - \varv_\rmn{a})^2 + \varv_\perp^2 = \rmn{const.}$, which is the mathematical representation of the velocity space circle in Fig.~\ref{fig:scattering}. The remaining magnetic field of the Alfv\'en wave is able to shift the pitch-angle of the CRs in an energy conserving manner, which leads to motion of CRs along the velocity space circle. The interaction of an Alfv\'en wave with a CR population that is randomly distributed in space leads to random phase shifts between the individual gyro motions and the wave (see Fig.~\ref{fig:Lorentz}). This causes a random change in pitch angle for each scattering event so that we are witnessing a random walk in $\mu$, implying that this can be described by a diffusion process along the equal energy surface in velocity space (see Fig.~\ref{fig:scattering}). CR particles with $\varv_\parallel>\varv_\rmn{a}$ decrease their parallel velocity component (or equivalently, decrease their pitch angle cosine $\mu=\varv_\parallel/\varv$, which corresponds to an increasing angle $\theta$) and particles with $\varv_\parallel<\varv_\rmn{a}$ increase their $\mu$. Formally, scattering across pitch angles of $\sim90^\circ$ (corresponding to $\mu\sim0$) would require wave modes with vanishing small wavelengths, $\lambda = 2\pi/k_\parallel\to0$. However, those modes do not exist because of fast damping of these waves. 

It turns out that there are several effects that circumvent this hypothetical problem, rendering it irrelevant in practice. First, the scattering coefficient for $\mu\to0$ remains finite by retaining $\omega$ in the resonance condition, implying a broadening of the resonance as a result of dielectric effects \citep{fedorenko_cosmic-ray_1983,schlickeiser_cosmic-ray_1989}. Second, the momentum of particles with $\mu\to0$ can be reversed by mirror interactions with long-wavelength compressible modes \citep{felice_cosmic-ray_2001}, thereby emphasizing the important role of non-resonant wave--particle interactions in the scattering process. Finally, resonance broadening as a result of non-linear effects of the interaction enables populating the other hemisphere of CR velocity space so that the system eventually approaches an isotropic CR pitch angle distribution \citep{shalchi_second-order_2005}. Mathematically, these collective CR pitch angle scatterings (representing the random walk in $\mu$) can described by a diffusion process for the gyro-averaged CR distribution $f=f(\bs{x},p,\mu)$ in $\mu$:
\begin{align}
\label{eq:mu_diffusion}
\left.\frac{\partial f}{\partial t} \right\vert_{\rm scatt} = \frac{\partial}{\partial \mu} \left[\frac{1-\mu^2}{2} \,\nu(p,\mu)\,  \frac{\partial}{\partial \mu} f\right],
\end{align}
where the time and the pitch angle are evaluated in the Alfv\'en wave frame, $\nu(p,\mu)$ denotes the scattering frequency and the factor $(1-\mu^2)/2=(\sin^2\theta)/2$ derives from gyro averaging the Lorentz force terms. A transparent derivation of quasi-linear theory of CR transport and pitch-angle scattering of CRs starting from elementary physics considerations can be found in \citet{thomas_timon_hydrodynamik_2022}.\\
\indent
The effective spatial scattering coefficient along the magnetic fields can be derived using the following heuristic random walk arguments. CRs can scatter on Alfv{\'e}n waves if the resonance condition~\eqref{eq:resonance} is met. Adopting the dispersion relation for Alfv{\'e}n waves $\omega_\rmn{r}=\varv_\rmn{a}k$ and realizing that those propagate at typical speeds of tens of km~s$^{-1}$, i.e., $\varv_\rmn{a}/c\ll1$, we drop $\omega_\rmn{r}\ll k_\parallel \varv$ (because CRs move relativistically, $\varv\sim c$) and obtain the simplified resonance condition $k_{\parallel}^{-1} = \mu r_{\rm L}$, where $r_\rmn{L}=p c/(qB)$ is the CR gyroradius. This defines a minimum CR momentum that is in resonance with the Alfv{\'e}n wave of a given $k_\parallel$,
\begin{align}
\label{eq:pmin}
    p_\rmn{min}=\frac{qB}{ck_\parallel}.
\end{align}
This corresponds to individual scattering events each lasting $\sim r_{\rm L}/c$, i.e., a gyration time $\sim\Omega^{-1}$. In the frame of the Alfv{\'e}n wave, Lorentz forces acting during a scattering event alter the CR pitch angle by $|\delta\alpha|\sim\delta B/B\ll 1$, where $\delta B$ is the magnetic field fluctuation associated with the wave and $B$ is the mean magnetic field, but do not change CR energy. Multiple and uncorrelated wave--particle interactions lead to the random walk of CRs in pitch angle, which approaches CR isotropization when $N_{\rm scatt}(\delta\alpha)^{2}\sim 1$, where $N_{\rm scatt}$ is the number of scattering events. The isotropization timescale $t_{\rm i}\sim N_{\rm scatt}\Omega^{-1}$ of the CR is thus $t_{\rm i}\sim \Omega_{\rm}^{-1}(\delta B/B)^{-2}$, and the corresponding spatial diffusion coefficient along the mean direction of the magnetic field is $\kappa\sim c^{2}t_{\rm i}\sim c r_{\rm L}(\delta B/B)^{-2}$ \citep{blandford_particle_1987}. In the following sections, we will improve upon this heuristic argument and distinguish between spatial CR transport owing to diffusion and streaming, the latter of which is the limiting case of CR transport of an almost isotropic CR distribution in the Alfv\'en frame.

\subsubsection{Cosmic ray driven plasma instabilities}
\label{sec:plasma_instabilities}

\begin{figure}[tbp]
\begin{center}
\includegraphics[width=0.9\textwidth]{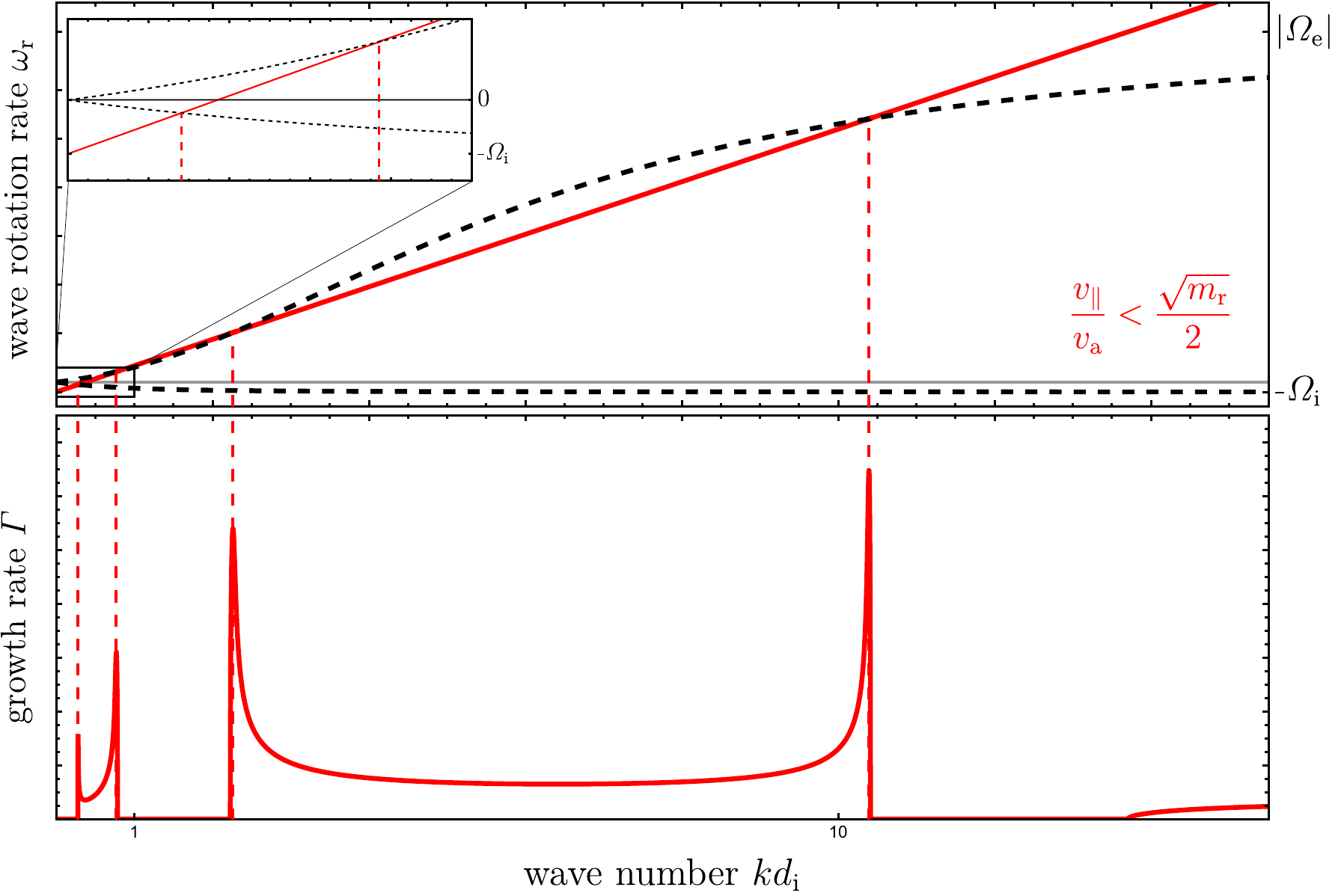}
\end{center}
\caption{Graphical visualization of resonant CR ion driven instabilities for reduced parameters $\varv_\rmn{a}=10^{-4}c$, $n_\rmn{cr}/n_\rmn{i}=10^{-6}$, $\varv_\parallel/\varv_\rmn{a}= 0.9\sqrt{m_\rmn{i}/m_\rmn{e}}=2.7$, $\varv_\perp=\varv_\rmn{a}$, and $m_\rmn{i}=36\,m_\rmn{e}$ for visual clarity. Shown are wave rotation rate $\omega_\rmn{r}$ (top panel) and instability growth rate in the linear regime (bottom panel) versus wave number times the ion skin depth $d_\rmn{i}$. Resonant CR driven instabilities are excited right at the intersection points (vertical red dashed lines) of the CR ion--cyclotron wave, $\omega_\rmn{r}=-\Omega_\rmn{i}+k \varv_\parallel$ (red), and the circularly polarized plasma waves (black dashed). At the gyroscale there are two peaks, which correspond to the crossing of the CR ion--cyclotron wave with backward and forward Alfv\'en waves (left to right), $\omega_\rmn{r}=\mp k\varv_\rmn{a}$, respectively (zoom-in panel). The electron cyclotron branch steepens at smaller scales to become a (parallel) whistler wave with $\omega_\rmn{r} \propto k^2$ until it levels off at $k d_\rmn{e}\gtrsim1$ to turn into an electron cyclotron wave with $\omega_\rmn{r}=|\Omega_\rmn{e}|$ (top panel). The bottom panel shows the growth rate of the gyro-resonant instability on large scales and the intermediate-scale instability toward smaller scales (larger $k$ values). Image from \citet{shalaby_deciphering_2023}; reproduced with permission from JPP.}
\label{fig:CR_driven_instabilities}
\end{figure}

In the previous section, we discussed the effect that Alfv\'en wave interactions with CRs have on their motion through phase space. We implicitly assumed that the waves have been driven by some unspecified process and have not accounted for a possible backreaction of the CRs on wave propagation. Here, we show that this backreaction naturally results from various \textit{resonant} CR-driven instabilities while we defer a discussion of the non-resonant hybrid instability \citep{bell_turbulent_2004} to Section~\ref{sec:acceleration}.\\
\indent
The nature of CR driven plasma instabilities can be understood by studying the response of a plasma to resonant driving provided by streaming (and gyrating) CRs. Perturbing the Vlasov equation for the CR distribution function and deriving a Fokker-Planck equation for CR transport leads to the picture of quasi-linear diffusion of CRs in phase space \citep{schlickeiser_cosmic_2002}. Assuming an equilibrium configuration with a uniform background magnetic field and no net electric field, we can analyze the behavior of small perturbations by linearizing the system. The resulting coupled system of equations describes electromagnetic modes in the plasma that are excited by anisotropically drifting CRs and consists of the dispersion tensor applied to electric perturbations, $T^{ij}\delta E^j=0$ with the boundary condition that these electric perturbations obey the linearized set of Maxwell's equations in Fourier space, $\bs{k} \bs\times \delta\bs{E}=\omega \delta\bs{B}$ and $\bs{k} \bcdot \delta\bs{B} = 0$. The dispersion tensor, $T^{ij}$, is complicated in general \citep[see Chapter 8 in][]{schlickeiser_cosmic_2002} but simplifies for parallel propagating wave modes:\footnote{For CR driven instabilities, modes that propagate at an oblique angle to $\bs{B}_0$ have typically a subdominant growth rate with respect to that of parallel modes \citep{zweibel_basis_2017,kulsrud_effect_1969}.}
\begin{eqnarray}
T^{ij}
=
\left(
\begin{array}{ccc}
T^{11} & 0 & 0 \\
 0 &T^{22} &T^{23} \\
 0 & -T^{23} &T^{22} \\
\end{array}
\right).
\end{eqnarray}
The explicit form of the matrix elements are given in \citet[][Appendix~B]{shalaby_new_2021} and we assume that the magnetic field and wave modes are aligned with the $x$ axis, i.e., $\bs{B}_0  = B_0 \bs{\hat{x}}$ and $\bs{k} = k\,\bs{\hat{x}}$. \\
\indent
The stability of electromagnetic plasma modes is determined by the linear dispersion relation in Fourier space, which assesses whether a perturbation of a certain wave mode decays or grows over time. The dispersion relation is the determinant of the dispersion tensor, $|T^{ij}| = T^{11} \left[ (T^{22})^2 + (T^{23})^2 \right]=0$. The solutions to $T^{11}=0$ represent electrostatic wave modes for which only the parallel electric field is finite while the solutions to $T^{22} \pm \rmn{i} T^{23} =0$ represent electromagnetic wave modes that are characterized by transverse electric and magnetic field components and a vanishing parallel electric field. Physically, the linear dispersion relation represents a relation between the complex wave frequency, $\omega=\omega_\rmn{r}+ \rmn{i} \Gamma$, and the wave number $k$ of waves propagating along the magnetic field:
\begin{align}
\label{eq:dispersion relation}
D^\pm \equiv T^{22} \pm \rmn{i} T^{23} = 1-\frac{k^2 c^2}{\omega^2} + \sum_s \chi_s^\pm = 0,
\end{align}
where $\chi_s^\pm$ is the linear plasma response for species $s$, which includes the background electrons and ions as well as the drifting CR electrons and ions \citep[e.g.,][]{shalaby_new_2021}. The circularly polarized eigenmodes of the background plasma relevant for CR scattering lie on the electron and ion cyclotron branches: on large scales with wave numbers $k d_\rmn{i}\ll1$, the electron cyclotron branch hosts forward-propagating Alfv\'en waves, which rotate at a rate $\omega_\rmn{r}= k\varv_\rmn{a}$. These scales are larger than the ion skin depth, $d_\rmn{i}=c/\omega_\rmn{i}$ (where $\omega_\rmn{i}$ is the ion plasma frequency) and represent the MHD limit. The Alfv\'en waves turn into whistler waves\footnote{The term whistler describes low-frequency plasma waves generated by lightning that emits for a few seconds in the radio band with a descending tone at the end, reminiscent of whistling. A whistler mode represents the continuation of the electron cyclotron branch on scales $k d_\rmn{i}> 1$. Like the Alfv\'en waves on larger scales, whistler waves are circularly polarized electromagnetic perturbations with a dispersion relation for wavelengths $k d_\rmn{e}\ll 1$ given by $\omega_\rmn{r}= \pm k_\parallel kd_\rmn{e}^2 \Omega_\rmn{e,0}=\pm k_\parallel k d_\rmn{i} \varv_\rmn{a,i}$, where $d_\rmn{e,i}=c/\omega_\rmn{e,i}=\varv_\rmn{a,e,i}/\Omega_\rmn{e,i,0}$ are the collisionless electron and ion skin depths, $\omega_\rmn{e,i} = (4\pi e^2 n_\rmn{e,i}/m_\rmn{e,i})^{1/2}$ are the electron and ion plasma frequencies, $\Omega_\rmn{e,i,0}$ are the non-relativistic electron and ion gyro frequencies, and $\varv_\rmn{a,e,i}=B_0/(4\pi n_\rmn{e,i}m_\rmn{e,i})^{1/2}$ are the electron and ion Alfv\'en speeds.} at smaller scales for $k d_\rmn{i}>1$ (with a dispersion relation $\omega_\rmn{r} = k^2 d_\rmn{i}\varv_\rmn{a}$ for parallel whistlers) in order to approach electron cyclotron waves at scales smaller than the electron skin depth, $k d_\rmn{e}=k d_\rmn{i}\sqrt{m_\rmn{e}/m_\rmn{i}}\gg1$. Those electron cyclotron waves show a constant wave rotation rate $\omega_\rmn{r} = |\Omega_\rmn{e}|$ (see the upper dashed black line in Fig.~\ref{fig:CR_driven_instabilities}). The ion cyclotron branch hosts backward-propagating Alfv\'en waves on scales $k d_\rmn{i}\ll1$, which directly turn into ion cyclotron waves on smaller scales with $\omega_\rmn{r} = -\Omega_\rmn{i}$ (see the lower dashed black line in Fig.~\ref{fig:CR_driven_instabilities}). By contrast, a population of gyrating and drifting CRs along the mean magnetic field excites a CR ion--cyclotron wave, which is a rotating and propagating electromagnetic wave with the rotation rate $\omega_\rmn{r}=-\Omega_\rmn{i}+k\varv_\parallel$ in the background plasma frame, i.e., its rotation rate equals the Doppler-shifted CR ion--cyclotron frequency. Hence, it is no coincidence that these gyrating CRs also obey the gyro-resonance condition~\eqref{eq:resonance} for parallel wave modes, which in this case is interpreted such that the Doppler-shifted plasma wave frequency resonates with the CR gyro frequency.

A resonant CR driven instability emerges if the rotating electromagnetic field of the CR ion--cyclotron wave as seen in the background frame equals one of the circularly polarized eigenmodes of the background plasma, namely those provided by the ion and electron cyclotron branches. As realized by \citet{shalaby_deciphering_2023}, this deep insight has a simple graphical interpretation, which is visualized in Fig.~\ref{fig:CR_driven_instabilities}: the locations at which the rotation rate of the CR ion--cyclotron wave $\omega_\rmn{r}=-\Omega_\rmn{i}+k\varv_\parallel$ intersects the wave frequency $\omega_\rmn{r}$ of the electron and ion cyclotron branches maximize the linear growth rate of a CR driven instability. Interestingly, CR ion--cyclotron modes are driven unstable not only at the points of intersection with the background circularly polarized eigenmodes but also when the rotation frequency of a given CR ion--cyclotron mode is smaller in magnitude than that of the closest background mode (in $\omega_\rmn{r}$) of either the ion or the electron cyclotron branch \citep[see Fig.~\ref{fig:CR_driven_instabilities};][]{shalaby_deciphering_2023}.

Starting on large scales, the first two intersection points result from a resonance of the CR ion--cyclotron wave with the backward- and then the forward-propagating Alfv\'en wave, with $\omega_\rmn{r}=\mp k\varv_\rmn{a}$, respectively. In consequence, the two peaks of the gyro-resonant instability \citep{kulsrud_effect_1969} emerge at wave numbers $kd_\rmn{i}=(\varv_\parallel/\varv_\rmn{a}\pm 1)^{-1}$, where $\varv_\parallel=\varv\mu$ is the parallel CR velocity component. This instability can be seen to the left in the growth rate plot in the bottom panel of Fig.~\ref{fig:CR_driven_instabilities}. On smaller scales, streaming CRs excite the intermediate-scale instability \citep{shalaby_new_2021}, with the shortest unstable wave mode at $k\varv_\parallel=|\Omega_\rmn{e,0}|=eB/(m_\rmn{e}c)$, where $\Omega_\rmn{e,0}$ denotes the non-relativistic electron gyro frequency. Interestingly, the steepening of the wave rotation rate in the whistler regime causes the emergence of the intermediate-scale instability that extends to the electron cyclotron scale, where the real part of the electron cyclotron branch flattens again and approaches $\Omega_\rmn{e}$.\\

Physically, the \textit{gyro-resonant streaming instability} describes the ability of CRs to excite resonant Alfv\'en waves \citep{kulsrud_effect_1969}. While Fig.~\ref{fig:CR_driven_instabilities} was computed for gyro-tropic CRs with a specific pitch angle, in practice the CR population shows a distribution in $\mu$. In this case, we can approximate the linear growth rate of the gyro-resonant instability for $n_\CR/n_\rmn{i}\ll1$ \citep{kulsrud_plasma_2005,thomas_timon_hydrodynamik_2022} in the frame that is comoving with the gas:
\begin{equation}
    \Gamma_\rmn{gyro} \approx \pi \int \mathrm{d}^3p \, \frac{1-\mu^2}{2} \frac{q^2 \varv}{p} \left[ p \frac{\partial f}{\partial p} + \left( \frac{\varv}{\varv_\mathrm{a}} - \mu \right) \frac{\partial f}{\partial \mu}\right] \delta_\rmn{D}\left[k (\mu \varv - \varv_\mathrm{a}) \mp \Omega_\rmn{i}\right], 
    \label{eq:gyroresonant_streaming_instability}
\end{equation}
where $\delta_\rmn{D}$ denotes Dirac's $\delta$ distribution and represents the resonance condition~\eqref{eq:resonance} for CR-Alfv\'en wave interactions. However, this condition is now interpreted from the viewpoint of the Alfv\'en waves and solely selects those CRs that meet this condition of resonantly interacting with parallely propagating Alfv\'en waves of wave number $k$ to excite or damp them, depending on the overall sign of the momentum-derivatives of the CR distribution function in the square bracket of equation~\eqref{eq:gyroresonant_streaming_instability}: a positive sign of the bracket signals instability while a negative sign indicates wave damping. Note that the combination of momentum derivatives in square bracket reduces to $\partial f/\partial\mu$ in the wave frame (see discussion surrounding Eq.~\ref{eq:mu_diffusion} and Fig.~\ref{fig:scattering}). Importantly, in the gyro-averaged CR rest frame, the CR gyro-motion needs to match the wave rotation frequency so that the CRs resonate with the Doppler-shifted wave oscillation. For a power-law momentum distribution of CRs, $f\propto p^{-\alpha}$, this growth rate can be simplified to yield \citep{kulsrud_plasma_2005}
\begin{align}
\label{eq:Gamma_gyro}
\Gamma_\rmn{gyro} \sim \frac{\pi}{4}\,\Omega_\rmn{i,0} C_\alpha\,\frac{n_\CR(>p_\rmn{min})}{n_\rmn{i}}\,\left(\frac{\varv_\rmn{d}}{\varv_\rmn{a}} - 1\right),
\end{align}
where $\Omega_\rmn{i,0}=qB/(m_\rmn{i}c)$ is the non-relativistic ion gyro frequency, $C_\alpha=(\alpha-3)/(\alpha-2)$ is a constant of order unity, $p_\rmn{min}$ is defined in equation~\eqref{eq:pmin}, $n_\CR(>p_\rmn{min})\propto p_\rmn{min}^{3-\alpha}$ is the number density of resonant CRs, $n_\rmn{i}$ is the number density of thermal ions, and $\varv_\rmn{d}$ is the CR drift velocity that defines a frame with respect to which the CR distribution is isotropic. Equation~\eqref{eq:Gamma_gyro} shows that only super-Alfv\'enic CRs can drive Alfv\'en waves unstable with a growth rate that depends on the resonant CR-to-thermal number density ratio, which decreases towards larger CR momenta due to the diminishing spectra at high energies for $\alpha>4$.\\
\indent
On smaller scales, CRs with a finite pitch angle resonantly excite parallel electromagnetic waves  on scales between the ion and electron gyro-resonances through the \textit{intermediate-scale instability} \citep{shalaby_new_2021}. As we can infer from Fig.~\ref{fig:CR_driven_instabilities}, CR ions drive background ion--cyclotron modes\footnote{An ion–cyclotron wave is a circularly polarized oscillation of ions in a magnetized plasma that is excited by the gyro motion of background ions in response to the Lorentz force with a dispersion relation $\omega_\rmn{r}=\pm\Omega_\rmn{i}$. It represents the small wavelength continuation of the backward propagating shear-Alfv{\'e}n wave and propagates along the magnetic field with the plasma.} unstable, which are comoving with the CR ions. While the electron modes have the wrong sense of rotation to interact with the CR ions in the background frame, Lorentz transforming to a frame that is comoving with the CR ions changes the sense of rotation of these wave modes to adopt that of CR ions and hence to enable resonance. Provided that $n_\CR\ll n_\rmn{i}$ and $\varv_\rmn{a}\ll c$, \citet{shalaby_new_2021} approximate the peak growth rate by 
\begin{eqnarray}
\label{eq:IntermediateRate}
\frac{ \Gamma_\rmn{inter} }{\Omega_\rmn{i,0} }
\approx 
\left(\frac{n_\CR}{n_\rmn{i}}\right)^{3/4}+ 
\left(\frac{n_\CR}{3n_\rmn{i}}\right)^{1/3}
\left( \frac{\varv_\parallel \varv_{\perp}}{\varv_\rmn{a}^2} \right)^{2/3},
\end{eqnarray}
where $\varv_\perp$ is the perpendicular CR velocity of a gyrotropic distribution. This interme{\-}diate-scale instability typically grows faster by more than an order of magnitude in comparison to the previously discussed gyro-resonant streaming instability. The two peaks of the intermediate-scale instability are approximately given by $kc/\omega_\rmn{i} = \varv_\parallel/\varv_\rmn{a}$ and $k c/ \omega_\rmn{i} = m_\rmn{r} \varv_{\rmn{a}}/\varv_\parallel -\varv_\parallel/\varv_\rmn{a} $, where $\omega_\rmn{i} = (4\pi e^2 n_\rmn{i}/m_\rmn{i})^{1/2}$ is the ion plasma frequency, $m_\rmn{r}=m_\rmn{i}/m_\rmn{e}$ is the ion-to-electron mass ratio, and $k c/ \omega_\rmn{i}  = m_\rmn{r} \varv_{\rmn{a}}/\varv_\parallel$ is the electron gyroscale. This amounts to a peak separation of
\begin{eqnarray}
\frac{ \Delta k c }{ \omega_\rmn{i} } \approx  m_\rmn{r}  \frac{\varv_\rmn{a}}{\varv_\parallel} - 2 \frac{ \varv_\parallel}{\varv_\rmn{a}} 
\qquad \Rightarrow \qquad 
\frac{ \Delta k \varv_\parallel  }{ \Omega_\rmn{i,0} } \approx  m_\rmn{r} - 2 \left( \frac{\varv_\parallel}{\varv_\rmn{a}}  \right)^2,
\label{eq:peak_position}
\end{eqnarray}
which is an excellent fit for low values of $\varv_\parallel/\varv_\rmn{a}$. For increasing $\varv_\parallel/\varv_\rmn{a}$, the peak separation decreases to the point were both peaks merge at $\varv_\parallel/\varv_\rmn{a}=\sqrt{m_\rmn{r}}/2$ (i.e., the factor 2 is replaced by 4 in equation~\ref{eq:peak_position}). For larger CR drift speeds, the instability ceases to exist because there is no intersection of the CR ion--cyclotron wave with the circularly polarized electron cyclotron branch at the whistler scales. Using the graphical representation in Fig.~\ref{fig:CR_driven_instabilities}, the slope of the 
CR ion--cyclotron wave is too steep to intersect and resonate with the overturning electron cyclotron branch towards small scales. While the gyro-resonant instability is governing CR transport in interstellar, circumgalactic, and intracluster media, the intermediate-scale instability may also play an important role in CR scattering and transport \citep{shalaby_new_2021} as well as in pre-accelerating electrons at collisionless shocks to enable them to participate in diffusive shock acceleration \citep[][see also Section~\ref{sec:acceleration_escape}]{shalaby_mechanism_2022}.

\subsection{Cosmic ray acceleration and escape from sources}
\label{sec:acceleration_escape}

Provided the CR distribution exhibits only a small degree of anisotropy, we can expand the CR distribution in a suitable system of basis functions (Legendre or Taylor polynomials) and truncate the expansion at finite order (e.g., after the dipolar or quadrupolar anisotropy) to derive a moment-based description of CR transport. This is governed by the wave--particle scattering rate, which itself depends on the amplitudes of resonant waves that result from the interplay of wave growth and damping. The resulting system of equations can be readily applied to the problem of galactic CR propagation, as we will detail in Section~\ref{sec:spatial_transport}. In situations where large CR fluxes (and associated anisotropies) are present, which is realized in the vicinity of shock fronts, one must solve the full kinetic equations and study the excitation of plasma instabilities in the regime of large CR fluxes, which scatter the particles and thereby reduce the CR flux, as will be discussed in this section.

\subsubsection{Cosmic ray sources}\label{CRsources}

The sources of Galactic CRs are generally assumed to be SNRs in the Milky Way. While there is an ongoing debate whether this source class is solely responsible for all CRs up to the ``knee'' in the CR momentum spectrum (at $E\approx 3\times 10^{15}$~GeV for CR protons while heavier CR ions have a break at larger energies) or whether other sources contribute some flux, energetic considerations clearly argue for a dominant contribution of SNRs for generating the pressure carrying (GeV--TeV) CRs. This was suggested in a visionary paper by \citet{baade_remarks_1934} and made quantitative by \citet{ginzburg_origin_1964}; here we will sketch the argument by estimating the luminosity required to supply all the Galactic CRs and balance their escape losses:
\begin{equation}
  \mathcal{L}_\CR=\frac{V \eps_\CR}{\tau_\rmn{esc}}\sim 3\times 10^{40}~ \rmn{erg~s}^{-1},
\end{equation}
where $V$ is the volume of the thick galactic disk (``the CR scattering halo'') with half-height $H\approx2$~kpc and radius $R\approx8$~kpc, $\eps_\CR\approx0.8\,\rmn{eV~cm}^{-3}$ is the CR energy density, and $\tau_\rmn{esc}\approx3\times10^7$~yr is the diffusive escape time (as inferred from the boron-to-carbon ratio of CRs and assuming a mean hydrogen density along the CR column in the scattering halo of $\bar{n}_\rmn{H}\approx0.1~\rmn{cm}^{-3}$, see Section~\ref{sec:B-to-C_ratio}). This power can be delivered by Galactic SNe if 10\% of their kinetic energy is transferred to CRs:
\begin{equation}
  \mathcal{L}_\CR\sim 0.1\mathcal{L}_\rmn{sn}\sim\frac{0.1E_\rmn{sn}}{\tau_\rmn{sn}}\sim
  \frac{0.1\times10^{51}\rmn{erg}}{100\,\rmn{yr}}\sim 3\times 10^{40}~ \rmn{erg~s}^{-1},
\end{equation}
where $\tau_\rmn{sn}$ is the average time scale between Galactic SNe. Within the Galaxy, various other (less prominent) sources of CRs exist. These include termination shocks found in stellar wind bubbles and superbubbles, shocks resulting from the collision of stellar winds in binary systems, bow-shocks generated by massive runaway stars, and pulsar wind nebulae. Time-averaging the mass loss rate times the square of the terminal velocity of stellar winds throughout different stages such as the main sequence, red supergiant, and Wolf-Rayet phases, the collective stellar wind luminosity from all massive stars in the Galaxy amounts to approximately $\mathcal{L}_\rmn{w}\approx1\times10^{41}~\rmn{erg~s}^{-1}$ \citep{seo_contribution_2018}, which is about 1/3 of the power of SN explosions, $\mathcal{L}_\rmn{sn}\approx3\times10^{41}~\rmn{erg~s}^{-1}$. Internal recollimation shocks in relativistic AGN jets as well as the termination or back-flow shocks are other important sites of CR acceleration, yet primarily for injecting CRs into the intergalactic and intracluster plasmas.

\subsubsection{Particle acceleration and magnetic amplification}
\label{sec:acceleration}

\begin{wrapfigure}{r}{0.5\textwidth}
  \vspace{-1em}
  \begin{center}
    \includegraphics[width=0.5\textwidth]{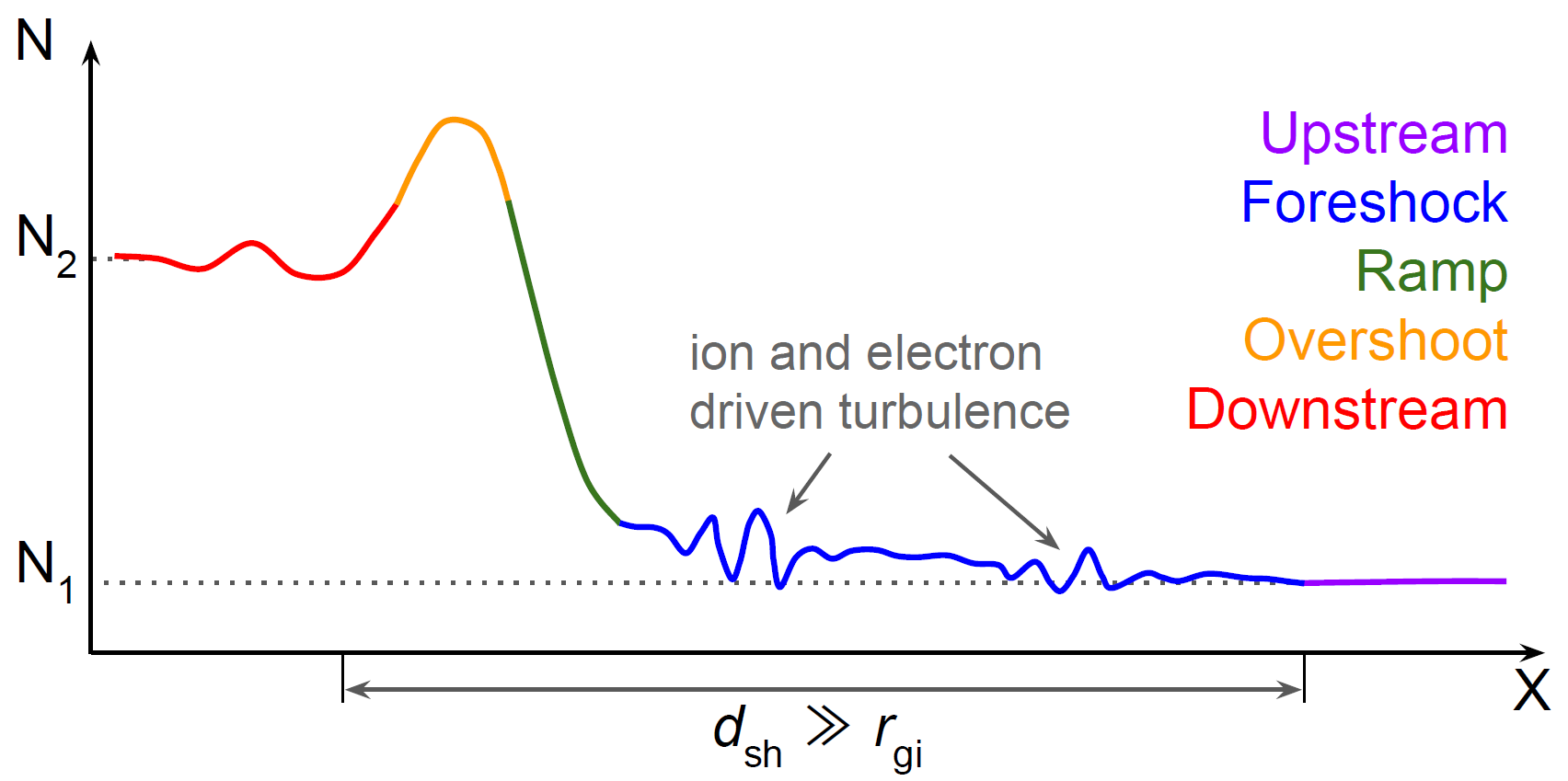}
  \end{center}
  \caption{Density structure of a quasi-parallel collisionless shock that accelerates CRs. The plasma is entering from the right and is gradually slowed down over the ``foreshock precursor'' where electron and ion driven plasma instabilities drive plasma turbulence that scatters ionized particles. As the plasma enters the shock transition of strong electromagnetic fluctuations that stretches across several ion gyro radii $r_\rmn{gi}$, frequent particle-wave scattering ensues. This slows down the incoming flow and steeply increases the plasma density in form of a ``ramp'' and an ``overshoot'', after which the plasma settles to a downstream state (Bohdan, private comm.).}
  \label{fig:shock}
  \vspace{-1em}
\end{wrapfigure}

\paragraph{General picture.} We now focus on the mechanism of \textit{diffusive shock acceleration} that energizes CR electrons and protons at astrophysical non-relativistic shocks \citep{krymskii_regular_1977,axford_acceleration_1978,bell_acceleration_1978-1,bell_acceleration_1978,blandford_particle_1978} and discuss the popular case of a SNR shock that causes the surrounding ISM to expand radially. Since the mean free path in the hot phase of the ISM is larger than the SNR, the shock wave cannot set the ISM in motion by momentum transfer from individual particle-particle collisions, but only by particles scattering at electromagnetic fluctuations in the surrounding plasma that are driven unstable by the expanding flux of plasma (e.g.,\ for negligible magnetization through the Weibel instability as shown by \citealt{medvedev_generation_1999}, while other electromagnetic plasma instabilities dominate the magnetized case). Such a shock wave is referred to as a collisionless shock wave. Essential to the CR acceleration process is the amplification of standing Alfv\'en waves by currents driven by high-energy ions \citep{bell_turbulent_2004} that outrun the shock towards the upstream region to generate a foreshock precursor (see Fig.~\ref{fig:shock} for a schematic drawing). This sources an (electron) return current in the 
thermal plasma that exponentially grows a spectrum of helical Alfv\'en waves with wavelengths smaller than the gyro radii of the high-energy protons. These waves act as a mediator that absorbs energy and can effectively scatter resonant protons and electrons at lower energies (Fig.~\ref{fig:Lorentz}) that are diffusing into the precursor. The scattering rate is close to, or below, the Bohm limit for a range of energies \citep{stage_cosmic-ray_2006}, i.e.\ they are scattered once per gyro orbit, implying that scattering in self-excited turbulence significantly reduces the diffusion coefficient \citep{reville_transport_2008}. This leads to isotropization of charged particles in the respective reference frames (before and after the shock front) so that they pick up momentum through head-on scatterings with the magnetic field without recoil to (re)cross the shock front.\footnote{The recoil of the magnetic field frozen into the surrounding plasma can be neglected because of the large inertia of the plasma.} In each of these cycles (downstream, upstream and back downstream), a part of the accelerated particles is advected with the plasma behind the shock front, so that they do not participate in the further acceleration process. Many acceleration cycles, i.e., repeated reflections on both sides of the shock wave, generate relativistic CR particles. As long as the gyroradius is smaller than the radius of the shock front, this scale invariant acceleration process produces a (non-thermal) power law in the CR momentum distribution with an efficiency that scales with $\varv_\rmn{s}/c$, where $\varv_\rmn{s}$ is the shock velocity in the laboratory frame (which is sometimes referred to as first-order Fermi acceleration, even though Fermi did not discover this process). Effectively, all particle acceleration at shocks is rooted in the electrostatic field $\mathbfit{E}$ of the shock, that slows down the incoming flow and causes the global velocity divergence.

As originally pointed out by \citet{fermi_origin_1949,fermi_galactic_1954}, particles can also be accelerated without a shock through interacting with externally driven turbulence. Resonant scattering off of moving magnetic irregularities, with $\mathbfit{E}=\mathbf{0}$ in the local rest frame causes isotropic and elastic scattering in the scattering center rest frame. As a result there is a momentum gain for head-on collision and a momentum loss for tail-on collisions in the laboratory frame. On average, the mean energy is conserved but because there are statistically more head-on than tail-on collisions due to the particle motion, this causes the width of the particle distribution to increase so that a small fraction of particles in the high-energy tail experiences acceleration. In the quasi-linear picture of particle transport in a bath of linear (Alfv\'en, magnetosonic) waves \citep{kennel_velocity_1966,skilling_cosmic_1971,skilling_cosmic_1975,schlickeiser_cosmic_2002,shalchi_nonlinear_2009}, this can be described by a diffusion process in momentum space, which models the energy gain through resonant interactions. This process is comparably inefficient because it scales as $(\varv_\rmn{w}/c)^2$, i.e. it is only second order in the dimensionless wave velocity of the plasma wave scattering centers $\varv_\rmn{w}$ in the laboratory frame. In the following, we will not consider this second-order Fermi process for particle acceleration (while it may play an important role in particle transport).\\
\indent
The (microscopic) processes at shocks are governed by plasma kinetics and can be studied by means of particle-in-cell (PIC) simulations that use macro particles to represent electrons and ions of the thermal plasma and the energetic CRs. According to Maxwell’s equations, moving charges source currents that perturb electromagnetic fields. These generate Lorentz forces that accelerate charged particles and hence modify the charge distribution and currents. The system is evolved by numerically iterating this loop on a fraction of the electron plasma timescale for macro-particles representing the individual elementary particles of a plasma. This methodology is ideal for exploring kinetic instabilities in the collisionless plasma around a shock. Prominent PIC codes include Tristan-MP  \citep{spitkovsky_simulations_2005}, VPIC \citep{bowers_ultrahigh_2008}, photo-plasma \citep{haugbolle_span_2013}, SHARP \citep{shalaby_sharp_2017,shalaby_new_2021}, and Warp-X \citep{vay_warp-x_2018}; see also \citet{pohl_pic_2020} for a review. However, this limits the physical length scale and total simulation time of PIC simulations to microscopic dimensions. In order to allow for longer run times of the simulations and/or multiple spatial dimensions, the electron timescale is integrated out in hybrid-PIC simulations, where the electron population is represented by an adiabatic fluid and the ions are treated as macro particles in the kinetic PIC model. Approaching even larger time and length scales (to study certain properties of CR transport) requires to treat the entire background either as an MHD fluid or as separate fluids for each particle species, which are coupled to the PIC component representing CR particles (see Section~\ref{sec:fluid-PIC_modeling_of_CR_transport}). This approach is however not well suited to study CR acceleration at shocks as this would require adopting a recipe for injecting non-thermal particles from the thermal (MHD) fluid, which then participate in the diffusive shock acceleration process. 
\begin{figure}[tbp]
\begin{center}
\includegraphics[width=0.95\textwidth]{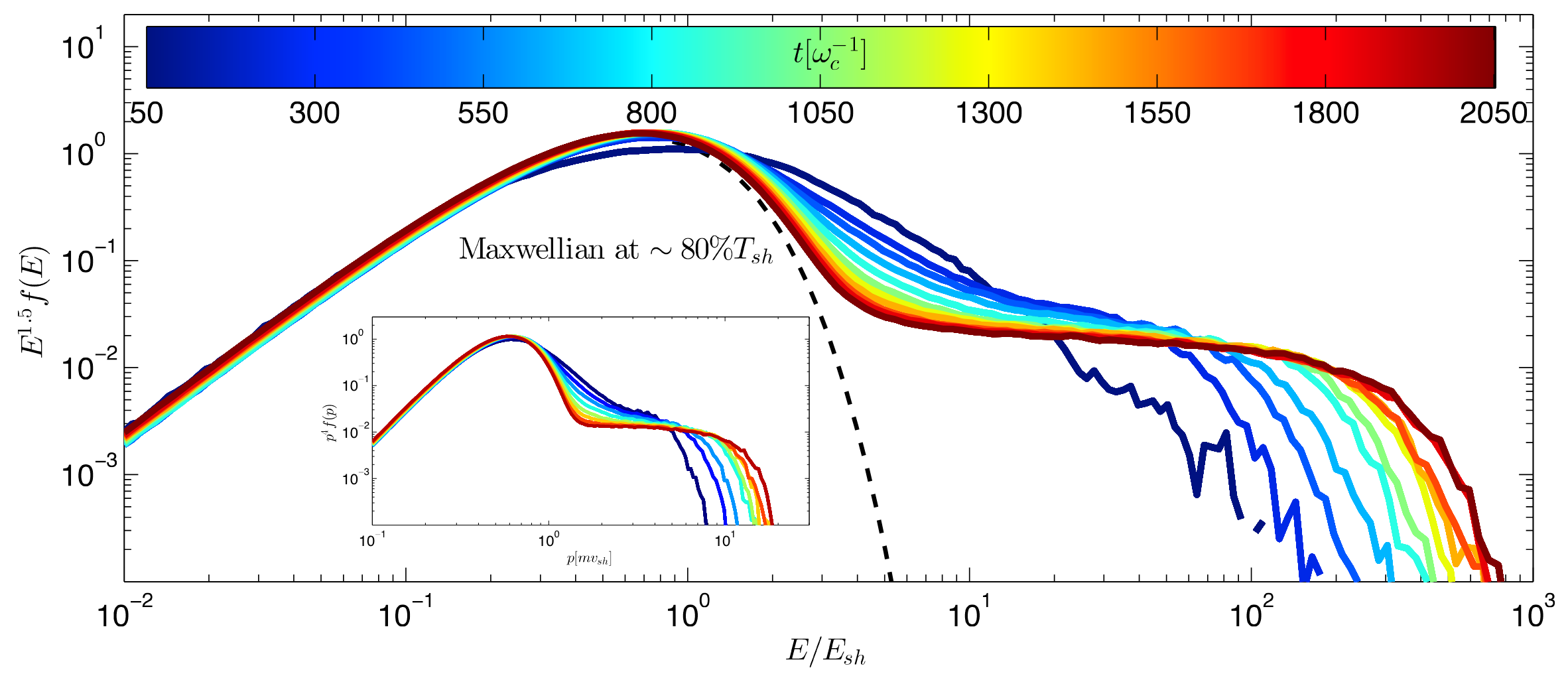}\\
\includegraphics[width=0.95\textwidth]{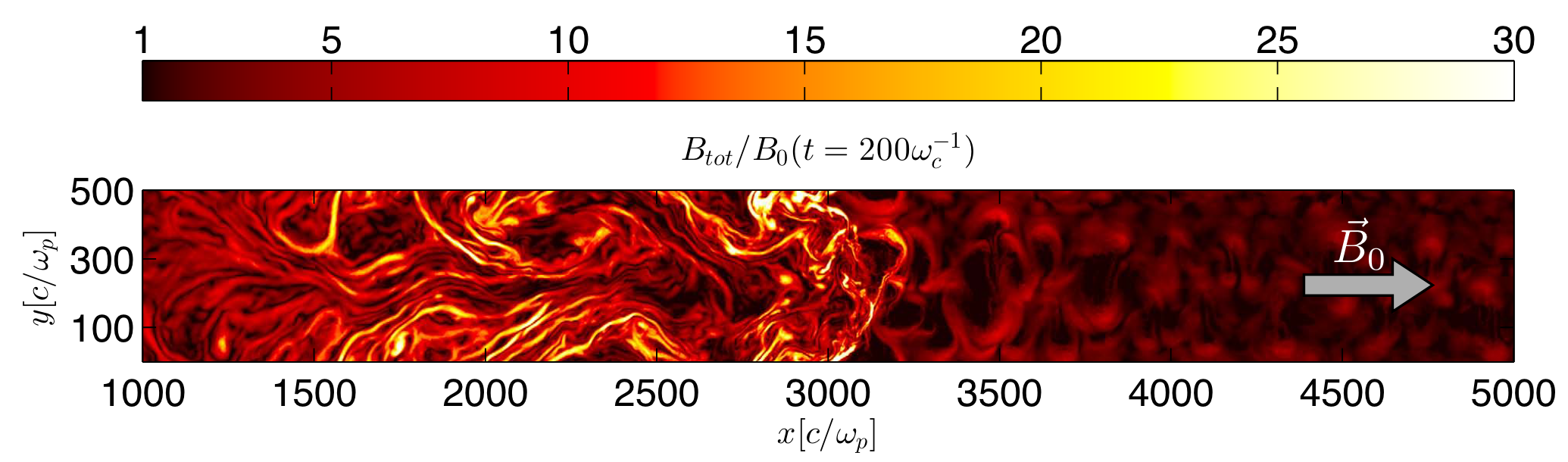}\\
\includegraphics[width=0.95\textwidth]{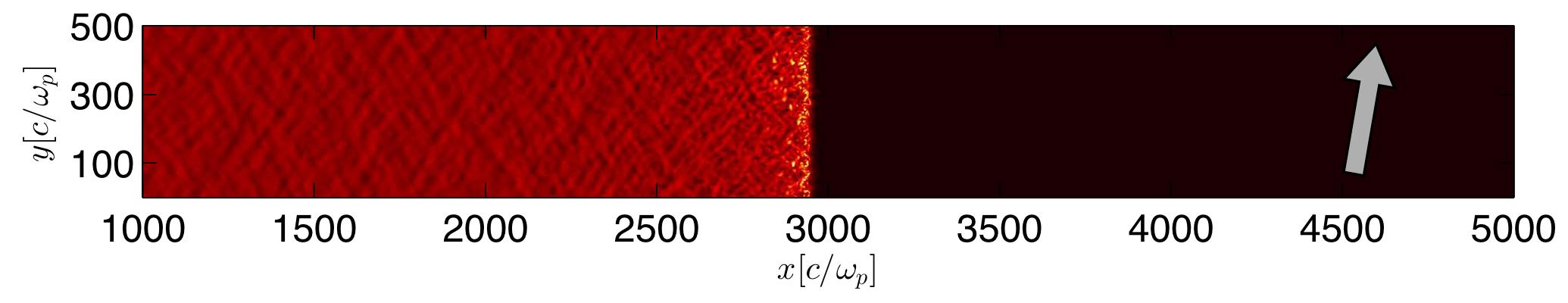}
\end{center}
\caption{Hybrid-PIC simulation of CR ion acceleration at a collisionless, non-relativistic strong shock. The top panel shows the downstream ion energy spectrum of a quasi-parallel shock, color coded by different times. The thermal distribution can be accurately described by a Maxwellian distribution with a temperature that is 80\% of the expected temperature for a shock (with a Mach number of 20) that does not accelerate particles (dashed line). The remaining energy is shared among magnetic turbulence and CRs that follow a power-law spectrum, which has an increasing maximum energy with time. To demonstrate the consistency with the theoretically predicted energy scaling in the non-relativistic regime of diffusive shock acceleration of test particles, the spectrum is scaled by $E^{1.5}$. This corresponds to a momentum spectrum $\propto p^{-4}$ (see inset). The bottom two panels show the magnitude of the total magnetic field for $\mathcal{M}=50$ shocks and a magnetic obliquity of $\theta_{Bn}=0^\circ$ and $80^\circ$, respectively (see gray arrows), implying that magnetic field amplification and thus CR acceleration is only efficiently at work in quasi-parallel shocks. Image from \citet{caprioli_simulations_2014-2}; reproduced with permission from ApJ.}
\label{fig:Caprioli}
\end{figure}

\paragraph{Ion acceleration at shocks.} Hybrid-PIC simulations in large computational domains and long run times demonstrate the formation of non-thermal tails at strong, non-relativistic shocks. While the efficiency of ion acceleration far into the relativistic regime approaches zero for quasi-perpendicular shocks (with magnetic fields nearly perpendicular to the shock normal, $\theta_{Bn}\approx90^\circ$), it is maximized for quasi-parallel shock geometries \citep[][see Fig.~\ref{fig:Caprioli}, bottom panels]{caprioli_simulations_2014-2,caprioli_simulations_2014-1} because of efficient magnetic field amplification through the Bell instability \citep{bell_turbulent_2004}.
\begin{wrapfigure}{r}{0.4\textwidth}
  \vspace{-1em}
  \begin{center}
    \includegraphics[width=0.4\textwidth]{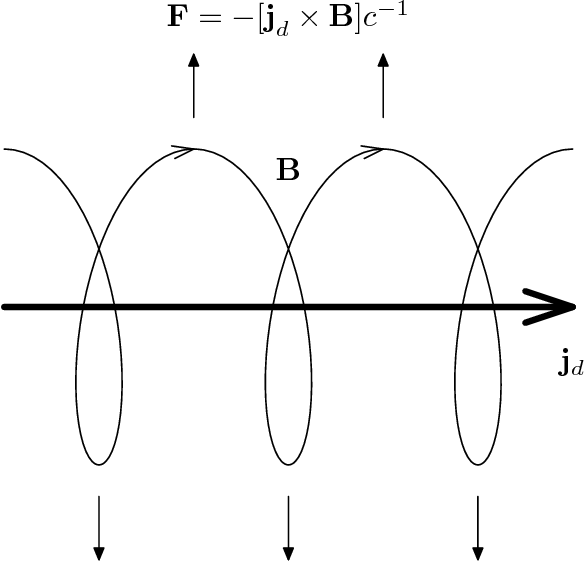}
  \end{center}
  \caption{Visualization of the underlying principle of Bell's non-resonant streaming instability. The CR current, $\mathbfit{j}_\rmn{d}$ induces a return current in the background electrons, $-\mathbfit{j}_\rmn{d}$, which amplifies a helical magnetic perturbation and stretches it via the Lorentz force $\mathbfit{F} = -(\mathbfit{j}_\rmn{d} \bs{\times} \mathbfit{B})\,c^{-1}$. Image from \citet{zirakashvili_modeling_2008}; reproduced with permission from ApJ.}
  \label{fig:Bell}
  \vspace{-1em}
\end{wrapfigure}
This nonresonant instability is the dominant CR streaming instability for sufficiently large CR fluxes, i.e., when the CR drift speed obeys \citep{shalaby_new_2021}
\begin{align}
\label{eq:Bell_criterion}
\varv_\rmn{d}> \left(\frac{3^{1/2}\gamma n_\rmn{cr}}{4 n_\rmn{i}}\right)^{-2/3}\varv_\rmn{a},
\end{align}
where $\gamma$ is the Lorentz factor of CR ions. This can be realized in quasi-parallel shock configurations where the acceleration process at the shock transition launches a strong CR flux towards the upstream along the mean magnetic field. The positive CR current induces a return current in the background electrons, which destabilized helical magnetic field fluctuations. The associated Lorentz force pulls the field lines outwards (Fig.~\ref{fig:Bell}), so that they start to approach each other, which thus exponentially amplifies the circularly polarized magnetic field \citep{bell_cosmic_2013}. Because CRs with a broad spectrum of momenta grow this field incoherently, there is a spectrum of Alfv\'enic turbulence generated on scales much smaller than the CR gyroradii. Clearly, this instability relies on a CR current that is aligned with the mean magnetic field. Thus, the magnetic amplification process saturates once the highest energy CRs become magnetized, i.e., if their trajectories start to be bent in the amplified field. This happens if the amplified magnetic energy density is comparable to the CR energy flux density at the shock divided by the light speed, yielding a saturated value for the amplified magnetic energy density  
\citep{bell_turbulent_2004,niemiec_production_2008,riquelme_nonlinear_2009,ohira_two-dimensional_2009,gargate_nonlinear_2010,blasi_high-energy_2015,zacharegkas_kinetic_2022}:
\begin{align}
    \eps_{B,\rmn{sat}} \sim \frac{1}{2}\frac{\varv_\rmn{s}}{c}\,\eps_\CR,
    \label{eq:eps_B,sat}
\end{align}
where $\varv_\rmn{s}$ is the shock speed, which is equal the CR drift velocity relative to the upstream plasma.

The average distance that energetic ions travel before experiencing pitch-angle scattering is similar to their gyro radii calculated in the turbulence they generate. In the case of moderately strong shocks, the magnetic field amplification can be characterized in the quasi-linear regime. This implies that all particles diffuse with their self-generated diffusion coefficient and undergo scattering once per gyro orbit (i.e., they diffuse in the Bohm limit). By contrast, in the case of very strong shocks, the magnetic field undergoes significant amplification, reaching non-linear levels. Most of the magnetic energy is concentrated in modes with wavelengths similar to the gyroradii of the highest-energy ions. As a result, only those particles undergo Bohm-like diffusion, while others scatter less efficiently \citep{caprioli_simulations_2014}. The evolution of Bell's (\citeyear{bell_turbulent_2004}) instability at high Mach number, quasi-parallel shocks is qualitatively similar in two and three-dimensional simulations \citep{vanmarle_three-dimensional_2019}.\\
\indent
Interestingly, quasi-parallel shocks are not stationary but they reform quasi-pe\-riodically on ion cyclotron timescales. This indicates that when ions encounter the steepest gradient of the shock discontinuity, they undergo specular reflection. This only happens for a quarter of the time. As a result, those ions are energized via shock-drift acceleration\footnote{Here, the term ``drift'' refers to the $\bs{E\times B}$ drift of charged particles, i.e., the constant motion of the particles' gyro centers perpendicular to the electric and magnetic fields. While shock-drift acceleration is most effective at perpendicular and quasi-perpendicular shocks, it does not work at a parallel shock \citep{ball_shock_2001}. For a quasi-perpendicular magnetic guide field, $\mathbfit{B}$, the induced electric field $\mathbfit{E}=-\bs{\varv\times B}/c$ (where $\bs{\varv}$ is the gyro-averaged particle velocity) is preferentially oriented perpendicular to the shock normal, so that the particles drift along the shock surface. As the particles' gyro orbits enter the shock transition, they experience an accelerating force from the somewhat stronger electric field there. Depending on the orientation of the gyro orbit, the field can either accelerate or decelerate the particle. After passing through the shock, the stronger post-shock magnetic field tightens the gyro orbits. If the gyro orbit is still large enough to enter the shock transition region, the increased electron field will continue to accelerate the particle.} and only a fraction of those gain enough energy to be injected into the process of diffusive shock acceleration \citep{caprioli_simulations_2015}. Hence, the popular thermal leakage model where all particles above a critical momentum in the tail of the Maxwell-Boltzmann distribution are accelerated is likely not applicable and too simplified to capture the correct physics. Energy transfer to a non-thermal (relativistic) particle population causes shocks to be modified and to assume an increased density jump at the shock transition due to an increased compressibility brought about by the decrease in the adiabatic index in the presence of the relativistic (CR) population. As a result, flatter CR spectra should emerge at high energies in comparison to the test-particle prediction $\propto p^{-4}$ \citep{drury_hydromagnetic_1981,malkov_nonlinear_2001}. This is in contradiction to the steeper slopes inferred from the non-thermal emission at SNRs implying momentum spectra $\propto p^{-4.3}$. Possible solutions of this spectral steepening include the insight that (i) the energy required to turbulently amplify magnetic field during the particle acceleration at shocks extracts energy from the CR population and steepens the CR energy spectrum \citep{bell_cosmic_2019}, (ii) that CRs potentially isotropize in the frame of Alfv\'enic turbulence downstream of the shock in a region known as the ``postcursor'' so that they would be advected downstream with this turbulence, implying a steepening of the CR momentum spectrum despite the density jump exceeding the canonical value for a non-relativistic gas of four \citep{caprioli_kinetic_2020}, and (iii) varying magnetic obliquity along the shock surface \citep{hanusch_steepening_2019}.

\paragraph{Electron acceleration at shocks.} By contrast, the gyroradii of electrons are smaller by the mass ratio relative to ions, $r_\rmn{e}/r_\rmn{i}=m_\rmn{e}/m_\rmn{i}$, so that they random walk through the shock transition and do not see a coherent electrostatic shock potential. This implies that thermal electrons cannot directly participate in the diffusive shock acceleration process that accelerates electrons to highly relativistic energies and requires a pre-acceleration process, which increases their momentum by a factor $\sim(m_\rmn{i}/m_\rmn{e})(c/\varv_\rmn{s})\sim3\times10^4$, thereby constituting the famous ``electron injection problem'' at shocks. Several processes have been suggested that differ depending on the magnetic obliquity in the upstream. Quasi-perpendicular shocks (with an obliquity $\theta_{Bn}\sim 90^\circ$) are characterized by a narrow shock transition so that particles cannot escape the shock and are (inefficiently) accelerated via shock-surfing \citep{shimada_strong_2000,hoshino_nonthermal_2002,bohdan_kinetic_2019-1,bohdan_kinetic_2019}, magnetic reconnection \citep{matsumoto_stochastic_2015,bohdan_kinetic_2020}, stochastic Fermi acceleration \citep{matsumoto_electron_2017,bohdan_kinetic_2019}, and compressional heating. In oblique shocks (with an obliquity $\theta_{Bn}\approx 50^\circ- 70^\circ$), electrons can escape the shock transition region to form the electron foreshock where they generate electrostatic electron-acoustic waves far upstream and whistler waves closer to the shock that can accelerate electrons \citep{xu_electron_2020,morris_preacceleration_2022,bohdan_electron_2022}. In addition, shock-surfing, magnetic reconnection, and stochastic shock drift acceleration also energize electrons \citep{matsumoto_electron_2017,katou_theory_2019,amano_observational_2020}.

However, by far the most efficient electron accelerators are quasi-parallel shocks ($\theta_{Bn}\lesssim 50^\circ$) where ions can escape the shock transition region forming the ion foreshock \citep{park_collisionless_2015,hanusch_electron_2020,arbutina_non-linear_2021}. In particular, the recently found intermediate-scale instability \citep{shalaby_new_2021} provides a natural way to produce large-amplitude electromagnetic fluctuations in parallel shocks. The instability drives ion–cyclotron waves unstable that are comoving with the upstream plasma at the shock front (see Section~\ref{sec:plasma_instabilities}). These unstable ion--cyclotron modes scatter the electrons parallel to the magnetic field so that some of them get accelerated to the required energies to be injected into diffusive shock acceleration \citep{shalaby_mechanism_2022}. However, these PIC simulations are numerically very challenging, which limits the achievable simulation time and/or dimensionality of the problem. Hence, we do not yet have a complete picture of the kinetic plasma physics that is responsible for particle acceleration.

\paragraph{Meso-scale models of shock acceleration.} There is an enormous range of scales between an entire SNR (with a typical radius of $r\sim3~\rmn{pc}\approx10^{19}$~cm) and typical scales of a plasma. In a cold plasma, the ion and electron distributions oscillate at their characteristic plasma frequencies, 
$\omega_\rmn{i,e} = (4\pi e^2 n_\rmn{i,e}/m_\rmn{i,e})^{1/2}$, which implies corresponding ion and electron skin depths of $d_\rmn{i} = c/\omega_\rmn{i}\sim 2\times 10^7~n_{\rmn{i,}0}^{-1/2}\rmn{cm}$ and $d_\rmn{e} = c/\omega_\rmn{e}\sim 5\times 10^5~n_{\rmn{e,}0}^{-1/2}\rmn{cm}$, where $n_{\rmn{e,}0} = n_{\rmn{i,}0} = 1~\rmn{cm}^{-3}$. Those length and time scales need to be resolved by PIC models of non-relativistic shocks at SN blast waves, which have typical shock transition widths of a few $d_\rmn{i}$. In consequence, these PIC simulations can only afford a limited simulation run time typically corresponding to $300~ \Omega_0^{-1}=300\, m_\rmn{p}c/(eB)\sim 9 B_{\mu\rmn{G}}^{-1}~\rmn{hours}$ of physical time and rarely model two or even three spatial dimensions, though hybrid PIC simulations are able to significantly expand upon these constraints. To bridge this range in scales, several different meso-scale approaches have been developed, which aim at answering different open problems of the CR shock acceleration problem. The answer to the question of whether SNR shocks are able to accelerate CRs to the ``knee'' at $\sim3$~PeV (for CR protons and greater energies for heavier ions) depends on the specific properties of the SNR shock, the CR diffusion coefficient, and eventually on the magnetic amplification upstream the shock.\\
\indent
To demonstrate this, we provide an order of magnitude argument that this is only possible if the magnetic field is amplified at the shock to levels of $100~\mu$G by balancing the acceleration time and the lifetime of SNR \citep{lagage_cosmic-ray_1983,lagage_maximum_1983,bell_cosmic_2013}. Assuming that the CR number density $n_\rmn{cr}$ in a small energy range has a uniform diffusion coefficient ahead of the shock, we can balance the CR diffusion flux $-\kappa \partial n_\rmn{cr}/\partial x$ with the advective flux $\varv n_\rmn{cr}$ to obtain an exponential precursor in the immediate upstream of the shock: $n_\rmn{cr}=n_\rmn{s}\exp(-x/L)$, where $x$ is the coordinate reaching from the shock to the upstream, $L=\kappa/\varv_\rmn{s}$ is the precursor scale height, $n_\rmn{s}$ is the CR number density at the shock discontinuity, and $\varv_\rmn{s}$ is the shock velocity. The CR number in the precursor region is given by $N=n_\rmn{s}LA$, where $A$ is the unit shock surface area. The rate at which CRs transition across the shock boundary from the upstream to the downstream and back to the upstream can be calculated from kinetic theory to yield $\dot{N}=n_\rmn{s}cA/4$. Combining these estimates enables us to estimate the average residency time of CRs ahead of the shock between crossings, $\Delta t=N/\dot{N}=4\kappa/(c \varv_\rmn{s})$. The CR acceleration rate is then given by $\rmn{d}E/\rmn{d}t\approx\Delta E/\Delta t=E \varv_\rmn{s}^2/(4\kappa)$, where we have used the average fractional energy gain of CRs per shock transition, $\Delta E/E=\varv_\rmn{s}/c$ \citep{bell_cosmic_2013}. This simple calculation only accounts for the time the CRs spend in the upstream, which is reasonable if the magnetic field in the downstream region is increased due to shock compression and if it is highly turbulent, implying a reduced diffusion coefficient. Assuming that additionally accounting for the time of CRs spent downstream doubles the total acceleration time yields $\tau_\rmn{acc}=8\kappa/\varv_\rmn{s}^2$, which can be equated to the SNR life time $\tau$ to obtain a maximum CR energy, $E_\rmn{max}$. We use the expression for the diffusion coefficient, $\kappa=c\lambda/3$, where $\lambda$ is the CR scattering free path measured in units of the CR gyroradius, $r_\rmn{L}= E/(ZeB)$. Hence, the maximum CR energy is 
\begin{align}
    \label{eq:Emax}
    E_\rmn{max}=\frac{3ZeB}{8c}
    \left(\frac{\lambda}{r_\rmn{L}}\right)^{-1}
    \tau\varv_\rmn{s}^2
    =3\times 10^{15}\,\left(\frac{\lambda}{r_\rmn{L}}\right)^{-1}
    \left(\frac{\varv_\rmn{s}}{5000~\rmn{km~s}^{-1}}\right)^2
    \,B_{2}\tau_{3}~\rmn{eV},
\end{align}
where $B_{2}$ is the magnetic field in units of $100\,\mu$G, $\tau_{3}$ is the SNR age in units of $1000$ years, and we adopted CR protons with $Z=1$. Because $\lambda$ cannot be smaller than $r_\rmn{L}$ and because $\tau_3\approx0.4$ for historic SNRs such as Cas~A, Kepler or Tycho, it is a challenge to reach the CR ``knee'' \citep{bell_cosmic_2013}. In particular, reaching this CR ``knee'' requires field strengths to be much larger than the $\mu$G fields of the average ISM. This provides circumstantial evidence for an efficient magnetic amplification mechanisms at SNR shocks such as Bell's instability.\\
\indent
Three-dimensional MHD-kinetic simulations, which employ a spherical harmonic expansion of the Vlasov--Fokker--Planck equation  \citep{bell_cosmic-ray_2013,reville_universal_2013} find that at the present time, historical SNRs such as Cas A, Tycho and Kepler appear to be expanding too slowly to accelerate CRs to the knee and argue for a larger maximum energy at an earlier stage of their evolution. Coupling the hydrodynamics to the transport equations for CRs and magnetic turbulence in one spherical dimension and zooming onto the shock transition enables one to resolve the scales of the fast growing Bell instability in the shock precursor and the associated Bohm diffusion while following diffusive acceleration of CRs \citep{brose_cosmic-ray_2020}. This yields the time evolution of the CR spectrum and facilitates accurate comparisons to observational data. The simulated $\gamma$-ray emission spectra \citep{brose_morphology_2021} of the different stages of the remnant's evolution such as free expansion, energy-conserving Sedov--Taylor, and early post-adiabatic expansion phases appear to match the observed time sequence of TeV $\gamma$-ray spectra \citep{funk_ground-_2015}.

\begin{figure}[tbp]
\begin{center}
\includegraphics[width=0.95\textwidth]{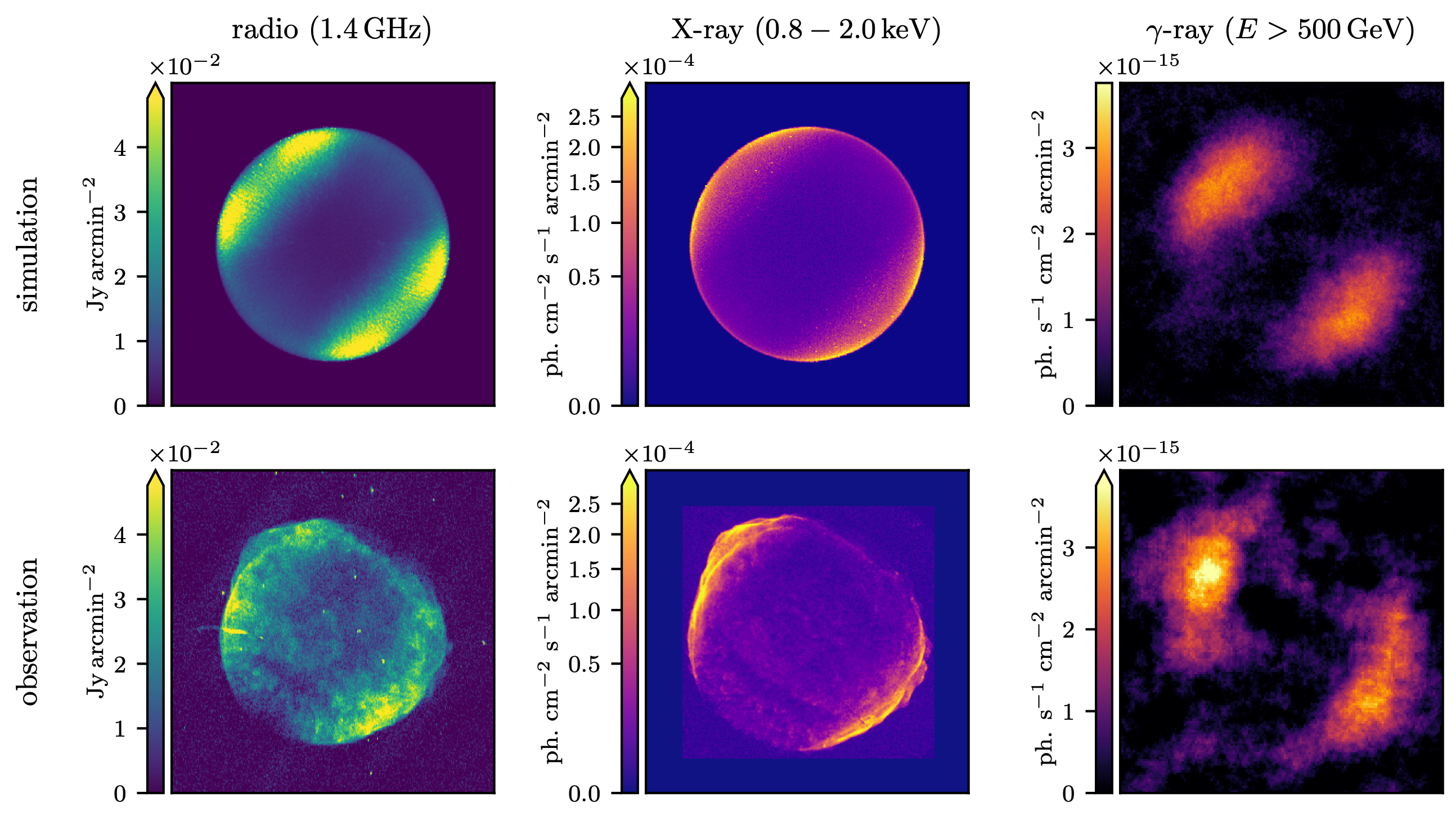}
\end{center}
\caption{Multi-frequency maps of the SNR SN 1006: the bottom row shows observations in the 1.4~GHz radio band \citep{dyer_14_2009}, the soft X-ray band  \citep{cassam-chenai_morphological_2008}, and at very-high energy $\gamma$ rays \citep{acero_first_2010}. Those are compared to simulated maps derived from global MHD simulations of Sedov explosions, which account for CR ion acceleration at the shock and evolve the CR electron spectrum \citep{winner_evolution_2019}. Initially, the magnetic field points from the bottom right to the top left. These simulations account for an acceleration efficiency that depends on magnetic obliquity, implying efficient acceleration at quasi-parallel shocks (i.e., when the shock travels at a narrow angle relative to the orientation of the magnetic field in the upstream region), and inefficient acceleration at quasi-perpendicular shock configurations. Hence, they explain the polar cap morphology seen in the maps. The simulated $\gamma$-ray map is convolved with the point-spread function of the H.E.S.S.\ instrument and incorporates Gaussian distributed noise, which matches the observed amplitude and correlation structure. The maps have a side length of 21~pc or 42.5' at a distance of 1.66~kpc. Image from \citet{winner_evolution_2020}; reproduced with permission from MNRAS.}
\label{fig:SN1006}
\end{figure}

\paragraph{Three-dimensional MHD models of shock acceleration.} A complementary line of research uses three-dimensional MHD simulation models, where a suitable algorithm finds the shock surface during the run-time of the simulation and characterises its jump conditions, Mach number, and magnetic obliquity \citep{pfrommer_simulating_tmp_2017,pais_effect_2018,dubois_shock-accelerated_2019,boss_crescendo_2023}. In the next step, CR ions are injected and transported either in the grey approximation \citep{pfrommer_simulating_tmp_2017} or by accounting for the Fokker-Planck evolution of the full CR ion spectrum \citep{yang_spatially_2017,girichidis_spectrally_2020,girichidis_spectrally_2022}, which enables us to account for the dynamical back reaction of CRs on the MHD flow. Additionally, a full spectrum of (primary, shock-accelerated) CR electrons can be injected according to a ``subgrid'' model, which is inferred from PIC simulations or motivated by theoretical considerations. The CR electron spectrum is transported with the flow while accounting for all re-acceleration and cooling processes that modify the spectrum \citep{winner_evolution_2019}. At every time step, we can compute the hadronic proton-proton reaction that generates pions, which decay into $\gamma$ rays and secondary electrons/positrons and neutrinos (see Section~\ref{sec:cooling_times}). Subsequently, the (primary and secondary) leptons radiate synchrotron emission from the radio to the X-ray band (provided the shock generated sufficiently energetic electrons and amplified magnetic fields). In addition, they undergo IC interactions, which results in emission from the hard X-ray to the $\gamma$-ray regime, which can be compared to observational data. \\
\indent
Comparison between observed and simulated emission maps, ranging from the radio, to X-rays to the $\gamma$-ray regime and the detailed multi-frequency spectrum of SN1006 argue in favor of a preferred quasi-parallel electron and ion acceleration for TeV electrons \citep[][see Fig.~\ref{fig:SN1006}]{winner_evolution_2020}. While simulated radial profiles in the equatorial sectors match X-ray data very well, there is some residual radio flux visible in the radio observations, which is missed by those simulations. This may suggest that the acceleration efficiency for GeV electrons is less strongly dependent on magnetic obliquity while the acceleration efficiency for TeV electrons, which radiate synchrotron X-rays, is suppressed by a factor of about 10 for quasi-perpendicular shocks in comparison to quasi-parallel shock geometries. By incorporating the obliquity-dependent CR ion acceleration model of \citet{caprioli_simulations_2014-2} into these comprehensive three-dimensional MHD models, we can account for the puzzling range of morphologies observed in TeV shell-type SNRs by adjusting the magnetic morphology and attribute regions exhibiting high (low) $\gamma$-ray intensity to quasi-parallel (quasi-perpendicular) shock configurations. Moreover, this enables constraining the magnetic coherence scale in the environment of several SNRs by relating the degree of $\gamma$-ray patchiness of SNRs to the correlation scale in the magnetic field \citep{pais_constraining_2020,pais_simulating_2020}.

\subsubsection{Particle escape from supernova remnants to the interstellar medium}
Diffusive shock acceleration is a self-regulating process: provided the level of upstream turbulence is low, many particles escape upstream, which sources a large current and a very efficient amplification of resonant and non-resonant Alfv\'en waves through streaming instabilities \citep{kulsrud_effect_1969,bell_turbulent_2004}. In turn, this causes the level of Alfv\'enic upstream turbulence to increase until the magnetic energy density in the amplified field is comparable to the CR energy flux density at the shock divided by the light speed; see equation~\eqref{eq:eps_B,sat}. The strong  Alfv\'enic turbulence scatters particles more efficiently and suppresses particle escape in the upstream \citep{schure_cosmic_2014,cardillo_cosmic_2015}. Besides regulating the shock itself, this phenomenon also has a nonlinear impact on the escape and confinement times of CRs in the immediate vicinity of their sources \citep[e.g.,][]{bell_cosmic-ray_2013, marcowith_cosmic_2021}. The self-generated magnetic turbulence increases the scattering rate of CRs, resulting in a reduction of the diffusion coefficient. Consequently, CRs accumulate within the environment of the sources and start to build up a significant pressure gradient. This gradient excavates a cavity around the source and cause to the formation of a CR-dominated bubble, so that the CRs can accumulate significant grammage $X(t)=\rho c t$ in the bubbles before moving into the ISM \citep{reville_transport_2008,malkov_analytic_2013,schroer_dynamical_2021, schroer_cosmic-ray_2022, recchia_grammage_2022}; the level of the increased CR scattering depends on the ISM properties in the immediate environment of the sources \citep{reville_cosmic_2007,reville_cosmic-ray_2021,nava_non-linear_2016,nava_non-linear_2019}.\\
\indent
A crucial point to consider when discussing CR feedback in galaxies is the presence of a spatially varying and reduced diffusion coefficient within the source regions in comparison to its galactic average value. This modification affects the phase-space structure of star-forming regions, leading to the accumulation of CRs. As a result, even in situations where the disk is unstable, the accumulation of CRs prevents local fragmentation and preserves a well-defined spiral structure \citep{semenov_cosmic-ray_2021}. Provided the level of electromagnetic fluctuations is small enough so that the quasi-linear theory of CR transport applies, advancing to the next level calls for the two-moment model of CR transport, which accounts for a temporarily and spatially varying diffusion coefficient in the self-confinement picture \citep{jiang_new_2018,thomas_cosmic-ray_2019,thomas_comparing_2021,thomas_finite_2021}. These hold the promise to elucidate the picture of how exactly CRs migrate from the sites of their increased confinement around sources to the galactic environment where transport is faster.

\subsection{Cosmic ray spatial transport}
\label{sec:spatial_transport}

\subsubsection{Theoretical background}
\label{sec:background}

To model CR transport in macroscopic systems such as galaxies, galaxy clusters, or AGN jets, and to study dynamical impact of CRs on the evolution of these systems, we need to coarse grain the kinetic physics and develop a fluid description for CRs that is coupled to MHD. Here, we sketch the derivation and various approximations of the different approaches. The CR distribution is defined in phase space, which is spanned by the momentum and spatial coordinates $\bs{p}$ and $\bs{x}$, respectively, and reads
\begin{align}
\label{eq:fp}
f\equiv f(\bs{x},\bs{p},t)=\frac{\dd^6N}{\dd{x^3}\dd{p^3}}.
\end{align}
Its time evolution is governed by the Vlasov equation in the semi-relativistic regime in the frame that is comoving with the fluid,
\begin{align}
\frac{\partial f}{\partial t} + (\bs{\varv} + \bs{\varv}_\CR) \bs{\cdot} \bs{\nabla}_{\bs{x}} f + \bs{F} \bs{\cdot} \bs{\nabla}_{\bs{p}} f = 0,
\label{eq:vlasov}
\end{align}
where the distribution function is advected by the total velocity that consists of the mean gas velocity $\bs{\varv}$ and the CR velocity $\bs{\varv}_\CR$. Note that $\bs{\varv}$ is measured in the laboratory frame while CR velocity and momentum $\bs{p}$ are measured in the comoving frame. The total force is represented by $\bs{F}$. Introducing the comoving frame introduces pseudo forces, which are denoted as $\bs{F}_\rmn{pseudo}$ and which come about because the momentum measured by a comoving observer changes as a result of a change of the frame velocity $\bs{\varv}$. Consequently, the momentum of CRs is altered by the Lorentz force, which can be divided into contributions from large-scale and small-scale electromagnetic fields denoted as $\bs{F}_{\rm macro}$ and $\bs{F}_{\rm micro}$, respectively (as derived with a covariant formalism by \citealt{thomas_cosmic-ray_2019}, Appendix C):
\begin{alignat}{4}
	\bs{F} &=\bs{F}_{\rm pseudo} &~+~& \bs{F}_{\rm macro} &~+~& \bs{F}_{\rm micro}\\
    	   &=-m\frac{\rmn{d}\bs{\varv}}{\rmn{d}t}-\bs{(\bs{p} \bs{\cdot} \bs{\nabla})} \bs{\varv}  &~+~& q\frac{\bs{\varv}_\CR \bs{\times} \bs{B}}{c} &~+~& q \left( \delta\bs{E} + \frac{\bs{\varv}_\CR \bs{\times} \delta\bs{B}}{c}\right),
           \label{eq:pseudo}
\end{alignat}
where $m$ and $q$ are the CR mass and charge, $\bs{B}$ is the local mean magnetic field, and ${\rm d}/{\rm d}t = \partial / \partial t + \bs{\varv} \bs{\cdot} \bs{\nabla}$ denotes the Lagrangian time derivative. The first pseudo force arises from the accelerating comoving frame itself. From the perspective of a comoving observer, a CR at rest in the laboratory frame appears to be accelerating. The second pseudo force stems from spatial variations in the velocity field of the background plasma. If a CR moves in the laboratory frame, any change in its position leads to a change in the frame velocity due to the inhomogeneities. This relationship between the laboratory and comoving frames results in acceleration in the comoving frame. The small-scale fluctuations, denoted as $\delta\bs{E}$ and $\delta\bs{B}$, represent electric and magnetic fluctuations, respectively. These fluctuations are generated by plasma waves, especially Alfv\'en waves on MHD scales and ion--cyclotron waves that are comoving with the CRs on smaller scales, and they serve as a source of scattering for CR (see Fig.~\ref{fig:Lorentz}):
\begin{align}
\left.\frac{\partial f}{\partial t} \right\rvert_{\rm scatt} = \bs{F}_{\rm micro} \bs{\cdot} \bs{\nabla}_{\bs{p}} f.
\end{align} 
The waves are either provided by external turbulence and cascading from large scales or generated by the CR-driven gyroresonant instability.

The CR proton gyroradius is $r_\rmn{g}=p_\perp c/(eB)=0.22~\rmn{AU}\, E_\rmn{GeV}\,B_{\mu\rmn{G}}^{-1}$ and is much smaller than any macroscopic scale in galaxies or galaxy clusters. Hence, we can project out the full phase dynamics of CRs by taking their gyroaverage to arrive at the {\em focused} transport equation \citep{skilling_cosmic_1971,zank_transport_2014}, which is expressed in the reference frame that moves with the average velocity of the gas
\begin{align}
\frac{\partial f}{\partial t} &+ (\bs{\varv} + \mu \varv_\CR \bs{b}) \bs{\cdot} \bs{\nabla} f + \left[ \frac{1-3\mu^2}{2} (\bs{b} \bs{\cdot} \bs{\nabla} \bs{\varv} \bs{\cdot} \bs{b}) - \frac{1-\mu^2}{2} \bs{\nabla} \bs{\cdot} \bs{\varv} \right] p \frac{\partial f}{\partial p}  \nonumber\\
&+ \left[ \varv_\CR \bs{\nabla} \bs{\cdot} \bs{b} + \mu \bs{\nabla} \bs{\cdot} \bs{\varv} - 3 \mu (\bs{b} \bs{\cdot} \bs{\nabla} \bs{\varv} \bs{\cdot} \bs{b}) \right] \frac{1-\mu^2}{2} \frac{\partial f}{\partial \mu}  = \left. \frac{\partial f}{\partial t}  \right\rvert_{\rm scatt}.
\label{eq:fpe_skilling} 
\end{align}
Here, the mean gas velocity $\bs{\varv}$ and the direction of the mean magnetic field on large scales, $\bs{b} = \bs{B} / B$ are measured in the laboratory frame while the magnitude of the CR momentum, $p=|\bs{p}|$, and the cosine of the pitch angle, $\mu = \bs{\varv}_\CR \bs{\cdot} \bs{b} / \varv_\CR$ are given in the comoving gas frame. As a result, the CR distribution function depends on 6 variables, $f=f(\bs{x}, p, \mu, t)$. If the CR distribution function is characterized by a high degree of anisotropy (which is the case for CR escape ahead of shocks), we would have to directly solve the focused transport equation \eqref{eq:fpe_skilling} without making further approximations. Here, we define the CR anisotropy as the deviation of the CR momentum-space distribution from an isotropic distribution, $\delta f/f_0$, where $f_0=f_0(\bs{x}, p, t)$ is the isotropic part of the CR distribution function.

\subsubsection{One-moment cosmic-ray hydrodynamics}

In our Galaxy, the  CR anisotropy at particle energies in the range of GeV to TeV is observed to be very small of order $\mathcal{O}(10^{-4})$ \citep{kulsrud_plasma_2005}. The simplest approximation for CR transport is the one-moment method, which approximates $f$ by the isotropic part of the CR distribution function and is derived in quasi-linear theory of CR transport that assumes electromagnetic fluctuations with a small amplitude relative to the mean field. Taking the zeroth $\mu$-moment of the focused transport equation~\eqref{eq:fpe_skilling}, yields a Fokker-Planck equation for CR transport \citep{skilling_cosmic_1971,skilling_cosmic_1975,schlickeiser_cosmic_2002},
\begin{equation}
  \label{eq:f0}
  \frac{\partial f_0}{\partial t} + (\bs{\varv} + \bs{\varv}_{\rmn{st}}) \bcdot \bs{\nabla} f_0
  = \bs{\nabla}\bcdot \left[\kappa \bs{b} \left(\bs{b}\bcdot\bs{\nabla} f_0\right)\right]
  + \frac{1}{3} p\frac{\partial f_0}{\partial p}\,\bs{\nabla}\bcdot(\bs{\varv} + \bs{\varv}_{\rmn{st}})
  + \frac{1}{p^2}\frac{\partial}{\partial p}\left[p^2\Gamma_\rmn{p}\,\frac{\partial f_0}{\partial p}\right],
\end{equation}
where $p=|\bs{p}|$ is the magnitude of the momentum, $\Gamma_\rmn{p}$ is the momentum diffusion rate related to second-order Fermi acceleration, and we have omitted CR sources and non-adiabatic CR losses. $\bs{\varv}_{\rmn{st}}$ and $\kappa$ are the CR streaming speed and spatial diffusion coefficient along the mean magnetic field, which will be introduced in detail below. This equation holds true only for time periods significantly longer than the relaxation time for pitch angle scattering $\tau\sim\mathcal{O}(D_{\mu\mu}^{-1})$ where $D_{\mu\mu}=D_{\mu\mu}(\bs{x},p,\mu)$ is the Fokker-Planck coefficient representing the frequency of pitch angle scattering of CRs by hydromagnetic waves. 

Multiplying the Fokker-Planck equation~\eqref{eq:f0} by the CR kinetic energy and integrating it over momentum space results in the evolution equation for the CR energy density, $\eps_\CR$: 
\begin{eqnarray}
\frac{\partial\varepsilon_{\rmn{cr}}}{\partial t} 
+ \boldsymbol{\nabla}\bcdot \left[ P_{\rmn{cr}} \bs{\varv}_{\rmn{st}}
+ \eps_{\rmn{cr}} (\bs{\varv} + \bs{\varv}_{\rmn{st}} + \bs{\varv}_{\rmn{di}})
  \right]
&=&
- P_{\rmn{cr}}\boldsymbol{\nabla}\bcdot\bs{\varv} + \bs{\varv}_{\rmn{st}}\bcdot\boldsymbol{\nabla}P_{\rmn{cr}},
\label{eq:ecr_1m}\\
\frac{\partial \eps}{\partial t} 
+ \boldsymbol{\nabla}\bcdot \left[(\eps+P_{\rmn{th}}+P_{\rmn{cr}}) \bs{\varv}\right]
&=&
\phantom{-}P_{\rmn{cr}}\boldsymbol{\nabla}\bcdot\bs{\varv} -\bs{\varv}_{\rmn{st}}\bcdot\boldsymbol{\nabla}P_{\rmn{cr}}.
\label{eq:eth}
\end{eqnarray}
Note that the CR equation~\eqref{eq:ecr_1m} resembles the evolution equation for the thermal and kinetic energy density, $\eps = \eps_{\rmn{th}}+\rho \varv^2/2$ shown in equation~\eqref{eq:eth}. The equations of state relate the thermal and CR pressures,  $P_\rmn{th}$ and $P_\CR$, to the corresponding energy densities via
\begin{alignat}{4}
\label{eq:eos1}
P_\rmn{th} &= (\gamma_{\rm th} &&- 1) \varepsilon_{\rm th}, \hspace{40pt} & \gamma_{\rm th} &= \frac{5}{3},\\
\label{eq:eos2}
P_\CR &= (\gamma_{\rm cr} &&- 1) \eps_\CR, & \gamma_{\rm cr} &= \frac{4}{3}.
\end{alignat}
CR streaming and diffusion are characterized by characteristic velocities, $\bs{\varv}_{\rmn{st}}$ and $\bs{\varv}_{\rmn{di}}$ (defined below). Note that here we neglected second-order Fermi acceleration and any non-adiabatic gain and loss terms for the thermal plasma and CRs for sake of transparency. Combining equations~\eqref{eq:ecr_1m} and \eqref{eq:eth} causes the terms on the right-hand side to vanish identically. This means that in the absence of non-adiabatic gain and loss terms, the total energy in form of CRs, thermal plasma and kinetic energy is conserved. This can be readily inferred by integrating the sum of equations~\eqref{eq:ecr_1m} and \eqref{eq:eth} over volume and applying Gauss' theorem. The physics of the CR energy equation \eqref{eq:ecr_1m} can be more transparently discussed by expanding the $P_{\rmn{cr}} \bs{\varv}_{\rmn{st}}$ term in the brackets to arrive at
\begin{eqnarray}
\label{eq:ecr1}
\frac{\partial\varepsilon_{\rmn{cr}}}{\partial t} 
+ \boldsymbol{\nabla}\bcdot \left[\eps_{\rmn{cr}} (\bs{\varv}+\bs{\varv}_{\rmn{st}} + \bs{\varv}_{\rmn{di}})\right]
=
- P_{\rmn{cr}}\boldsymbol{\nabla}\bcdot (\bs{\varv}+\bs{\varv}_{\rmn{st}}).
\end{eqnarray}
This equation states that $\varepsilon_{\rmn{cr}}$ is advected with the total CR velocity, $\bs{\varv}+\bs{\varv}_{\rmn{st}}+\bs{\varv}_{\rmn{di}}$. CRs can either adiabatically gain or lose energy, depending on the sign of the total velocity divergence: for adiabatic compression, $\boldsymbol{\nabla}\bcdot (\bs{\varv}+\bs{\varv}_{\rmn{st}})<0$, so that CRs gain energy.

\begin{figure}[tbp]
\begin{center}
\includegraphics[width=0.95\textwidth]{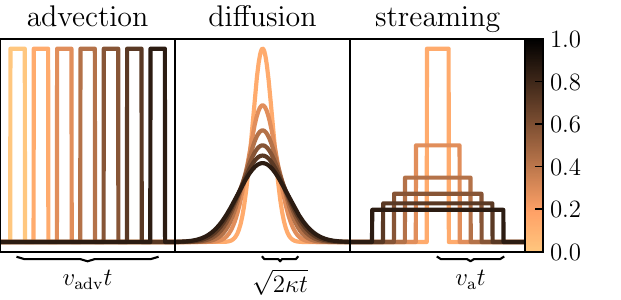}
\end{center}
\caption{\textit{Left:} The magnetic fields, which are frozen-in, are carried along with the plasma at an average velocity, $\varv=\varv_\mathrm{adv}$. Hence, CRs orbiting individual field lines are also advected alongside the plasma. \textit{Middle:} When the amplitude of Alfv\'en waves is strongly damped, CRs are weakly scattered and (after an initial time) diffuse away from the source. They are transported with a characteristic root-mean-square velocity of $\sqrt{2\kappa/t}$, where $t$ is the time and $\kappa$ denotes the diffusion coefficient along the magnetic field. \textit{Right:} If Alfv\'en waves are weakly damped, their amplitude remains large so that CRs are effectively scattered and stream at the Alfv\'en speed, $\varv_\rmn{a}$, along the orientation of the magnetic field. These different transport modes of CR fluids are not realized in their pure forms in Nature as they are instead superposed onto each other. Image from \citet{thomas_probing_2020}; reproduced with permission from ApJ. }
\label{fig:CR_transport}
\end{figure}

This surprising simplicity of CR transport can be understood by considering the kinetic physics of CR-wave interactions. Because the magnetic field is flux frozen into the ionized plasma in the MHD approximation, CRs are advected with the field lines as they are themselves advected with the fluid velocity $\bs{\varv}$. When CRs propagate along the magnetic field, their microphysical transport depends crucially on the scattering frequency. A small but non-negligible scattering rate implies diffusive transport along the magnetic field. In addition to diffusion, CR particles can collectively drift at an average velocity: when CRs move at a speed exceeding the local Alfv\'en speed, they generate gyroresonant Alfv\'en waves through a phenomenon known as the streaming instability \citep{kulsrud_effect_1969}. In turn, these waves increase the CR scattering (Fig.~\ref{fig:Lorentz}), which decelerates the CR population along the mean magnetic field, effectively transferring CR energy to further growing resonant Alfv\'en waves. This self-reinforcing feedback loop is stopped when CRs flow on average at the speed of Alfv\'en waves because there is no more net energy transfer from CRs to Alfv\'en waves.\\
\indent
Wave damping processes transfer wave energy originally borrowed from the CRs to the background thermal plasma. This means that CRs exert pressure on the thermal plasma by scattering of off Alfv\'en waves. In the presence of weak wave damping, the high (remaining) wave amplitudes enable a strong coupling, causing CRs to stream alongside Alfv\'en waves. Conversely, when wave damping is strong, wave amplitudes are reduced to the level that they cannot maintain a frequent scattering that isotropizes CRs in the wave frame, so that the CRs preferentially diffuse. This self-limiting picture of CR transport is confirmed by recent simulations using the PIC technique \citep{holcomb_growth_2019,shalaby_new_2021}, hybrid PIC \citep{weidl_three_2019,haggerty_hybrid_2019} as well as MHD PIC models \citep{lebiga_kinetic-MHD_2018,bai_magnetohydrodynamic_2019,plotnikov_influence_2021,bambic_mhd-pic_2021}. The highly idealized pure CR transport modes advection, diffusion and streaming are visualized in Fig.~\ref{fig:CR_transport}.

To concretize this qualitative discussion, we define the pitch-angle averaged CR scattering frequencies of right- and left-ward propagating Alfv\'en waves using their energy densities $\eps_{\rmn{a},\pm}$ \citep{thomas_cosmic-ray_2019}:
\begin{align}
\bar{\nu}_\pm
=\frac{3 \pi\, \Omega}{16} \frac{\eps_{\rmn{a},\pm}}{\varepsilon_B}
=\frac{c^2}{3\kappa_\pm},
\label{eq:closure_scatt_coeff}
\end{align}
where $\eps_B=B^2/(8\pi)$ is the magnetic energy density and $\kappa_\pm$ denote the CR diffusion coefficients associated with the corresponding scattering frequencies. Hence, a larger level of Alfv\'enic fluctuations increases the scattering rate. CRs stream down their own pressure gradient relative to the background plasma along the local direction of $\mathbfit{B}$ with a velocity \citep{skilling_cosmic_1975}
\begin{align}
\label{eq:v_st}
\bs{\varv}_{\rmn{st}}=\bs{\varv}_{\rmn{a}}\,
    \frac{\bar{\nu}_+ - \bar{\nu}_-}{\bar{\nu}_+ + \bar{\nu}_-}
    \to -\bs{\varv}_{\rmn{a}}\,\rmn{sgn}(\bs{B}\bcdot\bs{\nabla}P_\CR).
\end{align}
The limit on the right-hand side is realized in the self-confinement picture of CR transport where CRs streaming down the CR gradient excite resonant Alfv\'en waves. Those waves are moving in the direction of the streaming CRs so that their wave amplitudes dominate over the counter propagating type of waves, which are efficiently damped through various collisionless damping processes, see Section~\ref{sec:streaming}. If $\bs{B}$ points in the opposite direction of $\bs{\nabla}P_\CR$, equation~\eqref{eq:v_st} implies streaming along the magnetic field, $\bs{\varv}_{\rmn{st}}=\bs{\varv}_{\rmn{a}}$ and a dominating scattering rate $\bar{\nu}_+$. Efficient scattering of CRs isotropizes them in the Alfv\'en wave rest frame (see Section~\ref{sec:CR-wave}). As discussed above, CRs can also diffuse in the wave frame along the local direction of $\mathbfit{B}$ due to less frequent pitch angle scattering by MHD waves \citep{thomas_cosmic-ray_2019}:
\begin{align}
\label{eq:v_di}
\bs{\varv}_{\rmn{di}}=-\kappa\mathbfit{b}\,
\frac{\mathbfit{b}\bcdot\bs{\nabla} \eps_{\rmn{cr}}}{\eps_{\rmn{cr}}},
\qquad
\mbox{where}
\qquad
\kappa=\frac{c^2}{3(\bar{\nu}_+ + \bar{\nu}_-)}
\end{align}
is the total CR diffusion coefficient, which decreases for an increased CR-wave scattering rate. Because $\bs{\varv}_{\rmn{di}}$ depends on the gradient of the CR energy density, this type of transport process spreads the initial CR distribution in time. As a result, the diffusion operator is characterized by a Laplacian (combining equations~ \ref{eq:ecr1} and \ref{eq:v_di} and assuming $\kappa=\rmn{const}$). In one dimension, the solution of the diffusion equation is given by convolving the initial condition with a Gaussian representing the Green's function of the diffusion operator. This causes the diffusive spread of CRs away from a local maximum of CRs along the magnetic field.

The first numerical simulations to include dynamical CR feedback on the hydrodynamics studied the problem of diffusive shock acceleration at a SN blast wave  \citep{dorfi_numerical_1984,dorfi_time-dependent_1985} and employed the so-called ``two-fluid'' model laid out in equations~\eqref{eq:ecr_1m} and \eqref{eq:eth}, which are closed by the equations of state with the corresponding ratio of specific heats, \eqref{eq:eos1} and \eqref{eq:eos2}, for the thermal and CR populations \citep{drury_hydromagnetic_1981,kang_diffusive_1990}. These early studies assumed simplified expressions for the spatial and momentum dependence of the CR diffusion coefficient and neglected the CR streaming term. Using a simplified analytic two-fluid model of the SN expansion, \citet{drury_simplified_1989} account for dynamical CR feedback as well as the influence of Alfv\'en wave heating on the diffusion coefficient. The first fully time-dependent and non-linear numerical simulations of spherical SN blast waves with the ``two-fluid'' model of CR-hydrodynamics find that the shock can transfer $\sim10\%$ of its kinetic energy to the CR population at the end of the adiabatic, Sedov--Taylor phase \citep{jones_time-dependent_1990,dorfi_evolution_1990}. The first CR hydrodynamical simulations coupled to an equation for the resonant Alfv\'en wave energy demonstrate the emergence of CR-driven galactic winds with one-dimensional flux-tube models \citep{breitschwerdt_galactic_1991,breitschwerdt_galactic_1993}.

In the new century, new computational capabilities enabled the study of the dynamical effect of CR pressure and heating in galaxy formation and galaxy clusters, using three-dimensional (magneto-)hydrodynamical simulations. Pioneering work was performed with the Eulerian MHD code PIERNIK that followed anisotropic CR diffusion \citep{hanasz_incorporation_2003} and succeeded by simulations of galaxies, clusters and cosmological volumes with the smoothed particle hydrodynamics code \textsc{Gadget}, which evolves a CR population described by a single power-law momentum spectrum while accounting for isotropic CR diffusion \citep{pfrommer_detecting_2006,enslin_cosmic_2007,jubelgas_cosmic_2008} and CR streaming \citep{uhlig_galactic_2012}. The promising results of galactic winds driven by CR streaming were confirmed by several other groups that modelled isotropic CR diffusion using adaptive mesh-refinement codes \textsc{Ramses} \citep{booth_simulations_2013}, \textsc{Enzo} \citep{salem_cosmic_2014}, and ART \citep{semenov_cosmic-ray_2021}. Studies of anisotropic CR diffusion coupled to MHD were made possible by CR implementations in the Eulerian \textsc{Pencil} code \citep{snodin_simulating_2006}, the adaptive mesh-refinement codes \textsc{Flash} \citep{yang_fermi_2012,Girichidis_anisotropic_2014} and \textsc{Ramses} \citep{dashyan_cosmic_2020}, and the unstructured moving-mesh code AREPO \citep{pakmor_semi-implicit_2016,pfrommer_simulating_tmp_2017}. Anisotropic CR streaming in MHD simulations was modelled by \citet[][with \textsc{Flash}]{ruszkowski_global_2017}, \citet[][with \textsc{Enzo}]{butsky_role_2018}, and \citet[][with \textsc{Ramses}]{dubois_shock-accelerated_2019} while employing the regularization of the anisotropic CR streaming term proposed by \citet{sharma_numerical_2010}.

\subsubsection{Two-moment cosmic-ray hydrodynamics}
\label{sec:two-moment CR hydro}

The previously discussed picture of one-moment CR hydrodynamics has the disadvantage that the (direction of the) CR streaming is ill-defined at the extremes of the CR distribution (see equation~\ref{eq:v_st}). This causes numerical instabilities of the CR streaming term, which can be cured by regularizing the sharp transition of the sign function and smoothing it out \citep{sharma_numerical_2010}, but at the expense of introducing a numerical parameter that dominates the solution in case of a CR source on top of a CR background \citep[see figure~6 of][]{thomas_cosmic-ray_2019}. Capitalizing on analogies of CR and radiation hydrodynamics, \citet{jiang_new_2018} propose a two-moment CR hydrodynamics scheme that assumes steady-state CR scattering approximation and additionally solves for the CR flux density as an independent and evolved quantity, thereby solving the problem of non-uniqueness of the CR streaming flux at the extremes of $f$ \citep[see also][for a similar development]{chan_cosmic_2019}. 

In an alternative approach, we can expand $f$ into a complete set of basis functions in  pitch-angle (such as the Legendre polynomials, see Section~\ref{sec:CR-wave}), which has a long history in CR transport \citep[see e.g.][]{klimas_foundation_1971, earl_diffusion_1973, webb_hydrodynamical_1987, zank_transport_2000, snodin_simulating_2006, litvinenko_numerical_2013,rodrigues_fickian_2019}. Using the Eddington approximation for the CR distribution, $f = f_0 + 3 \mu f_1$, and requiring that the isotropic term dominates over the anisotropy, $f_0 \gg f_1$, \citet{thomas_cosmic-ray_2019} compute the zeroth and first $\mu$-moments of the 
focused transport equation \eqref{eq:fpe_skilling} to arrive at evolution equations for the isotropic and anisotropic parts of the CR distribution, $f_0$ and $f_1$. Multiplying those with the particle energy and the energy flux, respectively, and integrating the equations for $f_0$ and $f_1$ over momentum space yields the evolution equations for the CR energy and momentum density, $\eps_\CR$ and $\bs{f}_\CR/c^2$. In the laboratory frame, the pressure, energy and momentum densities transform as follows (neglecting the relativistic corrections to the CR energy and pressure):  
\begin{align}
\label{eq:lab}
\mat{P}_{\CR,\rmn{lab}} = P_\CR \mat{1},  
\quad
\eps_{\CR,\rmn{lab}} = \eps_\CR,
\quad
\bs{f}_{\CR,\rmn{lab}} = \bs{f}_\CR + \bs{\varv}\left(\eps_\CR + P_\CR\right),
\end{align}
so that we obtain the governing equations for the CR energy and momentum density as measured in the laboratory frame \citep[][see Appendix E]{thomas_cosmic-ray_2019}:
\begin{alignat}{6}
\label{eq:ecr_lab}
&\frac{\partial \eps_{\CR,\rmn{lab}}}{\partial t} + \bs{\nabla} \bcdot \bs{f}_{\CR,\rmn{lab}}
  = && - \bs{\varw}_+ \bcdot & \frac{\bs{b}\bs{b}}{3\kappa_+} \bcdot \left[   \bs{f}_{\CR,\rmn{lab}} - \bs{\varw}_+ (\eps_{\CR,\rmn{lab}} + P_{\CR,\rmn{lab}})\right]&&\\
   &&& - \bs{\varw}_- \bcdot & \frac{\bs{b}\bs{b}}{3\kappa_-} \bcdot \left[ \bs{f}_{\CR,\rmn{lab}} - \bs{\varw}_- (\eps_{\CR,\rmn{lab}} + P_{\CR,\rmn{lab}})\right]
& -\bs{\varv}\bcdot\bs{g}_{\rmn{Lorentz}} &+ S_\eps,
\nonumber\\
\label{eq:fcr_lab}
\frac{1}{c^2}&\frac{\partial \bs{f}_{\CR,\rmn{lab}}}{\partial t} + \bs{\nabla} \bs{\cdot} \mat{P}_{\CR,\rmn{lab}}
  = &&\, - &\frac{\bs{b}\bs{b}}{3\kappa_+} \bcdot \left[ \bs{f}_{\CR,\rmn{lab}} - \bs{\varw}_+ (\eps_{\CR,\rmn{lab}} + P_{\CR,\rmn{lab}})\right]&&\\
&&&\, - &\frac{\bs{b}\bs{b}}{3\kappa_-} \bcdot \left[ \bs{f}_{\CR,\rmn{lab}} - \bs{\varw}_- (\eps_{\CR,\rmn{lab}} + P_{\CR,\rmn{lab}})\right]
& - \bs{g}_{\rmn{Lorentz}} &+ \bs{S}_f,
\nonumber
\end{alignat}
where the Alfv\'en wave velocity in the laboratory frame is given by $\bs{\varw}_\pm = \bs{\varv} \pm \bs{\varv}_{\rmn{a}}$, $S_\eps$ and $\bs{S}_f$ are non-adiabatic source terms of CR energy and flux, respectively, and the Lorentz force density in the ultra-relativistic limit is $c^2\bs{g}_{\rm Lorentz} = \bs{f}_\CR\bs{\times}\bs{\Omega}$, where $\bs{\Omega} = \Omega \bs{b}$.

These equations also contain the previously discussed three CR transport modes: advection, streaming, and diffusion. In the absence of sources and along the magnetic field (where $\bs{g}_{\rm Lorentz} = \bs{0}$), we recover CR streaming and diffusion. In the presence of efficient CR-wave scattering, i.e., for a large CR scattering frequency $\bar{\nu}_\pm = c^2/(3\kappa_\pm)$, we approach a homogeneous CR flux with a steady-state CR energy density, implying $\partial_t \eps_{\CR,\rmn{lab}} + \bs{\nabla} \bs{\cdot} \bs{f}_{\CR,\rmn{lab}} = 0$. In consequence, the CR enthalpy streams with the Alfv\'en wave frame, $\bs{f}_{\CR,\rmn{lab}} = \bs{\varw}_\pm (\eps_{\CR,\rmn{lab}} + P_{\CR,\rmn{lab}})$, where the sign of $\bs{\varw}_\pm$ determines the CR propagation direction. If CR scattering is inefficient but finite, we also approach a steady state CR flux ($\partial_t \bs{f}_{\CR,\rmn{lab}} = \bs{0}$) and recover the CR diffusion velocity, $\bs{\varv}_\rmn{di} \eps_{\CR,\rmn{lab}} = -\kappa_\pm \bs{b}\bs{b}\bcdot \bs{\nabla}\eps_{\CR,\rmn{lab}}$, where we have identified the non-vanishing bracket on the right-hand side with $\bs{\varv}_\rmn{di} \eps_{\CR,\rmn{lab}}$ and used the ultra-relativistic equation of state for CRs, $P_\CR = \eps_\CR/3$. Finally, we recover CR advection in steady state, $\bs{f}_{\CR,\rmn{lab}} = \bs{\varv} (\eps_{\CR,\rmn{lab}} + P_{\CR,\rmn{lab}})$ due to the balance of forces parallel and perpendicular to the magnetic field: while the perpendicular pressure force balances the Lorentz force acting on CRs, $\bs{\nabla}_\perp P_{\CR,\rmn{lab}} = -\bs{g}_{\rmn{Lorentz}}$ (which may however act on kinetic time scales), we have forces due to CR streaming and diffusion balancing the pressure forces parallel to the magnetic field. 

These equations closely resemble the lab-frame equations for radiation energy and momentum density, $\eps$ and $\bs{f}/c^2$ \citep{mihalas_foundations_1984,lowrie_coupling_1999}
\begin{alignat}{4}
&\frac{\partial \eps}{\partial t} + \bs{\nabla} \bs{\cdot} \bs{f} 
  &&= - \sigma_{\rmn{s}}\,\bs{\varv}\, \bcdot\, & \left[ \bs{f} - \bs{\varv} \bcdot (\eps \mat{1} + \mat{P})\right]
&+ S_{\rmn{m}},
\\
\frac{1}{c^2}&\frac{\partial \bs{f}}{\partial t} + \bs{\nabla} \bs{\cdot} \mat{P}
  &&= - \sigma_{\rmn{s}} &\left[ \bs{f} - \bs{\varv} \bcdot (\eps \mat{1} + \mat{P})\right]
&+ S_{\rmn{m}}\bs{\varv},
\end{alignat}
where $\mat{P}$ is the radiation pressure tensor, $\sigma_\rmn{s}$ is the scattering coefficient, and $S_\rmn{m}$ describes photon absorption and emission processes, thus coupling radiation to matter. The similarities of the CR and photon transport equations are apparent and result from the same underlying physics: while efficient photon scattering causes the photon enthalpy to be advected with the gas, efficient CR-wave scatterings causes the CR enthalpy to stream with the Alfv\'en waves. However, there are two main differences: (i) the propagation direction of photons is given by the path of least resistance unlike for CRs, whose gyrotropic motion is directed along the local mean magnetic field and (ii) in contrast to photon transport, the CR lab-frame equations require resolving rapid gyro kinetics owing to the Lorentz force. As pointed out by \citet{thomas_cosmic-ray_2019}, transforming those lab-frame equations into the comoving frame enables us to project out the fast gyro kinetics so that the CR hydrodynamics theory can be applied to macroscopic astrophysical scales.

CR scattering and hence their spatial transport depends on the amplitudes of resonant Alfv\'en waves, which are excited by the gyro-resonant instability and lose energy due to various collisionless damping processes as well as second-order Fermi acceleration. Hence, a complete theory of two-moment CR hydrodynamics not only needs to couple the equations of CR energy and flux density to the MHD equations but also to the equations for resonant Alfv\'en wave energy \citep{dewar_interaction_1970,jacques_momentum_1977,breitschwerdt_galactic_1991,zweibel_basis_2017} to account for spatial and temporal variations of wave growth and damping. The non-linearly coupled subsystem of four CR-wave equations in the comoving frame reads as follows \citep{thomas_cosmic-ray_2019}:
\begin{align}
&\frac{\partial \eps_\CR}{\partial t} + \bs{\nabla} \bs{\cdot} [\bs{\varv} (\eps_\CR + P_\CR) + \bs{b} f_\CR] = \bs{\varv} \bs{\cdot} \bs{\nabla} P_\CR  \nonumber\\ 
&\quad\quad - \frac{\varv_{\rm a}}{3\kappa_+} \left[ f_\CR - \varv_{\rm a} (\eps_\CR + P_\CR) \right] + \frac{\varv_{\rm a}}{3\kappa_-} \left[ f_\CR + \varv_{\rm a} (\eps_\CR + P_\CR) \right],
\label{eq:ecr}\\[1em]
&\frac{\partial f_\CR/c^2}{\partial t} + \bs{\nabla} \bs{\cdot} \left( \bs{\varv} f_\CR/c^2  \right) + \bs{b} \bs{\cdot} \bs{\nabla} P_\CR =  - (  \bs{b} \bs{\cdot} \bs{\nabla} \bs{\varv}) \bs{\cdot} (\bs{b}  f_\CR/c^2) \nonumber\\
&\quad\quad-\frac{1}{3\kappa_+} \left[ f_\CR - \varv_{\rm a} (\eps_\CR + P_\CR) \right]  - \frac{1}{3\kappa_-} \left[ f_\CR + \varv_{\rm a} (\eps_\CR + P_\CR) \right],\
\label{eq:fcr}\\[1em]
&\frac{\partial \eps_{\rmn{a},\pm}}{\partial t} + \bs{\nabla} \bs{\cdot} \left[ \bs{\varv} (\eps_{\rmn{a},\pm} + P_{\rmn{a},\pm}) \pm \varv_{\rm a} \bs{b} \eps_{\rmn{a},\pm} \right] = \bs{\varv} \bs{\cdot} \bs{\nabla} P_{\rmn{a},\pm}  \nonumber\\ 
&\quad\quad\pm \frac{\varv_{\rm a}}{3\kappa_\pm} \left[f_\CR \mp \varv_{\rm a} (\eps_\CR + P_\CR)\right] - S_{{\rm a},\pm},
\label{eq:ew}
\end{align}
where $P_{\rmn{a},\pm}=\eps_{\rmn{a},\pm}/2$ are the 
ponderomotive pressures of Alfv\'en waves, which arise due to the nonlinear force experienced by a charged particle in an oscillating electromagnetic field with spatial variations. These Alfv\'en waves have wave lengths comparable to the gyroradii of pressure-carrying CRs, well below the resolved scales in the simulation. $S_{{\rm a},\pm}$ are wave energy loss terms due to damping processes.\\
\indent
In contrast to the laboratory frame equations of CR transport \eqref{eq:ecr_lab} and \eqref{eq:fcr_lab}, here the CR flux density is aligned with the local mean magnetic field, $\bs{f}_\CR = \bs{b} f_\CR$ to reduce the dimensionality of the system of equations. Clearly, the terms of the laboratory frame equations are also apparent in this comoving frame. In addition, there are pseudo forces that result from transforming into the non-inertial comoving frame as discussed in Section~\ref{sec:background}. Notably, the adiabatic term on the right-hand side of equation~\eqref{eq:ecr} is the manifestation of a pseudo force in this two-moment method. Like all hydrodynamical theories that are derived from a moment hierarchy and are truncated at finite order, a closure relation is required that relates the highest-order moment to lower order moments in order to arrive at a consistent and closed system of evolution equations. Unlike the case of radiation transport, where there are significant differences in the solutions with the various closure relations, the different closures for the CR transport models produce the same results for frequent CR-wave scattering, which is typically expected in the (warm and hot phases of the) interstellar, circumgalactic or intracluster media \citep{thomas_comparing_2021}. This is because CR transport is primarily constrained to magnetic field lines, allowing CRs to propagate only along or against the direction of the magnetic field. This is not the case for radiation, because its transport may occur in arbitrary directions.

\begin{figure}[tbp]
\begin{center}
\includegraphics[width=0.3\textwidth]{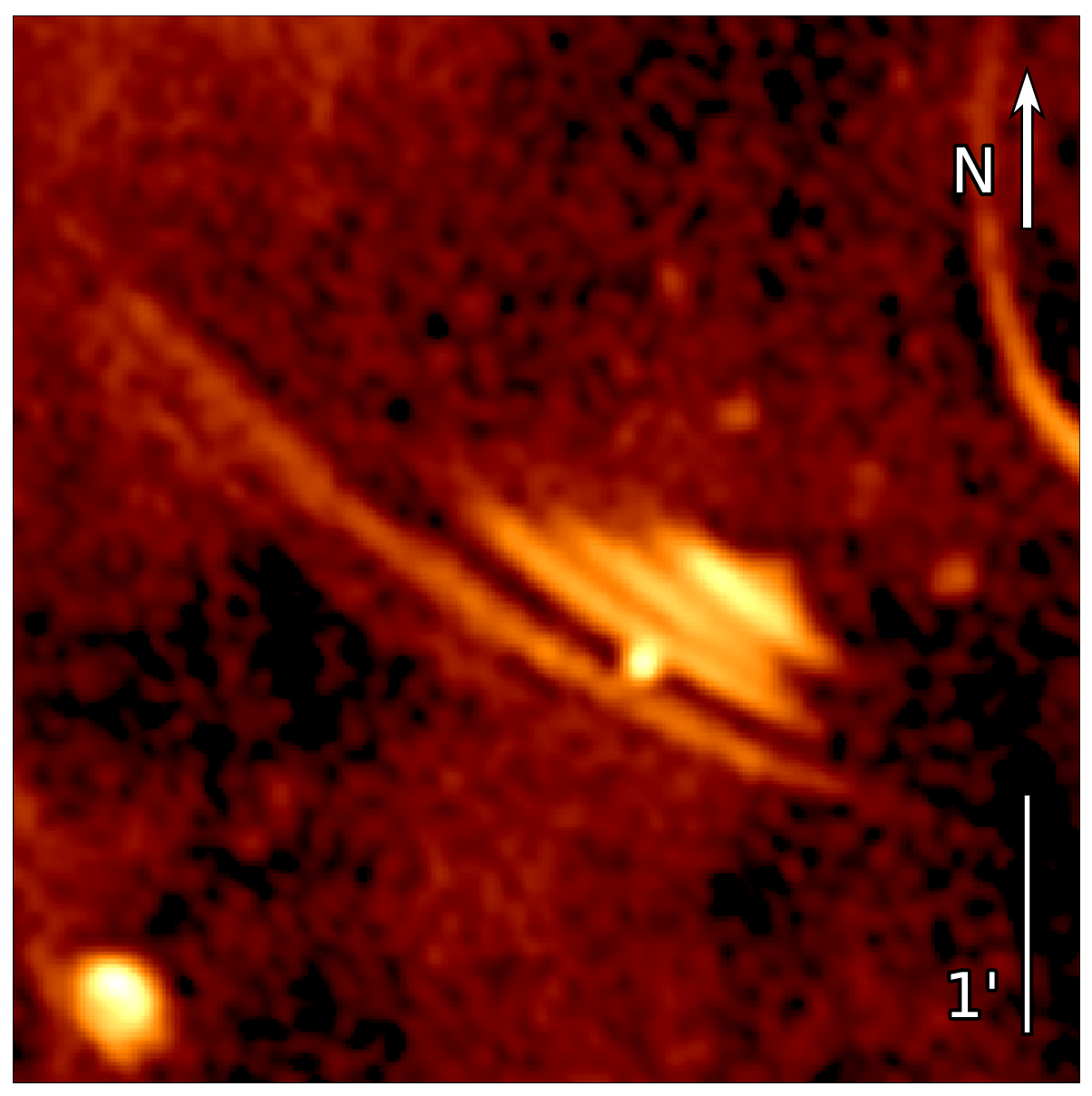}
\includegraphics[width=0.65\textwidth]{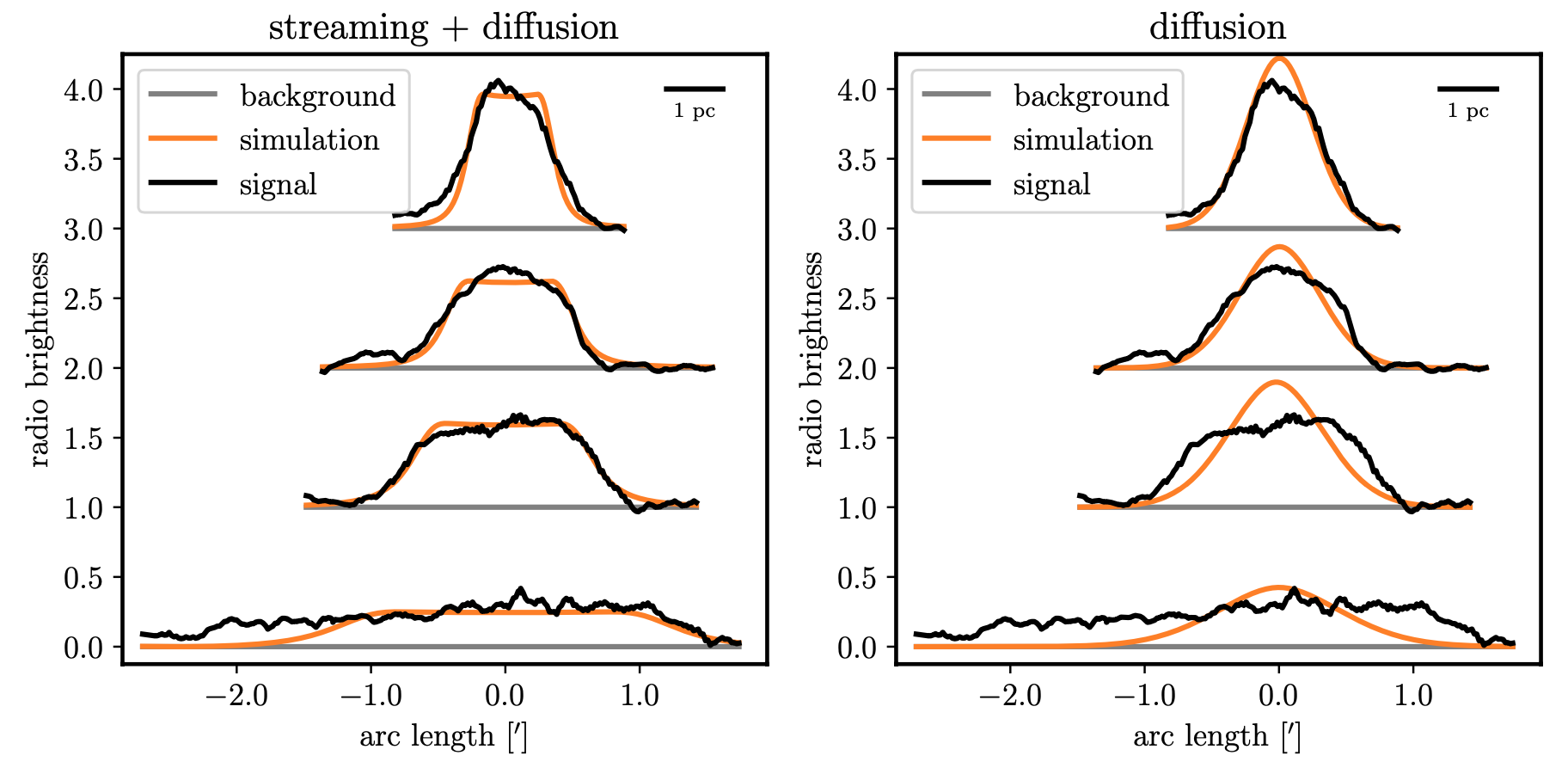}
\end{center}
\caption{A radio harp in the MeerKAT observation by \citet{heywood_inflation_2019} at the Galactic center (left). The observed radio brightness profiles along the individual radio-emitting filaments, the ``harp strings'' are compared with model calculations (orange) of streaming and diffusion (middle) and pure diffusion (right). In the lower two profiles, where the CR electrons had more time to propagate, the models show significant differences and favor the streaming model. Image from
\citet{thomas_probing_2020}; reproduced with permission from ApJ. }
\label{fig:harps}
\end{figure}

Are there methods to observationally validate these theoretical considerations? The observed level of CR anisotropy is an indirect measure of the degree of isotropization, which should be of order $\varv_\rmn{a}/c \sim10^{-4}$ in the ISM if CRs are indeed streaming close to the Alfv\'en speed \citep{kulsrud_plasma_2005}. Observing the time sequence of CR profiles as they are propagating from a source would provide a strong insight into the physics. In recent radio observations of the center of our Milky Way using the MeerKAT telescope, radio-emitting structures with nearly parallel filaments have been discovered. These filaments are lined up by their length, resulting in a morphology reminiscent of a harp, with the filaments resembling radio-synchrotron emitting ``strings'' \citep[][see left-hand panel of Fig.~\ref{fig:harps}]{heywood_inflation_2019}. These structures are formed when massive stars or pulsars fly through ordered magnetic fields of the ISM, discharging CR electrons along their paths into these magnetic fields either (i) via magnetic reconnection of shocked ISM fields with the magnetic field of pulsar wind nebulae or (ii) turbulent mixing of the shocked ISM and shocked stellar wind, which contains CRs accelerated at a strong stellar wind termination shock. The particles propagate along the ISM magnetic field lines, usually transverse to the star's orbit, causing the magnetic fields to be illuminated in the radio through the synchrotron process, thus explaining their appearance in form of strings of a harp (with the different string lengths corresponding to the timing of the discharge). If this propagation was a diffusion process, the radio profiles should have a rounded bell shape (see right-hand panel of Fig.~\ref{fig:harps}). Instead, the radio profiles have a flat-top shape, which can only be explained by streaming CRs: propagating relativistic electrons excite the magnetic fields of the ``strings'' to resonantly oscillate with their gyro orbits, which then amplifies Alfv\'en waves through the streaming instability \citep{kulsrud_effect_1969}. This in turn decelerates CRs by the described scattering processes so that CRs can be described by a streaming fluid \citep[plus a small amount of diffusion,][]{thomas_probing_2020}. This validates the predictions of the theory of self-limiting hydrodynamic transport of CRs by streaming and diffusion. Recently, this picture obtained observational support from angular associations of compact radio and infra-red objects with non-thermal filaments that both share similar latitude distributions, suggesting that they both co-exist spatially \citep{yusef-zadeh_population_2022}.\\
\indent
The two-moment method of CR transport cures the numerical problems of the streaming term and provides a more faithful representation of CR transport by including a spatially and temporally varying diffusion coefficient. However, it comes with new challenges. In addition to Alfv\'en characteristics, the system of equations \eqref{eq:ecr} to \eqref{eq:ew} also contains light-like characteristics with eigenvalues $\pm c/\sqrt{3}$ (which is not equal to $c$ because of the truncation of the expansion at linear order). Resolving time steps associated with this velocity would not only make the solutions numerically very expensive, but also increase the numerical diffusivity at fixed numerical resolution \citep{thomas_finite_2021}, suggesting the use of a reduced speed of light. This has only a minor impact on the transport of CR momentum and energy because there are no large mass fluxes associated with these light-like characteristics, unlike in the case of radiation transport, where this velocity impacts the propagation speed of ionization fronts. Most importantly, if the magnetic dynamo is not fully resolved and the magnetic field does not saturate at its true value, the resulting CR transport would be too slow and bias the solution. In addition, if a plasma instability is not modelled in the macroscopic fluid model or if the adopted damping rates are too strong/weak (perhaps because of an incomplete modelling of the multi-phase nature of the ISM), the CR transport speed would also be wrong and may not supply the correct amount of feedback. These considerations make a strong case for an increased effort in micro- and meso-scale studies of CR transport.

\subsubsection{Cosmic ray energy vs.\ entropy methods}

The one- and two-moment methods of CR hydrodynamics both evolve the CR energy. However, this is not a conserved quantity because only the total energy composed of kinetic, magnetic, thermal and CR energies is conserved. As a result,
the CR energy equation~\eqref{eq:ecr1} contains an adiabatic source term that establishes a coupling between CRs and the thermal gas. This problem is also present in the two-moment approach of CR hydrodynamics because the one-moment equation emerges in the limit of a steady CR flux density. This can cause problems for different discretizations of this adiabatic term at shocks, where kinetic energy is dissipated into thermal energy while the CR energy should only be adiabatically compressed in the picture of grey CR hydrodynamics. In practice, because the energy formulation does not ensure the conservation of CR entropy, there could be artificial numerical CR entropy generated that would render the solution inaccurate, thereby defining the ``non-uniqueness problem'' of the two-fluid equations. While \citet{gupta_numerical_2021} find that the formulation where the adiabatic source term of equation~\eqref{eq:ecr1} is integrated over the cell volume and Gauss' theorem is applied does minimize the problem, \citet{semenov_entropy-conserving_2022} instead argue in favor of the entropy formulation of CR hydrodynamics. \\
\indent
However, there are several points to consider as pointed out by \citet{weber_comparing_2023}: (i) in the limit of low numerical resolution (which is the typical case for galaxy-scale simulations), both schemes perform well and are largely unaffected by the shock Mach number; (ii) however, the absolute truncation error is substantially larger for fixed-grid Eulerian methods because of the increased numerical diffusion in comparison to moving mesh formulations of MHD as, e.g., employed in the one-moment CR hydrodynamics formulation \citep{pfrommer_simulating_tmp_2017,pakmor_semi-implicit_2016} in the AREPO code \citep{springel_e_2010,pakmor_improving_2016}, and (iii) a true collisionless shock can accelerate CRs because of self-regulating plasma kinetic processes, which cause CRs to stream into the upstream, to generate magnetic turbulence, which ensures CR scattering, acceleration and a non-linearly modified shock structure as discussed in Section~\ref{sec:acceleration_escape}. Hence, only accounting for adiabatic compression of CRs at a shock is an inaccurate academic representation of the underlying plasma physics and must be accompanied by a subgrid model that accounts for CR acceleration \citep{pfrommer_simulating_tmp_2017}. Moreover, when accounting for CR acceleration in low-resolution simulations, the entropy formulation shows artificial density oscillations in the post-shock regime, which are sourced by numerical noise as a result of converting injected  CR energy to entropy. This causes the shock to propagate with a much faster velocity in the entropy formulation in comparison to the CR energy formulation  \citep{weber_comparing_2023}.

\subsubsection{Streaming instability, wave damping mechanisms and cosmic ray self-confinement}
\label{sec:streaming}

\paragraph{Setting the stage.} CR-driven instabilities play an essential role in CR propagation in galaxies and galaxy clusters \citep{kulsrud_effect_1969,shalaby_new_2021}. On the one hand, these instabilities amplify magnetic field fluctuations at the scale of CR gyro radii, which is explained in great detail in \citet{kulsrud_plasma_2005} and \citet{thomas_timon_hydrodynamik_2022}. On the other hand, collisionless wave damping processes reduce the wave power. This modulates CR-wave scattering and tightly couples the intrinsically collisionless CR population to the thermal plasma to enable dynamical feedback on macroscopic scales. Among the various (collisionless) wave damping mechanisms, there are linear and non-linear Landau damping, turbulent damping, and ion--neutral damping, which are explained in the following paragraphs. 

{\em Landau damping} describes the energy exchange between an electromagnetic wave with phase velocity $\varv_\rmn{ph}$ and charged particles with velocity along the mean magnetic field close to $\varv_\rmn{ph}$. Provided the particle velocities are  slightly less than $\varv_\rmn{ph}$, the electric field of the wave will accelerate the particles to move at $\varv_\rmn{ph}$, while particles with velocities slightly greater than $\varv_\rmn{ph}$ experience a decelerating Lorentz force so that they lose energy to the wave. As a result, particles tend to synchronize with the wave. {\em Non-linear Landau damping} occurs because of the interaction of two circularly polarized Alfv\'en waves with similar wave numbers $k_1$ and $k_2$ that propagate in the same direction. The interaction generates a beat wave \citep{lee_damping_1973,volk_characteristics_1981,miller_magnetohydrodynamic_1991,kulsrud_plasma_2005,wiener_cosmic_2013}, which propagates at the group velocity
\begin{align}
	\varv_{\rm beat} = \frac{\omega_1 - \omega_2}{k_1 - k_2}.
\end{align}
The magnetic mirror force associated with this beat wave accelerates thermal particles travelling at similar velocities, implying a net extraction of wave energy by the particles and leading to efficient wave damping. Physically, the interaction between the two waves occurs via their beat wave at the Landau resonance with thermal particles, $\varv_\rmn{th}\approx \varv_\rmn{beat}$.

Magneto-hydrodynamic Alfv\'enic turbulence is anisotropic on spatial scales much smaller than the turbulent injection scale \citep{goldreich_toward_1995} so that the elongated ``eddies'' are aligned with the mean magnetic field. {\em Turbulent damping} is not a classical wave damping process but occurs because of the shearing of two counter-propaga{\-}ting Alfv\'en wave packets, which causes field-line wandering. While propagating along the perturbed field lines of the colliding partner, one wave packet undergoes transverse distortion with respect to the average magnetic field on a time comparable to the eddy turnover time \citep{lithwick_compressible_2001}. This causes cascading of energy to higher wave numbers $k_\parallel$, which decreases the wave energy at the resonant scale and makes CR transport more diffusive \citep{farmer_wave_2004,lazarian_damping_2016,lazarian_damping_2022}.

{\em The ion--neutral damping} of Alfvén waves occurs as a result of the frictional forces between ions and neutrals within a partially ionized medium \citep[see Appendix C of][]{kulsrud_effect_1969}. Ion--neutral collisions equilibrate the temperature of these two species to approach $\varv_\rmn{i}^2/\varv_\rmn{n}^2=m_\rmn{n}/m_\rmn{i}$, where neutrals refer to hydrogen and helium \citep[for damping rates of this three-component fluid, see][]{soler_role_2016}. Additionally, ions are accelerated by the Lorentz force exerted by the Alfv\'en waves, which are self-generated by streaming CRs. However, this force is opposed by friction between neutral and ionized particle species, which damps the waves. In summary, the energy loss by the waves is converted to thermal energy of ions and neutrals.

Furthermore, there are other processes that quench the CR streaming instability via CR trapping by magnetic bottles \citep{holcomb_growth_2019}, as well as pressure anisotropy \citep{zweibel_role_2020} or streaming bottlenecks as CRs are propagating in a multi-phase plasma with warm, dense clouds embedded in a hot phase in (approximate) pressure equilibrium: provided the magnetic field does not substantially differ between these phases (which can be obtained via thermal instability and collapse along the magnetic field), those cool clouds are characterized by a decreasing Alfv\'en velocity. As CRs propagate into the dense cloud, they are decelerated by efficient scattering to adjust to the smaller Alfv\'en speed. As a consequence, a reservoir of stationary CRs accumulates ahead of the cloud and a pressure gradient across the cloud, which in turn gets accelerated as a result of this CR pressure gradient \citep{wiener_interaction_2017,wiener_cosmic_2019,thomas_finite_2021}. In all these processes, inhomogeneities of the magnetic field and/or the density of the background plasma cause additional confinement of the CRs, which modifies their transport speed and hence their momentum and energy transfer to the thermal plasma.

\paragraph{CR scattering and transport in the kinetic picture.} The PIC technique allows for the computation of the nonlinear development of the gyroresonant streaming instability. It also enables the study of the concurrent evolution of electron and ion distributions along with the self-generated wave spectra; albeit only in one-dimensional setups due to the numerical complexity. \citet{holcomb_growth_2019} perform such simulations of the CR streaming instability and show that the initial instability growth of Alfv\'en waves agrees with the predictions from linear physics. However, the behavior of the instability during the non-linear saturation stage differs depending on the degree of anisotropy of the initial CR distribution: CR distributions that are highly anisotropic cannot efficiently isotropize in the Alfv\'en frame due to the reduced generation of left-handed resonant modes. Most importantly, because the CR gyroradius is required to be sufficiently small so that the resonant modes are well resolved within the computational domain, numerical limitations enforce the choice of a rather high value of the Alfv\'en speed of $\varv_\rmn{a} = 0.1 c$. This causes the streaming instability to saturate via particle trapping in magnetic bottles rather than via turbulent or non-linear Landau damping, which should be the dominating damping mechanisms in the ionized interstellar plasma, where $\varv_\rmn{a} \sim 10^{-4} c$. Nevertheless, streaming CRs efficiently couple momentum and energy to the background plasma, causing the emergence of bulk flows and confirming the microphysical basis of CR-driven winds. If the CRs have a non-zero pitch angle, the nature of the dominant instability changes and instead CRs excite background ion--cyclotron modes in the frame that is comoving with the CRs \citep{shalaby_new_2021}. The associated growth rate is typically much larger in comparison to that of the resonant streaming instability at the ion gyroscale. Because this new instability grows waves on intermediate scales between the gyroradii of CR ions and electrons, lower-energy (resonant) CRs should get efficiently scattered and more strongly coupled to the background plasma (see Section~\ref{sec:plasma_instabilities}).  

\paragraph{Fluid-PIC modeling of CR scattering and transport.}\label{sec:fluid-PIC_modeling_of_CR_transport} In order to access inhomogeneities on larger scales, simulate multi-dimensional effects, or approach realistic ISM parameters (with a large scale separation of Alfv\'en and light speeds and a CR-to-thermal background density ratio of $\sim10^{-9}$), the background needs to be treated as a fluid, of which the simplest description is provided by MHD models while the CR component is modeled in the kinetic picture with the PIC method \citep{bai_magnetohydrodynamic-particle--cell_2015}. MHD-PIC simulations of the CR streaming instability show that the CR distribution can be fully isotropized in the wave frame as a result of non-linear wave--particle interactions \citep[rather than mirror reflections,][]{lebiga_kinetic-MHD_2018,bai_magnetohydrodynamic_2019}. In order to reduce the Poisson noise inherent to the PIC method, \citet{bai_magnetohydrodynamic_2019} employ the $\delta f$ method, where the CR distribution is split into an (analytically known) isotropic part $f_0$ and the difference from the full distribution function $\delta f = f-f_0$, which is evolved using individual particles as Lagrangian PIC particles.

Adopting ion--neutral drag to damp Alfv\'en waves in a portion of the simulation domain, the simulations of \citet{bambic_mhd-pic_2021} show the emergence of spatial CR gradients across the fully ionized region, which is directed opposite to the CR flux. This supports predictions of CR hydrodynamics, in which the combination of a CR energy density gradient and time-dependent energy flux balances wave--particle scattering throughout the domain. As the ion--neutral damping rate increases, Alfv\'en wave are efficiently damped and CRs are not any more efficiently isotropized \citep{plotnikov_influence_2021}. A systematic MHD-PIC study of CR scattering rates in steady state where the CR streaming instability balances ion--neutral damping yields momentum scalings consistent with quasi-linear theory, but with a reduced normalization; thus offering the promise to calibrate CR hydrodynamic models with kinetic simulations \citep{bai_toward_2022}. MHD-PIC simulations of charged dust and CRs embedded in magnetized gas show the growth of resonant drag instabilities, that cause the formation of charged dust concentrations \citep{ji_numerical_2022}. Those excite Alfv\'en waves that efficiently scatter the initially perfectly streaming CRs, which eventually become fully isotropic and are decelerated to drift at the Alfv\'en speed. The associated momentum transport to the background plasma may be responsible for outflows in the dusty CGM around quasars or superluminous galaxies. 

All these models use the MHD to describe the background plasma, which precludes Landau damping into separate electron and ion fluids and integrates out the electron scale. Moreover, the MHD limit operates on scales larger than the ion skin depth, $kd_\rmn{i}\ll1$, which also makes it impossible to resolve the faster growing inter{\-}mediate-scale instability (see \citealt{shalaby_deciphering_2023} and Fig.~\ref{fig:CR_driven_instabilities}) and thus precludes studying the complete plasma system. A more promising approach to include these physical processes is to model the background with separate fluids for the electron, ion and neutral components (which are equipped with Landau closures for the electron and ion populations). Those fluids are then coupled to the CR population that is represented by PIC particles via Maxwell's equations. Simulations of one spatial dimension and three velocity space dimensions with this approach show that the CR streaming instability indeed saturates via processes that remove wave energy from the resonant scale, namely non-linear Landau damping as well as (inverse) cascading of wave energy towards larger scales in the regime of small Alfv\'en velocities $\varv_\rmn{a}\lesssim 10^{-2}c$ \citep{lemmerz_coupling_2023}.

\subsubsection{Cosmic ray scattering on MHD turbulence -- external confinement by turbulence} 
\label{sec:CR_scattering_turbulence}

\paragraph{CR interactions with Alfv\'en waves and fast and slow magnetosonic waves.} 
Provided MHD turbulence is injected on scales much larger than the CR gyro radii, Alfv\'en and slow modes are less efficient in pitch-angle scattering particles \citep{chandran_convective_2006,yan_cosmic-ray_2004,maiti_cosmic-ray_2022,xu_resonance-broadened_2018}. This is a consequence of the critical balance condition in the inertial-range between linear wave periods and non-linear turnover timescales in MHD turbulence \citep{goldreich_toward_1995}, which leads to elongated ``eddies'' along the magnetic field on small spatial scales with a wave number scaling $k_\perp \propto k_\parallel^{3/2}$. While
the Alfv\'enic cascade perpendicular to the mean magnetic field exhibits Kolmogorov scaling with $E(k_\perp)\dd k_\perp\propto k_\perp^{-5/3}\dd k_\perp$, the parallel scaling is significantly steeper,
\begin{align}
E(k_\parallel) \dd k_\parallel
= E\left[k_\perp(k_\parallel)\right]\frac{\dd k_\perp}{\dd k_\parallel}\dd k_\parallel \propto k_\parallel^{-2}\dd k_\parallel.
\end{align}
Because CRs resonate with the parallel wave vectors (see equation~\ref{eq:resonance} and Fig.~\ref{fig:Lorentz}), the steep spectrum leaves little wave energy to scatter CRs on the resonant scale. In configuration space, this result can be intuitively understood because of the alignment of the elongated eddies along the direction of the mean magnetic field so that the gyroradius of a CR encloses numerous eddies. These incoherently aligned neighboring eddies exert incoherent Lorentz forces on the CR and cause it to random walk during a gyro orbit. This phenomenon attenuates and broadens the gyro resonance, resulting in a decrease in the efficiency of CR scattering. Provided the compressible fast-mode cascade is isotropic and independent of the strength of the large-scale field with respect to the turbulence \citep{cho_compressible_2003,makwana_properties_2020}, and provided it has a flat spectral slope similar to the Iroshnikov-Kraichnan \citep{iroshnikov_turbulence_1963,kraichnan_inertial-range_1965} cascade ${\propto~}k^{3/2}$ \citep{zakharov_spectrum_1970,cho_compressible_2003,makwana_properties_2020}, fast modes may dominate CR propagation for the typical interstellar conditions \citep{yan_cosmic-ray_2004,maiti_cosmic-ray_2022}. Nonetheless, fast modes experience significant kinetic damping on collisionless scales \citep{klein_using_2012,told_comparative_2016}. The precise details of their cascade and spectral slope remain uncertain.

CR interactions with compressible modes can also accelerate these particles via a second-order Fermi process. In the scenario where CRs are only advected with the gas, any adiabatic energy gain through compression will be completely lost during the process of rarefaction. Additionally accounting for diffusion changes this picture: in a single compression event, the CRs adiabatically gain energy as they develop a peaked distribution while diffusing outwards conserves CR energy. Thus, there is a net CR energy gain in a compression event while there is a net energy loss in a rarefaction event. Because interactions with compressible waves are more probable than with expanding waves, the gain in mean CR energy is second order in the wave velocity divided by the light speed \citep{ptuskin_cosmic-ray_1988}. This acceleration efficiency depends on the balance between advective and 
diffusive timescales, $\kappa/(c_\rmn{s}L)$, where $c_\rmn{s}$ is the sound speed and $L$ is the size of the compressible perturbation. The net CR energy gain is reduced if we account for CR streaming in addition to CR diffusion because (i) CRs drain energy from gas motions at a reduced rate that is modified by the factor $1-\varv_\rmn{a}/c_\rmn{s}$ and (ii) the excitation of the streaming instability increases the wave energy at the extend of CR energy \citep[which is less important because the CR pressure gradients are typically misaligned with the magnetic field;][]{bustard_turbulent_2022}. This removal of compressible wave energy leads to a steepening of the compressive turbulent power spectrum when the time required for wave damping ($t_\rmn{damp}\sim \rho\varv^2/\dot{\eps}_\rmn{cr} \propto \eps_\rmn{cr}^{-1}$) becomes similar to the timescale of the turbulent cascade process \citep{bustard_cosmic_2023}.

Waves in strong MHD turbulence are not long-lived but have a decay time that is comparable to their eddy turn-over time. Hence wave--particle interactions are not efficiently mediated through linear resonances, but are instead mediated through non-linearly broadened resonances \citep{yan_cosmic-ray_2008, lynn_resonance_2012}. This is particularly important for particles with perpendicular pitch angles with respect to the local magnetic field orientation since those particles would otherwise find no scattering modes that meet the resonance condition on the correspondingly tiny length scales because those modes are subject to efficient damping. 

\paragraph{MHD and thermal effects on CR transport.}  Using a characteristic ISM volume that is initially in a thermally unstable state, \citet{commercon_cosmic-ray_2019} study CR propagation within the turbulent and magnetised interstellar plasma. They identify a clear transition in the ISM dynamics where the thermal instability is suppressed for CR diffusion coefficients below a critical value of $\kappa_\rmn{crit}\sim 10^{24}$--$10^{25} \rmn{cm}^2 \rmn{s}^{-1}$ or in regions where the CR pressure is at least ten times larger than the thermal pressure. This is because of efficient trapping of CRs in these regions and because of the substantially larger cooling times of CR ions with relativistic energies (see Section~\ref{sec:cooling_times}) compared to the quickly cooling thermal plasma \citep{jubelgas_cosmic_2008}. The transport of streaming CRs depends critically on the level of ionic Alfv\'en fluctuations. \citet{beattie_ion_2022} study compressible MHD turbulence simulations and find that for sub-Alfv\'enic turbulence, the probability density function of the ionic Alfv\'en velocity only depends on the density fluctuations that result from shocks forming parallel to the magnetic field. By contrast, for super-Alfv\'enic turbulence, the correlations between magnetic and density fluctuations are more complex. \citet{sampson_turbulent_2023} use a large ensemble of MHD turbulence simulations to quantify how plasma properties affect the transport of streaming CRs. They find that the macroscopic CR transport can be described by a combination of streaming along the mean field and superdiffusion along and across it. The Alfv\'en Mach number $\mathcal{M}_\rmn{a}$ (that characterizes the strength of the large-scale field with respect to the turbulence) sets the anisotropy between parallel and perpendicular diffusion and the ionization fraction modulates the magnitude of the diffusion coefficient. CR transport does not depend on the compressibility except in the sub-Alfv\'enic ($\mathcal{M}_\rmn{a}\lesssim 0.5$) regime.

\subsection{Radiative and non-radiative cosmic ray processes and their cooling times}
\label{sec:cooling_times}

\subsubsection{Overview}

\begin{figure}[tbp]
\begin{center}
\includegraphics[width=0.8\textwidth]{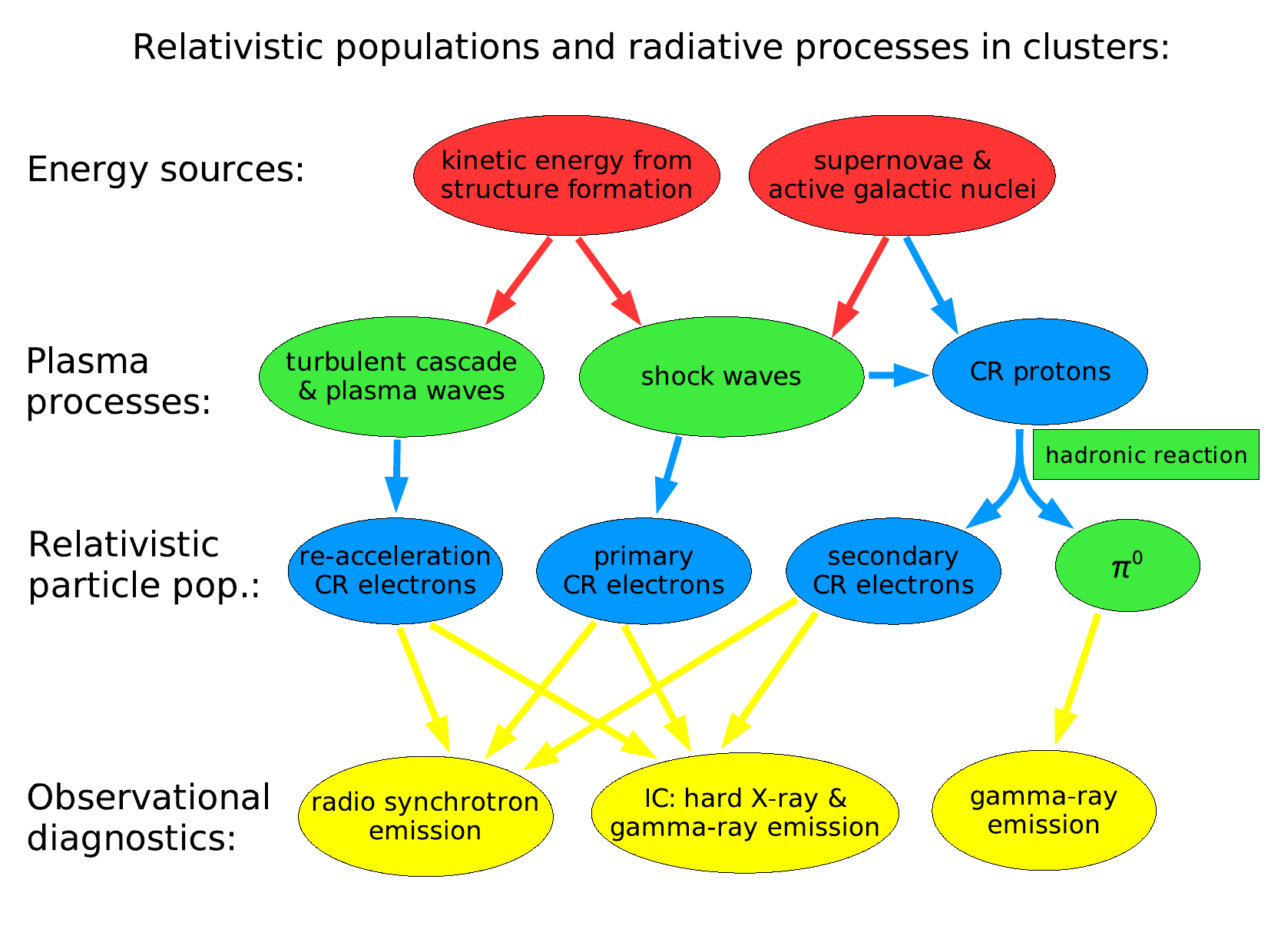}
\end{center}
\caption{Schematic overview of relativistic particle populations and  non-thermal radiative processes in galaxies and galaxy clusters. Cosmologically growing galaxies and groups accrete matter and merge with other halos to assemble larger structures, thus releasing gravitational binding energy in form of kinetic energy. Other sources of kinetic energy injection are SNe and AGNs as the most important non-gravitational energy sources (in red). These energy sources give rise to plasma processes at shocks and interactions with unstable electromagnetic plasma modes (in green), which accelerate relativistic particle populations, so-called CRs (in blue). Non-thermal observables (in yellow) link these CR populations to the underlying physical acceleration process. However, there is a degeneracy because the radio synchrotron and inverse Compton (IC) radiation could be emitted by any of the primary, secondary, and re-accelerated CR electron populations. This degeneracy can be (partially) broken by observing the characteristic spectral pion-decay feature in the $\gamma$-ray spectrum that is associated with hadronic CR interactions with gas protons. This characteristic serves as a distinctive indicator of the presence of a population of CR protons and has been observed in SNRs and in the diffuse emission in the Milky Way. The decay of charged pions that are also produced in hadronic CR interactions yields neutrinos at a very low flux (not shown). Image from \citet{pfrommer_simulating_2008}; reproduced with permission from MNRAS. }
\label{fig:radiative_processes}
\end{figure}

Figure~\ref{fig:radiative_processes} provides an overview of the various relativistic particle populations and radiative processes in galaxies and galaxy clusters. Shocks driven by cosmologically growing structures, AGN jets, and galactic winds can directly accelerate {\em primary CR electrons and ions}. The hadronic interaction between CR protons and protons of the surrounding gas results in the production of \textit{secondary relativistic electrons and positrons} (see the decay chain in equation~\ref{eq:hadronic}). These two CR electron populations quickly cool at high energies via synchrotron emission in the ubiquitous magnetic fields in galaxies and clusters and by means of inverse Compton (IC) interactions with the radiation fields provided by the CMB or by stellar light to settle down at Lorentz factors $\gamma_\rmn{e}=E_\rmn{e}/(m_\rmn{e} c^2)\sim100$--$300$. This makes them invisible in our accessible observational windows in the radio and at $\gamma$ rays. Continuous in-situ re-acceleration of these CR electrons by means of interactions with turbulent MHD waves leads to the emergence of a distinct population of \textit{re-accelerated relativistic electrons}.

\begin{figure}[tbp]
\begin{center}
\includegraphics[width=0.95\textwidth]{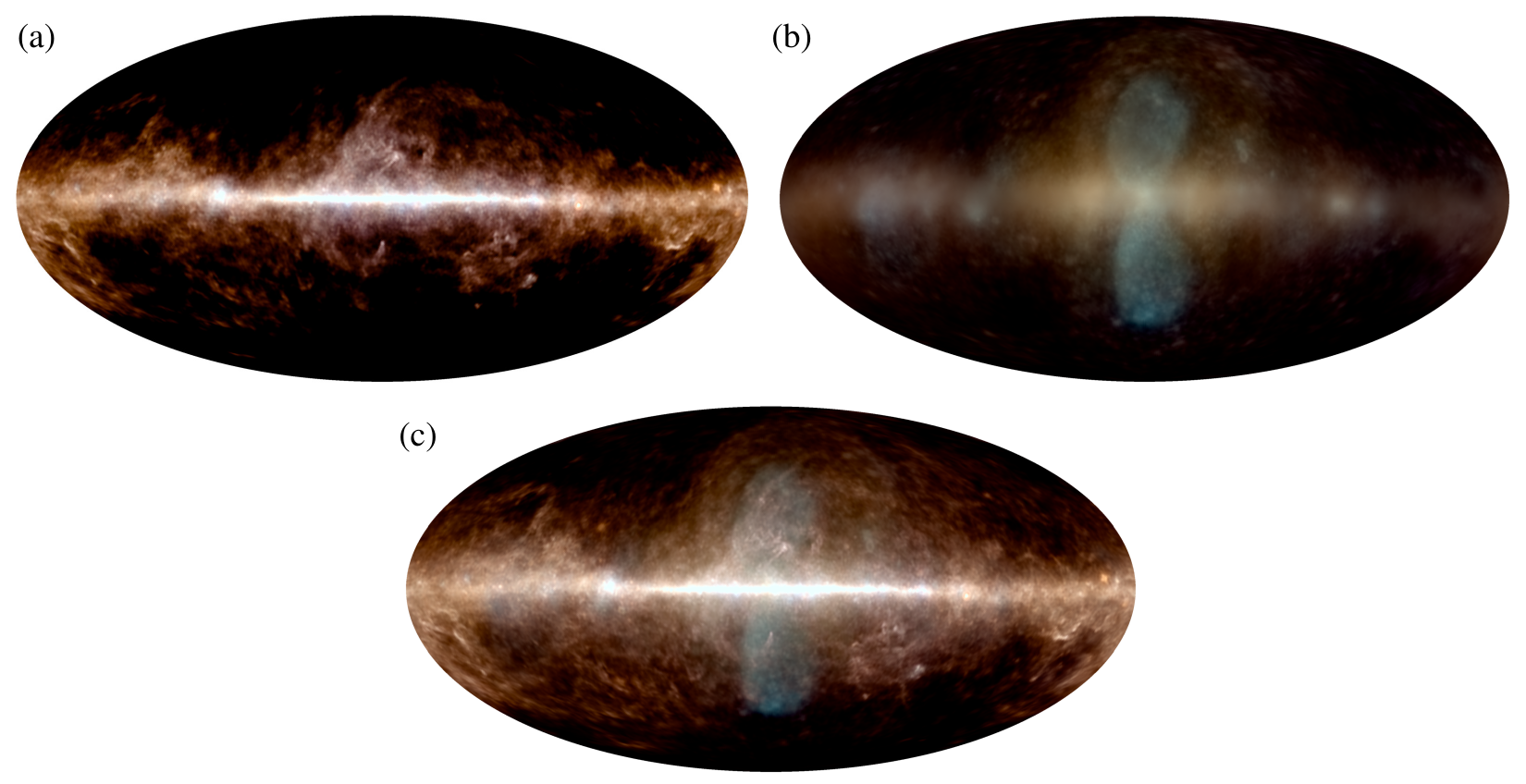}
\end{center}
\caption{Decomposition of the diffuse $\gamma$-ray sky observed by the \textit{Fermi} $\gamma$-ray space telescope in a Mollweide projection. This physics-informed model reconstructs one point source component (not shown) and two hadronic pion decay components, one of which uses the \textit{Planck} dust map as a modifiable template and probes CR interactions with the cold ISM phase that is narrowly distributed around the Galactic midplane (panel a) while the other one traces the more extended warm phase of the ISM (with a soft spectral index, visualized with an orange color in panel b). The third diffuse component exhibits a harder spectral index, characteristic of the leptonic IC component (denoted by blue colors in panel b), and also includes the \textit{Fermi Bubbles}. The sum of all three diffuse components is shown in panel c. Image from \citet{platz_multi-component_2022}; reproduced with permission from A\&A. }
\label{fig:Fermi_sky}
\end{figure}

All these three CR electron populations contribute to the observed radio synchrotron emission (which are described in detail below) and should Compton upscatter CMB and starlight photons into the X-ray and $\gamma$-ray regime. Owing to the uncertainty in the distribution of magnetic field strengths, the predictive ability of radio synchrotron emission alone is limited. In the case of clusters, the observation of the conceptually simpler IC emission is challenging due to the intense radiation background in the soft and hard X-ray range.
In galaxies, IC interactions with the intense starlight photon fields generates a significant level of $\gamma$ rays and rivals the pion-decay $\gamma$-ray emission resulting from hadronic CR-proton interactions. Generally, it is hard to distinguish these leptonic and hadronic emission components. The ``pion bump'', i.e., the spectral decay signature at half the neutral pion's rest mass at around 67.5~MeV in the spectral photon density, which originates from the two photons leaving the interaction site back to back, is a unique electromagnetic signature of hadronic interactions. It has been detected at old SNRs \citep{ackermann_detection_2013} and can be analytically modeled \citep{pfrommer_constraining_2004}. The \textit{Fermi} Gamma-ray Space Telescope (hereafter called \textit{Fermi}) produced a $\gamma$-ray map of the entire sky, which is composed of resolved and unresolved point sources, and the diffuse and compact emission from various source classes in the Milky Way and nearby galaxies. Using information field theory \citep{enslin_information_2009}, the $\gamma$-ray map can be decomposed into several independent emission components while simultaneously employing correlations of the diffuse $\gamma$-ray flux in angular and energy space. Reconstructing one point source component and three diffuse components allows one to separate spectrally and spatially distinct diffuse components that resemble the leptonic IC emission and two pion-decay components that trace the cold and warm phases of the ISM in the Milky Way, respectively (see Fig.~\ref{fig:Fermi_sky}, \citealt{platz_multi-component_2022}, improving upon an earlier analysis by \citealt{selig_denoised_2015}). Alternatively, the accompanying neutrinos provide another unique signal of hadronic CR proton reaction.

To quantify these considerations, in the following we introduce the physics of hadronic and leptonic CR interactions and pay special attention to the cooling timescales of these relativistic particle species. The reader is referred to \citet{rybicki_radiative_1979} for a more detailed exposition of the physics of radiative processes or to \citet{sarazin_energy_1999}, for solutions of the energy spectrum of primary CR electrons under simplified assumptions.

\subsubsection{Cosmic ray ion interactions}
\label{sec:CR_ion_interactions}

\paragraph{Streaming instability losses.} As CRs stream down their gradient, they resonantly excite Alfv\'en waves through the streaming instability \citep{kulsrud_effect_1969}. The associated transfer of energy to Alfv\'en waves is given by the last term of equation~\eqref{eq:ecr_1m} and causes CRs to lose their energy density at a rate
\begin{align}
\label{eq:streaming_losses}
\dot\eps_\rmn{st} =
-\left|\bs{\varv}_\rmn{st}\bcdot\bs{\nabla}P_\CR\right|
\quad\Rightarrow\quad
\tau_\rmn{st} = \frac{\eps_\CR}{\left|\dot\eps_\rmn{st}\right|},
\end{align}
where $\tau_\rmn{st}$ is the CR loss timescale due to CR streaming and $\bs{\varv}_\rmn{st}$ is the CR streaming velocity in the steady-state streaming limit defined in equation~\eqref{eq:v_st}. This definition of the energy density loss rate ensures that the CR cooling goes to zero in the limit of balanced turbulence, i.e., for $\bar{\nu}_+=\bar{\nu}_-$, while it is maximized for imbalanced turbulence, $\bar{\nu}_\pm\gg\bar{\nu}_\mp$, which can be realized in the self-confinement picture for strong CR fluxes. As detailed in Section~\ref{sec:streaming}, the wave energy is dissipated through various collisionless wave damping processes, thereby heating the background plasma.

\paragraph{Hadronic interaction.} \label{hadronic} Of particular relevance for deciphering the CR proton population is the hadronic reaction of a CR proton with a thermal proton: when the momentum of CR protons surpasses the kinematic threshold of approximately $0.78~\mbox{GeV}/c$, the interaction generates pions that subsequently decay, producing secondary electrons, positrons, neutrinos, and gamma rays:\footnote{The kinematic threshold for the hadronic reaction is $2\gamma' m_\rmn{p}=2m_\rmn{p}+m_{\pi^0}$ in the center-of-momentum system. After Lorentz transforming to the laboratory frame, this amounts to a total energy of $\gamma'^2(1+\beta'^2)\,m_\rmn{p}c^2=(2\gamma'^2-1)\,m_\rmn{p}c^2\simeq1.22~\rmn{GeV}$ (where $\gamma'$ denotes the Lorentz factor and $\beta'=\varv'/c$ in the center-of-momentum system), which corresponds to a CR kinetic energy of 280~MeV or a CR momentum of $0.78~\rmn{GeV}/c$.} 
\begin{eqnarray}
\label{eq:hadronic}
  \pi^\pm &\rightarrow& \mu^\pm + \nu_{\mu}/\bar{\nu}_{\mu} \rightarrow
  e^\pm + \nu_{e}/\bar{\nu}_{e} + \nu_{\mu} + \bar{\nu}_{\mu}\nonumber\\
  \pi^0 &\rightarrow& 2 \gamma \,.
\end{eqnarray}
Thus, only CR protons with momentum exceeding this kinematic threshold are observable through the detection of their decay products either directly in form of $\gamma$-ray and neutrino emission\footnote{The physics of this reaction and the spectral decay signatures are at best mildly affected when considering heavier CR nuclei because the nuclear MeV binding energies are small in comparison to the involved CR energies that exceed GeV. Hence, the overall increased emissivity of heavier CR compositions can be parametrized by a moderate nuclear enhancement factor.} or indirectly via radiative processes such as synchrotron and IC emission of secondary electrons and positrons, making them observationally detectable. The cooling timescale due to hadronic processes above the kinematic threshold for pion production is given by
\begin{align}
\label{eq:tau_pp}
\tau_\rmn{pp} =
\frac{1}{0.5\,n_\rmn{n} \varv_\rmn{cr}\sigma_\rmn{pp}}
\approx 6.6\,\left(\frac{n_\rmn{n}}{10^{-2}~\rmn{cm}^{-3}}\right)^{-1}\,\rmn{Gyr}.
\end{align}
Here, $\sigma_\rmn{pp} \approx 32~$mbarn represents the inelastic cross section of protons with an inelasticity of approximately 0.5. Additionally, $n_\rmn{n}=\rho/m_\rmn{p}$ denotes the number density of target nucleons for the hadronic reaction (assuming gas of mass density $\rho$ that is mainly composed of hydrogen and helium) and $\varv_\rmn{cr}\approx c$ is the CR proton velocity that approaches the light speed $c$ for relativistic CR energies.

\begin{figure}[tbp]
\begin{center}
\includegraphics[width=0.49\textwidth]{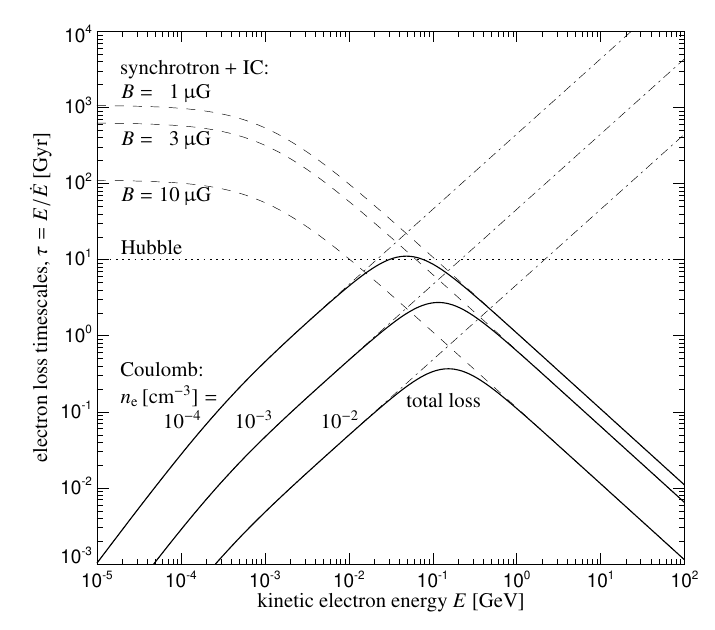}
\includegraphics[width=0.49\textwidth]{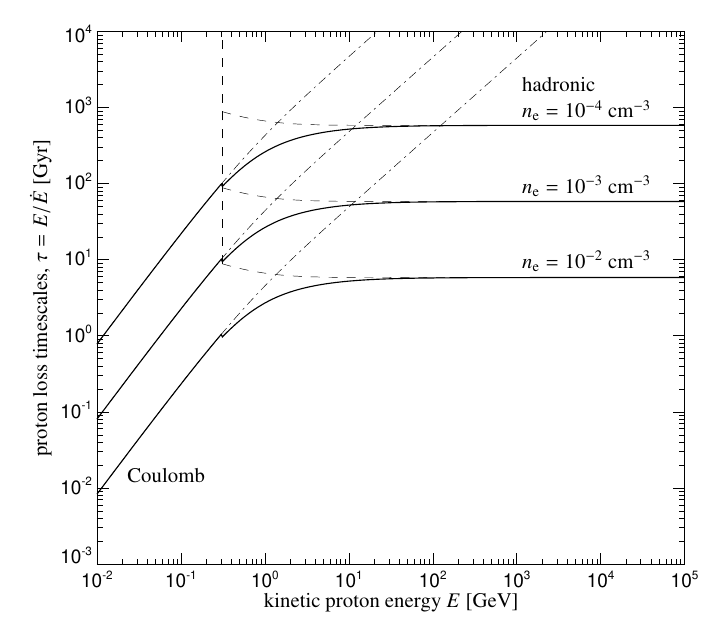}
\end{center}
\caption{Cooling timescales of CRs in various astrophysical plasmas as a function of kinetic energy. \textit{Left:} CR electron cooling times resulting from Coulomb and IC/synchrotron interactions for densities and magnetic field strengths as indicated. \textit{Right:} CR proton cooling times as a result of Coulomb and hadronic interactions for the same parameters as in the left panel. It is evident that CR protons with energies above 10 GeV have a lifespan that is at least 60 times longer than that of CR electrons at any energy. CR electrons can only persist for a Hubble time without re-acceleration in the dilute outer regions of the CGM and the ICM. Electrons emitting radio waves at a frequency of 1.4 GHz possess an energy of approximately 5 GeV in microgauss magnetic fields, resulting in a lifespan of 0.2 Gyr or less. If these electrons are generated through hadronic interactions of CRs, their parent CR protons would have had energies around 80 GeV, so that they can be injected over a considerably longer lifetimes. Image from \citet{enslin_cosmic_2011}; reproduced with permission from A\&A. }
\label{fig:cooling_times}
\end{figure}

\paragraph{Coulomb interactions of CR ions with a thermal plasma.} The Coulomb field of electrons of the background plasma can deflect a CR ion. The resulting momentum and energy transfer from the CR ion to background electrons (i.e., the stopping power) decelerates the ion. The calculation is most easily done in the center-of-momentum reference frame, which nearly coincides with the CR ion rest frame. In this frame (that is denoted by primed quantities), the electron is deflected by an angle $\theta_\rmn{d}'$ and attains a perpendicular momentum change of $\Delta p_\rmn{e}'=m_\rmn{e}^{} \varv_\rmn{e}'\theta_\rmn{d}'$ (in the non-relativistic limit). Because this change of momentum occurs perpendicular to the boost direction from the lab to the CR ion rest frame, we have $\Delta p_\rmn{e}'=\Delta p_\rmn{e}$. Hence in the lab frame, the associated energy gain of the electron corresponds to the energy loss of the ion of 
\begin{align}
    \label{eq:DeltaE_Coulomb}
    \Delta E=\frac{\left(\Delta p_\rmn{e}'\right)^2}{2 m_\rmn{e}}
    =\frac{m_\rmn{e}}{m_\rmn{i}}\,\theta_\rmn{d}'^2 E,
\end{align}
where $E=\frac{1}{2}m_\rmn{i}^{}\varv_\rmn{i}^2$ is the CR ion energy in the lab frame.\footnote{Note that the CR ion velocity in the lab frame equals the negative electron velocity in the CR ion rest frame, $\varv_\rmn{i}=-\varv_\rmn{e}'$.} Note that the impact of ion-ion scattering rate on the energy loss of the incoming fast ion is suppressed because the inertia of the background ions is much larger than that of the background electrons. If the impact parameter of the interaction is less than a critical impact parameter (at which the electron's kinetic energy balances the electrostatic potential energy on average in the CR ion frame),
\begin{align}
  b_0=\frac{2 Z e^2}{m_\rmn{e} \varv_\rmn{i}^2}
  =\frac{2 r_0 c^2}{\varv_\rmn{i}^2},
\end{align}
we observe a (rare) large-angle scattering event. In the above expression $Ze$ and $\varv_\rmn{i}$ are the charge and velocity of the CR ion, $r_0=Ze^2/(m_\rmn{e} c^2)$ denotes the classical electron radius, $m_\rmn{e}$ is the electron mass and $e$ is the elementary charge. Because there are many more electrons at distances larger than $b_0$, small-angle deflections at large impact parameters up to the Debye length (which characterizes the scale at which the charge of a plasma particle is screened) dominate the Coulomb scattering rate by a factor of $2\ln\Lambda$, where $\ln\Lambda\sim35$--$40$ is the Coulomb logarithm. The timescale at which the average squared deflection angle $\langle\theta_\rmn{d}'^2\rangle$ in equation~\eqref{eq:DeltaE_Coulomb} becomes approximately equal to unity corresponds to the deflection time, $\tau_\rmn{d}^\rmn{ei}$. The Coulomb cooling timescale ($\tau_\rmn{Coul,i}$) for a CR ion as it moves through a plasma is determined by dividing the particle's energy by its rate of energy loss. In other words, $\tau_\rmn{Coul,i}$ can be calculated as the deflection timescale divided by the average relative energy transfer to the background plasma:
\begin{align}
  \label{eq:tau_Coul}
  \tau_\rmn{Coul,i}
  = \left.\frac{E}{|\dot{E}|}\right|_\rmn{Coul,i}
  \approx\tau_\rmn{d}^\rmn{ei}\, \frac{m_\rmn{i}}{m_\rmn{e}}
  = \frac{m_\rmn{i}}{m_\rmn{e} n_\rmn{e} \varv_\rmn{i} \sigma_{\rmn{ei}}}
  = \frac{m_\rmn{i}}{m_\rmn{e} n_\rmn{e} \varv_\rmn{i} \pi b_0^2\, 2\ln\Lambda}
  = \frac{m_\rmn{i}^{}\varv_\rmn{i}^3}{8\pi m_\rmn{e} n_\rmn{e} r_0^2 c^4\ln\Lambda},
\end{align}%
where $n_\rmn{e}$ is the electron number density and $\sigma_{\rmn{ei}}$ is the Coulomb cross section. In the second step, we adopted the non-relativistic limit for simplicity. While the hadronic and Coulomb cooling times both scale with the inverse gas density, the strong velocity dependence of $\tau_\rmn{Coul}\propto \varv_\rmn{i}^3$ implies that for proton energies $\lesssim1$~GeV, Coulomb interactions are more effective than the hadronic reaction in removing energy from the CR proton (see right-hand panel of Fig.~\ref{fig:cooling_times}). A more precise calculation for this process by \citet{gould_energy_1972-1} also takes into account quantized plasma oscillations which slightly modifies the Coulomb logarithm.

\paragraph{Ionization interactions of CR ions.} Ionization losses are important both for low energy CRs and for the ISM. Ionization losses for CR protons that move with velocity $\varv_\rmn{p}$ can be obtained from the Bethe-Bloch equation \citep{groom_passage_2000,enslin_cosmic_2007}. The associated ionization timescale in the non-relativistic limit reads
\begin{align}
  \label{eq:tau_ion}
  \tau_\rmn{ion}
  = \left.\frac{E}{|\dot{E}|}\right|_\rmn{ion}
  \approx \frac{m_\rmn{p}^{}\varv_\rmn{p}^3}{\displaystyle8\pi m_\rmn{e} r_0^2 c^4\sum_Z Z n_Z\ln \left(2 m_\rmn{e} \varv_\rmn{p}^2/I_Z\right)},
\end{align}
where $n_Z$ represents the density of atomic species characterized by an electron number $Z$, $I_Z$ is the ionization potential ($I_Z=13.6$ and 24.6~eV for hydrogen and helium, respectively), and we ignore a density correction factor of order unity. The similarity of ionization and Coulomb loss timescales (equation~\ref{eq:tau_Coul}) is not a coincidence but is rooted in the interaction physics that only differs by the ionization process.

\subsubsection{Cosmic ray lepton interactions}
\label{sec:CR_lepton_interactions}

\paragraph{Coulomb interactions of CR electrons with a thermal plasma.} Those interactions can be derived analogously to the case of CR ions, but now we consider the scattering of a CR electron in the Coulomb field of an electron of the background plasma. This increases the energy transfer by a factor of the mass ratio $m_\rmn{i}/m_\rmn{e}$ 
(compared to the energy transfer rate for ion-electron collisions)
because of the identical masses of the scattering partners, thus making this process more efficient than electron-ion scattering while the scattering rates of both processes are nearly identical. Hence, we obtain from equation~\eqref{eq:tau_Coul}:
\begin{align}
  \label{eq:tau_Coul_e}
  \tau_\rmn{Coul,e}
  = \left.\frac{E_\rmn{e}}{|\dot{E}_\rmn{e}|}\right|_\rmn{Coul,e}
  \approx \frac{1}{n_\rmn{e} \varv_\rmn{e} \sigma_{\rmn{e e}}}
  = \frac{1}{n_\rmn{e} \varv_\rmn{e} \pi b_0^2\, 2\ln\Lambda}
  = \frac{\varv_\rmn{e}^3}{8\pi n_\rmn{e} r_0^2 c^4\ln\Lambda},
\end{align}
where $\sigma_{\rmn{ee}}$ is the Coulomb cross section and $b_0=2 r_0 c^2/\varv_\rmn{e}^2$ is the impact parameter where the electron's kinetic energy balances its electrostatic potential energy. Interestingly, the Coulomb cooling timescale of CR electrons also decreases steeply towards low electron energies as for the CR ions (see left-hand panel of Fig.~\ref{fig:cooling_times}). A more precise calculation by \citet{gould_energy_1972}
results in an Coulomb energy loss timescale for CR electrons which slightly differs from that of CR ions because of two primary factors: firstly, the presence of exchange effects (which slightly modifies the Coulomb logarithm), and secondly, the fact that CR electrons and positrons have the potential to relinquish a significant portion of their energy in a single interaction with a plasma electron.

\paragraph{Inverse Compton interactions.} At highly relativistic energies, CR electrons experience synchrotron interactions with the magnetic field and IC interactions with the ambient photon field. In the intergalactic medium and in the bulk of the intracluster plasma, the photon energy density is dominated by the CMB while the energy density of starlight photons (in the infrared-to-ultraviolet regime) exceeds that of the CMB in and around galaxies. In general, the photon energy density is the sum of the CMB and starlight, $\eps_\rmn{ph}=\eps_\rmn{cmb}+\eps_\star$. 

To derive the energy loss rate of electrons due to IC interactions, we transform to the electron rest frame. In this frame, we can assume elastic photon scattering with a relativistic electron of energy $E_\rmn{e}=\gamma_\rmn{e} m_\rmn{e} c^2$ in the Thomson regime, i.e., when the Lorentz-boosted photon energy is much less than the electron rest mass, $\gamma_\rmn{e}\bra E \ket \ll m_\rmn{e}c^2$, where $\bra E \ket$ is the average photon energy before scattering. Hence, after Lorentz boosting the photon into the electron rest system, reversing the normal component of the photon momentum as a result of elastic scattering, and Lorentz de-boosting it into the lab system, we pick up two Lorentz factors and find a net photon energy gain of $\bra E_1\ket=\frac{4}{3}\,\gamma_\rmn{e}^2\,\bra E\ket$, where the factor $4/3$ derives from averaging over an isotropic photon field. This photon energy gain corresponds to the IC energy loss rate of a CR electron of
\begin{equation}
\dot{E}_\rmn{e}
= -\sigma_\rmn{T}c n_\rmn{ph} \bra E_1 \ket 
= -\frac{4}{3} \sigma_\rmn{T}c \varepsilon_\rmn{ph} \gamma_\rmn{e}^2
= -\frac{\sigma_\rmn{T}c}{6\pi}\,  B_\rmn{ph}^2 \gamma_\rmn{e}^2,
\end{equation}
where $\sigma_\rmn{T}=2\pi r_0^2= 2\pi Z^2 e^4/ (m_\rmn{e}^2 c^4)$ is the Thomson cross section and $\eps_\rmn{ph}=\bra E\ket n_\rmn{ph}=B_\rmn{ph}^2/(8\pi)$ is the photon energy density in the laboratory frame with an equivalent magnetic field strength $B_\rmn{ph}$. In general, the Thomson cross section of a charged particle of mass $m$ interacting with a photon scales as $\sigma_\rmn{T}\propto m^{-2}$ so that the IC interaction rate of ions in comparison to that of electrons is suppressed by the square of the electron-to-ion mass ratio.

\paragraph{Synchrotron interactions.} The Lorentz force associated with the magnetic field causes a CR electron or positron to gyrate and hence to emit synchrotron radiation. Formally, this can be described by a scattering process, which obeys the same Feynman diagram as the IC interaction: while the IC process evokes an electron scattering with a real photon, in a synchrotron interaction, the electron borrows a ``virtual photon'' from the magnetic field. Hence, the total energy loss rate of a CR electron at high energies is given by
\begin{equation}
    \dot{E}_\rmn{e}=
    -\frac{\sigma_\rmn{T}c}{6\pi}\, \left(B_\rmn{ph}^2+B^2\right) \gamma_\rmn{e}^2.
    \label{eq:E_sync,IC}
\end{equation}
The first term in equation~\eqref{eq:E_sync,IC}, $\dot{E}_\rmn{e}\propto B_\rmn{ph}^2$, represents the energy loss resulting from IC scattering with photons from the radiation field. The second term $\propto B^2$ represents the energy loss due to synchrotron emission. The magnetic field strength equivalent to the energy density of the CMB is $B_\rmn{cmb}\simeq3.2\,(1+z)^2~\mu\rmn{G}$ at a redshift $z$. Thus, for synchrotron emission to dominate over the IC process, the magnetic field must exceed either $B_\rmn{cmb}$ or $B_\rmn{ph}$ if the energy density of the stellar radiation field exceeds that of the CMB. The cooling time $\tau_\rmn{cool}=E_\rmn{e}/\dot{E}_\rmn{e}$ of a relativistic electron due to synchrotron and IC interactions can be calculated as follows:
\begin{equation}
    \label{eq:t_cool}
    \tau_\rmn{cool}=\frac{E_\rmn{e}}{|\dot{E}_\rmn{e}|}=
    \frac{6\pi m_\rmn{e} c}{\sigma_\rmn{T}\,\left(B_\rmn{ph}^2+B^2\right)\gamma_\rmn{e}}
    \approx200\,\rmn{Myr},
\end{equation}
for $B=1~\mu$G and $\gamma_\rmn{e}=10^4$ and we assume a negligible starlight contribution. A CR electron population that was injected at one epoch and cools for a time $t=\tau_\rmn{cool}$ shows an exponentially suppressed electron spectrum above the energy/Lorentz factor that corresponds to $\tau_\rmn{cool}$. In practice, CR electrons are generated over a finite time interval so that the cooled spectrum probes a range of cooling times and associated spectral energy cutoffs. Hence, if we observed such a CR electron population at time $t\gtrsim \tau_\rmn{cool,\,init}$ (the cooling time of the initially injected CR electron population), we would expect to observe a considerably steepened power-law spectrum in comparison to the injected spectrum. The synchrotron frequency, $\nu_\rmn{syn}$, in the monochromatic approximation \citep{enslin_synchrotron_2002, pfrommer_simulating_2022} is given by
\begin{eqnarray}
  \label{eq:nu_s}
  \nu_\rmn{syn} &=& \frac{3 e B}{2\pi\, m_\rmn{e} c}\,\gamma_\rmn{e}^2 \simeq 
  1 \, \left(\frac{B}{\mu\mbox{G}}\right)\, 
  \left(\frac{\gamma_\rmn{e}}{10^4}\right)^2 \mbox{ GHz}.
\end{eqnarray}
By combining equations~\eqref{eq:t_cool} and \eqref{eq:nu_s} and eliminating the Lorentz factor $\gamma_\rmn{e}$, we can obtain the cooling time of electrons emitting at frequency $\nu_\rmn{syn}$,
\begin{eqnarray}
  \tau_\rmn{cool} =
  \frac{\sqrt{54\pi\, m_\rmn{e} c\, e B \nu_\rmn{syn}^{-1}}}
       {\sigma_\rmn{T}\,(B_\rmn{ph}^2+B^2)}
       \lesssim190\,\left(\frac{\nu_\rmn{syn}}{1.4\,\rmn{GHz}}\right)^{-1/2}\rmn{Myr},
  \label{eq:t_cool,max}
\end{eqnarray}
The cooling time $t_\rmn{cool}$ is then bound from above and -- in the case of negligible starlight contribution -- attains its maximum cooling time at $B=B_\rmn{cmb}/\sqrt{3} \simeq 1.8\,(1+z)^2\mu\rmn{G}$, independent of the magnetic field.
  
Figure~\ref{fig:cooling_times} shows the cooling timescales of CR electrons (left) and CR protons (right) as a function of their kinetic energy. It is evident that CR protons with energies above 10 GeV have a lifespan at least 60 times longer than that of CR electrons at any energy. CR electrons can only persist for a Hubble time without undergoing re-acceleration within the low-density regions of galaxy clusters or in the outer CGM. But in this case, they cool down to Lorentz factors $\gamma_\rmn{e}\sim100$--$300$ with kinetic energies $E_\rmn{e}\sim(50$--$150)$~MeV where they cannot be observed on Earth because the ionospheric plasma cutoff precludes radio waves to propagate through the Earth's atmosphere at frequencies below 1--10~MHz.\footnote{At frequencies below the plasma frequency $\omega_\rmn{e}$ the refractive index becomes imaginary so that any plasma wave is exponentially attenuated and does not propagate through the plasma.} The electrons that emit in the GHz radio range have an energy of approximately 5 GeV in $\mu$G magnetic fields (as indicated by equation~\ref{eq:nu_s}). Consequently, their lifespan is estimated to be 0.2 Gyr or shorter. If they are of hadronic origin, their parent CR protons had energies of about $E_\rmn{p}\approx 16 \bra E_{\rmn{e}^\pm}\ket \approx80$~GeV, which have considerably longer lifetimes.

\subsubsection{Equilibrium electron distribution}
\label{sec:equilibirum}

We now discuss the connection between the radio synchrotron spectrum and the radiating CR electron population. In particular, we have to distinguish between freshly accelerated electrons and an equilibrium distribution of electrons where injection is balanced by radiative losses. CR electrons are either directly accelerated at shocks (driven by structure formation in galaxy clusters or by SNe in galaxies) or injected in hadronic CR proton interactions. This implies a CR electron source function $s_\rmn{e}=C_\rmn{inj}E_\rmn{e}^{-\alpha_\rmn{inj}}$, with a spectral index $\alpha_\rmn{inj}\simeq2.1$--$2.4$. Note that the test particle limit of diffusive shock acceleration yields a spectral index $\alpha_\rmn{inj}=2$ in case of a strong shock. In steady state, the acceleration and injection of CR electrons are balanced by the cooling effects of synchrotron and IC processes:
\begin{align}
  \frac{\partial }{\partial E_\rmn{e}} \left[ \dot{E_\rmn{e}}(E_\rmn{e}) f_\rmn{e} (E_\rmn{e})
    \right] = s_\rmn{e}( E_\rmn{e}),
\end{align}
where the electron energy loss rate, $\dot{E_\rmn{e}}$, is given by equation~\eqref{eq:E_sync,IC}.  For $\dot{E_\rmn{e}}(E_\rmn{e}) < 0$,
the solution to this equation is
\begin{align}
  \label{eq:fe_cooled}
  f_\rmn{e} (E_\rmn{e}) &= \frac{1}{|\dot{E_\rmn{e}}(E_\rmn{e})|} \int_{E_\rmn{e}}^\infty
  \dd E_\rmn{e}'  s_\rmn{e}( E_\rmn{e}') = \frac{C_\rmn{inj}}{(\alpha_\rmn{e}-1)\,|\dot{E_\rmn{e}}(E_\rmn{e})|}\,E_\rmn{e}^{1-\alpha_\rmn{inj}}
  \propto E_\rmn{e}^{-\alpha_\rmn{inj}-1},
\end{align}
assuming that the dominant processes are synchrotron and IC losses in the last step (see equation~\ref{eq:E_sync,IC}). Hence, the electron spectral index steepens by unity in steady state, $\alpha_\rmn{e}=\alpha_\rmn{inj}+1$. CR electrons with a power-law spectrum, $f_\rmn{e}={C_\rmn{e}}E_\rmn{e}^{-\alpha_\rmn{e}}$, radiate synchrotron emission with a power law in frequency,
\begin{align}
  \label{eq:jnu}
  j_\nu \propto C_\rmn{e} B^{\alpha_\nu+1}\nu^{-\alpha_\nu},
\end{align}
where $\alpha_\nu\equiv\dd{\log}j_\nu/\dd\log\nu=(\alpha_\rmn{e}-1)/2$. Observationally, the spectral index is determined by comparing radio surface brightness maps at two different frequencies $\nu_1$ and $\nu_2$,
\begin{align}
  \alpha_{\nu_1}^{\nu_2}\equiv\frac{\log(S_{\nu_2}/S_{\nu_1})}{\log(\nu_2/\nu_1)}.
\end{align}

\begin{figure}[tbp]
\begin{center}
\includegraphics[width=0.49\textwidth]{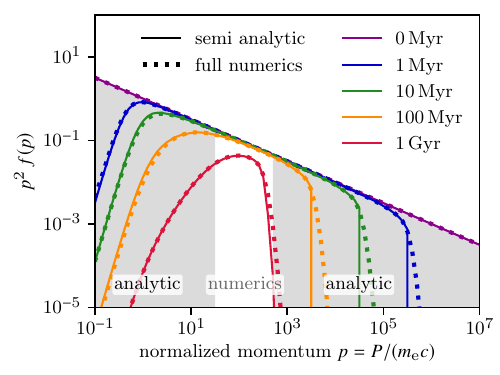}
\includegraphics[width=0.49\textwidth]{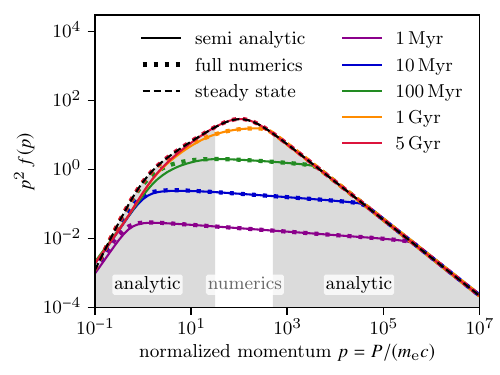}
\end{center}
\caption{Cooling processes shaping the CR electron spectrum. The left-hand panel shows a freely cooling power-law momentum spectrum with spectral index $\alpha_\rmn{e}=2.5$ where Coulomb cooling at low energies and IC/synchrotron cooling at high energies progressively narrow down the spectrum. The right-hand panel illustrates the accumulation of a steady-state spectrum resulting from continuous injection (with $\alpha_\rmn{inj}=2.1$) and cooling. In the cooled regime on the right-hand side, the spectrum steepens so that the spectral index increases by unity ($\alpha_\rmn{cool}=3.1$). The break energy from the unimpeded injection to the cooled spectrum moves to lower energies with time. The solid line represents the semi-analytical solution, while the dotted lines represent the fully numerical solutions to the Fokker-Planck equation of electron transport. The gas density, magnetic field strength and photon energy density are given by ${n_\mathrm{gas} = 10^{-3}\,\mathrm{cm}^{-3}}$,  $B = 5~\mu\rmn{G}$ and $\varepsilon_\rmn{ph} = 6\,\varepsilon_\mathrm{cmb}$, respectively. Image from \citet{winner_evolution_2019}; reproduced with permission from MNRAS. }
\label{fig:cooled_spectra}
\end{figure}

Hence, for a steady-state CR electron population that has been accelerated by a strong shock, we expect $\alpha_\rmn{e}=\alpha_\rmn{inj}+1=3$ and $\alpha_\nu=(\alpha_\rmn{e}-1)/2=1$ in the test particle limit. If we instead were to resolve the freshly accelerated CR electron population directly at the shock, we would obtain $\alpha_\nu=(\alpha_\rmn{inj}-1)/2=0.5$ at the shock provided we can neglect radiative cooling losses. Observed spectral indices at SNR shocks of $\alpha_\nu\simeq0.65$ imply $\alpha_\rmn{inj}\simeq2.3$, which may require a revision of the theory of diffusive shock acceleration (see discussion in Section~\ref{sec:acceleration}). In steady state and for a negligible starlight contribution, the synchrotron emissivity can be obtained by combining equations~\eqref{eq:fe_cooled} and \eqref{eq:jnu}, yielding
\begin{align}
  \label{eq:jnu_cooled}
  j_\nu \propto \frac{C_\rmn{inj}B^{\alpha_\nu+1}\nu^{-\alpha_\nu}}{B_\rmn{cmb}^2+B^2},
\end{align}
which is nearly independent of $B$ in the synchrotron cooling regime, $B>B_\rmn{cmb} \simeq 3.2\, (1+z)^2\,\mu$G in stready state. While IC emission only depends on the amount of CR electrons and the photon energy density, the synchrotron emission depends on $B$ in the IC cooling regime, $B<B_\rmn{cmb}$. Figure~\ref{fig:cooled_spectra} shows the impact of the various leptonic cooling processes discussed in this section on the CR electron distribution while neglecting any re-acceleration processes. On the left-hand side, we show a freely cooling electron spectrum that develops a cutoff at low and high energies as a result of Coulomb and IC/synchrotron cooling, respectively. After 1~Gyr, only CR electrons with Lorentz factors $\gamma_\rmn{e}\sim100$--$300$ survive, as expected from our discussion of Fig.~\ref{fig:cooling_times}. The right-hand side of Fig.~\ref{fig:cooled_spectra} shows the build up of a steady-state spectrum where continuous injection (with spectral index 2.1) balances cooling due to Coulomb and IC/synchrotron interactions. At early times $t\lesssim1$~Gyr, the transition from the acceleration/injection spectrum to the cooled spectrum with an index of $\alpha_\rmn{cool}=\alpha_\rmn{inj}+1=3.1$ is clearly visible. With time, this break moves to lower energies and nearly vanishes for $t\gtrsim2$~Gyr.

\subsection{Cosmic ray spectral transport}
\label{sec:spectral_transport}

\subsubsection{Momentum-dependence of spatial cosmic ray transport}\label{self-confinement}

\paragraph{Self-confinement of CRs.} Following our discussion of the various radiative and non-radiative cooling processes that shape the CR electron and ion spectra, we now consider the momentum dependence of spatial CR transport. Provided the self-excited Alfv\'en waves are only weakly damped, CRs are efficiently scattered and stream close to the local Alfv\'en velocity. In the opposite regime when waves are efficiently damped, CRs are no longer confined to the frame of the local Alfv\'en waves and diffuse with an effective velocity that exceeds the Alfv\'en velocity, $\varv_\rmn{di}\gg\varv_\rmn{a}$. The value of the CR drift speed $\varv_\rmn{d}$ is determined by the equilibrium between the wave damping rate and the growth rate of the CR streaming instability, $\Gamma_\rmn{gyro}\propto n_\CR(>p_\rmn{min})/n_\rmn{i} \times(\varv_\rmn{d}/\varv_\rmn{a} - 1)$, as shown in equation~\eqref{eq:Gamma_gyro}. The CR streaming instability growth rate indirectly depends on the CR momentum because it depends on the number density of CRs, $n_\CR(>p_\rmn{min})$, with a resonant minimum momentum $p_\rmn{min}$. Because the CR spectra are typically soft, there are much fewer high-momentum particles, which decreases the wave growth rate towards high CR momenta. The damping rates scale differently with CR particle momentum \citep{wiener_cosmic_2013}, depending whether we consider a linear or non-linear wave damping process \citep{kempski_reconciling_2022} and on the specifics of the wave damping process: e.g., ion--neutral damping is essentially independent of wavelength while damping by background turbulence scales as $\Gamma\propto k^{1/2}$ \citep{farmer_wave_2004}. Provided the prevailing damping process depends weaker on $n_\CR$ than the wave growth rate, the equilibrium of wave growth and damping evolves towards a dominant CR diffusion regime at high CR momenta, which implies that high-momentum CRs drift faster than their lower momentum analogues. Hence, this transport physics introduces an energy dependence of the self-generated CR diffusion coefficient, $\kappa\propto E^\alpha$ with $\alpha>0$. Depending on which process dominates wave damping, the correction term to the CR drift velocity is neither truly diffusive- nor streaming-like in nature \citep{wiener_cosmic_2013}. 

 \paragraph{External confinement of CRs.} In this case, CR (ion or electron) scattering is dominated by externally driven turbulence, and the specifics of the scattering modes (Alfv\'enic vs.\ fast mode turbulence) and their spectral slopes (see Section~\ref{sec:CR_scattering_turbulence}) may imprint an energy dependence on the CR diffusion coefficient as we briefly sketch here. A CR scattering with a parallel propagating Alfv\'en wave obeys the resonance condition $r_\rmn{g}=2\pi/k_\parallel$ (see equation~\eqref{eq:resonance} or Fig.~\ref{fig:Lorentz}), where the gyroradius of a CR particle of mass $m$ and charge $q$ in a magnetic field of strength $B$ is given by
\begin{align}
\label{eq:rg}
r_\rmn{g} = \frac{p_\perp c}{q B} = \frac{\beta_\perp c}{\Omega} 
\to\frac{E}{qB}.
\end{align}
Here, $\Omega=qB/(\gamma mc)$ denotes the relativistic gyro frequency, $p_\perp$ and $\beta_\perp$ are the perpendicular momentum and relativistic $\beta$ factor, respectively, and the limit in the last step applies to the relativistic regime. The resonant interaction of a CR with a parallel propagating Alfv\'en wave with energy density $\eps_{\rmn{w},\pm}$, thus generates a CR diffusion coefficient (equations~\ref{eq:v_di} and \ref{eq:closure_scatt_coeff})
\begin{align}
\label{eq:kappa}
\kappa=\frac{c^2}{3(\bar\nu_+ + \bar\nu_-)}
\propto\frac{c^2}{3\Omega}\,\frac{\eps_B}{\eps_{\rmn{w},+} + \eps_{\rmn{w},-}}
\to \frac{c r_\rmn{g}}{3}\,\frac{\eps_B}{\eps_{\rmn{w},+} + \eps_{\rmn{w},-}},
\end{align}
where the limit in the last step applies to the relativistic regime. Depending on the wave spectrum and degree of anisotropy, we can distinguish three important cases:
\begin{equation}
\label{eq:eps_w}
\eps_\rmn{w} = \eps_{\rmn{w},+} + \eps_{\rmn{w},-}
  \propto \left\{
\begin{array}{ll}
k_\parallel^{-1}, & \mbox{anisotropic Alfv\'enic turbulence}, \\[.5em]
k^{-2/3}, & \mbox{isotropic Kolmogorov turbulence}, \\[.5em]
k^{-1/2}, & \mbox{isotropic Iroshnikov-Kraichnan turbulence}. \\
\end{array}
\right.
\end{equation}
Inserting the different wave spectra into the expression for the CR diffusion coefficient in equation~\eqref{eq:kappa}, and adopting the resonance condition, enables us to derive its energy dependence,
\begin{equation}
\label{eq:kappa_E}
\kappa\propto \frac{1}{k_\parallel \eps_\rmn{w}}
  \propto \left\{
\begin{array}{lll}
k_\parallel^{0}&\propto E^0, & \mbox{anisotropic Alfv\'enic turbulence}, \\[.5em]
k^{-1/3}&\propto E^{1/3}, & \mbox{isotropic Kolmogorov turbulence}, \\[.5em]
k^{-1/2}&\propto E^{1/2}, & \mbox{isotropic Iroshnikov-Kraichnan turbulence}. \\
\end{array}
\right.
\end{equation}

\subsubsection{Numerical methods for evolving the cosmic ray momentum spectrum in space}
\label{sec:numerical_methods_CR_spectrum}

Because CR cooling processes and spatial transport depend on CR momentum, this calls for the development of numerical methods that evolve the CRs distribution function simultaneously in space and momentum. While the time evolution of the CR electron spectrum is essential for understanding observational signatures at different frequencies, following the CR ion spectrum may be additionally important for improving our modeling of CR energy and momentum feedback in galaxies. As shown in Fig.~\ref{fig:cooling_times}, and as discussed in Section~\ref{sec:cooling_times}, the electron cooling times are substantially shorter than those of CR ions, which suggests the need for different numerical treatments for CR electrons and CR ions.

Spectral CR propagation codes, such as GALPROP \citep{strong_propagation_1998,moskalenko_production_1998}, USINE \citep{maurin_cosmic_2001,putze_markov_2010}, DRAGON \citep{evoli_cosmic_2008,evoli_cosmic-ray_2017,maccione_dragon_2011}, PICARD \citep{kissmann_picard_2014}, and SPINNAKER \citep{heesen_radio_2018} aim at numerically solving the CR transport equation~\eqref{eq:f0} in the one-moment approximation for a given magnetic field, gas density, source distribution, and stationary background flow. Another approach is to transform the CR transport equation~\eqref{eq:f0} into a set of equivalent stochastic differential equations, partially equipped with advection fields \citep{kopp_stochastic_2012,merten_crpropa_2017,merten_propagation_2018}. Coupling this approach with a Monte Carlo code for simulating the propagation of ultra-high-energy CRs was achieved in the \textsc{CRPropa} code \citep{armengaud_crpropa_2007,alves_batista_crpropa_2022}, which uses a set of stochastic differential equations for propagating CRs in its latest release. Because these codes aim at understanding CR and $\gamma$-ray observables in the Milky Way or radio maps of nearby galaxies, the magnetic field and density distributions are typically inferred from other observations that adopt certain simplifying assumptions \citep{boulanger_imagine_2018}. While adequate for these scientific goals, the background state does not necessarily represent a self-consistent solution of the MHD equations and as such cannot be used to study the dynamical impact of CRs. Moreover, these codes face difficulty in modeling a scenario in which the transport of CRs transitions from predominantly streaming to predominantly diffusion as a function of CR energy. To improve upon these issues, a set of stochastic differential equations suited to follow CR transport in simulations of MHD turbulence was implemented in the CRIPTIC code \citep{krumholz_cosmic_2022} and used to determine an effective transport theory for CRs that stream through a turbulent MHD plasma \citep{sampson_turbulent_2023}. However, common to all those approaches is that they do not allow for the back-reaction of CR pressure forces or heating on the \mbox{(magneto-)}hydrodynamics. As such, those approaches cannot be used to study CR feedback in galaxies and galaxy clusters.

Early works that numerically solve the coupled time-dependent CR transport equation~\eqref{eq:f0} and hydrodynamic equations in one spatial dimension (assuming planar or spherical geometry) address the problem of non-linear self-regulation of CR acceleration at shocks \citep{falle_time-dependent_1987,bell_non-linear_1987,kang_numerical_1991}. These studies use piece-wise constant values to discretize the CR momentum spectrum. However, when the CR transport equations are coupled to the system of MHD equations, this discretization requires approximately 40--80 bins per momentum decade to produce accurate results of the numerical integration \citep{winner_evolution_2019}. A more efficient discretization is given by a piece-wise power-law representation of the CR momentum distribution on a logarithmically spaced momentum grid, which is evolved by assuming continuity of the momentum spectrum and employing CR number conservation, where the backreaction of CRs on the MHD is either not included \citep{jones_simulating_1999} or included \citep{yang_spatially_2017}.

Unfortunately, there is a numerical instability associated with this approach for a localized energy injection in momentum space (e.g. as a result of energy-dependent spatial diffusion or CR shock acceleration). \citet{girichidis_spectrally_2020} show that the continuity assumption enforces changes of the local logarithmic slope that leads to a non-physical, oscillatory behavior across the entire spectrum with alternating convex and concave regions of the spectrum. Generally, the piece-wise power-law representation of the CR distribution exhibits two degrees of freedom, the normalization and spectral slope in every bin. Ensuring CR energy and number conservation (in the absence of CR sources) while evolving the momentum distribution and abandoning spectral continuity is a very promising approach, which is similar in spirit to the discontinuous Galerkin method of hydrodynamics. This so-called ``coarse-grained momentum finite volume'' method is employed for evolving the CR electron spectrum on Lagrangian particle trajectories in MHD simulations, where the CR pressure does not back-react on the gas dynamics \citep{miniati_cosmocr_2001,mimica_spectral_2009,vaidya_particle_2018,boss_crescendo_2023}.

In order to study CR-modified shocks or CR-driven galactic winds, we need to account for the non-linear feedback of the CR pressure on the MHD while simultaneously solving for the CR transport equation~\eqref{eq:f0}. To this end, advection and diffusion processes in physical and momentum spaces are followed with the coarse-grained momentum finite volume method on Eulerian fixed meshes \citep{jones_efficient_2005,girichidis_spectrally_2020,ogrodnik_implementation_2021} or on unstructured moving Voronoi meshes \citep{girichidis_spectrally_2022}. The very small cooling times at low CR electron and ion momenta as well as high CR electron momenta would substantially slow down the computations, rendering fully three-dimensional MHD simulations of galaxies or SN blast waves with spectral CRs prohibitively expensive. To improve the efficiency and accuracy of these computations, analytical solutions in the low- and high-momentum regimes are connected to numerical solutions in the intermediate momentum range, thus enabling computations at numerically affordable costs \citep{winner_evolution_2019,girichidis_spectrally_2020}. This allows us to study the impact of obliquity-dependent CR shock acceleration at SN blast waves and to generate multi-frequency emission maps from radio to $\gamma$ rays \citep{winner_evolution_2020}, the dynamical impact of spectral CR ions during galaxy formation \citep{girichidis_spectrally_2023}, and non-thermal processes in galaxy clusters such as radio halos and relics \citep{miniati_cosmic-ray_e_2001,miniati_numerical_2003,pfrommer_simulating_2007,pfrommer_simulating_2008,pfrommer_simulating_tmp_2008,donnert_rise_2013,pinzke_simulating_2010,pinzke_giant_2013,pinzke_turbulence_2017,boss_crescendo_2023}. The CR spectral algorithm has also found application in galaxy simulations with star formation and feedback that employ the spatial two-moment method for CR hydrodynamics to yield CR spectra of electrons, positrons, (anti)protons, and heavier nuclei \citep[][using a meshless finite-mass discretization of the underlying MHD]{hopkins_first_2022}. This novel approach allows studying the size of the CR scattering halo reaching into the CGM in realistic galaxies and to address which CR spectral features arise from local structures rather than CR transport physics.

\vspace{0.25in}

\clearpage

\section{Astrophysical systems}\label{Astrophysical_Systems}
\begin{figure}[tbp]
\begin{center}
\includegraphics[width=1.0\textwidth]{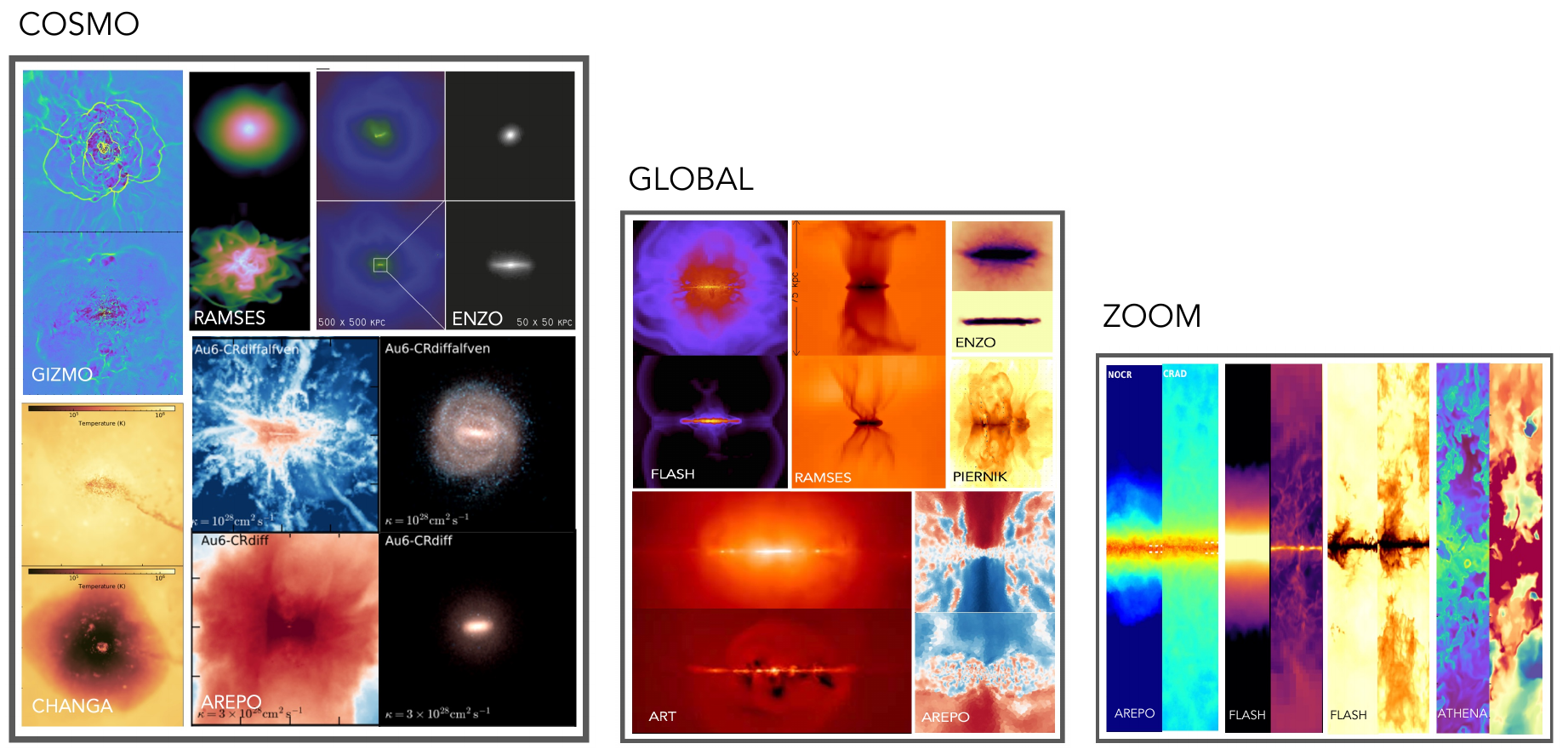}
\end{center}
\caption{A collage of simulations of CR-driven winds performed by different groups using various approaches, physics, and numerical techniques: (i) cosmological simulations (\textit{left panel};
\citet{ji_virial_2021,liang_column_2016,salem_role_2016,butsky_impact_2020,buck_effects_2020}; GIZMO, RAMSES, ENZO, CHANGA, AREPO, respectively), (ii) global galactic disks (\textit{middle panel}; \citet{ruszkowski_global_2017,booth_simulations_2013,butsky_role_2018,hanasz_cosmic_2013,semenov_cosmic-ray_2021,pakmor_galactic_2016}; FLASH, RAMSES, ENZO, PIERNIK, ART, AREPO, respectively), (iii) stratified boxes, each representing a zoom-in on the galactic disk (\textit{right panel};   \citet{simpson_role_2016,farber_impact_2018,
girichidis_cooler_2018, armillotta_cosmic-ray_2021}; AREPO, FLASH, FLASH, ATHENA, respectively). In all cases, groups of images corresponding to a given code illustrate the differences arising due to varying CR physics.}
\label{fig:collage}
\end{figure}
As argued in the Introduction (Section~\ref{Introduction}), CRs may shape in fundamental ways how the astrophysical feedback processes operate in nature. We now begin the discussion of the applications of the physical principles governing CR interactions with plasma waves and matter (Section~\ref{sec:physics}) to these feedback processes. Our general philosophy is to separate the discussion by the relevant astrophysical scales, where the discussion of feedback on small scales informs that on larger scales. This allows us to first focus on the in-depth introduction of the detailed impact of CR physics on the dynamical state of the ISM before transitioning to progressively larger scales of galactic halos, where we connect the processes operating on different scales in order to synthesize the emerging picture of CR feedback. To this end, we discuss the role of CR physics in SNe, cold ISM clouds, multiphase ISM, star formation, galactic wind launching, interactions of galactic winds with the CGM and cosmological infall, and the largest galaxies and galaxy clusters. As we progress from smaller to larger scales, the preferred approach to modeling feedback tends to change from one-dimensional simulations, simulations of individual gas clouds and multiphase medium, small gravitationally stratified boxes, global models of galaxies, to fully cosmological simulations. Examples of images from simulations based on some of these approaches are shown in Fig.~\ref{fig:collage}. In many instances, these approaches are to some extent complementary. Consequently, as an example, we may resort to discussing and interpreting the results from cosmological simulations also using one-dimensional analytical models in order to help build intuition. 

\subsection{Cosmic ray ionization} 
\label{CR_ionization}
The subject of low-energy CRs is extensively discussed in recent reviews by \citet{padovani_impact_2020} and \citet{gabici_low-energy_2022}. Here we present a limited discussion of this topic focusing on key concepts that have implications for CR feedback. In particular, we discuss the impact of CRs by concentrating on the effect of CR-induced ionization in the ISM. Broadly speaking, this discussion is motivated by the fact that this type of ionization is essential for maintaining the coupling of the magnetic fields with the ISM plasma and for facilitating complex ISM chemistry. Furthermore, these processes have important implications for the dynamical interactions of CRs with the ISM, which we discuss in more detail in Section~\ref{multiphase}.\\
\indent
The rate of star formation in giant molecular clouds is controlled by the competition between gravity and non-thermal pressure due to turbulence and magnetic fields \citep{crutcher_magnetic_2012}. The degree to which magnetic forces can participate in this process, and couple to the ISM plasma, depends on the level of ionization in the gas. 
While the observed ionization fractions in molecular clouds are very low, they are nevertheless sufficient to at least partially couple the magnetic fields to the gas, which allows them to, e.g., enable magnetic braking of interstellar clouds, slow down star formation, and enable magneto-rotational instability to operate in protoplanetary disks.
Interestingly, these low ionization levels significantly exceed those expected from photoionization due to UV stellar radiation because of the very large column densities of the clouds \citep{mckee_photoionization-regulated_1989}. This strongly suggests that the additional ionization may come from CRs that can penetrate the clouds. The kinetic energy of the electrons generated via CR ionization of the ISM gas also provides the heating needed to maintain the molecular clouds at the observed temperatures. Both the ionization and heating of the gas are due to low-energy CRs with energies up to $\sim$1 GeV \citep[e.g.,][]{field_cosmic-ray_1969}.
This picture is also consistent with the observations of complex chemistry in molecular clouds. The self-shielding of the clouds prevents the photodissociation of the complex molecules and allows them to form. However, the formation rates of these molecules due to the interaction between neutral species, or neutral-neutral reactions, are too slow. This implies that additional ionization sources, again likely due to low-energy CRs, could lead to ion--neutral reactions that are much faster \citep[e.g.,][]{bergin_cold_2007}.\\
\indent 
In addition to playing a pivotal role in shaping the properties of molecular clouds in the ISM, low-energy CRs are also responsible for producing light elements -- lithium (Li), beryllium (Be), and boron (B) -- in the Universe. These elements are easily destroyed in thermonuclear reactions in stellar interiors. The abundances of these light elements in the solar system are much smaller than those of other elements characterized by comparable atomic numbers, e.g., carbon, nitrogen, and oxygen \citep[e.g., figure~1 of][]{tatischeff_particle_2018}. Curiously, when the comparison is restricted to the composition of CRs, the relative abundances of Li, Be, and B and the neighboring elements in the periodic table, the abundances are comparable, thus underscoring the role of low-energy CRs in the synthesis of these elements. The primary channel for the formation of these elements is the spallation of CNO nuclei by low-energy CRs -- a process in which the heavier nuclei (C, N, O) are split as a result of collisions with the low-energy CRs or when heavier CR species, such as carbon, collide with ISM hydrogen and eject nucleons to form lighter CR species \citep[e.g.,][see also Section~\ref{sec:B-to-C_ratio}]{meneguzzi_production_1971}.\\
\indent
The physical origin of low-energy CRs is not fully understood. While most of the CR energy may be generated in SN explosions, superbubbles, and stellar wind termination shocks \citep[e.g.,][see also Section~\ref{CRsources}]{bykov_nonthermal_2014} followed by Coulomb and ionization losses (see Section~\ref{sec:CR_ion_interactions}), low-energy CRs could also be accelerated in other environments, e.g., along bipolar protostellar jet shocks or at surfaces of protostars, where the matter from the accretion disk is channelled by strong magnetic fields and hits protostellar surface \citep{padovani_impact_2020}. The life cycle of molecular gas in star-forming regions appears to rely significantly on protostellar outflows. The interaction between jets and magnetic fields could potentially enable prolonged star formation across multiple dynamic periods. This process will indirectly impact on CR propagation, as the outflow properties will determine the turbulent characteristics of the cold phase \citep{wang_outflow_2010}. Below we briefly discuss direct measurements and indirect constraints on low-energy CRs.

\subsubsection{Direct and indirect constraints on low-energy cosmic rays}\label{lowEconstraints}
Unravelling the origin of low-energy CRs is complicated by the fact that their spectra are attenuated due to the interactions of low-energy CRs with the solar wind. When measured close to Earth, these interactions lead to the modulation of CR flux at low particle energies \citep[below $\sim$~10 GeV;][]{vos_new_2015} with a period of $\sim$~11 years. During the solar maximum, the attenuation is the strongest as the low-energy CRs are unable to penetrate the strong turbulent and magnetized solar wind, while the high energy CRs remain unaffected. This effect is termed solar modulation and it prevents us from directly measuring the flux of low-energy CRs on Earth, though the flux can be approximately de-modulated by considering poorly constrained CR propagation models. However, thanks to recent data from the \textit{Voyager} probes that have  travelled past the heliopause, we now have direct measurements of the low-energy CR flux that is nearly constant in time and unaffected by the modulation. The CR spectra shown in Fig.~\ref{fig:fig1} include the data from the \textit{Voyager} missions \citep{cummings_galactic_2016,stone_cosmic_2019} in addition to high-energy CR data from other missions and instruments that is unaffected by the solar modulation (see caption of that figure for details). The spectra unaffected by the solar modulation can be translated into reliable measurements of total energy density of CR ions and electrons in the solar vicinity (i.e., outside the heliopause).\\
\indent
In addition to direct measurements, low-energy CR spectra can be constrained via indirect means based on $\gamma$-ray emission observations. CR protons colliding with interstellar protons can produce neutral pions that decay into $\gamma$-ray photons (see also Section~\ref{hadronic} for a more detailed discussion of hadronic processes). This reaction is possible as long as the kinetic energy of a CR proton exceeds a threshold value of $\sim$~280 MeV. Similarly, heavier CR nuclei can also produce $\gamma$-ray emission via collisions with the ISM gas. Since the $\gamma$-ray photons can easily penetrate the ISM, $\gamma$-ray observations open up a window for observations of CRs including mildly relativistic, low-energy CR nuclei. In fact, the diffuse $\gamma$-ray spectrum from the Milky Way disk above $\sim$~100 MeV is dominated by this process \citep[][see also Fig.~\ref{fig:Fermi_sky}]{ackermann_fermi-lat_2012,selig_denoised_2015,platz_multi-component_2022}.\\
\indent
The observed $\gamma$-ray emission comes from a convolution of the spatial distribution of CRs and the ISM. Thus, if the spatial distribution of the diffuse gas can be reliably modeled, then the spatial distribution of CR energy density can be obtained. This approach reveals that, in the local ISM, the CR spectra derived using this approach exceed the directly measured spectra from AMS-02 (at high CR energies, where the effects of solar modulation are negligible) by only a few tens of percent \citep{strong_local_2016,orlando_imprints_2018}. Similarly, when the comparison is made between the CR energy density measured directly in the local ISM and at other galacto-centric radii, the scatter in the typical values is less than a factor of two \citep{acero_development_2016}. 
In a similar vein, giant molecular clouds (GMCs) can be used as ``CR barometers.'' A key difference compared to the diffuse ISM case discussed above is that the GMC can serve as a probe of very localized CR energy density. The appeal of using GMCs as CR probes is their large density, which can make them bright $\gamma$-ray sources \citep{black_production_1973}. Consequently, they can be detected at both nearby and remote locations in the Milky Way. The expected $\gamma$-ray flux from GMCs depends on a number of factors such as the ability of diffuse CRs to penetrate the clouds or the cloud masses, both of which are poorly constrained. Measurements of CR energy densities based on $\gamma$-ray emission from GMCs yield similar results to those based on diffuse $\gamma$-ray emission in the Galactic disk \citep{aharonian_probing_2020,peron_probing_2021}.\\
\indent
In addition to hadronic $\gamma$-ray emission associated with the decay of neutral pions, $\gamma$-rays can also be produced via leptonic processes: IC scattering of low energy photons and synchrotron emission\footnote{Relativistic bremsstrahlung can also contribute to $\gamma$-ray emission but this contribution is subdominant compared to that from the hadronic and IC processes for photon energies above and below $\sim 100$ MeV, respectively \citep{strong_interstellar_2011,de_angelis_science_2018}.}\citep{blumenthal_bremsstrahlung_1970}. One key advantage of relying on the leptonic emission signatures (see Section~\ref{sec:CR_lepton_interactions} for the discussion of leptonic processes) to constrain CR spectra is that they probe CR electron/positron energies that are not accessible to direct observations from the \textit{Voyager} missions and AMS-02, which can reliably detect CR electrons only with energies $\lesssim 100$ MeV and $\gtrsim 10$ GeV (where solar modulation is unimportant), respectively. \\
\indent
It is instructive to estimate typical photon energies or frequencies corresponding to these processes. For IC scattering, a typical photon energy is $E_{\rm ph}\sim 5 E_{1}^{2} \epsilon_{1}$~MeV, where $E_{1}$ and $\epsilon_{1}$ are the CR electron energy in GeV and seed photon energy in eV. In the order of increasing characteristic seed photon energy $\epsilon$, the primary sources of seed photons for the IC upscattering process are: CMB ($\epsilon\sim 6\times 10^{-4}$~eV), dust, and interstellar radiation field ($\epsilon\sim 1$~eV). Thus, for GeV electrons (i.e., in the middle of the energy range unaccessible to direct observations), the observed energies of the upscattered photons span a wide range from soft X-rays to soft $\gamma$ rays. This range of CR electron energies can also be accessed via radio observations, as the characteristic synchrotron emission frequency of GeV electrons is $\nu_{\rm ph}\sim 320~E_{1}^{2}B_{10}$~MHz (see equation~\ref{eq:nu_s}), where $B_{10}$ is the magnetic field in $10~\mu$G. Spectral energy distributions of CR electrons derived from the radio and $\gamma$-ray observations of the local ISM are in agreement with direct measurements from the \textit{Voyager} 1 (for CR energies $\lesssim 100$ MeV) and AMS-02 (for CR energies $\gtrsim 10$ GeV), and smoothly interpolate  between these regimes \citep{orlando_imprints_2018}. However, the measurement of CR electrons in the Milky Way beyond the local ISM is challenging. In this case, indirect arguments based on the modelling of CR propagation suggest that the spatial variation in the CR electron distribution are also likely relatively small \citep[e.g.,][]{di_bernardo_diffuse_2015}.

\subsubsection{Low-energy cosmic ray ionization rates}\label{ionization_rates}
The spatial distribution of low-energy CRs can also be probed by searching for signatures of ionization of molecular gas by CRs. While photoionization by UV photons dominates throughout most of the ISM volume, hydrogen molecules in clouds are effectively shielded by atomic hydrogen in their outer layers because the atomic hydrogen ionization potential is lower than that of the molecular hydrogen. Consequently, the dominant source of ionization of H$_{2}$ in molecular clouds is low-energy CRs \citep[e.g.,][]{padovani_cosmic-ray_2009}. 
Ionization of a hydrogen molecule H$_{2}$ by a CR leaves behind a singly ionized H$^{+}_{2}$, a free electron, and a CR. Subsequent reaction of H$^{+}_{2}$ with another hydrogen molecule releases a hydrogen atom and leads to the formation of protonated hydrogen H$^{+}_{3}$ \citep[e.g.,][]{dalgarno_galactic_2006}. Since the latter (ion--neutral) reaction is much faster than the former CR ionization reaction, the formation rate of H$^{+}_{3}$ is determined by the rate at which CRs ionize molecular hydrogen. In other words, the rate of protonated hydrogen formation is $\zeta n(\rm{H}_{2})$, where $\zeta$ is the CR ionization rate and $n(\rm{H}_{2})$ is the number density of molecular hydrogen.\\
\indent
In an equilibrium situation, the formation rate of H$^{+}_{3}$ is balanced by its destruction rate. The latter can proceed via recombination of H$^{+}_{3}$ ions with free electrons, which produces either atomic hydrogen or molecular and atomic hydrogen. These reactions dominate the destruction of H$^{+}_{3}$ as long as the electron fraction is sufficiently high ($\gtrsim 10^{-4}$), which is possible in diffuse clouds. 
The main source of free electrons in diffuse clouds is ionization of carbon. This is possible thanks to the high abundance of carbon and the fact that its ionization threshold is below that of hydrogen. Thus, UV photons with energies below the hydrogen ionization threshold can penetrate the diffuse clouds and ionize carbon, thereby maintaining a small albeit significant electron fraction in these clouds.\\
\indent
In dense molecular clouds, the electron fractions are much lower due to significant absorption of UV flux and the dominant H$^{+}_{3}$ destruction channel involves reactions of protonated hydrogen with carbon monoxide molecules. 
The net result is that, irrespectively of cloud density, the balance between production and destruction of H$^{+}_{3}$ permits the determination of the CR ionization rate $\zeta$ as long as the abundances of H$^{+}_{3}$, H$_{2}$, and CO, and the relevant destruction reaction rates are known. 
Measurements of CR ionization rates based on similar principles (i.e., those that involve quantifying the abundances of ionized molecules such as H$^{+}_{3}$, OH$^{+}$, H$_{2}$O$^{+}$, 
H$_{3}$O$^{+}$) demonstrate that $\zeta$ can range from $10^{-17}$s$^{-1}$ for high column density clouds \citep[$N(\rm{H}_{2})\sim 10^{23}$cm$^{-2}$, e.g., ][]{sabatini_survey_2020} to $10^{-15}$s$^{-1}$ for low column density clouds \citep[$N(\rm{H}_{2})\sim 10^{23}$cm$^{-2}$, e.g., ][]{indriolo_herschel_2015}.

\subsubsection{Penetration of molecular clouds by low-energy cosmic rays}\label{gmc_penetration}
As stated above, CR ionization rates vary widely and tend to decrease with the column density of molecular clouds. This suggests that the level of ionization may depend on the ability of CRs to penetrate the clouds. The probability that a CR impinging on a cloud will ionize the cloud molecules increases with the time the CR can spend inside the cloud. While frequent scattering of CRs in the cloud gas increases this probability, it also affects how deep the CRs can penetrate the cloud. The penetration depth is maximized by assuming that impinging CRs propagate along straight lines. However, the frequent scattering of CRs on Alfv{\'e}n waves in the turbulent and magnetized plasma precludes that possibility. While the principles behind such interactions, and the related CR transport processes, are described in more detail in Section~\ref{sec:CR-wave} and Section~\ref{sec:spatial_transport}, respectively, here we lay out in broad strokes the key points relevant to the discussion of the CR penetration of molecular clouds.\\
\indent
In the quasi-linear regime, CRs scatter on Alfv{\'e}n waves characterized by typical wavelengths comparable to CR gyroradii and multiple scattering events isotropize CRs in pitch angle. The likely origin of the magnetic field fluctuations on which the CRs can scatter is the streaming of CRs along the magnetic fields \citep[e.g.,][see also Section~\ref{sec:plasma_instabilities}]{kulsrud_effect_1969,shalaby_new_2021}. The fundamental reason why CRs can trigger the instability and amplify the waves is the following. Suppose that CRs propagate at a given net drift speed and encounter slower Alfv{\'e}n waves propagating in the same direction. While the wave--particle scattering does not change CR energy in the wave frame, the effective streaming speed of CRs along the mean magnetic field adjusts to the wave speed in the laboratory frame on the isotropization timescale (see Section~\ref{sec:CR-wave}). Thus, CRs experience effective momentum loss in the laboratory frame as their net drift speed is reduced. By momentum conservation, this momentum loss corresponds to the gain in the wave momentum. Alfv\'en waves do not have an intrinsic momentum in the MHD approximation like normal electromagnetic waves would. But they are carried by the background gas particles which absorb the momentum loss of CRs. This effectively accelerates the background gas. Associated with this exchange of momentum is the energy transfer. The kinetic energy of CRs is directly converted into the magnetic energy of the Alfv\'en waves. This process thus grows the magnetic field perturbations on which CRs can scatter. The growth rate of the gyroresonant streaming instability is given by equation~\eqref{eq:Gamma_gyro}. In dense molecular clouds, where neutral gas dominates the mass budget, wave damping due to ion--neutral friction can efficiently damp Alfv{\'e}n waves and, thus, net CR propagation speeds can approach the speed of light \citep[e.g.,][]{everett_interaction_2011,ivlev_penetration_2018} provided there is no other instability driving plasma waves unstable that could dominate CR-wave scattering.\\
\indent
\begin{figure}[tbp]
\begin{center}
\includegraphics[width=1.0\textwidth]{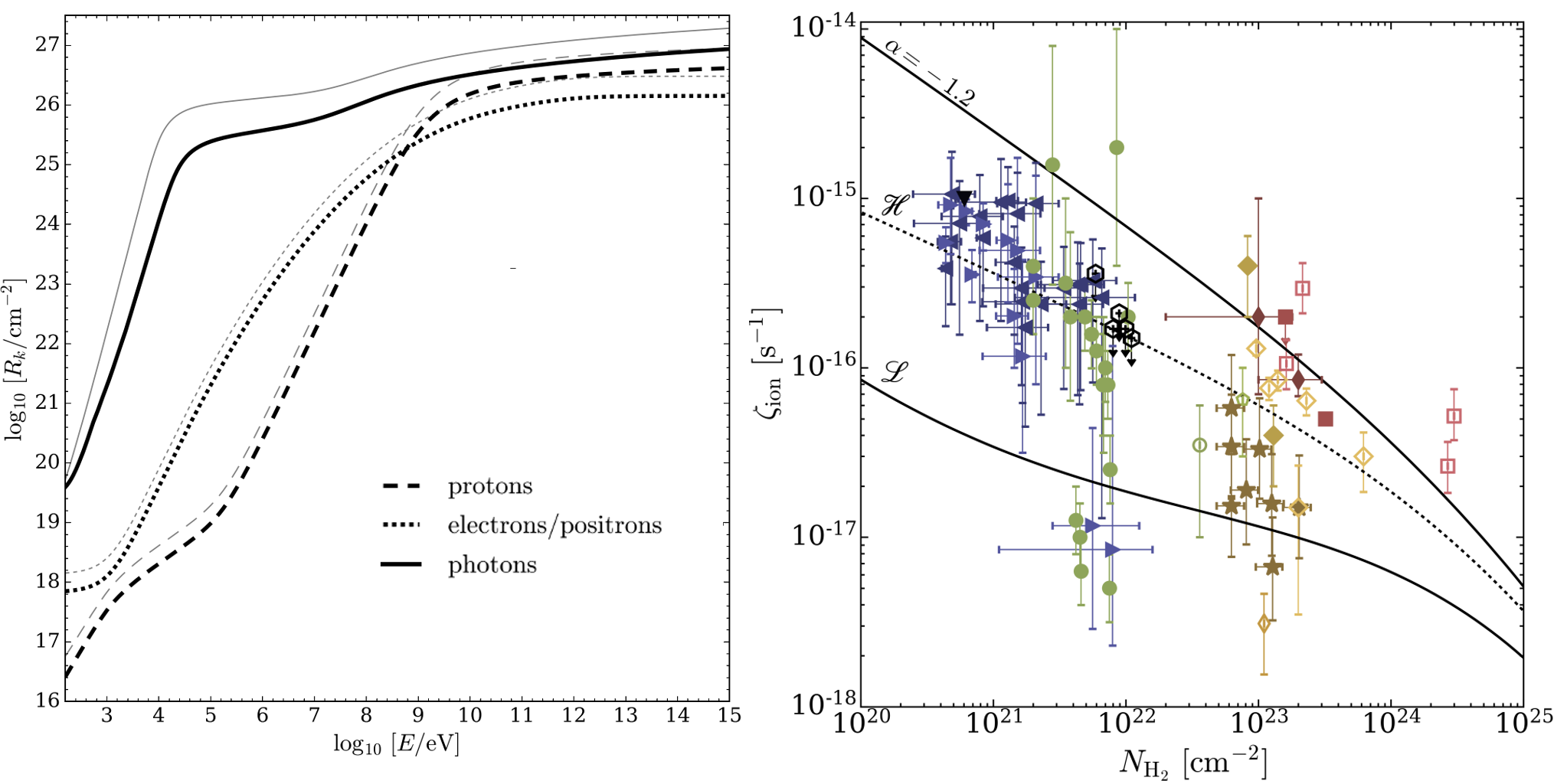}
\end{center}
\caption{\textit{Left:} CR energy dependence of the column density that CRs can travel before losing their energy (dotted and dashed lines correspond to CR electrons and CR protons, respectively) and the analogous quantity for photons (solid lines). Thick and thin lines correspond to CRs and photons traversing (i) molecular ICM clouds characterized by solar abundance and (ii) clouds made of atomic hydrogen, respectively. Modified figure from \citet{padovani_cosmic-ray_2018}; reproduced with permission from A\&A. \textit{Right:} Observationally-inferred CR ionization rates as a function of molecular cloud column density (data points) and free-streaming CR ionization models \citep[see][and references therein]{padovani_cosmic_2022}. Figure from \citet{padovani_cosmic_2022}; reproduced with permission from A\&A.}
\label{fig:cionization}
\end{figure}
In addition to scattering on Alfv{\'e}n waves, the ability of CRs to penetrate molecular clouds is further impeded by inelastic interactions with the cloud gas. These interactions are dominated by ionization losses for CR energies up to GeV and by pion production and bremsstrahlung at higher energies. The left panel in Fig.~\ref{fig:cionization} shows the characteristic column densities of gas, that CRs can traverse before being attenuated, as a function of CR energy \citep{padovani_cosmic-ray_2018}. Plotted also is the same quantity for high-energy photons. This figure illustrates the fact that only higher energy CRs are capable of penetrating the clouds; low energy CRs can be easily absorbed in diffuse clouds or outer layers of dense clouds. Similarly, only high-energy X-ray and $\gamma$-ray photons can penetrate cloud interiors. This can help to explain the observed trend for the CR ionization rate $\zeta$ to decrease with the molecular cloud column density shown in the right panel of Fig.~\ref{fig:cionization} because high-energy CRs have a lower ionization rate \citep[][see also Section~\ref{ionization_rates} above]{padovani_cosmic_2022}.\\
\indent
Significant energy losses of CRs entering clouds are expected to create CR pressure gradients and force CRs to propagate down their pressure gradient toward the cloud centers. As described above, this is expected to trigger the streaming instability. The efficient scattering of CRs on self-generated Alfv{\'e}n waves would then prevent CRs from penetrating deep into the dense molecular clouds, thus limiting their ability to explain the ionization rates in seen in the data. Cloud ionization models that include (i) CR ionization losses, (ii) scattering due to the streaming instability in the ionized envelopes of clouds, as well as (iii) ion--neutral damping inside the clouds, where the gas is largely neutral and CRs can free-stream through the gas, fall significantly short of the observed ionization levels \citep[e.g.,][]{phan_what_2018}. However, the upper envelope of CR ionization rates can be estimated by considering a limiting case whereby CRs free-stream without scattering through the clouds on straight trajectories allowing them to penetrate deep into the clouds. 
Under these assumptions, the ionization rates computed assuming a low-energy CR spectrum slope matching \textit{Voyager} 1 and 2 data underpredict the observed ionization rates in diffuse clouds (see bottom solid line in the right panel in Fig.~\ref{fig:cionization}). However, for steeper CR slopes, the models relying on the free-streaming approximation can match typical CR ionization values across the full range of molecular cloud column densities \citep[][see dotted line in the right panel in Fig.~\ref{fig:cionization}]{padovani_cosmic_2022}.\\
\indent 
Current CR ionization rate data exhibit significant scatter (see, e.g., right panel in Fig.~\ref{fig:cionization}), which limits the constraining power of these measurements. The likely culprits responsible for the scatter include uncertainties in the chemical network, large-scale and local (e.g., due to the irradiation of the gas by CRs from nearby SNe) variations in the CR flux in the ISM, and a possible presence of CR sources inside the molecular clouds mentioned earlier. Direct probes of the CR ionization rate that will circumvent these complications, and thus allow us to better constrain CR propagation models, may include future observations of rovibrational lines of H$_{2}$ with the \textit{James Webb Space Telescope} \citep{bialy_cold_2020} or possible detection of MeV lines with \textit{Compton Spectrometer and Imager} \citep[COSI;][]{de_angelis_science_2018}.

\subsection{Cosmic ray-driven galactic winds}\label{galactic_winds}

Large-scale outflows from late-type galaxies are ubiquitously detected in the Universe \citep[e.g., see reviews by][]{rupke_outflows_2005,alsabti_galactic_2017,veilleux_cool_2020}. These galactic winds play an essential role in addressing some of the fundamental issues in the galaxy formation field such as quenching of star formation, enriching the CGM and intergalactic medium with metals, and the ``missing baryon problem,'' i.e., reducing the galactic baryon fraction below the cosmic mean \citep[e.g.,][]{somerville_physical_2015,naab_theoretical_2017,tumlinson_circumgalactic_2017}. Several mechanisms have been proposed to launch the winds. Broadly speaking, they fall into the categories of energy-driven winds, where SN explosions provide thermal energy to power the outflow \citep[e.g.,][]{chevalier_wind_1985}, or momentum driven winds, where the outflow is accelerated by radiation pressure  \citep[e.g.,][]{murray_maximum_2005}. While these mechanisms may help to drive winds in extreme environments of starburst galaxies and AGNs, they are less likely to be the primary mechanisms responsible for the wind acceleration in normal disk galaxies, where the thermal and radiation pressure forces are weaker. They may also be prone to the problem of overcooling, in which the injected thermal energy is radiated away before the wind can be accelerated, or to problem of the lack of sufficient coupling of radiation pressure to the multiphase gas clouds (see Section~\ref{CR_and_star_formation}).\\
\indent
An alternative solution to the problem of wind driving may involve gas acceleration by dynamically important CR pressure gradients associated with CRs accelerated in SN shock waves.
One of the most appealing features of this mechanism is the fact that inelastic CR energy losses tend to be much slower than radiative cooling of the thermal gas. Furthermore, compared to thermal gas, the relatively small adiabatic index of the CR fluid results in slower decrease of CR pressure with the height above the disk as the gas expands, which further helps to drive the outflow. In the self-confinement model of CR propagation (see Section~\ref{CR_scattering}), CRs scatter on self-excited Alfv{\'e}n waves. The scattering reduces CR anisotropy in the wave frame (thus essentially coupling CRs to the gas) and transfers CR momentum and energy to the waves, and eventually to the gas via wave damping. CR transport out of the galactic disk then establishes a CR pressure gradient that can accelerate the gas. The pioneering work describing and applying these principles in the context of spherically-symmetric and steady-state galactic wind models was first presented by \citet{ipavich_galactic_1975}, who demonstrated that galactic outflows can be launched by CRs even when the gas temperature vanishes. It was also realized early on \citep{jokipii_consequences_1976} that the long CR confinement times in the Galaxy, inferred from CR spallation arguments (see Section~\ref{sec:B-to-C_ratio}), imply that CRs should, on average, interact with gas of much smaller density than the typical density of the ISM. This suggests that CRs may escape from the dense disk, and thus establish CR pressure gradients and a ``dynamical halo.''

\subsubsection{One-dimensional models} \label{1D_models}
Given the above general motivation for including CRs in galactic wind models, we now systematically discuss the dynamical role of CRs. We intentionally begin with relatively simple semi-analytic and one-dimensional models before discussing more sophisticated multi-physics and multi-dimensional models applicable to progressively larger physical scales. Broadly speaking, the one-dimensional winds are modeled either as magnetic flux tube outflows or as spherically-symmetric winds, and either assuming steady state or allowing for time dependent solutions. Below we briefly describe the main results and the advantages and disadvantages of these one-dimensional approaches. 

\paragraph{Magnetic flux tube models of galactic winds.}
The physical motivation for approximating the outflow geometry as an ensemble of flux tubes anchored in the disk and pointed away from it comes from the realization that SNe and superbubbles injecting energy in the galactic midplane will tend to open up the magnetic field lines in the vertical direction. As the gas is accelerated away from the disk, the cross-sectional area of the flux tubes will increase to maintain lateral pressure balance with neighboring flux tubes. Despite its simplicity, this model can capture some of the essential aspects of the wind launching problem such as (i) adiabatic cooling of the gas as it is lifted above the disk, (ii) departure from spherical symmetry by modelling an essentially cylindrically symmetric configuration, and (iii) effects of rotation. However, it does not capture the outflow from the entire disk, focusing rather on small patches thereof.\\
\indent 
The majority of one-dimensional CR wind models in the literature adopted the magnetic flux tube approximation, most notably the influential early work of \citet{breitschwerdt_galactic_1991}.
In this steady-state model, the gas is assumed to flow parallel to the magnetic field and the magnetic stresses are neglected. Thus, the field is decoupled from the flow and the cross-sectional area of the funnel-like magnetic tubes is prescribed manually. The model tracks the evolution of thermal, CR, and forward-propagating Alfv{\'e}n wave pressures, and CR transport occurs via streaming at the Alfv{\'e}n speed, while wave damping is neglected in most cases. Radiative losses of the thermal gas and inelastic CR losses are also neglected. The gravitational potential, $\varPhi$, includes contributions from bulge, disk, and halo. Under these assumptions, the MHD equations take a particularly simple form. In particular, the \textit{wind equation} reads:
\begin{equation}
\left(\varv^{2}-c_{\rm eff}^{2}\right)\frac{\dd\ln\varv}{\dd\ln z} = c_{\rm eff}^{2}\frac{\dd\ln A}{\dd\ln z} + z\, g_z(z),
\label{eq:delaval}
\end{equation}
where $\varv$, $A$, $g_z$, and $c_{\rm eff}$ are the gas velocity, cross-sectional area ($\dd A/\dd z>0$) of the magnetic tube, magnitude of gravitational acceleration in vertical direction ($g_z\equiv -\partial\varPhi/\partial z<0$), and the effective sound speed of the composite fluid that includes pressure contributions from thermal gas, CRs, and waves excited by CRs, respectively. This equation bears close similarity to the \textit{Parker solar wind equation} and to the \textit{de Laval nozzle equation}, which describes gas flow through a pinched tube and the resulting conversion of thermal to kinetic energy and thus acceleration of the gas to supersonic speeds. The right hand side of equation~\eqref{eq:delaval} plays the role of the vertical slope of the effective cross-sectional area of the de Laval nozzle. As the accelerating ($\dd \varv/\dd z>0$) subsonic flow passes the point where the ``nozzle'' is narrowest (i.e., where the right hand side of equation~\eqref{eq:delaval} vanishes), it must become supersonic. Past that critical point, the ``nozzle'' is diverging and the surface term dominates over the gravitational acceleration term and the gas continues to accelerate \citep[e.g.,][]{recchia_cosmic_2017}.\\
\indent 
Equation~\eqref{eq:delaval} illustrates well a fundamental problem faced by thermally driven winds. As the adiabatic thermally driven wind expands and the density drops, the effective sound speed decreases as $c_{\rm eff}^{2}\propto\rho^{2/3}$. Consequently, the first term on the right hand side of equation~\eqref{eq:delaval} decreases faster than the effective gravitational potential $g\sim-\varv^{2}_{\rm circ}/r$ because the circular velocity $\varv_{\rm circ}$ as a function of spherical radius $r$ in the halo is essentially flat.\footnote{This can be seen by considering the limiting case $r\gg R_{0}$, where $R_{0}$ is the cylindrical galactocentric distance at which the flux tube is anchored to the disk, i.e., $r\sim z$. In this case, the right hand side of equation~\eqref{eq:delaval} becomes $2c_{\rm eff}^{2}-\varv_{\rm circ}^{2}$ for commonly assumed profile $A(r)\propto r^{2}$ \citep[e.g.,][]{recchia_cosmic_2017}.}
This prevents the outflow from reaching the critical point and accelerating past the sound speed. On the other hand, in the case of CRs and for sub-Alfv{\'e}nic flows (an assumption well justified for fiducial parameters), CRs streaming at the Alfv{\'e}n speed $\varv_\rmn{a}\propto B\rho^{-1/2}$, where $B$ is the strength of the magnetic field, results in CR pressure varying with gas density along the flux tube as $P_{\rm cr}\propto (\varv_\rmn{a}A)^{-4/3}\propto\rho^{2/3}$ (where the first proportionality derives from the CR transport equation~\eqref{eq:ecr_1m} in the strong coupling limit with $\kappa\approx0$, small gas velocities $\varv<\varv_\rmn{a}$, and without any source terms, see Appendix~B of \citealt{breitschwerdt_galactic_1991}, while the second proportionality takes into consideration magnetic flux conservation), thus implying $c_{\rm eff}^{2}\propto\rho^{-1/3}$. Consequently, the right hand side of equation~\eqref{eq:delaval} can become positive and the CR accelerated flow can become supersonic (see \citet{mao_galactic_2018}, where a careful derivation of the flow streamlines and $A(z)$ corroborates this conclusion).\\
\indent 
The key conclusion from this model is that stationary wind solutions are possible for a wide range of parameters as long as the pressure of the ambient intergalactic medium is sufficiently low. For Milky Way conditions, mass loss rates are on the order of $\rmn{M}_{\odot}~\rmn{yr}^{-1}$. Importantly, the results are consistent with the main mass loss mechanism being indeed the acceleration of the gas by CRs rather than thermal driving as can be seen from the following simple argument. On the one hand, the mass loss rate expected in the limiting case where CRs are the sole agent driving the wind, can be estimated by comparing the kinetic power of the accelerated gas to the CR energy injection rate
\begin{equation}
\frac{1}{2}\rho_{0}\varv_{0}\varv_{\rm esc}^{2}A_{\rm disk} \sim \frac{\eps_{\rm cr}V_{\rm disk}}{\tau_{\rm esc}},
\label{eq:CRflux}
\end{equation}
where $\rho_{0}$ and $\varv_{0}$ are reference values of gas density and velocity at a reference height close to the disk, and where it is assumed that the velocity at the reference height is on the same order as the escape speed $\varv_{\rm esc}$ \citep{breitschwerdt_galactic_1991}. The remaining symbols $A_{\rm disk}$, $\eps_{\rm cr}$, $V_{\rm disk}$, and $\tau_{\rm esc}$ are the surface area of the galactic disk, CR energy density in the disk, volume of the disk, and a typical escape time of CR from the disk, respectively. Rearranging terms in the above equation leads to the expression for the mass loss rate per unit area
\begin{equation}
{\dot \Sigma}_{\rm cr} \approx \frac{2\eps_{\rm cr}h_{\rm disk}}{\varv_{\rm esc}^{2}\tau_{\rm esc}},
\end{equation}
where the disk thickness $h_{\rm disk}$ and other terms can be estimated from observations. 
On the other hand, assuming no significant heating and cooling of the gas and CRs (other than energy exchange mediated via the streaming instability) and that the terminal wind velocity is of the same order as $\varv_{\rm esc}$, the mass loss rate per unit area predicted by the model can be estimated from the mass conservation and the Bernoulli equation as
\begin{equation}
{\dot \Sigma}_{\rm th} \approx \frac{\gamma_{\rm cr}}{\gamma_{\rm cr}-1}\frac{P_{\rm cr0}\varv_{\rm a0}}{\varv_{\rm esc}^{2}},
\end{equation}
where $\varv_{\rm a0}$ and $P_{\rm cr0}$ are the Alfv{\'e}n speed and CR pressure at the reference level of $z_{\rm 0}=1$~kpc, respectively. Assuming $\tau_{\rm esc}\sim 3\times 10^{7}$~yr (based on B-to-C ratio measurements; see Section~\ref{sec:CRprop}), $h_{\rm disk}\sim 100$~pc (conservatively), midplane CR pressure $P_{\rm cr}=\eps_{\rm cr}/3\sim 3.3\times10^{-1}$~eV cm$^{-3}$, and adopting reference values $\varv_{\rm a0}\sim 70$ km s$^{-1}$ and $P_{\rm cr0}\sim6.2\times 10^{-2}$~eV \citep{breitschwerdt_galactic_1991}, the ratio ${\dot \Sigma}_{\rm cr}/{\dot \Sigma}_{\rm th}\sim1.5\,(P_{\rm cr}/P_{\rm cr0})\,[h_{\rm disk}/(\varv_{\rm a0}\tau_{\rm esc})]\sim\mathcal{O}$$(1)$, thus demonstrating that the outflow may be CR driven.\\
\indent
The original magnetic flux tube wind model of \citet{breitschwerdt_galactic_1991} can be generalized in a number of ways. For example, the effects of rotation can be incorporated by modifying the effective gravitational acceleration. When magnetic tension forces are also included, this can result in angular momentum loss and enhanced mass loss rate from the disk \citep{zirakashvili_magnetohydrodynamic_1996}. An important limitation of the original \citet{breitschwerdt_galactic_1991} model is an ad hoc prescription for the cross-sectional area of the vertically-flared magnetic flux tubes. This limitation can be overcome by considering flow streamlines aligned with the gradient of the effective gravitational potential that includes realistic gravitational and centrifugal terms \citep{mao_galactic_2018}. While unlike in the original model, this approach neglects Alfv{\'e}n wave pressure, assumes an isothermal equation of state, and considers very weak gas pressure, it leads to the wind equation very reminiscent of that given by equation~\eqref{eq:delaval}. This model makes a specific prediction for the scaling of the mass loading $\eta_\rmn{m}$, defined as the ratio of the mass loss rate in the wind to the star formation rate, with the circular velocity $\varv_{\rm circ}$ of the halo such that $\eta_\rmn{m}\propto\varv_{\rm circ}^{-5/3}$, which is very close to that observed \citep{chisholm_mass_2017}. Interestingly, the predicted slope of this scaling relation lies in between that expected from energy- ($\eta_\rmn{m}\propto\varv_{\rm circ}^{-2}$) and momentum-driven ($\eta_\rmn{m}\propto\varv_{\rm circ}^{-1}$) outflow models \citep[e.g.,][]{somerville_physical_2015}.\\
\indent
The basic framework of the steady-state magnetic flux tube model can also be used to make predictions for the spectra of CRs accelerating galactic winds. Building on the model of \citet{breitschwerdt_galactic_1991}, but assuming instead that the Alfv{\'e}n waves are damped by non-linear Landau damping which reduces their amplitudes, one can eliminate the need to follow the dynamical contribution from the wave pressure, which becomes subdominant. The transport of CRs can be followed by solving the advection-diffusion equation for the isotropic part of the momentum-dependent CR distribution function $f_0(p)$ (see equation~\eqref{eq:f0} in the steady-state case and neglecting the second-order Fermi acceleration term). In this approach, the diffusion coefficient can be self-consistently computed by balancing the streaming instability growth rate (that depends on $f_0(p)$) with the non-linear Landau damping rate, while the effective CR advection velocity is approximately $\bs{\varv} + \bs{\varv}_{\rmn{a}}$ because, for the parameters relevant to ionized gas in the wind, non-linear Landau damping is not strong enough to result in super-Alfv{\'e}nic drift speeds. Under the above assumptions, CR spectra in the wind can be self-consistently computed  
\citep{ptuskin_non-linear_2008,recchia_cosmic_2016,recchia_cosmic_2017}. However, these studies generally find that CR spectra are inconsistent with the observations. In particular, very fast advection leads to CR spectra harder than observed for energies $\lesssim200$ GeV, though this conclusion depends sensitively on the conditions close to the disk, where ion--neutral friction may significantly alter CR transport and non-equilibrium and/or three-dimensional CR transport effects could significantly modify the results \citep{thomas_cosmic-ray-driven_2023}. This warrants future studies to elucidate the exact role of momentum-dependent damping mechanisms and their impact on launching winds and spectra of CR driving them.\\
\indent
The realism of the magnetic flux tube models can be further improved by considering time-dependent wind solutions \citep{dorfi_time-dependent_2012,dorfi_time-dependent_2019}. In realistic situations, feedback from SNe and superbubbles in the disk will be variable in time. In the context of the flux tube model, this can be modeled via time-variable inner boundary conditions. Multiple episodes of stellar feedback will therefore imprint pressure fluctuations steepening into shock waves in the accelerating wind that will subsequently merge to form a large-scale shock. The interaction of that shock with the ambient wind will also produce a reverse shock travelling inward.  An appealing consequence of the formation of the forward and reverse shocks in the outflow already ``polluted'' with CRs is that the shocks can efficiently re-accelerate CRs in situ via the first-order Fermi process. This mechanism thus circumvents the common problem of the injection of seed CRs needed for the acceleration to start. Furthermore, since these large-scale shocks are expected to significantly outlast SNRs, CR acceleration lifetimes can be much longer than in the case of SNe making it possible to accelerate CRs to very high energies. This line of reasoning suggests that CRs could be accelerated to energies past the ``knee''  ($\sim3\times10^{15}$~eV) and possibly as high as the ``ankle'' ($\sim10^{18}$~eV) in the CR spectrum
\citep[see Fig.~\ref{fig:fig1}; ][see also \citealt{jokipii_ultra--high_1987,volk_cosmic_2004,bustard_cosmic_2017}, who reached similar conclusions regarding efficient CR acceleration in the context of wind termination shocks]{dorfi_time-dependent_2019}. 

\paragraph{Spherically-symmetric wind models.}\label{Spherically-symmetric wind models}
One of the disadvantages of magnetic flux tube models is that, by construction, they model outflows only from a patch of the disk. An alternative approach is to consider global spherically-symmetric models albeit at the expense of sacrificing the possibility of including rotation or probing the disk potential. This is in fact the original approach adopted by \citet{ipavich_galactic_1975}, who considered steady-state, non-radiative CR-driven winds in gravitational potentials determined by a central point mass. Generalization of this approach to include spatially distributed dark matter potential and radiative cooling demonstrated that CRs can drive strong winds even if cooling is very efficient \citep{samui_cosmic_2010}. This steady-state model predicts mass loading factor $\eta_\rmn{m}\propto\varv_{\rm circ}^{-2}$ \citep[cf.,][]{mao_galactic_2018}.\\
\indent
Time-dependent simulations can reveal qualitatively different results than those assuming exact steady states. We illustrate this by discussing recent results based on spherically-symmetric time-dependent simulations of CR-driven winds in the presence of diffusion \citep{quataert_physics_2022-1} or Alfv{\'e}nic streaming \citep{quataert_physics_2022}. In both cases, one can proceed by starting with analytic steady-state solutions in spherically-symmetric geometry and compare them to the corresponding solutions from time-dependent simulations as described below.\\
\indent
In the diffusion case, the analytic treatment can be simplified substantially by specializing to the case of fast diffusion. In this limit, the advection timescale vastly exceeds the diffusion timescale, $\tau_{\rm adv}/\tau_{\rm diff}\sim \kappa/(r\varv)$, where $\kappa$ is the diffusion coefficient, and $\varv$ is the outflow speed at distance $r$. For large $\kappa \gtrsim \kappa_{\rm crit}\sim r_{\rm 0}c_{\rm 0}$ (where  $r_{\rm 0}$ and $c_{\rm 0}$ are the radius of the wind base and the sound speed at that distance, respectively) and close to the center, the advection term in the equation for CR energy density can be dropped. The mass conservation and Euler equations can be combined to form a wind equation similar to equation~\eqref{eq:delaval} above. Below the critical point the gas flow is subsonic and so the density profile can be approximated reasonably well assuming hydrostatic equilibrium. The mass loss rate is then completely specified by the conditions at the sonic radius. Importantly, this approach leads to results that are validated by time-dependent spherically-symmetric simulations. In this fast diffusion limit, the flux of CR energy supplied at the center is almost entirely converted to the kinetic flux at infinity. The mass loss rate predicted by this model is $\dot{M}_\rmn{w}\propto \varv^{-2}_{\rm circ}$ \citep{quataert_physics_2022-1}. Interestingly, the scaling of the mass loading $\eta_\rmn{m}\propto \varv^{-2}_{\rm circ}\kappa^{-1}$ with the diffusion coefficient exhibits the same overall decreasing trend as seen in full three-dimensional non-radiative simulations of CR-driven winds \citep{salem_cosmic_2014}. This behavior is also intuitively correct as faster diffusion implies lower CR pressures and thus weaker outflow. In the opposite regime of low $\kappa$, time-dependent simulations demonstrate that wind acceleration is very slow and CR flux supplied at the wind base is expended on doing work against gravity.\\
\indent
The physical outcomes in the case of CR streaming transport at the Alfv{\'e}n speed are diametrically different from those in the CR diffusion case described above. As in the fast diffusion case, a steady-state analytic solution can be obtained below the critical point in the sub-Alfv{\'e}nic regime, i.e., when CR transport is relatively fast. However, the solutions obtained this way are linearly unstable as confirmed by time-dependent spherically-symmetric numerical simulations. The instabilities trigger formation of shocks, and the density peaks associated with the shocks produce local minima in the Alfv{\'e}n speed. This leads to the formation of CR ``bottlenecks'' and ``staircase'' distribution of CRs, where CR gradients are flat in between successive density peaks \citep[see also][we discuss CR bottlenecks in more detail in Sections~\ref{Interaction of CR with dense clouds}, \ref{porous}, and \ref{CR Eddington limit}\footnote{We note in passing that time-dependent simulations of \citet{dorfi_time-dependent_2012} and \citet{dorfi_time-dependent_2019}, which include both streaming and diffusion as well as shocks driven by SN explosions, do not exhibit the bottlenecks.}]{tsung_cosmic-ray_2022}. One implication of this effect is that, in the time-average sense, the CR pressure distribution is significantly flatter and better approximated as $P_{\rm cr}\propto\rho^{1/2}$ rather than $P_{\rm cr}\propto\rho^{2/3}$ expected in the sub-Alfv{\'e}nic regime for the usual steady-state streaming case (see above). The wind mass loss rate is significantly altered by these effects and is reduced by two orders of magnitude, while the scaling of $\dot{M}_{\rm w}\propto\varv_{\rm circ}^{-4}$ is much steeper than in the diffusion case (\citealt{quataert_physics_2022}, in the streaming case with isothermal equation of state; cf., \citealt{modak_cosmic-ray_2023}, who find that low power winds characterized by $\dot{M}_{\rm w}\propto\varv_{\rm circ}^{-2}$ develop in the streaming case with radiative losses included).\\
\indent
Overall, this illustrates the point that the evolutionary outcomes of galactic winds are highly sensitive to the physics of CR transport. While one-dimensional models are valuable tools that help to build physical intuition, interpret simulation results or validate codes, and may be useful in semi-analytic models of galaxy formation, they do rely on multiple approximations. These considerations therefore strongly motivate further research to elucidate the role of CR transport in time-dependent, multi-dimensional, and multi-physics high-resolution simulations. Below we begin a systematic discussion of these topics in increasingly complex systems.

\subsubsection{Dynamical impact of cosmic rays on multiphase medium}\label{multiphase}

In this section we primarily focus on the dynamical role of CRs in the ISM. We start by discussing the dynamics of CRs on very small scales near CR sources and then progress on to the topics related to systematically larger scales such as the dynamical interaction of CRs with individual clouds and CR dynamics in porous multiphase medium.

\begin{wrapfigure}{r}{0.5\textwidth}
  \begin{center}
    \includegraphics[width=0.45\textwidth]{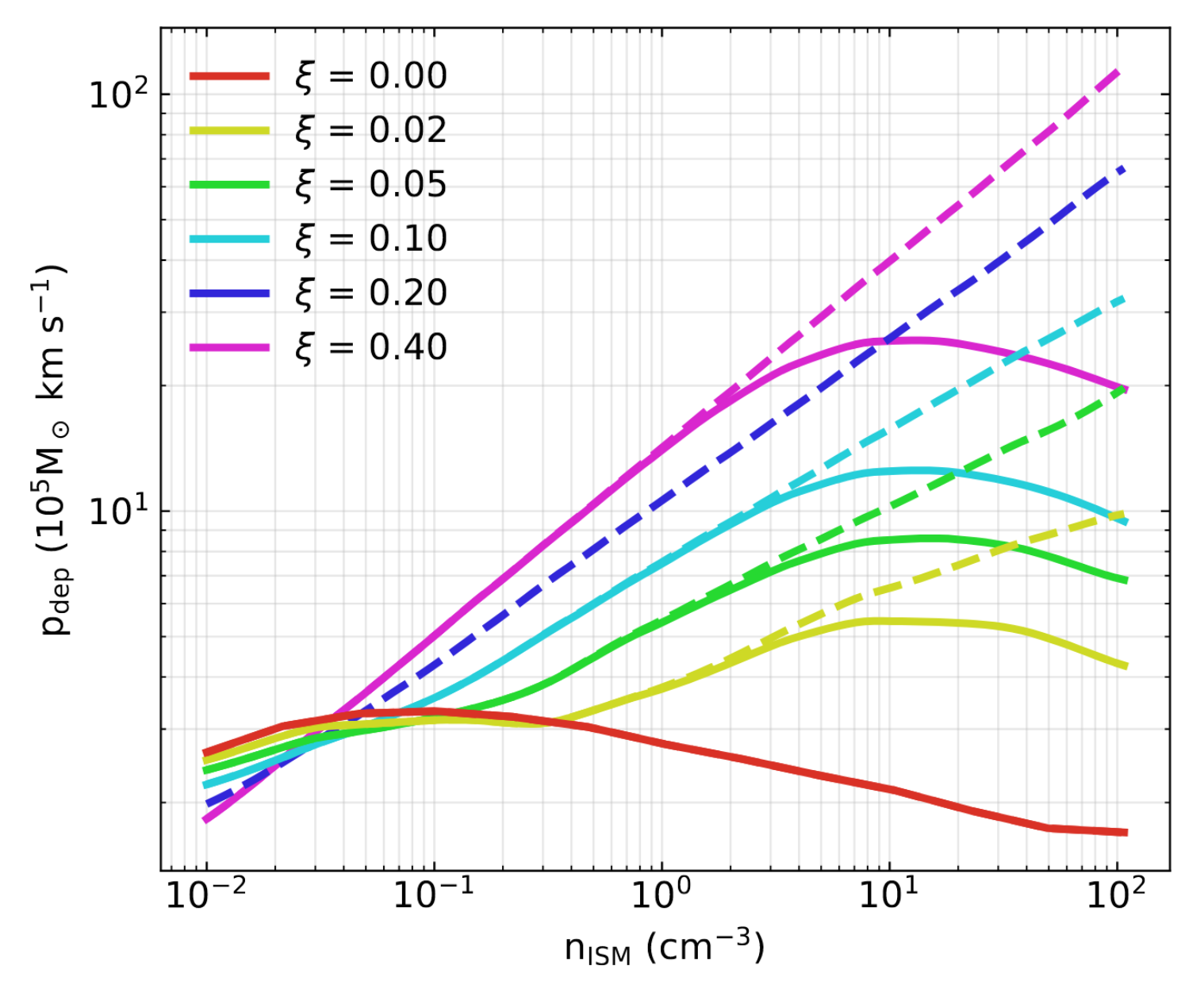}
  \end{center}
  \caption{Momentum $p_{\rm dep}$ deposited in the ISM by a SN as a function of the ISM density for different values of CR acceleration efficiency $\xi$. Solid an dashed lines correspond to the cases with (i) hadronic plus Coulomb CR cooling and (ii) without CR cooling, respectively. Note that the relative boost of CR momentum deposition at high gas densities is smaller when CR cooling losses are present. Image from \citet{diesing_effect_2018} reproduced with permission from PRL.  }
  \label{fig:diesing_caprioli}
\end{wrapfigure}

\paragraph{Dynamics of cosmic rays near their sources.}\ \label{dynamics_of_CR_near_sources}
CRs are efficiently produced in SNR shock waves (e.g., \citealt{blandford_particle_1987}; see \citealt{caprioli_cosmic-ray_2016} for a recent review), as well as winds from massive young stars \citep{bykov_nonthermal_2014} and jet termination and accretion shocks in young stellar objects \citep[e.g.,][]{padovani_cosmic-ray_2015, padovani_protostars_2016}. CRs affect the dynamical evolution of SNRs in two distinct ways. First, the adiabatic index of the CR fluid is smaller than that of the thermal gas and, consequently, the pressure inside the remnant is dominated by CRs late in the evolution. Second, CR energy losses are much smaller than those of the thermal gas during the snowplow phase, when the gas density becomes high enough and the temperature low enough for the SNR to enter the radiatively efficient stage of expansion. This has important consequences for the strength of stellar feedback in the ISM as CRs are able to help the SNRs expand into the ISM and impart more momentum onto the surrounding medium. This underscores the need to study the dynamical impact of CRs on SN feedback from first principles. In the remainder of this section we discuss the role of CRs in simulations of individual SNe and defer the discussion of CRs in shaping star formation to Section \ref{CR_and_star_formation}.\\
\indent
Using three-dimensional MHD simulations, \citet{pais_effect_2018} investigated the role of magnetic obliquity-dependent in situ acceleration of CRs at strong shocks in the early Sedov--Taylor stages of SN expansion. They demonstrated that at the locations of quasi-parallel shocks (i.e., where the shock normal is approximately parallel to the local magnetic field) efficient injection of CRs increases gas compressibility, which results in ellipsoidal shapes of SNRs. The average CR acceleration efficiency was about 30\% of the maximum efficiency expected from kinetic simulations independently of the coherence length of the magnetic field in the ISM. The impact of CRs at the later snowplow expansion stage was considered by \citet{diesing_effect_2018}, who also studied deposition of momentum from SN explosions in the presence of in situ acceleration of CRs at SNR shocks. Using one-dimensional spherically symmetric models and working in the thin shell approximation, they demonstrated that CR injection results in an effective momentum boost of a factor of a few beyond what is expected in a purely hydrodynamic case without CRs. This is quantified in Fig.~\ref{fig:diesing_caprioli}, which shows the momentum deposited in the ISM by a SN 
as a function of the ISM density for different values of CR acceleration efficiency. Interestingly, for extreme values of the CR acceleration efficiency, the boost factor increases by over an order of magnitude in the densest regions, where radiative losses of the gas are strongest, despite CRs experiencing hadronic and Coulomb losses. This work was extended by \citet{rodriguezmontero_momentum_2022}, who used three-dimensional MHD simulations of SN explosions in uniform media characterized by different gas densities, CR injection fractions (while not considering in situ acceleration), and strengths and topologies of the magnetic field. This work also included anisotropic CR diffusion and streaming. As expected, including CRs resulted in the enhancement in the momentum deposition late in the evolution of SNRs when CR pressure dominated the expansion. However, this enhancement was only at the level of 50\%. Interestingly, the strength of the effect was more pronounced for lower gas densities in contrast to the results from semi-analytic model of \citet{diesing_effect_2018}. Since the simulations included CR transport, the escaping CRs were able to accelerate the gas beyond the remnant. However, the question of the effective transport of CRs near sources is still open and the speed with which CRs escape SNRs is likely to affect the dynamics of the expanding gas. \\
\indent
Gamma-ray emission from the vicinity of star clusters \citep{aharonian_massive_2019} and enhanced $\gamma$-ray fluxes from molecular clouds located close to SNRs \citep{casanova_molecular_2010, hanabata_detailed_2014} suggests that CR diffusion could be suppressed near these sources. In the case of SNRs, the diffusion times of GeV CRs, that contribute most to the CR pressure, are indeed larger than the characteristic lifetimes of the remnants \citep{caprioli_cosmic-ray_2012}. Moreover, when the flux of CRs streaming away from their sources exceeds the local magnetic pressure, CRs can excite the non-resonant streaming Bell instability \citep{bell_turbulent_2004} and grow magnetic turbulence that scatters and effectively reduces the CR transport speed. This in turn can create inflated and overpressured bubbles that distort the magnetic fields in the direction perpendicular to the original CR flux. The ensuing turbulent motions in the bubble and the Bell instability amplify the fields, thereby effectively confining CRs to their sources as demonstrated using hybrid kinetic simulations \citep{schroer_cosmic-ray_2022, schroer_dynamical_2021}. This strong suppression of CR transport could result in the accumulation of significant grammage by the CRs and thus change the observed B/C and secondary-to-primary ratios \citep{cowsik_positron_2010, lipari_spectral_2018}, and possibly account for the excess $\gamma$-ray emission mentioned above. The characteristic sizes of these over pressurized bubbles can be as large 100 pc. Even in the absence of the Bell instability, the CRs streaming away from their sources could be subject to self-confinement due to the pressure anisotropy instability \citep{zweibel_role_2020}. 

\begin{figure}[tbp]
\begin{center}
\includegraphics[width=0.95\textwidth]{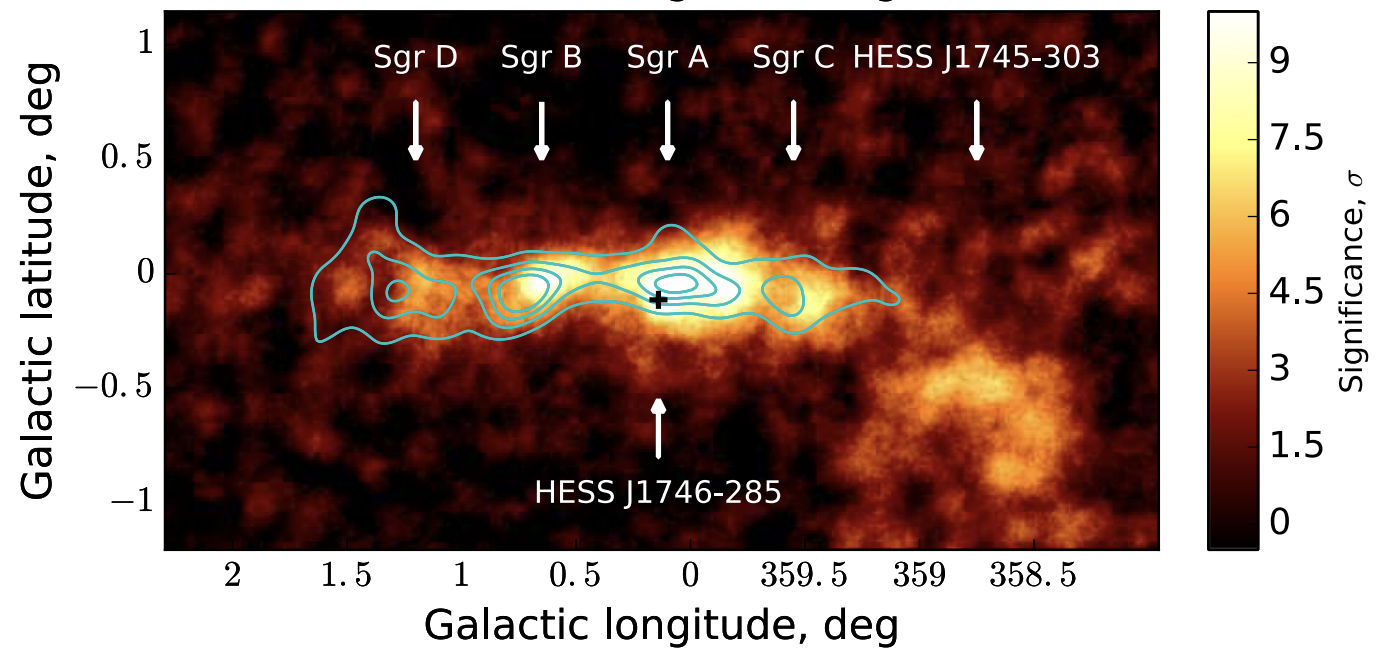}
\end{center}
\caption{Residual $\gamma$-ray image of the Galactic Center after subtraction of point sources. The cyan contours indicate molecular gas emission. Note the spatial coincidence of the emission components. Image from \citet{hess_collaboration_characterising_2018}; reproduced with permission from A\&A.}
\label{fig:HESS_clouds}
\end{figure}

\paragraph{Dynamical interaction of cosmic rays with dense clouds.}
\label{Interaction of CR with dense clouds}
As the CR distribution in the Galaxy is broader than the distribution of their likely sources \citep{strong_propagation_1998}, CRs must eventually break out from the vicinity of these sources and permeate the multiphase ISM. These escaping CR can then dynamically interact with the ISM. In particular, they can heat and impart momentum to cold clouds. Figure~\ref{fig:HESS_clouds} shows an example of $\gamma$-ray and molecular emission from ISM clouds \citep{hess_collaboration_characterising_2018}. The fact that these two emission components coincide spatially strongly indicates that CRs interact with the cold ISM clouds. \\
%
\indent
The process of CR--cloud interaction can be associated with the formation of so-called \textit{CR bottlenecks} that we discuss next. When CRs stream down their pressure gradient and encounter a cold cloud, the local Alfv{\'e}n speed at the cloud location will be reduced in comparison to the ambient medium provided that the magnetic field strength does not increase with the ion mass density faster than $\rho^{1/2}$. There is no consensus in the literature regarding the exact slope of the relationship between the strength of the magnetic field and gas density: numerical \citep{li_magnetized_2015, mocz_moving-mesh_2017} and observational \citep{crutcher_magnetic_2010, tritsis_magnetic_2015, myers_magnetic_2021} studies suggest a range of values that lead to the Alfv{\'e}n speed either increasing or decreasing with the density. While the question of the slope is not settled, the scatter in the relationship may nevertheless result in local dips in the Alfv{\'e}n speed. As long as CRs are not prevented from entering the clouds due to magnetic draping, they can penetrate the clouds. CRs that stream down their pressure gradient and encounter smaller values of the Alfv{\'e}n speed begin to slow down, provided they are efficiently coupled to the gas via frequent scatterings. As a result, the CR pressure builds up in the upstream region \citep[thus forming a bottleneck; ][]{{skilling_cosmic_1971}}.
Since CRs cannot stream up their pressure gradient, their distribution adjusts so that the pressure becomes flat. This in turn implies that CRs dynamically decouple from the gas upstream of the cloud and can no longer heat and exert pressure on it. However, the coupling is restored downstream of the minimum of the Alfv{\'e}n speed \citep{wiener_interaction_2017,jiang_new_2018}. As a corollary, if a cloud has an inverted Gaussian Alfv\'en-speed profile, it will start to disperse as the CR pressure force only acts on the back side of the cloud. However, for an inverted top hat Alfv{\'e}n-speed profile, a constant CR pressure gradient develops 
across the cloud, which does not get dispersed and instead experiences a spatially uniform accelerating force and collisionless heating of the gas
\citep{thomas_finite_2021}.\\
\indent
The above considerations have important implications for the launching of multiphase outflows from galaxies. While we discuss this topic in more detail in Section~\ref{wind_launching}, here we merely argue that CR forces imparted on to the clouds can be sufficiently strong so as to render them a promising agent for accelerating cold clouds away from the disk. More broadly, there are a few classes of competing ideas for explaining fast multiphase outflows, which we now list before briefly discussing some of their limitations: 
(i) \textit{in situ formation}, where the cold clouds form by cooling out of the hot wind emerging from a galaxy \citep[e.g.,][i.e., in this scenario, the cold gas is \textit{not} accelerated but instead forms in an already moving hot medium]{thompson_origin_2016, scannapieco_production_2017, schneider_production_2018}, 
(ii) acceleration by a \textit{hot wind}, where cold gas clouds exposed to the hot galactic wind are accelerated by the wind ram pressure \citep[e.g.,][]{klein_hydrodynamic_1994},
(iii) acceleration by \textit{magnetic draping}, where the cold gas clouds are draped by magnetic fields from the hot phase, a process that transfers the momentum from the hot wind to the draped cold gas \citep{lyutikov_magnetic_2006,dursi_draping_2008},
(iv) acceleration by \textit{radiation pressure}, where cold gas clouds exposed to intense radiation from massive stars are accelerated as a result of scattering and absorption of radiation by dust grains embedded in the clouds \citep[][the efficiency of this process may be reduced due to relatively weak coupling of radiation to gas; see Section \ref{CR_and_star_formation}]{murray_radiation_2011,hopkins_self-regulated_2011}, 
(v) acceleration due to \textit{CR bottlenecks}, where the acceleration is due to CR pressure forces inside the cold clouds \citep[e.g.,][]{wiener_interaction_2017}.\\
\indent 
One fundamental problem with the acceleration models is that they are plagued by instabilities that tend to disrupt the clouds. In the hot wind scenario, the characteristic timescale for the acceleration of the clouds is longer than the ``cloud crushing time'' over which the clouds are shattered \citep{klein_hydrodynamic_1994}. The severity of this problem can be reduced by invoking supersonic flows past the clouds \citep{scannapieco_launching_2015} or thermal conduction that forms an evaporative interface that helps to shield the clouds from destruction \citep{bruggen_launching_2016}. 
Magnetic fields could also suppress the instabilities operating at the interfaces of the clouds and the ambient medium \citep{dursi_bubble_2007}, but their impact on the survivability of the clouds is unclear \citep{mccourt_magnetized_2015, cottle_launching_2020}. 
Alternatively, the shattering problem could be avoided if radiative cooling is fast enough to lead to recondensation of the clouds \citep[e.g.,][]{gronke_how_2020, farber_survival_2021}. While larger clouds exposed to the wind can shatter into smaller cloudlets via the Kelvin-Helmholtz instability \citep{mccourt_characteristic_2018,sparre_physics_2019}, these cloudlets mix with the hot wind to form a thermally unstable warm phase that quickly cools and joins the cool phase thus re-forming the clouds. The net effect is that the momentum of the hot wind is transferred to the cold phase that survives the exposure to the hot wind \citep{gronke_growth_2018,gronke_how_2020,sparre_interaction_2020,li_survival_2020,girichidis_situ_2021,farber_survival_2021}.\\
\indent
As mentioned above, cold clouds can be accelerated by CR pressure gradients. While an accelerating cold cloud exposed to a CR gradient can also undergo shattering, the tendency of a cloud to disrupt can be suppressed if the Alfv{\'e}n speed profile across the cloud has sharp edges at the cloud interface \citep{thomas_finite_2021}. Furthermore, even when the Alfv{\'e}n speed profiles do not resemble inverted top hat profiles, two-dimensional simulations with radiative cooling demonstrate that the clouds can survive, and the survival is even possible despite CR heating operating at cloud-ISM interfaces \citep{wiener_cosmic_2019}. In the simulations of this process without CR heating, the cloud destruction is primarily driven by CR streaming on the sides of the cloud, which results in a significant shear at the boundary layer \citep{bruggen_launching_2020}. However, the cloud destruction rates can be slow enough to allow for cloud survival and for cold cloud acceleration to hundreds of km s$^{-1}$ as seen in the observed galactic winds.

\begin{figure}[tbp]
\begin{center}
\includegraphics[width=0.95\textwidth]{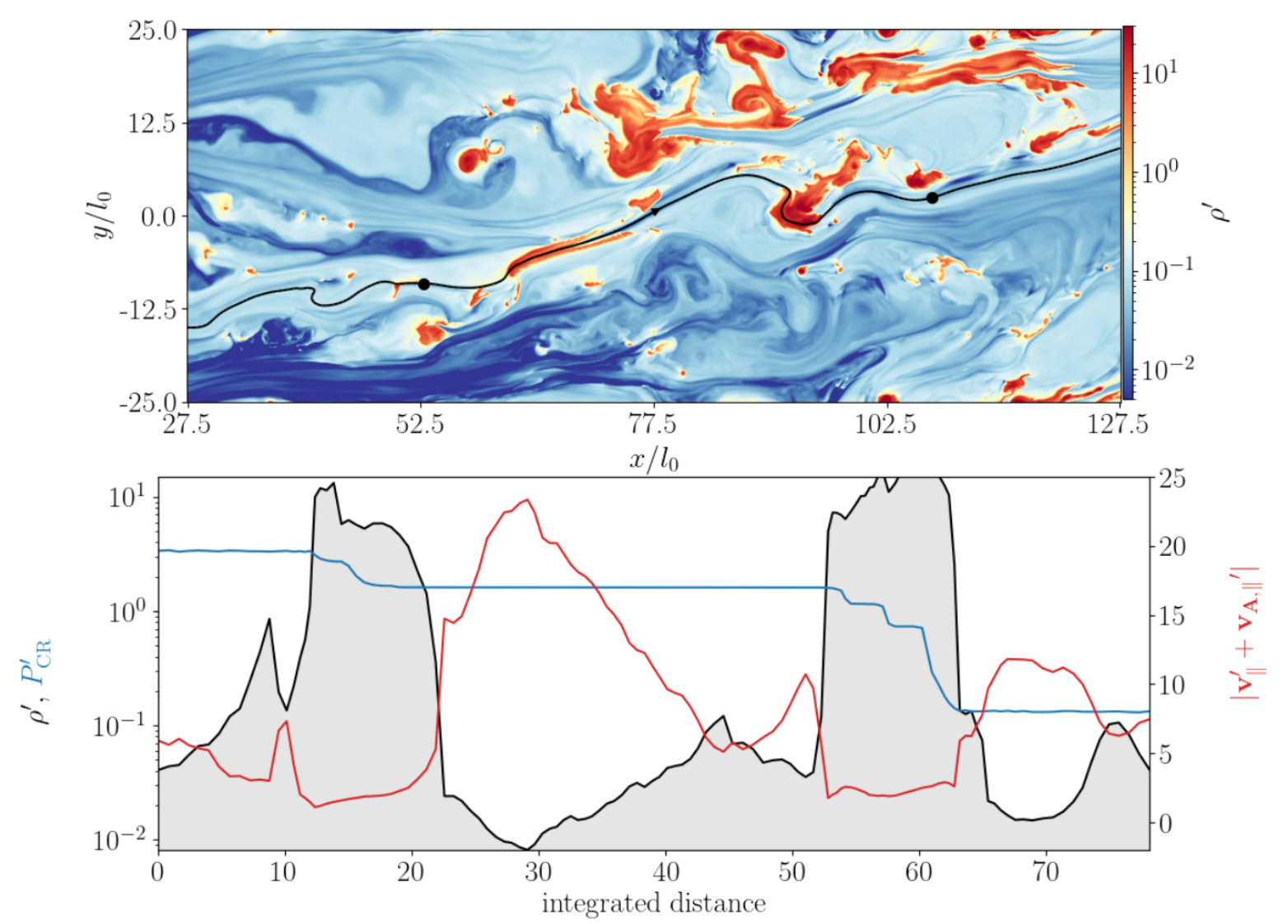}
\end{center}
\caption{\textit{Top panel:} thermally unstable gas density distribution exposed to CR flux. 
\textit{Bottom panel:} gas density, CR pressure, and projected gas velocity along a magnetic field line 
(magnetic field line is denoted as a solid line in the top panel). Formation of flat CR plateaus and bottlenecks coincident with gas density maxima is evident in the bottom panel. Image from \citet{huang_cosmic-ray-driven_2022} reproduced with permission from ApJ.}
\label{fig:huang}
\end{figure}

\paragraph{Dynamical impact of cosmic rays in porous multiphase media.}\label{porous}
In realistic situations, CRs interact with multiphase gas characterized by a range of densities, temperatures, and ionization fractions. Additional physical processes shape the dynamical interactions of CRs with such porous media. Specifically, in the mostly neutral gas phase, the effective CR streaming speed can vastly exceed the speed in the fully ionized gas because, in the self-confinement model in the fully coupled regime, CR transport proceeds at the ion Alfv{\'e}n speed, i.e., $\varv_\rmn{a}=B/\sqrt{4\pi\rho}$, where $B$ is the magnetic field strength and $\rho$ is the ion mass density. Furthermore, in the weakly ionized phase, ion--neutral friction can very efficiently damp self-generated Alfv{\'e}n waves on which CRs scatter \citep{kulsrud_effect_1969}. Consequently, in the mostly neutral gas, CR can propagate much faster than in the fully ionized medium. These processes have interesting implications not only for the CR transport but also for the dynamical coupling of CRs to the gas and heating of the gas by CRs. \\
\indent
The inclusion of ion--neutral damping leads to subtle yet important modifications to the bottleneck process described above. As CRs stream down their pressure gradient toward an overdense cloud, they propagate in a progressively denser medium, and as long as the magnetic field remains constant, the Alfv{\'e}n speed decreases. As before, this creates a bottleneck effect and the CR pressure flattens in the upstream region and drops significantly past the minimum in the transport speed. However, the transport speed eventually increases significantly as CRs propagate deeper into the cloud because the ionization fraction drops and the ion--neutral friction damps the confining waves. This effect can overcompensate for the initial drop in the transport speed. Consequently, this very fast transport inside the cloud leads to the reduction in the CR pressure. As the CRs traverse the cloud, they eventually encounter the far side of the cloud, where the density decreases, ionization fraction increases, and ion--neutral damping becomes ineffective again. The latter of these three factors is decisive in controlling the effective CR transport speed and CRs reestablish strong coupling with the gas on the far side of the cloud. This results yet again in the reduction of the CR transport speed and, thus, a second bottleneck \citep{bustard_cosmic-ray_2021}. Consequently, the reduced CR pressure remains flat inside the cloud just as it did in front of the leading edge of the cloud, which leads to a stair-step distribution of the CRs across the cloud.
It is worth pointing out that the CR pressure profile in and near the cloud is quite sensitive to how well the cloud-intercloud interface is resolved.\\
\indent
Substantial CR collisionless losses can occur at the locations of sharp CR pressure gradients at cloud edges. These losses are further increased because of the ionization fraction dependence of the collisionless heating, $|\bs{\varv}_\rmn{a}\bs{\cdot} \bs{\nabla} P_{\rm cr}|$. In addition to collisionless losses, CR also experience substantial hadronic and Coulomb losses. In objects that are CR calorimeters, by definition, these energy losses are comparable to the CR energy production rate in SNe \citep{lacki_-ray_2011, yoast-hull_cosmic_2015, yoast-hull_equipartition_2016, krumholz_cosmic_2020}. Despite the complexities introduced to CR transport by the varying ionization fraction, the overall CR losses may be only weakly dependent on this factor. In the low ionization case, the decrease in the CR residence time in the clouds, and the associated reduction in the collisional losses, can be compensated for by the CR probing denser medium in which the losses are larger and by the fact that collisionless losses can occur on both cloud-ISM interfaces rather than just one \citep{bustard_cosmic-ray_2021}. These considerations have implications for the $\gamma$-ray emission which probe hadronic interactions of CRs with the gas, which we comment on in Section~\ref{Non-thermal emission from galaxies}.\\
\indent
The dynamical impact of CRs can also be influenced by the ionization structure of the multiphase medium. In terms of the effective momentum imparted on to the gas by CRs, the outcomes can be bracketed by two extreme limits. When the medium is very close to being neutral, ion--neutral damping can be very effective throughout the clouds thus completely flattening CR pressure profiles in the cold and dense phase and making CR driving of outflows from the galaxy ineffective \citep{everett_interaction_2011}. On the other hand, if the medium is fully ionized, CR bottlenecks form and, as mentioned above, the clouds can be accelerated to astrophysically relevant speeds \citep{wiener_interaction_2017, wiener_cosmic_2019,bruggen_launching_2020, thomas_finite_2021}. Interestingly, in the intermediate case of partial ionization, cloud acceleration rates remain largely unaffected by the level of ionization and the clouds can be accelerated to velocities still comparable to those found for full ionization cases \citep{bustard_cosmic-ray_2021}.\\
\indent
Controlled and highly idealized experiments of individual clouds are helpful for developing physical intuition and interpreting simulations in which the formation of the multiphase gas is computed from first principles. Simulations of thermal instability in magnetized media with CRs naturally predict phase space distributions of such cold clouds 
\citep[e.g.,][see also Fig.~\ref{fig:huang}, which shows density, velocity, and CR pressure distributions in thermally unstable gas exposed to CR flux]{sharma_thermal_2010, huang_cosmic-ray-driven_2022}.
Such simulations reveal that CRs reduce the density contrast between the clouds and the hot ambient medium and that the clouds can have very filamentary morphology. In a purely hydrodynamical cases, when clouds are exposed to an incoming wind, they experience the strongest acceleration on the side exposed to the wind. However, when the clouds are instead exposed to CR flux, bottleneck formation can help to accelerate the clouds. Assuming an Alfv\'en-speed profile that follows an inverted Gaussian, acceleration tends to be stronger on the far side of the clouds with respect to the CR source, thus producing a differential force stretching the clouds, which can make the clouds filamentary. The net acceleration of the multiphase gas formed via thermal instability tends to be suppressed compared to the expectations based on idealized experiments described above. This can be attributed to two factors. First, because in such thermally unstable environments magnetic fields are tangled, and CR transport is anisotropic, the forces associated with the bottlenecks tend to push the gas both in the direction parallel and perpendicular to the overall wind direction. Second, the strength of the bottlenecks is now reduced because, in the process of thermal instability development, the decrease of the Alfv{\'e}n speed in the clouds due to elevated density is partially compensated for by the increased magnetic field strength in the cold phase. While the momentum imparted to the gas via CR pressure forces is reduced, the velocity dispersion of the gas increases. Interestingly, the impact of the CR flux on the acceleration of the cold gas is nevertheless significant because CR heating in the hot gas increases thermal pressure that in turn is able to accelerate the cold gas to the observed velocities \citep{huang_cosmic-ray-driven_2022}. 

\subsubsection{Parker instability and cosmic rays}
\begin{figure}[tbp]
\begin{center}
\includegraphics[width=0.8\textwidth]{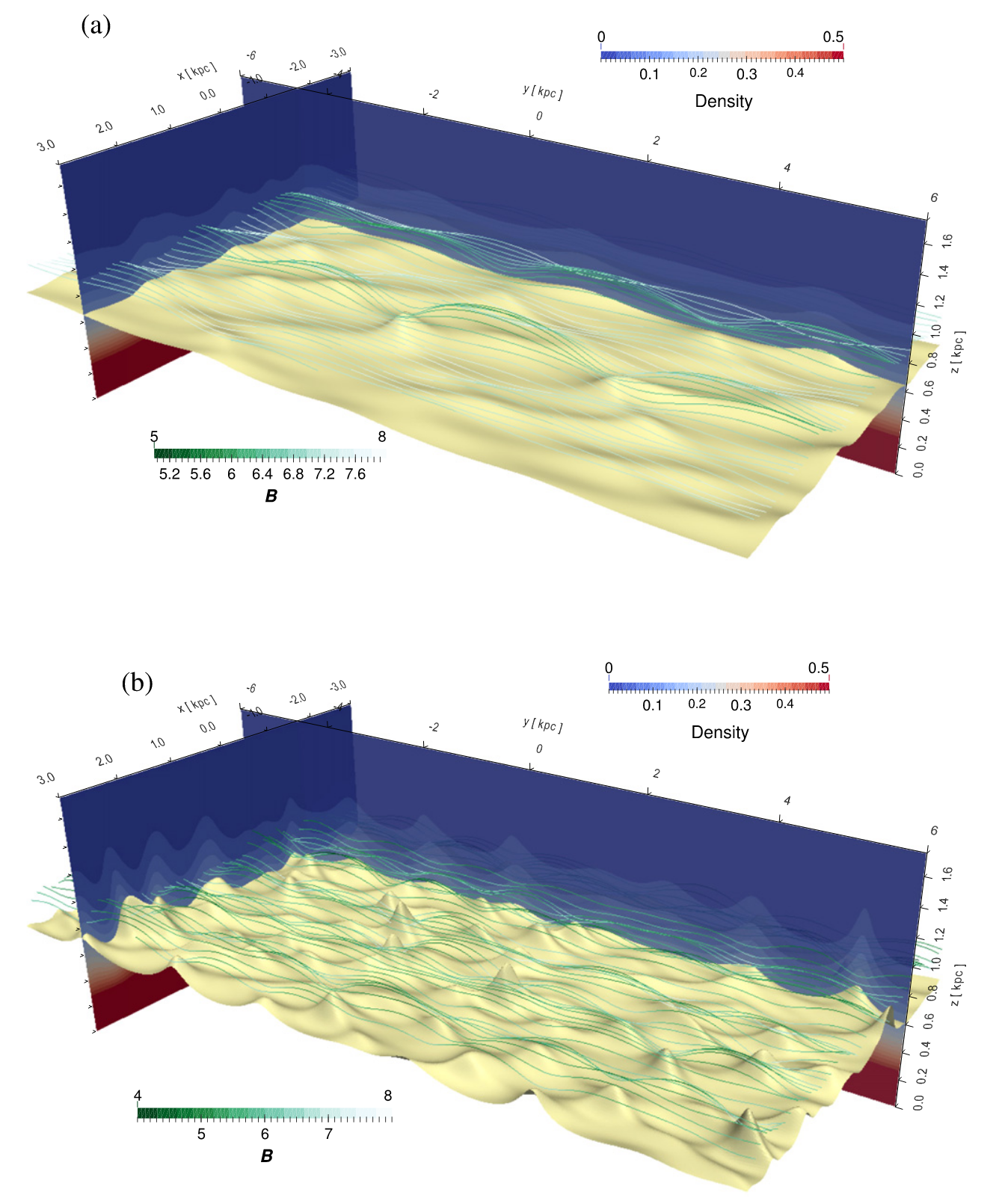}
\end{center}
\caption{Iso-density surfaces in a gravitationally stratified patch of a galactic disk (yellow surfaces), density slices (vertical planes), and magnetic fields (green lines) in simulations with and without CRs (top and bottom panel, respectively; the simulation with CRs includes anisotropic diffusion).  
Note that including CR physics makes the gas more unstable and leads to prominent magnetic ``Parker loops.'' Image from \citet{rodrigues_parker_2016}; reproduced with permission from ApJ.}
\label{fig:rodriguez}
\end{figure}

We now turn our attention to larger length scales and discuss the impact of CRs on the stability of gravitationally stratified patches of the ISM. These considerations are highly relevant for a range of topics including regulation of star formation, magnetic dynamos, or polluting of the CGM with magnetic fields and CRs, where the latter may dominate pressure support. While we defer the discussion of these topics to later sections, here we focus specifically on the triggering and development of the Parker instability and the role that CR physics plays in it. \\
\indent
In the solar neighborhood, typical values of the energy densities of CRs, magnetic fields, radiation field, and kinetic energy of the gas are comparable and at the level of 1 eV cm$^{-3}$ \citep[e.g.,][]{boulares_galactic_1990, badhwar_hydrostatic_1977}. One consequence of this apparent equipartition is that CRs and magnetic fields can play a dynamically important role in setting up gravitational equilibrium in the gaseous component of galactic disks \citep{badhwar_hydrostatic_1977,ghosh_role_1983,chevalier_cosmic-ray_1984}. 
We note in passing that, while equipartition may be a reasonable assumption in normal star forming galaxies (but see the discussion in Section~\ref{sec:magnetic equipartition}), it may not hold in starburst galaxies \citep{yoast-hull_equipartition_2016}. In the latter case, high gas densities and mass loss due to winds, can lead to significant CR energy density losses. Consequently, magnetic fields need to be much stronger in order for the synchrotron emission to match the observed radio luminosities. This also implies that magnetic fields must be strongly correlated with the sites of significant star formation. In such situations, magnetic energy density may in general be out of equipartition and may significantly exceed the CR energy density.\\
\indent
Since the magnetic fields and CRs are essentially weightless components of the ISM, a gravitationally stratified ISM may become unstable to a buoyancy instability that is analogous to the Rayleigh-Taylor instability that is triggered when a heavier fluid is placed on top of a lighter one. In the context of magnetized medium with CRs, this instability is known as the Parker instability \citep{parker_dynamical_1966}. In a Parker unstable system, the gas begins to flow down along a perturbed magnetic field. In the process of infall, gravitational potential energy is released and spent on compressing the gas and counteracting magnetic tension. 
Parker's instability has been demonstrated to play a role in the magnetic field amplification by dynamo action and in the transport of magnetic fields and CRs out of the disk \citep{hanasz_cosmicray_2000,hanasz_building_2004,hanasz_amplification_2004}, which could aid magnetic field growth via the small-scale dynamo \citep{brandenburg_astrophysical_2005,pakmor_magnetic_2017,pfrommer_simulating_2022} and a large-scale $\alpha$-$\Omega$ dynamo \citep{ntormousi_dynamo_2020}. While it was originally envisioned that Parker's instability could be responsible for the formation of molecular clouds, in the modern picture, supersonic ISM turbulence driven by gravitational collapse and stellar winds or supernova explosions regulate star formation \citep{mac_low_control_2004, krumholz_general_2005}, and GMCs are only transient objects that form as a result of converging turbulent flows and only survive a few crossing time scales, which may explain the small star formation efficiency.\\
\indent
In the classic treatment of the Parker instability, \citet{parker_dynamical_1966} assumed that the CR pressure is constant along all magnetic field lines. This assumption is justified when the speed of sound of the composite fluid is much larger than any characteristic velocities in the problem and the CR fluid is not affected by gravity. These assumptions can be relaxed in several ways by considering general CR pressure perturbations, which leads to several interesting consequences that we discuss next. In brief, while treating CRs as a compressible fluid described by an adiabatic index $\gamma_{\rm cr}>0$ makes the gas less susceptible to the Parker instability, while including non-adiabatic effects renders the gas more unstable.\\
\indent
First, the classic treatment can be extended by approximating the CR fluid as a relativistic fluid with $\gamma_{\rm cr}=4/3$, which makes the gas less compressible. This in turn implies that more work needs to be done by gravity to compress the gas, which makes it less susceptible to the development of the instability \citep{zweibel_stabilizing_1975,boettcher_testing_2016,heintz_parker_2018}. 
This could explain why, in simulations of stratified disks involving adiabatic CRs, the atmospheres can remain puffed up by CR pressure rather than become unstable to the Parker instability \citep[e.g.,][see also Section~\ref{wind_launching} below]{uhlig_galactic_2012,ruszkowski_global_2017}. \\
\indent
Second, including non-adiabatic effects, such as radiative cooling, CR diffusion, streaming, and thermal conduction have destabilizing effect.
The impact of radiative cooling is easiest to grasp. Radiative losses make the gas more compressible, which makes it easier for the gas to flow deeper into the valleys formed by the magnetic field and accelerate the instability.
CR transport via anisotropic diffusion along the magnetic fields also increases the growth rate of the Parker instability. As the coupling of CRs to the gas is reduced (i.e., diffusion is faster), so is the steepness of the CR pressure gradient that opposes gravity, which allows the gas to fall down more easily. Consequently, the growth rate of the instability increases with the parallel diffusion coefficient as demonstrated both by linear stability analysis and numerical simulations (e.g., 
\citealt{ryu_effect_2003,kuwabara_nonlinear_2004,rodrigues_parker_2016}; but see \citealt{hanasz_incorporation_2003}, who reported the opposite trend). 
This is evident in Fig.~\ref{fig:rodriguez} that illustrates the stark differences between the simulations with and without CR physics. In particular, the emergence of prominent magnetic ``Parker loops'' is clearly visible in the bottom panel that shows the results from CR simulations that include anisotropic CR diffusion. 
%
This study was further extended to include self-gravity of the gas and it corroborated earlier findings regarding the impact of diffusion on the growth rates \citep{kuwabara_parkerjeans_2006,kuwabara_dynamics_2020}. This work also demonstrated that CRs can effectively oppose gravity and suppress the Jeans instability. For low parallel diffusion, large CR gradients can be established that prevent the gas from collapsing, which leads to dense filaments aligned with the local direction of the magnetic field. In the opposite case of fast diffusion, the gas slides along the magnetic field and forms filaments perpendicular to it. 
Including CR transport via streaming further destabilizes the system. By invoking similarities to overstability of acoustic waves driven by CR streaming \citep{drury_stability_1986,begelman_acoustic_1994}, and considering non-radiative simulations of the Parker instability with streaming, \citet{heintz_parker_2018} identified the CR streaming instability heating as an important destabilizing process. They noted that capturing the instability in this case requires (volume-filling) numerical resolution better than 100 pc, which is higher than in the case of classic Parker instability. However, since the instability growth rates can be comparable for both the diffusion and streaming cases \citep{heintz_role_2020}, the destabilizing impact of CR streaming heating and the exact nature of transport may be secondary for the development of the Parker instability, i.e., the instability can develop more easily as long as \textit{some} form of transport is present that allows CRs to escape from the compressing gas regions.\\
\indent
The ISM may be even more unstable to the Parker instability in the presence of efficient anisotropic thermal conduction. The origin of this instability can be understood by considering a horizontal magnetic field in a gravitationally stratified medium in which temperature increases in the direction of gravity. As the field is perturbed in the vertical direction, the heat begins to flow along the field in the upward direction. If the process is isobaric, then the displaced parcel of gas becomes hotter and less dense, and thus more buoyant. While this instability was considered by \citet{balbus_stability_2000,balbus_convective_2001} in the context of high plasma $\beta$ medium, such as the ICM, it has been used to generalize the Parker instability in the ISM by \citet{dennis_parkerbuoyancy_2009}, who demonstrated that the ISM is more prone to the Parker instability in the presence of anisotropic conduction.\\
\indent
We note that the regime of applicability of the above models is limited by the fact that they neglect driving of supersonic turbulence by SNe and superbubbles in the ISM. In the environments where this driving is important, slow buoyant motions are subdominant compared to fast advection, and the Parker instability is unlikely to operate. However, in the coronal hot phase regions of the ISM, the gas may be Parker unstable. The outcome will therefore depend on a combination of uncertain factors such as the ratio of the time between superbubble explosions and the instability growth timescale, spatial separation between superbubble locations, and the properties (temperature, velocity dispersion) of the phase ISM relatively unaffected by SN activity. If these complications can be neglected, in the saturated state of the Parker instability, large magnetic loops extending to significant heights above the disk midplane can be formed. These loops could transport both the magnetic fields and CRs away from the disk, possibly leading to CR-dominated galactic halos. These processes are closely connected to the issue of galactic wind launching that we discuss in the next section.

\subsubsection{Wind launching and simulations of disk patches and stratified boxes}\label{wind_launching}
In this section we begin the discussion of the impact of CRs on the scales intermediate between the ISM scales responsible for star formation in molecular clouds and the large scales of the CGM. Specifically, we discuss what role CRs can play in shaping the smooth transition between the disk-like distribution of the ISM and the more spherically distributed CGM surrounding the galaxy. One of the key processes shaping this transition is the process of galactic fountain and wind launching. We start by discussing the basic mechanisms behind CR wind driving and the crucial role that CR transport plays in this process. We then examine the consequences of the physics of wind launching for star formation, mass and energy loading in the wind, and the phase space distribution of the gas in the ISM and the wind immediately above the disk.

\paragraph{CR Eddington limit and CR staircases.} \label{CR Eddington limit}
As discussed above, CRs can exchange momentum with the gas via pitch angle scattering off of Alfv{\'e}n waves and thus accelerate the gas. In a pioneering work,  \citet{socrates_eddington_2008} suggested that if the CR energy generation rate in a galaxy (i.e., CR luminosity) exceeds a critical value, then the gas can be unbound. This argument is analogous to the Eddington luminosity argument with the difference being that photons are replaced by different relativistic particles, i.e., by CRs. 
The critical CR luminosity $L_{\rm Edd, cr}$ above which CR forces exceed gravitational forces can be easily obtained in the diffusion limit \citep[cf.][similar expressions to the ones presented below can be obtained for the streaming-dominated case]{huang_launching_2022}. In this case, balancing CR flux with the critical flux needed to offset gravity, and assuming spherical symmetry, we have
\begin{equation}
\frac{L_{\rm Edd, cr}}{4\pi r^{2}}\bs{\hat{r}} = -\frac{\kappa}{\gamma_{\rm cr}-1}\bs{\nabla} P_{\rm cr} = -\frac{\kappa}{\gamma_{\rm cr}-1}\rho \bs{g},
\end{equation}
where $\kappa$ is the CR diffusion coefficient, $P_{\rm cr}$ is CR pressure, $\rho$ is the gas density, and $\bs{g}$ is the gravitational acceleration. For a point source of gravity, the Eddington luminosity is
\begin{equation}
L_{\rm Edd, cr} = \frac{4\pi GM \kappa \rho}{\gamma_{\rm cr}-1}. \label{edd}
\end{equation}
Analogous expression can be derived for radiation, in which case $\kappa\rho/(\gamma_{\rm cr}-1)=m_{\rm p}c/\sigma_{\rm T}$, for Thomson scattering. If $\kappa$ is constant (as is often assumed in simple CR propagation models), then equation~\eqref{edd} suggests that lower density gas should be more easily accelerated. This is in contrast to the radiation case, whereby photons tend to transfer little momentum to the gas as they escape the atmosphere through low density channels.
This comparison can be put on a more quantitative footing by expressing the Eddington limit in terms of mean free path $\lambda_{\rm mfp}$ of CRs or photons. In both cases, respective $L_{\rm Edd}\propto \rho \lambda_{\rm mfp}$, and consequently the ratio of CR and radiative Eddington limits is $L_{\rm Edd,cr}/L_{\rm Edd,\gamma}\sim\lambda_{\rm mfp,cr}/\lambda_{\rmn{mfp,}\gamma}$. 
Based on the spallation argument presented in the Introduction (and detailed in Section~\ref{sec:B-to-C_ratio}), the relatively long residence time $\tau_{\rm esc}\sim 3\times10^{7}$~yr of CRs near the disk midplane (compared to the light crossing time $\sim H/c\sim 2\times 10^{4}$ yr; assuming a height of the CR scattering halo of $H\sim 5$~kpc) translates to a short mean free path of CRs $\lambda_\rmn{mfp,cr}\sim 3H^{2}/c\tau_{\rm esc}\sim 8$~pc. Since this mean free path is much shorter than the Thomson mean free path for photons ($\sim 500$ kpc for typical ISM density of 1 cm$^{-3}$), the CR Eddington limit is much lower than the usual limit for radiation. Above this CR limit, the gas cannot exist in hydrostatic equilibrium and a wind should develop. Since the production rate of CRs is closely linked to the star formation rate, this limit corresponds to a critical star formation rate above which winds should be launched.
We note that, just as in the case of the classic derivation of the Eddington limit for radiation, the CR Eddington limit is not strict due to a number of simplifying assumptions such as spherical symmetry, neglect of inelastic energy losses, spatial variation in opacity that can flatten pressure profiles, etc.\\
\indent
Whether the wind is actually launched depends on a variety of factors. Using one-dimensional steady state calculations, \citet{heintz_galaxies_2022} generalized the above argument to include gravitational potential from a spherical mass distribution at the galactic center and the dark matter in the halo. By resorting to the solar wind analogy \citep{parker_dynamics_1958}, where the solar wind is expected to be launched when the asymptotic pressure in the solar corona exceeds that of the interstellar space, they compared asymptotic CR pressure to the CGM pressure to determine the conditions for CR wind launching. By examining the role of different CR transport mechanisms, \citet{heintz_galaxies_2022} found that solutions were strongly dependent on the physics of CR transport and that winds were only likely to develop in the presence of streaming. Moreover, while they confirmed that the Eddington limit does exist, it was not likely to be exceeded in realistic galaxies unless gas densities were reduced below the observed average values.\\
\indent
Nevertheless, in realistic multiphase environments, it is plausible that CRs could be transported to lower density regions and drive the winds for a wider range of parameters. More broadly, the work of \citet{heintz_galaxies_2022} therefore underscores the need for considering the role of multidimensional and time-dependent effects and the rich physics of CR transport in launching the winds. The work of \citet{huang_launching_2022} makes a step in that direction by performing two-dimensional MHD simulations of stratified atmospheres including radiative cooling and CR transport. In their simulations, CRs are injected at the midplane of the atmosphere and propagate either via diffusion or streaming. They find that the Eddington limit can be reached for a broad range of star formation environments and that winds can be launched when CR transport is governed by diffusion or streaming. In simulations with diffusion, streaming heating is absent and all of the CR energy can be used to increase the kinetic energy of the gas, thus making the gas acceleration more efficient. In simulations with streaming, they observed growth and saturation of the CR acoustic instability. This instability has the potential to considerably alter the picture of CR wind acceleration. \\
\indent
In general, the CR acoustic instability arises as a result of a relative displacement due to streaming or diffusion of the CR fluid with respect to the gas during the propagation of a sound wave. In the diffusion-dominated regime, this instability only occurs in rare cases when the CR pressure scaleheight is shorter than the CR diffusion length \citep{drury_stability_1986} and, consequently, it is not seen in the simulations of \citet{huang_launching_2022}. In the streaming-dominated regime, the instability can occur even in the presence of weak CR background gradient \citep{begelman_acoustic_1994}. In this case, the relative shift between CRs and the gas, combined with the CR streaming heating, leads to effective forcing of the gas that may either amplify the oscillations in the gas density, if the forcing is in phase with the waves, or damp them in the opposite case. As demonstrated by \citet{tsung_cosmic-ray_2022} via one-dimensional time-dependent simulations, as the waves steepen, they form sequences of shocks in the gas. The density jumps associated with these shocks may result in the formation of CR bottlenecks. This implies that the CR distribution develops plateaus interspersed by sharp jumps at the locations of the shocks. Consequently, CRs are dynamically decoupled throughout most of the volume but exert strong forces at the locations of the jumps in CR pressure. While the gas is not heated by CRs in the plateau regions, it is subject to intense heating at the shock locations. Furthermore, since radiative cooling is not offset by CR heating at the plateaus, the gas can be severely out of thermal equilibrium, which can facilitate the formation of additional shocks. Overall, the transfer of energy and momentum from CRs to the gas is expected to be highly spatially and temporally dependent. The formation of the bottlenecks and CR decoupling at stair-like plateaus in the distribution of CRs was also observed by \citet{huang_launching_2022}. They also determined that the CR heating of the gas is dominated by that occurring near the shocks rather than the more volume-filling regions characterized by shallower CR gradients. Despite these intricacies of CR interactions with the gas, the numerical experiments by \citet{huang_launching_2022} show that CR and the gas are nevertheless well coupled dynamically -- the efficiency of momentum transfer was found to be nearly perfect irrespectively of whether CR transport occurred via diffusion or streaming.  Overall, these simulations suggest that CR fluxes can achieve super-Eddington levels. The impact of CR staircase formation and CR transport on launching highly variable outflows is further discussed in Section~\ref{1D_models}.\\
\indent
The level of sophistication of the models presented above can be increased by considering a number of factors. First, the ability of bottlenecks to impede CR transport is likely to be reduced in three-dimensional simulations compared to those performed in one and two dimensions because the volume-filling factor of the plateaus would be significantly reduced. Second, the severity of the impact of the bottlenecks can be further affected by considering self-consistent simulations of thermal instability in which the magnetic field is allowed to increase with the gas density. Allowing for this effect would reduce the frequency with which the bottlenecks occur because Alfv{\'e}n speed minima would be less deep. Third, the level of decoupling of CRs from the gas can be further affected by, e.g., ion--neutral damping. Further work is needed to ascertain the impact of these factors on CR propagation and triggering of outflows from multiphase gravitationally stratified atmospheres. All of these issues fall under the broad category of the impact of CR transport on wind launching, which is the topic we discuss in the next section. We return to the issue of the CR Eddington limit in Section \ref{CR_and_star_formation}, where we more closely examine the link between this limit and star formation. 

\begin{figure}[tbp]
\begin{center}
\includegraphics[width=0.8\textwidth]{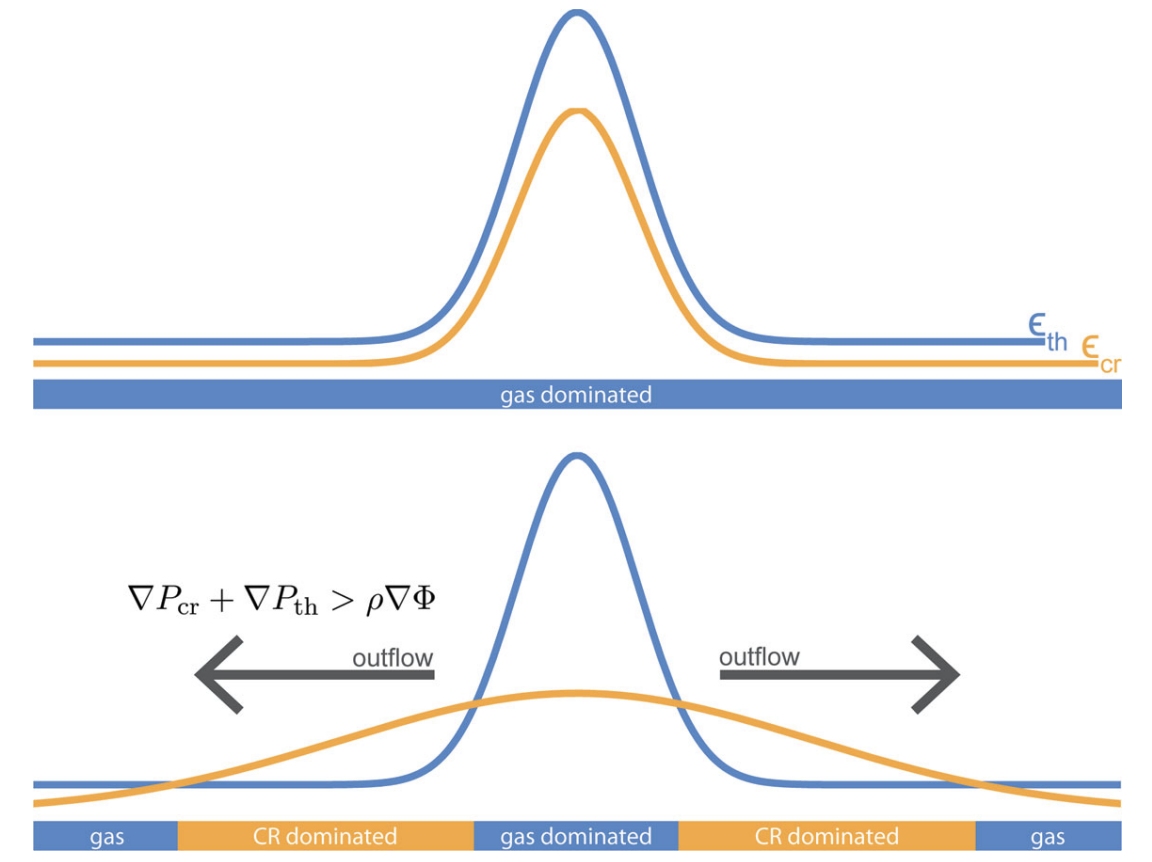}
\end{center}
\caption{Impact of CR transport on wind launching. The blue and yellow lines correspond to hypothetical gas density and CR energy density profiles, respectively. The top panel shows these distributions early in the evolution, and the bottom panel corresponds to a time when CR got transported away from the disk midplane. Image from \citet{salem_cosmic_2014} reproduced with permission from MNRAS.}
\label{fig:salem_bryan}
\end{figure}

\paragraph{Impact of CR transport on wind launching.} \label{Impact of CR transport on wind launching}
In this section, we discuss in more detail the role of CR transport in triggering the winds in gravitationally stratified atmospheres. We begin by considering the dependence of transport on environment and then discuss how CR pressure builds up in the disk and launches winds. From the perspective of wind launching, the key reason why CR transport is important is that it regulates the magnitude of the outward-pointing gas acceleration. It does so either via (i) shaping the CR pressure gradient and/or (ii) by allowing for the CR momentum to be deposited in lower density gas that is easier to accelerate away from the disk. These issues are central to the question of whether an adequate CR pressure gradient can be built up to accelerate the gas and launch the wind.\\
\indent
As mentioned in Section \ref{sec:CR_scattering_turbulence}, CRs can either scatter on Alfv{\'e}n waves generated via the streaming instability or on extrinsically generated waves that are due to turbulent cascade. For CRs with energies less than $\sim 10$ GeV, which contribute most of the CR pressure support, self-confinement due to the streaming instability is expected to dominate over extrinsic confinement \citep{evoli_origin_2018}. The effective drift speed of CRs with respect to the bulk flow of the gas, either dominated by streaming or diffusion, can then be determined by comparing, under the assumption of steady state, the streaming instability growth rate of the scattering waves to their damping rates. The damping can be due to ion--neutral damping, turbulent damping, linear or non-linear Landau damping. Depending on the environment, the effective speed of CR transport will be dominated by advection, streaming, or diffusion. By neglecting CR-gas dynamical coupling, \citet{armillotta_cosmic-ray_2021} focused on the physics of transport and considered CR propagation in the presence of non-linear Landau damping and ion--neutral damping by post-processing TIGRESS simulations \citep[Three phase ISM in Galaxies Resolving Evolution with Star formation and Supernova feedback,][]{kim_three-phase_2017,kim_numerical_2018} of galactic disk patches typical of the solar neighborhood. They found that transport is highly spatially variable and the scattering coefficient varies over four orders of magnitude. Near the disk plane, where most of the mass of the ISM is in neutral form, ion--neutral damping is extremely important, and consequently, diffusion is much faster than advection. While in this environment streaming is also expected to be very fast due to very low ionization, and thus large ion Alfv{\'e}n speed, diffusion is nevertheless more important than streaming for $\sim$~GeV CRs. The coupling of CRs to the gas is restored in the layers adjacent to the disk midplane, where ionization is much higher so that non-linear Landau damping is the dominant damping process, significantly slowing down CR transport. \\
\indent
Due to the sudden drop in the effective CR transport speed in that region, CRs can be effectively trapped in the cold gas. As a result, their actual transport speed may depend more strongly on the conditions in the volume filling phase above the disk rather than in the region where ion--neutral damping leads to very fast transport \citep{hopkins_testing_2021,bustard_cosmic-ray_2021,huang_launching_2022,armillotta_cosmic-ray_2021}. While the CR pressure can build up near the midplane, the actual impact of CRs on the gas acceleration depends on the gradient of pressure rather than its magnitude. In the cold gas, where transport is very fast, pressure gradients along the magnetic field will be very small and the ability of CR to accelerate the gas will be diminished \citep[e.g.,][]{armillotta_cosmic-ray_2021, bustard_cosmic-ray_2021}. Consequently, the impact of CRs on suppressing star formation on very small scales or on launching winds in such environments may be reduced. However, above the disk midplane, or perpendicular to the local magnetic field orientation, or at the interfaces between individual cold clouds and the hotter ISM, the coupling between CR and the gas can be restored, and large CR pressure gradients can be present.\\
\indent
Significant CR pressure gradients can affect the gas distribution and the impact of these gradients can be larger when they act on less dense gas. Figure \ref{fig:salem_bryan} illustrates schematically the impact of CR transport on triggering the development of a galactic outflow \citep{salem_cosmic_2014}. The blue and yellow lines correspond to hypothetical gas density and CR energy density profiles, respectively. Gravitational field is in
the horizontal direction and points toward the density peak. The top panel shows the gas and CR distributions in the initial state. In the absence of CR transport relative to the gas such as diffusion and streaming, CR pressure forces can thicken the disk \citep[e.g.,][]{girichidis_launching_2016,simpson_role_2016}. When active CR transport is present, CRs can diffuse or stream away from the midplane and their pressure gradients can begin to act on less dense gas at larger heights above and below the disk. This effectively breaks the global hydrostatic balance\footnote{In this argument, we neglect turbulent pressure and velocities that, by definition, are not hydrostatic.} and allows the gas to begin to flow out of the potential well. In the context of zoom-in simulations of galactic disk patches, this was seen in the simulations of \citet{girichidis_launching_2016} and \citet{simpson_role_2016}, who considered anisotropic CR diffusion with a constant coefficient and a detailed chemical network that followed the abundances of key ions and molecules. In terms of the treatment of CR transport, this was extended by \citet{farber_impact_2018}, who considered a spatially-dependent anisotropic transport prescription in which diffusion was boosted in the low-temperature ISM. This approach was motivated by the desire to emulate the effects of ion--neutral damping combined with the increased Alfv{\'e}n speed in the cold gas. They demonstrated that despite the effective CR decoupling from the gas in the low ionization environment, CRs can still drive strong winds. While this work was performed under the assumption that diffusion dominates transport, in a similar setup, \citet{holguin_role_2019} showed that the results carry over to the case of CR streaming. They showed that streaming regulated by turbulent damping dependent on local ISM magnetization leads to effectively super-Alfv{\'e}nic CR drifts. Stronger damping resulted in more extended CR and gas distributions and stronger winds.\\
\indent
Due to their ability to remove gas from galaxies, CRs have a potentially significant bearing on star formation, mass and energy loading of the outflows, and the gas phase-space distribution in galactic winds. Below we continue the discussion of these topics and focus on role of CRs in star formation feedback. 

\paragraph{CR feedback and star formation.} \label{CR_and_star_formation}
Stellar feedback plays a critical role in regulating star formation. In order to place the impact of CRs in the appropriate context, we first list well established stellar feedback mechanisms. The most important contributions to the feedback come from massive stars in their late evolutionary stages. These stars interact with the surrounding ISM via: (i) radiation pressure associated with ultra-violet and IR photons interacting with dust
\citep[e.g.,][]{krumholz_dynamics_2009,fall_stellar_2010,murray_disruption_2010,raskutti_numerical_2016,thompson_sub-eddington_2016}, (ii) photoionization heating
\citep[e.g.,][]{geen_feedback_2016,gavagnin_star_2017,dale_ionizing_2012,walch_clumps_2013,dale_before_2014,haid_silcc-zoom_2019,kim_duration_2021}, 
(iii) stellar winds 
\citep[e.g.,][]{castor_radiation-driven_1975,weaver_interstellar_1977,lancaster_star_2021}, and
(iv) SNR shock waves 
\citep[e.g.,][]{rogers_feedback_2013,calura_feedback_2015}.\\
\indent
Each of these processes comes with its limitations in terms of the ability to regulate star formation. For example, the question of the importance of radiation pressure feedback is actively debated. While it has been argued that radiation pressure can be very effective in regulating star formation in high density environments 
\citep{krumholz_dynamics_2009,murray_disruption_2010,kim_modeling_2018}, claims to the contrary have also been presented 
\citep{rosdahl_scheme_2015,rosdahl_galaxies_2015,tsang_radiation_2018}, arguing that radiation can escape low-density channels without depositing a large fraction of the photon momentum. In particular, \citet{menon_infrared_2022} suggested that IR radiation pressure should have only very mild impact on star formation and is unlikely to drive galactic winds. 
Similarly, \citet{reissl_spectral_2018} demonstrated that UV and optical photons from young stars 
do not impart significant momentum to the ambient gas as these photons are efficiently absorbed and re-emitted in the far-IR, where the gas is optically thin.
Heating resulting from photoionization of the surrounding gas due to ultra-violet photon absorption is also an important feedback mechanism, but its importance is dependent the environment. For example, \citet{fukushima_radiation_2021} argue that photoionization feedback is inefficient in giant molecular clouds with surface densities above 100 M$_{\odot}$~pc$^{-2}$. Similarly, 
\citet{haid_relative_2018}, demonstrated that photoionizing radiation is important in the cold neutral medium, but it is subdominant compared to stellar winds in the warm ionized medium. 
Stellar winds can help to disperse gas clouds surrounding star formation sites \citep[e.g.,][]{castor_radiation-driven_1975,wunsch_twodimensional_2008,wunsch_evolution_2011,toala_radiation-hydrodynamic_2011,dale_ionizing_2012,rogers_feedback_2013}. However, the efficiency of this form of feedback can be limited by the tendency of the hot wind gas to escape through low-density channels and by fast cooling at the mixing layers between the hot and cold gas \citep{rogers_feedback_2013,rosen_gone_2014,lancaster_star_2021}.
The last of the feedback processes mentioned above involves deposition of the momentum and energy of expanding SNR shocks in the ISM.
SN are the most violent of these feedback processes. While energetically very important, SN explosions are delayed with respect to the onset of star formation and other processes can disperse GMCs before SN feedback can act 
\citep{fall_stellar_2010,geen_feedback_2016,grudic_dynamics_2022}.
Nevertheless, SN have a dramatic impact on driving turbulence in the ISM 
\citep{mckee_theory_1977,mac_low_control_2004,padoan_supernova_2016,kim_three-phase_2017}, especially, if the SNe are clustered in space \citep{girichidis_silcc_2016}.\\
\indent
CRs injected by SNe add a new ``dimension'' to the feedback problem and possess a number of appealing features that help to maximize the impact of SN feedback.
In most numerical simulations of SN feedback, the resolution is insufficient to resolve individual SNe and the SN energy is quickly radiated away. This catastrophic energy loss is known as the overcooling problem 
\citep[e.g.,][]{walch_silcc_2015,naab_theoretical_2017}. From the numerical point of view, several approaches have been proposed to avoid this problem including: temporarily switching off radiative cooling \citep[e.g.,][]{stinson_star_2006, teyssier_cusp-core_2013}, distributing SN energy injection sites stochastically \citep[e.g.,][]{dalla_vecchia_simulating_2012, simpson_role_2016}, including injection of momentum in addition to thermal energy \citep[e.g.,][]{martizzi_supernova_2015,keller_empirically_2022}, and including natal kicks \citep[e.g.,][]{steinwandel_impact_2022}. As discussed in Section \ref{dynamics_of_CR_near_sources}, CRs are injected by SNe and the pressure inside SNRs may be dominated by CRs later in their evolution because the adiabatic index of CRs is lower than that of the thermal gas. Moreover, in situ injection of CRs at the SN shock fronts tends to result in larger momentum transfer to the ambient ISM \citep{diesing_effect_2018}. This effect, combined with the relatively slower CR energy losses compared to radiative losses of the thermal gas \citep[see e.g., Figure 5 in][]{jubelgas_cosmic_2008}, 
makes CRs an attractive element of the feedback process in which the issues of overcooling could be mitigated. \\
\indent
The impact of CR feedback can be amplified in superbubble environments, where multiple shocks can accelerate CRs.
Because most massive stars form in giant molecular clouds %
and live for relatively short times before they explode as SNe, the feedback due to SNe is expected to be strongly clustered in space and time. Multiple SN explosions then lead to the formation of hot and low density superbubbles.  
Consequently, rather than expending most of the SN energy in high density environments, where catastrophic radiative cooling losses would quickly radiate the energy away \citep{gatto_modelling_2015}, 
the SN energy can be injected in lower density environment of overlapping SNRs \citep[e.g.,][]{mac_low_superbubbles_1988}, where the effect of SN explosions can be sustained over longer periods. Indeed, the impact of SN feedback is enhanced if they detonate in low-density environments both in terms of their momentum transfer to the surrounding ISM  
\citep{walch_silcc_2015, martizzi_supernova_2015, gentry_enhanced_2017} and the ability to drive winds  
\citep{creasey_how_2013,fielding_how_2017,fielding_clustered_2018}. The dynamical impact of CRs could be further increased in clustered SNe as the re-acceleration of pre-existing CR populations could result in an elevated CR acceleration efficiency \citep[e.g.,][]{caprioli_diffusive_2018,vieu_cosmic_2022}.

\paragraph{CRs and energy and mass loading of the winds.} \label{Energy_and_mass_loading}
As discussed above, CRs can play an important role in launching galactic winds and regulating star formation. The efficiency of this feedback can be quantified in terms of either energy $\eta_\rmn{e}$ or mass $\eta_\rmn{m}$ loading factors. The energy and mass loading factors are typically defined as the ratio of the kinetic plus thermal luminosity in the outflow to the energy injection rate from SNe, and the ratio of the mass outflow rate to the star formation rate, respectively. In simulations, star formation and outflow rates are typically measured simultaneously. However, the instantaneous star formation rate does not directly control the outflow at some height, because the outflow was launched by momentum deposition associated with earlier star formation events. The situation is even more complicated in observations. The star formation rate is indirectly inferred through various estimators such as the rest-frame UV or FIR emission, which probe an instantaneous star formation event, while the outflow rate measured at a finite height above the disk is typically inferred from spectral line shifts excited some time after the corresponding star formation event that triggered the wind near the disk midplane. These are difficult to localize in space and hence, cannot be unambiguously attributed to outflowing gas at a specific height above the disk, which complicates direct comparison to simulations.

Observations suggest that the mass loading factor $\eta_\rmn{m}$ varies widely and ranges from $\sim$~0.01 to $\sim$~10 \citep[e.g.,][]{veilleux_galactic_2005}
Values broadly falling within this range have been seen in numerical simulations of small stratified boxes. For example, strong outflows with mass loading factor values up to 10 were reported from small-box SILCC simulations \citep[SImulating the Life Cycle of molecular Clouds,][]{girichidis_silcc_2016} and values ranging from 0.01 to 10 were seen in TIGRESS simulations \citep{kim_numerical_2018}. Furthermore, 
\citet{li_quantifying_2017} found that $\eta_\rmn{m}$ decreases as a function of galactic surface density with 50~$\rmn{M}_{\odot}~$pc$^{-2}$ marking the transition where $\eta_\rmn{m}=1$. However, particular mass loading values depend on the distance from the galactic midplane where the measurements of gas flows are made and on a particular gas phase. An insightful unification of these type of small-box simulations in terms of the wind loading factors has been presented by \citet{li_simple_2020}. They conducted a systematic literature search and showed that hot outflows are much more powerful than cool outflows. This analysis revealed a unifying trend that the specific energy density of the hot phase in the wind is $\propto \eta_\rmn{e}/\eta_\rmn{m}$ over four orders of magnitude in star formation surface density. However, it has been argued \citep[e.g.,][]{martizzi_supernova_2016} that simulations of Cartesian boxes with a uni-directional gravitational field may not fully capture the outflow physics as they neglect the correct global geometry and galactic gravitational potential. This omission tends to suppress the mass loading factors and reduces wind speeds. We return to this question in Section \ref{Role of CR in global models of isolated galaxies}, where we discuss the role of CRs in global models of isolated galaxies. \\
\indent 
How do CRs change this picture? The fundamental difference between the SN-powered winds and those also assisted by the injection of CRs is that the latter can experience continuous acceleration due to CR pressure gradients while the former are more episodic in nature. Consequently, a generic feature of CR-driven outflows is that their speeds increase as a function of the distance from the disk, which is seen not only in stratified box simulations \citep[e.g.,][]{girichidis_cooler_2018} but also in one-dimensional models \citep[e.g.,][]{dorfi_time-dependent_2012} discussed in Section~\ref{1D_models} and global models \citep[e.g.,][]{pakmor_galactic_2016} discussed in Section \ref{Role of CR in global models of isolated galaxies}. 
Importantly, this offers a natural way to explain observational trends \citep{steidel_structure_2010}.
However, other models not invoking CRs can also explain these trends (see Section~\ref{CGM line diagnostics of CR feedback: kinematics}).\\
\indent
The impact of CRs on wind driving is best illustrated via controlled experiments designed to isolate  differences in results that arise solely due to CR physics. For example, \citet{simpson_role_2016} 
\begin{figure}[tbp]
\begin{center}
\includegraphics[width=0.95\textwidth]{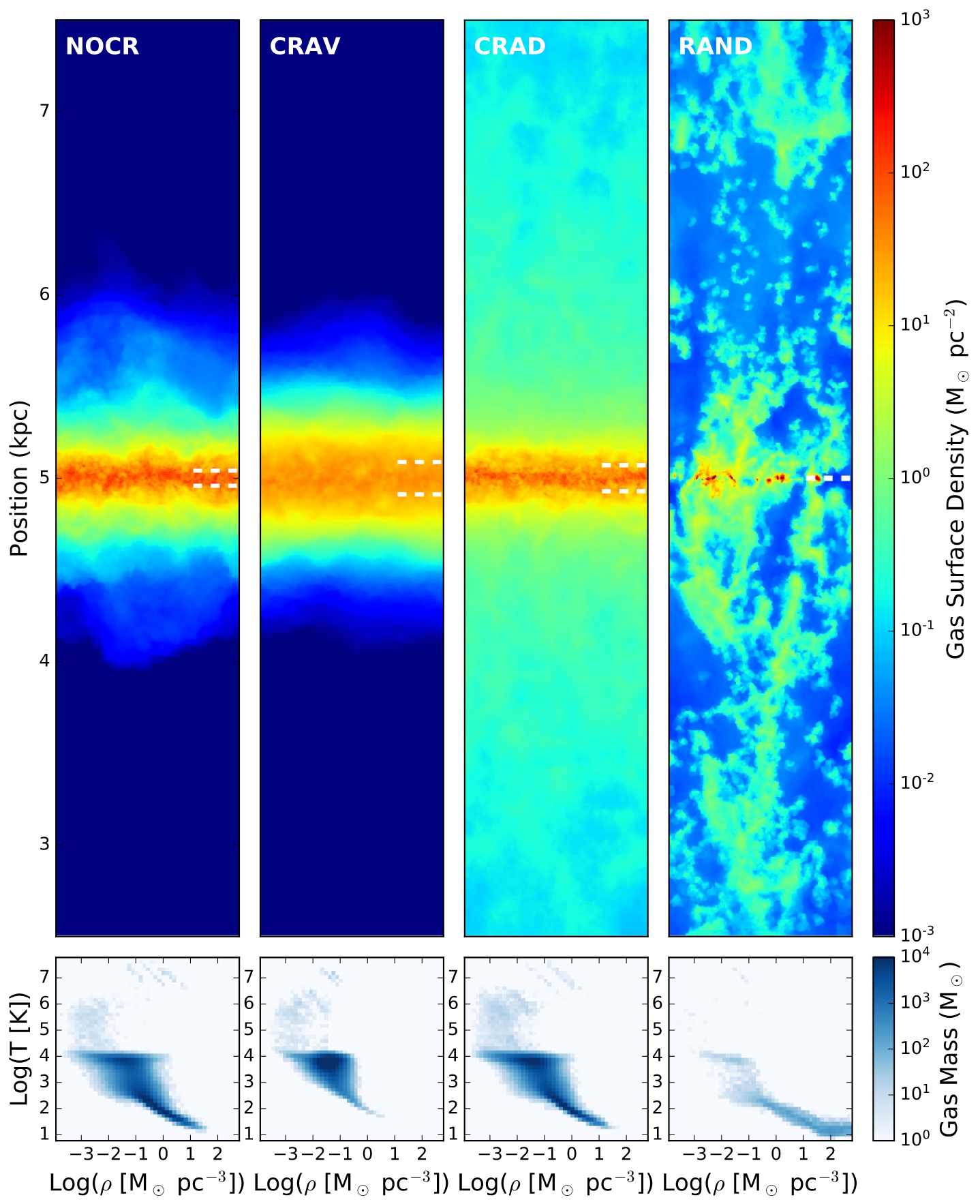} 
\end{center}
\caption{Gas density distributions (top row) in the plane perpendicular to a galactic disk and the corresponding phase space gas distributions (bottom row). From left to right shown are the cases corresponding to no CRs, CRs with advection only, CRs with advection and anisotropic diffusion, and random placement of SNe without CR physics. Dashed horizontal lines indicate the height that contains half the gas mass in the initial conditions. Image from \citet{simpson_role_2016} reproduced with permission from ApJ.}
\label{fig:simpson}
\end{figure}
performed small stratified box experiments in which SNe and CRs were injected at density peaks and CRs were allowed to diffuse out either isotropically or along the magnetic fields. These cases were then compared to non-CR runs in which SNe were placed randomly in space near the disk midplane (see Fig.~\ref{fig:simpson}). These experiments revealed that in the former case, despite the fact that SN feedback coincided with the densest regions where the radiative losses were the strongest, the mass loading of the wind was comparable to that corresponding to the latter case in which SNe were injected at random locations and CRs were absent. However, unlike the peak-driving cases (with and without CRs) that produced a self-regulated ISM, the random-placement model resulted in a unphysical cooling-induced formation of overdense ISM clouds.  
In an independent study of non-CR SN feedback in stratified boxes, \citet{girichidis_silcc_2016} confirmed that indeed 
only when SNe were placed randomly with respect to the density peaks could the SN energy be injected in sufficiently low density regions that the excessive radiative losses could be avoided and the winds launched with significant mass loading factors. 
Furthermore, by comparing non-CR runs to those that incorporate CR pressure and diffusion, \citet{girichidis_launching_2016} showed that including CR physics results in loading factors around unity, which were much larger compared to those seen in non-CR simulations. In the CR simulations, the dominant source of acceleration above a height of $\sim1$~kpc was CR pressure gradients \citep{girichidis_cooler_2018}. Moreover, emulating the effect of ion--neutral decoupling of CRs from the gas in low temperature ISM regions reduces the mass loading factors of the winds only moderately \citep{farber_impact_2018}. Similarly, \citet{holguin_role_2019} demonstrated that the increase in turbulent damping led to faster CR streaming that helped to drive winds with progressively larger mass outflow rates while allowing for increasingly larger star formation rates (due to reduced CR pressure support in the disk). Overall, these small-box studies demonstrate that CRs can play an important dynamical role in launching galactic winds.

\paragraph{Phase space structure.} \label{Phase_space_structure}
The observed galactic outflows are multiphase (e.g., molecular: 
\citealt{leroy_multi-phase_2015,chisholm_molecular_2016}; neutral: 
\citealt{heckman_absorptionline_2000,rupke_outflows_2005,contursi_spectroscopic_2013};
ionized: 
\citealt{steidel_structure_2010,erb_galactic_2012,heckman_systematic_2015}).
As discussed in Section \ref{Interaction of CR with dense clouds}, the origin of the multiphase nature of these outflows is uncertain, but it is either due to (i) in situ formation of the cold phase via local thermal instability in hot outflows or (ii) acceleration of multiphase gas from the vicinity of the galactic disks. \\
\indent
Given that the outflows are launched from the disk that is occupied by multiphase ISM, it is not surprising that the winds themselves could be multiphase. The phase-space distribution of the gas in the midplane and in the outflow regions above it depend on a number of factors including the details of stellar feedback briefly discussed in Section \ref{CR_and_star_formation} above. In this regard, CRs in particular have the ability to significantly alter these phase space distributions. For example, by systematically varying the assumptions about the environments in which SNe explode, \citet{simpson_how_2023} showed in stratified box experiments that the ISM either evolves to a state where warm gas is volume filling, when SN are preferentially placed at the locations of gas density peaks, or to a ``thermal runaway'' state, which occurs when SNe are placed randomly in space so that their energy deposition cannot prevent the mass-carrying cold and warm phases from collapsing to very high densities and the hot gas becomes volume filling.
Interestingly, dynamically important CR pressure and CR diffusion have the ability to transform one solution to another. In a related work, \citet{commercon_cosmic-ray_2019} considered CR propagation in a bi-stable ISM and discovered a critical diffusion level that separates the regime where CRs can propagate freely from the one in which CRs are trapped to produce strong CR pressure gradients that suppress thermal instability and thus, affect star formation rates. Interestingly, some of the most complete models of the ISM that in addition to including a myriad of processes mentioned above (e.g., SN clustering, pre-SN feedback, radiation pressure) also incorporate CR physics \citep{rathjen_silcc_2021}, result in the best matches to the observed gas phase space structure, energy density partition in the ISM, and star formation rates. 
In particular, \citet{rathjen_silcc_2023} directly compare non-CR and CR simulations of stellar feedback and find that while the simulations without CRs result in single or two-phase medium, a realistic galactic outflow consisting of three-phase medium develops in CR simulations.\\
\indent
A generic feature of CR-driven winds is the tendency of the CR pressure to become progressively more important compared to the kinetic, thermal, and magnetic pressures as a function of the distance from the disk (e.g., \citealt{hanasz_cosmic-ray-driven_2009, simpson_role_2016,girichidis_cooler_2018, armillotta_cosmic-ray_2021}; see also Section \ref{Role of CR in global models of isolated galaxies} below). One consequence of this dynamically important CR pressure support is that thermal gas can be significantly cooler because thermal pressure is replaced by CR pressure and different gas phases no longer have to be in pressure equilibrium with each other. Indeed, simulations demonstrate that CR driven outflows are denser and much cooler compared to those that are thermally driven (e.g., \citealt{girichidis_launching_2016, girichidis_cooler_2018}; see also bottom row in Fig.~\ref{fig:simpson}). The addition of CRs also generally tends to make the outflows slower and smoother in terms of density distribution (cf.\ third and fourth panel in the top row of Fig.~\ref{fig:simpson}) while reducing the volume filling factor of the hot gas phase provided half of the SNe are placed randomly in the ISM  \citep[see figure~18 of][]{simpson_how_2023}, though details of CR transport may affect these findings \citep{farber_impact_2018}. These overall trends may also help to explain the properties of the warm ionized gas layers above the disks of 
the Milky Way and other disk galaxies \citep{haffner_warm_2009}. This warm gas is most likely not expelled from the disk by very hot SN-driven thermal outflows, and cooler CR-driven outflows may offer a solution to the problem of lifting the gas without overheating it. While the warm ionized medium is most likely heated by photoionization due to massive OB stars 
\citep{reynolds_power_1990,reynolds_what_1990}, additional heating mechanisms may be required to provide the necessary heating to explain the [S II]/H$\alpha$ and [N II]/H$\alpha$ line intensity ratios \citep{reynolds_evidence_1999}.
Interestingly, this supplementary heating could be at least partially provided by the streaming CRs \citep{wiener_cosmic_tmp_2013} that are already present in the wind.

\paragraph{Impact of CRs on star formation and winds in varying galactic environments.}
We now turn our attention to the role of CRs in shaping star formation across a range of galactic environments. 
As discussed in Section \ref{CR Eddington limit}, whether a particular galaxy, or region within a galaxy, can launch a CR-driven wind depends on the relative value of (i) the CR flux supplied by SNe compared to (ii) the critical CR flux needed to unbind the gas \citep{socrates_eddington_2008}. Under the assumption of negligible CR losses, the former is proportional to the surface density of the star formation rate, $\dot{\Sigma}_\star$. The latter increases with the strength of disk gravity and the speed of CR transport, and both of these factors help to stabilize the disk against disruption. In particular, for faster transport of CRs away from the disk, larger CR fluxes are needed to unbind the gas in order to compensate for weaker coupling of CRs to the gas. Depending on the details of CR transport physics, this competition between outward-pointing CR pressure forces and downward-pointing disk gravity leads to a relationship between the surface density of the disk $\Sigma_{\rm gas}$ and the critical star formation rate density $\dot{\Sigma_\star}$ above which the gas can be unbound by CR forces. Predictions from such models can be compared to the classic Kennicutt-Schmidt law \citep{kennicutt_star_1989,de_los_reyes_revisiting_2019,kennicutt_revisiting_2021}, i.e., the observed relationship between $\dot{\Sigma_\star}$ and $\Sigma_{\rm gas}$. Based on this argument, CRs could drive winds over a wide range of star formation environments \citep{huang_launching_2022} as long as CR losses can be neglected. However, the impact of CR losses is generally expected to be higher for larger $\Sigma_{\rm gas}$. There is an emerging consensus that the observed star formation rate may be sufficient to drive winds at least in lower $\Sigma_{\rm gas}$ environments \citep{crocker_cosmic_I_2021,crocker_cosmic_II_2021,heintz_galaxies_2022}, but at higher surface densities the impact of CRs, and their pressure compared to the gas pressure, may be reduced due to a combination of catastrophic hadronic losses \citep{lacki_physics_i_2010, lacki_-ray_2011,krumholz_cosmic_2020,crocker_cosmic_II_2021} and/or fast transport away from the disk \citep{armillotta_cosmic-ray_2021,armillotta_cosmic-ray_2022}. \\
\indent
The relative importance of CR hadronic losses is closely tied to the details of modeling of the physics of CR transport. Consequently, a particular shape of the relationship between critical star formation rate needed to launch winds and the surface density of the gas likely depends on this physics. At high surface densities, galaxies may be good CR calorimeters (and powerful $\gamma$-ray sources), meaning that most of the energy injected by SNe in the form of CRs is radiated away. However, where exactly the transition to calorimetry occurs as a function of $\Sigma_{\rm gas}$ depends on CR transport physics and the ability to reliably model the small-scale porosity and magnetic connectivity of the different phases of the ISM. Moreover, it has been suggested that CR transport in the diffuse medium surrounding the cold ISM needs to be very fast in order to limit hadronic losses and prevent overproduction of $\gamma$-ray emission \citep{chan_cosmic_2019,hopkins_testing_2021,bustard_cosmic-ray_2021} but claims to the contrary have also been presented  (\citealt{nunez-castineyra_cosmic-ray_2022,simpson_how_2023}; see also \citealt{girichidis_cooler_2018,heintz_galaxies_2022}). Further detailed and fully self-consistent modeling of these processes for a wide range of relevant ISM parameters is needed to establish the conditions for the onset of CR wind driving and the consistency of the models with the observed far-infrared (FIR)--radio correlation (\citealt{condon_radio_1992}; see Section \ref{Non-thermal emission from galaxies} for further discussion). 

\subsubsection{Role of cosmic rays in global models of isolated galaxies} \label{Role of CR in global models of isolated galaxies} 
Certain aspects of the feedback process cannot be fully captured in small-scale ISM simulations \citep[e.g.,][]{girichidis_silcc_2016} or zoom-in simulations of galactic patches \citep[e.g.,][]{simpson_role_2016} discussed above. Most notably, zoom-in simulations are, by construction, performed in planar geometry and cannot include realistic galactic potentials. One consequence of adopting planar geometry is that the wind solutions cannot transition from subsonic to supersonic regime as is expected in the standard \citet{chevalier_wind_1985} galactic wind theory \citep{martizzi_supernova_2016}. As a result, outflows are slower, typical mass loading factors appear to be substantially below those seen in rapidly star-forming galaxies, and fountain flows are more likely to develop (cf.\ zoom-in TIGRESS simulations of \citealt{kim_numerical_2018} and CGOLS global simulations of \citealt{schneider_physical_2020}). By contrast, the effect of the ``dilution’’ of the gravitational potential in global models of galactic winds naturally predicts conical outflows and larger mass loading factors. Global models can also be used to study various aspects of the interactions of the outflows with the ambient medium, e.g., by including radiatively cooling and collapsing gaseous halos, that source turbulence and drive a small-scale magnetic dynamo \citep[e.g.,][]{pfrommer_simulating_2022}, and gas streams, that feed the disk and interact with the galactic wind \citep{peschken_angular_2021}, or considering the interplay between winds and ram pressure associated with the bulk motions of galaxies \citep[e.g.,][]{bustard_cosmic-ray-driven_2020,farber_stress-testing_2022}. In this section, we mostly focus on those aspects of CR feedback that are either unique to the global models or that can be more faithfully modeled in the context of global models of galaxies in non-cosmological settings. We discuss several intricately inter-related processes such as the magnetic field amplification, CR losses due to hadronic and Coulomb processes, and CR transport in global galactic models.  

\begin{wrapfigure}{r}{0.5\textwidth}
  \begin{center}
    \includegraphics[width=0.5\textwidth]{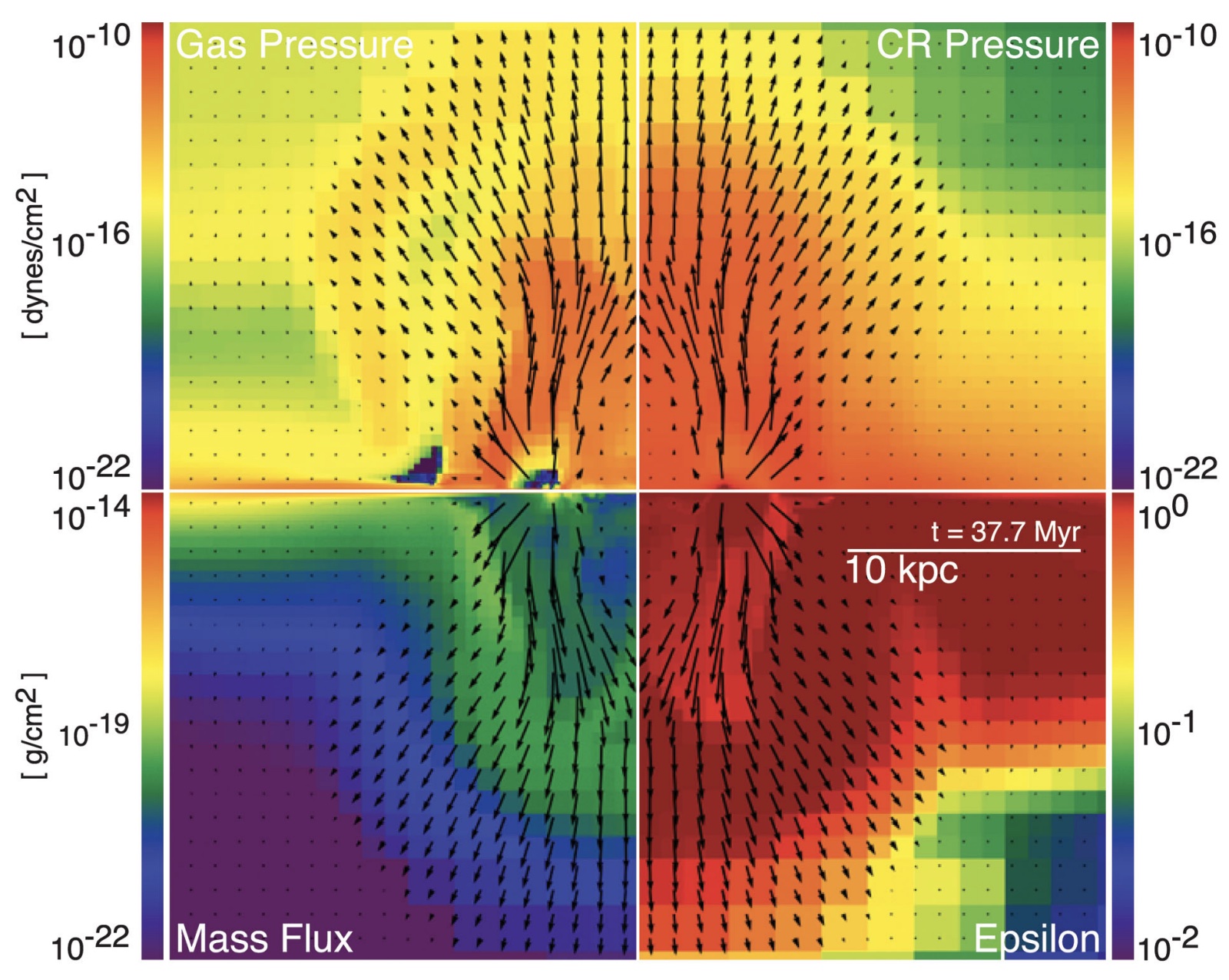}
  \end{center}
  \caption{Slices of mass flux, thermal gas pressure, CR pressure, and the ratio of the CR-to-thermal pressures in the plane perpendicular to the disk; from a global simulation of a disk galaxy from \citet{salem_cosmic_2014}; reproduced with permission from MNRAS.}
  \label{fig:salem_bryan_wind}
\end{wrapfigure}

\paragraph{Impact of CR transport on wind driving.}
As argued above in the context of one-dimensional models and zoom-in simulations of galactic patches, CR transport plays central role in shaping the properties of galactic outflows. These outflow properties can differ substantially depending on whether CRs are advected or actively transported via diffusion and/or streaming, and how exactly these processes are modeled. Despite these differences, some generic features emerge from these models including cooler and slower winds (compared to simulations where winds are driven by thermal feedback), potentially significant CR pressure support in the halo, substantial suppression of star formation, and thickening of the galactic disks.\\
\indent
Global three-dimensional simulations of galactic winds including CR diffusion were performed by a number of authors \citep[e.g.,][]{ hanasz_cosmic_2013,booth_simulations_2013,pakmor_galactic_2016,ruszkowski_global_2017}. Below we contrast two examples of early simulations in order to isolate the impact of non-adiabatic CR energy losses.
The first example is the global three-dimensional hydrodynamical simulations of galactic winds including isotropic CR diffusion presented by \citet{salem_cosmic_2014}. These simulations demonstrated that strong winds with mass loading factors exceeding unity can be driven by CR feedback in Milky Way mass halos. Figure~\ref{fig:salem_bryan_wind} shows
slices of mass flux, thermal gas pressure, CR pressure, and the ratio of the CR-to-thermal pressures (lower right panel) in the plane perpendicular to the disk. It is evident from these results that the outflow is conical and that CR pressure dominates over thermal pressure in the halo. As a result of the additional pressure support due to CRs, the disk is inflated and star formation is reduced. Since the CR content near the disk midplane is regulated by the speed of CR transport, the CR pressure support decreases with increasing CR diffusion coefficient $\kappa$, which leads to the trend for star formation rate to increase with $\kappa$. The second example is the extension of the above work by \citet{booth_simulations_2013}, who incorporated the effect of CR energy losses due to hadronic Coulomb collisions in dense regions and studied the impact of CR feedback in a dwarf and Milky Way-mass galaxies. Despite CR losses, the impact of CR pressure is still significant in these simulations because CR diffusion allows for the energy to be transferred from the densest regions to more tenuous gas before CRs cool completely. As a result, the buildup of CRs provides a robust and gentle wind acceleration mechanism. In agreement with observations, this gentle and continuous acceleration of the gas results in slower winds compared to those accelerated by SN feedback without CRs. In the case of dwarfs, the boost in the mass loading factor in these simulations is an order of magnitude larger than in the case of non-CR SN feedback. In agreement with predictions from smaller scale simulations of zoom-in patches of galactic disks discussed earlier, global simulations also predict significantly cooler winds compared to the cases where winds are powered by SNe only. This effect is due to the fact that a significant fraction of the energy density in the wind goes to the CR component rather than the thermal component. This shift reduces the gas temperature, which further increases the efficiency of radiative cooling as the gas is subject to cooling closer to the peak of the cooling function.\\
\indent  
The impact of CR streaming is similar to that of CR diffusion in the sense that the presence of this CR transport process facilitates the wind launching. While in the absence of any active CR transport mechanism the disk only puffs up and star formation is severely reduced, when CR streaming (or diffusion) is present the wind can develop. First global three-dimensional hydrodynamical simulations that included CR streaming were performed by \citet{uhlig_galactic_2012}. In addition to transport via streaming, CRs were subject to hadronic and Coulomb losses and cooling associated with the excitation of Alfv{\'e}n waves (though the process was approximated by assuming that streaming occurs at the local sound speed in these purely hydrodynamical simulations). The inclusion of streaming resulted in powerful and sustained winds with wind speeds and mass loading factors in agreement with simplified analytical estimates that are based on the Bernoulli equation. Moreover, CR streaming losses led to the excitation of Alfv\'en waves, and the associated heating of the halo. This work was further extended by \citet{ruszkowski_global_2017}, who included streaming CR in MHD simulations and also observed development of winds with substantial mass loading factors and suppression of the star formation rates. \\
\indent
Despite these similarities, there are important differences between winds enabled by streaming and diffusion of CRs. Generally speaking, the mass loading factor of the wind launched by CR feedback depends on the speed of CR transport. When the transport speed is vanishingly small, CRs are concentrated near the disk midplane, reduce star formation, and are unable to lift the dense gas. As the transport speed is increased, progressively larger CR pressure gradients can build up in the lower density gas above the disk midplane, which leads to the wind launching and the increase of the mass loading factor with CR transport speed. In the other extreme, very fast transport of CRs allows them to leave the midplane on a short timescale so that they spend systematically less time accelerating this gas, which implies a decreasing mass loading factor with increasing transport speed. While the latter regime corresponds to the diffusion-dominated wind driving (e.g., \citealt{salem_cosmic_2014}), the former regime regime is applicable to winds driven by CR streaming (e.g., \citealt{ruszkowski_global_2017}). This argument can be made more quantitative by considering characteristic transport speeds in the diffusion and streaming cases. For typical conditions operating near star forming regions (e.g., cloud size of $L=100$ pc, magnetic field of 5 $\mu$G, diffusion coefficient $\kappa=10^{28}$~cm$^{2}$~s$^{-1}$, and gas density of $n=10$ cm$^{-3}$), the characteristic diffusion speed is $\kappa/L=320\,(\kappa/10^{28}~{\rm cm}^{2}~{\rm s}^{-1})(L/100~ \rm pc)^{-1}$km s$^{-1}$, while the streaming speed is the Alfv{\'e}n speed $\sim3.5\;(B/5~\mu{\rm G}) (n/10\;{\rm cm}^{-3})^{-1/2}$ km s$^{-1}$. This implies that the streaming speed is generally much smaller than the diffusion speed. In fact, if CR streaming alone is to drive the winds, then the streaming speed needs to be boosted by collisionless wave damping mechanisms such as turbulent damping, non-linear Landau damping, or ion--neutral damping, and/or by the increase in the Alfv{\'e}n speed in the low ionization regions, for the gas to be more easily accelerated \citep[e.g.,][]{farber_impact_2018,farcy_radiation-magnetohydrodynamics_2022}.
In a realistic situation, the effective CR drift speed arises from a combination of streaming and diffusion and is sufficiently fast to launch winds.\\
\indent 
In addition to the above CR transport speed arguments, additional complexity is introduced by the fact that, unlike in the extrinsic confinement picture where CR transport is dominated by diffusion, in the self-confinement picture the streaming CRs also excite Alfv{\'e}n waves on which they scatter, and collisionless damping of these waves effectively cools CRs and heats the thermal gas, which subsequently quickly radiates this energy away. Consequently, this energy is not efficiently used for driving the winds and the winds are expected to be weaker as lower CR pressure gradients are available to accelerate them \citep{wiener_cosmic_2017}. On the other hand, in the purely diffusive regime, CRs only experience adiabatic changes of their energy because spatial diffusion itself is energy conserving (neglecting hadronic and Coulomb losses that are present in both transport regimes) and the effect of CRs on wind driving can be overestimated. Thus, pure streaming and pure diffusion cases likely bracket the range of outcomes in terms of wind driving.\\
\indent
While the exact CR transport mechanisms are still uncertain, for CR energies below $\sim$ 100 GeV that contribute most significantly to the CR energy density, CR scattering is likely to be dominated by self-generated turbulence \citep[e.g.,][]{aloisio_nonlinear_2015,evoli_origin_2018}, rather than scattering on extrinsic turbulence \citep{yan_cosmic-ray_2004,yan_cosmic-ray_2008}. Hydrodynamical simulations aimed at bracketing the range of possibilities in the scenarios where CR transport is dominated by these two regimes \citep{wiener_cosmic_2017} demonstrate that diffusive transport leads to mass loss rates that are an order of magnitude larger in comparison to those inferred in the streaming scenario. These findings are broadly consistent with those obtained using MHD simulations \citep{dashyan_cosmic_2020}. Specifically, while the simulations employing isotropic and anisotropic diffusion with typical values of the diffusion coefficient $\sim 3\times 10^{28}-10^{29}$ cm$^{2}$s$^{-1}$ result in efficient wind driving, the simulations with streaming resemble those with pure CR advection. Including streaming speed boost helps to launch the wind but the outflow rates are nevertheless smaller than in the diffusion case and smaller than those typically observed. This is consistent with the findings of \citet{chan_cosmic_2019} and \citet{hopkins_but_2020} performed in the cosmological context (for cosmological simulations see Section \ref{Cosmological effects}) and including boosted streaming, though significant differences exist in the typical levels of magnetic field strength, and consequently the baseline Alfv{\'e}n speed, between these simulations and those of \citet{dashyan_cosmic_2020} making these comparisons uncertain (see below). \\
\indent
The relationship between the mass loading and transport speed discussed above can be significantly altered in the presence of CR energy losses. Specifically, the results of \citet{dashyan_cosmic_2020} differ from those of \citet{ruszkowski_global_2017}, who reported significant outflows in the case of boosted streaming. A possible resolution to these differences lies in the fact that \citet{dashyan_cosmic_2020} include additional CR cooling due to hadronic and Coulomb processes. Since the timescale for CR cooling can be shorter than that associated with the CR escape timescale due to streaming at speeds moderately boosted beyond the Alfv{\'e}n speed, CR may lose energy before they can exert sufficient pressure forces on the gas to launch the wind. However, larger streaming boost due to damping processes could perhaps help to alleviate this problem. Similarly, the trend for the mass loading to decrease with the value of the diffusion coefficient \citep{salem_cosmic_2014} discussed above is reversed in \citet[][see also \citealt{farcy_radiation-magnetohydrodynamics_2022}]{dashyan_cosmic_2020}. In the latter work this dependence of the mass loading on transport speed is also attributed to the fact that, for higher $\kappa$, CRs spend less time in the denser regions where CR cooling times are shorter.\\
\indent 
Irrespective of whether CR cooling losses are included, recent studies report that faster transport speeds lead to more star formation \citep[e.g.,][]{jubelgas_cosmic_2008,salem_cosmic_2014,ruszkowski_global_2017,farber_impact_2018,chan_cosmic_2019}. However, the corresponding wind mass loading does not have to be monotonic with the transport speed. Based on global hydrodynamical simulations that include diffusion and CR losses, \citet{jacob_dependence_2018} report that while the mass loading decreases with the diffusion coefficient for low halo masses, at larger masses the relationship is non-monotonic. In the same regime of low halo masses, \citet{farcy_radiation-magnetohydrodynamics_2022} find the opposite trend but the comparison is made difficult by the fact that the latter work also includes radiation pressure. Irrespectively of the halo mass, total outflow rates increased with $\kappa$ up to a peak and then either saturated or began to decrease near very high values of $\kappa\sim 3\times 10^{29}$~cm$^{2}$~s$^{-1}$, where CRs did not spend enough time accelerating the gas before escaping from the galaxy. These trends for the mass outflow rate to peak at a certain $\kappa$ were more evident for cold gas component. For the most massive $10^{12}$ M$_{\odot}$ halo, the increase in $\kappa$ led to larger total outflow rates and hotter winds. On a qualitative level, this is consistent with the trends seen in zoom-in simulations where the transport speed is locally increased in cold gas due to the combination of ion--neutral friction and the increase in the ion Alfv{\'e}n speed \citep{farber_impact_2018}. 

\begin{figure}
  \begin{center}
    \includegraphics[width=0.9\textwidth]{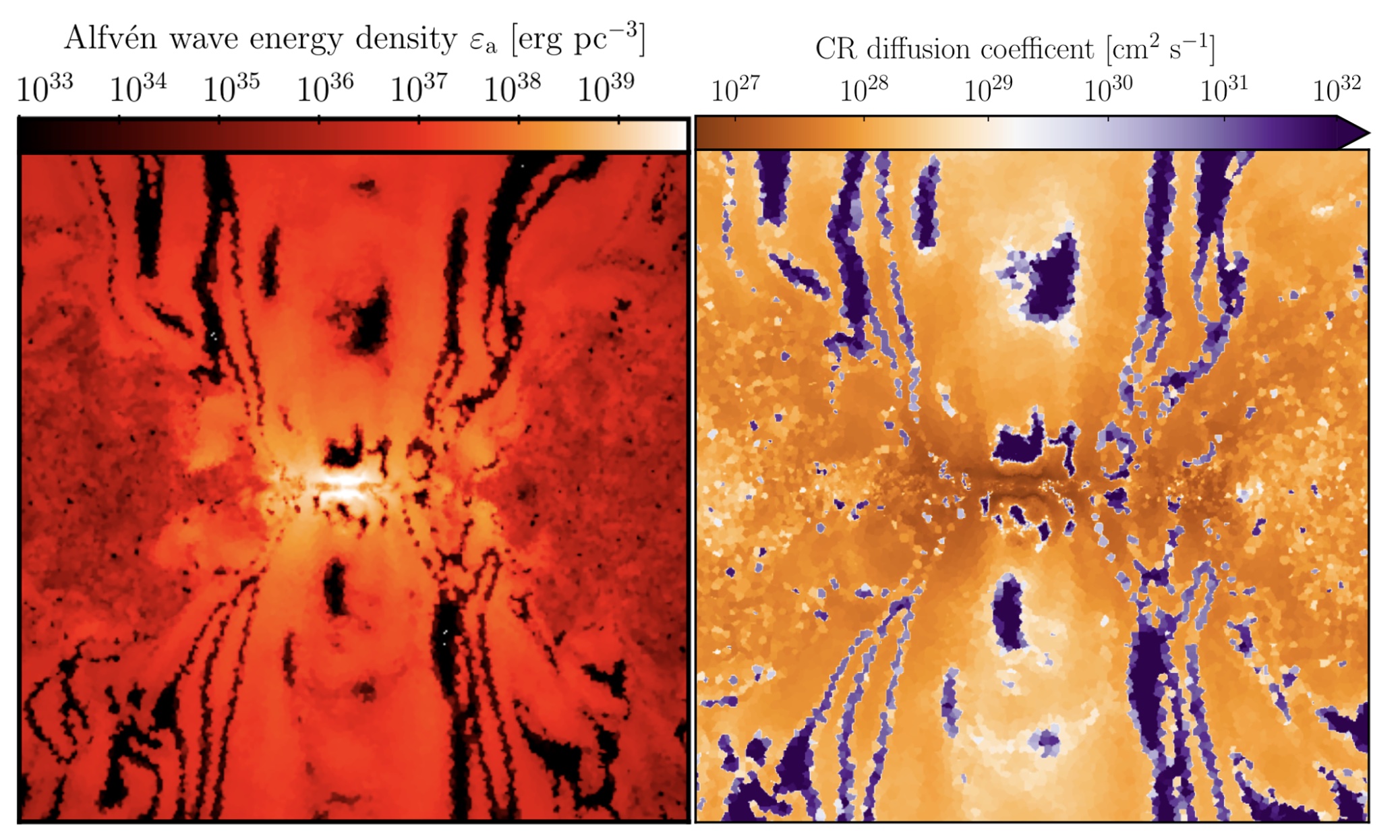}    
  \end{center}
  \caption{Slices perpendicular to the disk of the Alfv{\'e}n wave energy density (left panel) and CR diffusion coeffcient (right panel). Notice the spatial coincidence between the regions devoid of Alfv{\'e}n wave energy and those characterized by very large values of the diffusion coefficient. Image from \citet{thomas_cosmic-ray-driven_2023} reproduced with permission from MNRAS.  }
  \label{fig:thomas}
\end{figure}

\paragraph{Impact of non-equilibrium processes on CR transport.}
Most models of CR feedback make simplifying assumptions regarding the steady state CR spectra and transport. These assumptions are only justified when the relevant timescales associated with the processes governing CR evolution are shorter than other relevant timescales. Below we discuss the impact of relaxing these assumptions on the evolution of outflows in global models of CR feedback.\\
\indent
Despite the fact that the energies of CR particles span twelve orders of magnitude and the CR transport speed depends on CR energy, most models of CR feedback consider a single momentum-integrated CR fluid (in the ``gray’’ approximation). In this approximation, the evolution of CRs depends on momentum-integrated CR cooling rates and transport coefficients. This assumption is justified when the CR spectrum can reach a steady state on sufficiently short timescales. The evolution of CR energy density at a particular location depends on the competition between (i) gains due to CR injection and advective, diffusive, and streaming transport into the region of interest, and (ii) losses due to advective, diffusive, streaming transport out of that region combined with hadronic and Coulomb cooling. Therefore, the assumption of steady state, i.e., of balance between gains and losses, is justified when the instantaneous timescales associated with these processes are shorter than the timescale on which the overall net change of CR energy occurs. The validity of the steady state assumption can be assessed a posteriori by post-processing results obtained under the gray approximation. This type of analysis reveals that, close to the disk, the steady state approximation is valid, but it breaks down in very dynamical regions especially above the disk \citep{werhahn_cosmic_i_2021}. This has important implications of the modeling of galactic emission, which we defer to Section \ref{Non-thermal emission from galaxies}.\\
\indent
In order to relax the steady assumption, CR spectra need to be modeled on-the-fly using multi-energy CR fluid approach coupled with MHD \citep[][see also Section~\ref{sec:numerical_methods_CR_spectrum} for a detailed discussion]{miniati_cosmocr_2001,yang_spatially_2017,girichidis_spectrally_2020,girichidis_spectrally_2022,ogrodnik_implementation_2021}. Irrespective of whether we model the full CR spectrum or whether we treat them in the gray approximation, star formation rates and galaxy morphologies are not significantly affected. However, mass loading factors of galactic winds are smaller by up to a factor of four in dwarf galaxies in the spectral CR model \citep{girichidis_spectrally_2023}. In addition, the spectral CR treatment produces substantially more cold gas in fountain flows in Milky Way-mass galaxies than grey models. The limited influence of spectral CR modeling on hydrodynamics can be attributed to the dominant contribution of GeV CRs to CR pressure. Additionally, the proportion of fast-diffusing high-energy CRs is too small to exert a significant impact on fluid dynamics of galactic winds, particularly in larger galaxies. These simulations show that a powerful CR-driven outflow in smaller galaxies (with halo masses $M\lesssim3\times10^{11}~\rmn{M}_\odot$) primarily advects CRs with the gas. By contrast, in Milky Way-mass galaxies ($M\approx1\times10^{12}~\rmn{M}_\odot$) the CR pressure gradient alone is not able to power a sustained outflow, implying that CR transport is entirely diffusive.
Interestingly, the energy-weighted CR diffusion coefficient varies spatially by up to two orders of magnitude. In the disk and in the outflow region, it assumes the canonical values of GeV CRs of $1\times10^{28}$ to $3\times10^{28}~\rmn{cm}^2\rmn{s}^{-1}$ and increases up to $3\times10^{29}~\rmn{cm}^2\rmn{s}^{-1}$ in the CGM, where CRs with energies exceeding TeV dominate the spectrum. As a result, the CGM is supported by the pressure from high-energy CRs, which enables a more structured gas density and lower temperatures in comparison to the gray CR models \citep{girichidis_spectrally_2023}. \\
\indent
In general, both the CR flux, that determines how fast CRs are transported, and the diffusion coefficient, that determines how fast CRs are scattered, are time-dependent quantities. The steady state of the diffusion coefficient is determined by the balance between the wave growth and damping, and is reached on characteristic timescales corresponding to these processes. On the other hand, the steady state of the CR flux is determined by both the CR scattering rates and the hydrodynamic adjustments to the CR pressure. The timescale for the establishment of the steady state of the CR flux is in general longer than that needed to reach the steady state of the diffusion coefficient \citep{thomas_cosmic-ray-driven_2023}. Whether or not the steady state of CR flux can be reached depends on the bulk CR speed. It is the CR pressure gradient along the magnetic field that drives a CR flux to propagate along the field lines. If CRs move faster than the Alfv{\'e}n speed, they are able to excite gyroresonant Alfv{\'e}n waves on which they scatter. Sufficiently frequent scattering is a necessary condition for reaching the steady state of the CR flux. Thus, in the opposite case, i.e., when the bulk CR speed is smaller than the Alfv{\'e}n speed, CRs damp the waves, thereby reducing scattering and thus increasing the dynamical CR timescales associated with the adjustments to the CR pressure so that a steady state of the CR flux is not reached. In this case, the CR pressure gradient along the magnetic field is small, which implies a small growth rate of the gyroresonant instability. In addition, the CR pressure gradient can be rearranged on MHD timescales in response to gas motions and magnetic forces, which can also cause the CR flux to deviate from its steady state value. \\
\indent
Galactic wind models that include CR feedback and CR interactions with gyroresonant Alfv{\'e}n waves make it possible to self-consistently compute the CR diffusion coefficient and CR transport speeds in the two-moment approximation of CR transport (see Section~\ref{sec:two-moment CR hydro}). These models demonstrate that CR transport does not reach steady state solutions describable solely in terms of diffusion, streaming, or a combination thereof \citep{thomas_cosmic-ray-driven_2023}. Moreover, while CR diffusion indeed quickly reaches steady state in most environments, there also exist regions where equilibrium diffusion coefficient is not established. In these regions, the CR gradient is almost perpendicular to the local magnetic field (and the CR pressure is almost constant along the magnetic field). Consequently, the CR speed is smaller than the local Alfv{\'e}n speed and CRs are unable to excite gyroresonant Alfv{\'e}n waves leading to very high values of the diffusion coefficient at these ``Alfv{\'e}n wave dark regions.'' This is illustrated in Fig.~\ref{fig:thomas} that shows cross sections perpendicular to the disk of the Alfv{\'e}n energy density (left panel) and CR diffusion coefficient values (right panel). 

\paragraph{Halo mass dependence of galactic wind mass loading.}
Observationally-determined wind mass loading factors can serve as a means of constraining CR feedback models. Ultra-violet absorption line measurements suggest that the mass loading depends weakly on galaxy mass, with studies hinting at either no significant scaling with mass, consistent with $\eta_\rmn{m}\propto M_{\rm vir}^{-\alpha_M}$ with $\alpha_M\sim1/3$ to $2/3$ within the large scatter \citep{heckman_systematic_2015}, or a weak decrease with stellar mass as $M_\star^{-0.4}$ \citep{chisholm_mass_2017}.
This slope of the relationship is consistent with the scaling of the mass loading factor $\eta_\rmn{m}$ with galaxy circular velocity $\eta_\rmn{m}\propto \varv_{\rm circ}^{-\alpha_\varv}$, where $\alpha_\varv\sim 1$ to $2$, that is necessary to explain the low-mass end of the stellar mass function of galaxies \citep[e.g.,][]{oppenheimer_feedback_2010}.  From the theoretical standpoint, the consensus that emerges from global galactic models of stellar feedback aided by CR processes is that they can drive stronger winds in smaller systems \citep{uhlig_galactic_2012, booth_simulations_2013, jacob_dependence_2018, dashyan_cosmic_2020, farcy_radiation-magnetohydrodynamics_2022} but are insufficient to drive winds in halos more massive than $\sim10^{12}$~M$_{\odot}$ \citep[e.g.,][]{fujita_cosmic_2018}. However, there are considerable differences between theoretical predictions for the dependence of the mass loading factor on halo mass both in terms of the normalization and slope of the relation.\\
\indent
We now discuss various global (non-cosmological) galactic simulations of CR feedback from the perspective of deriving galactic wind mass loading factors. There are two popular setups for this type of simulations that potentially differ in the resulting wind speeds and mass loading factors, especially if measured at larger heights: (i) ``disk models'' that start from an initial configuration of a stellar bulge and an exponential disk of stars and gas embedded in a dark matter halo, and (ii) a ``collapsing halo'' setting that starts from a rotating dark matter halo filled with gas, which starts to cool at the beginning of the simulation. In the latter scenario, the densest gas at the center cools fastest, conserves its specific angular momentum, and starts to form a disk inside out. This disk finds itself embedded in a turbulent and hot CGM that has been heated by the outwards propagating accretion shock \citep[e.g.,][]{springel_cosmological_2003,pfrommer_simulating_2022}. Moreover, the infalling CGM provides inertia against which the galactic wind has to do volume work.
We caution that while many trends that emerge from these models are broadly self-consistent, exact comparisons of models with different physics is complicated by the fact that no unique definition of the mass loading factor exist with various authors using custom definitions of this quantity. Similarly, the adopted definitions of the mass loading factor are in general not identical to those considered in observational studies. Nevertheless, clear general trends emerge from the simulations and illustrate the role of CR feedback in galactic wind driving. \\
\indent
Early hydrodynamical simulations of CR feedback using the ``collapsing halo'' setup demonstrated that strong winds can develop in disk galaxies less massive than $\sim\!10^{11}$M$_{\odot}$ \citep{uhlig_galactic_2012}. General patterns that transpire from these simulations are that in dwarf galaxies CR feedback leads to significant suppression of star formation (up to a factor of $\sim5$) and large mass loading factors (also up to a factor of $\sim5$). While the wind speed increases with the halo mass, the mass loading factor exhibits the opposite trend becoming negligibly small at  $\sim 10^{11}$~M$_{\odot}$. However, even in massive systems, where streaming did not lead to significant outflows, strong fountain flows were present. In these simulations, CR transport is modeled as streaming at the local sound speed and, similarly, the heating of the gas associated with streaming is proportional to the CR pressure gradient and local sound speed. Consequently, hotter regions are preferentially heated, which leads to a significant sink of CR energy \citep[cf.,][]{wiener_cosmic_2017} that may limit the impact of CR wind driving due to fast radiative cooling of the gas heated by CRs.\\
\begin{figure}
  \begin{center}
    \includegraphics[width=0.9\textwidth]{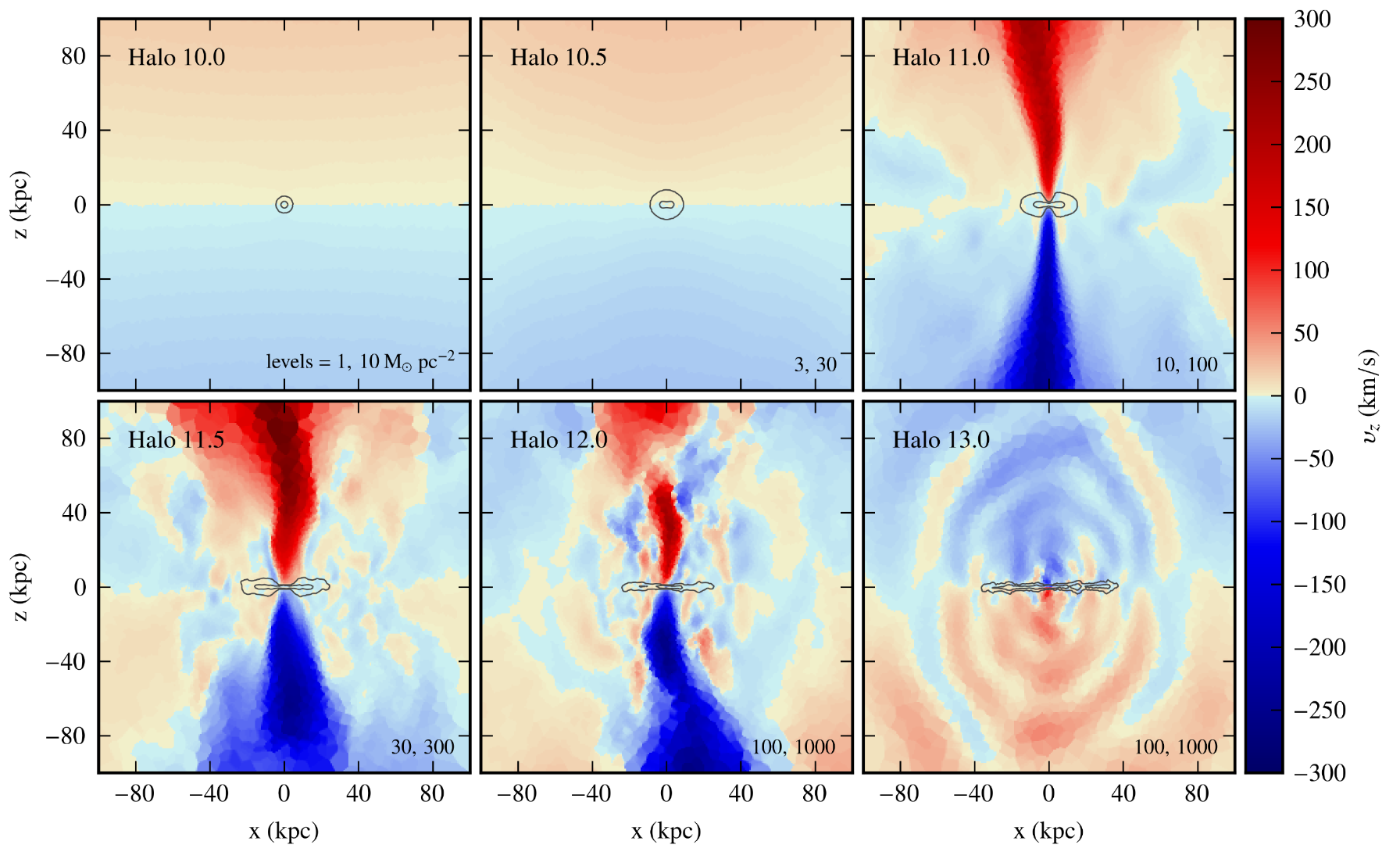}
  \end{center}
  \caption{Projections of the vertical component of the gas velocity for halos ordered according to mass (logarithm of mass is shown in the upper left corner of each panel). Note that quasi-spherical CR-driven outflows exist in dwarf galaxies; they transition to biconical outflows at intermediate mass galaxies ($\gtrsim10^{11}$~M$_{\odot}$) and fail to exist in halos more massive than that of the Milky Way. Image from \citet{jacob_dependence_2018} reproduced with permission from MNRAS.  }
  \label{fig:jacob}
\end{figure}
\indent
Despite the above limitations, other models that do not rely on these simplifying assumptions find broadly consistent trends albeit with significant quantitative differences. For example, global hydrodynamical simulations using the ``disk model'' that incorporate CRs via isotropic diffusion also lead to wind velocities increasing with circular velocity of the halo \citet{booth_simulations_2013}. These simulations by construction do not include streaming losses as CR transport is treated within the framework of the extrinsic CR confinement model. In the simulations by \citet{booth_simulations_2013} that include CR feedback, the wind velocity is substantially reduced (compared to the hydrodynamic-only cases) and, unlike in the case of thermal SN feedback, it matches the slope and normalization of the observed relation \citep{schwartz_keckhires_2004,rupke_outflows_2005}. Similarly to \citet{uhlig_galactic_2012}, the mass loading in these simulations was found to decrease with the halo mass. In dwarf galaxies, the magnitude of the mass loading was significantly larger when CR feedback was included as opposed to the case where only thermal SN feedback was considered. Surprisingly, the mass loading factor of $\sim 0.3$ for the Milky Way-size halo matched that reported by \citet{salem_cosmic_2014}, who considered similar models but without hadronic and Coulomb losses. The above models were further generalized to include magnetic field and anisotropic CR diffusion \citep{pakmor_galactic_2016,ruszkowski_global_2017, pfrommer_simulating_2017, jacob_dependence_2018}. Projections of the vertical component of the gas velocity for halos ordered according to mass are shown in Fig.~\ref{fig:jacob} and reveal that only intermediate mass galaxies show biconical outflows \citep{jacob_dependence_2018}. 
Similarly to earlier studies, the mass loading factor in these simulations was found to decrease with the halo mass with CRs unable to drive winds for halos more massive than $\sim10^{12}$~M$_{\odot}$. However, the scaling of the mass loading factor with the halo mass $M_\rmn{vir}^{-\alpha_M}$, with $\alpha_M\sim 1$ to 2 independently of whether CR diffusion was isotropic or magnetic-field-aligned, was steeper than found in observations and other galactic wind simulations \citep[e.g.,][]{muratov_gusty_2015}. However, despite the above differences, the star conversion efficiency predicted by the model of  \citet{jacob_dependence_2018} agrees with the empirically derived one \citep{vogelsberger_model_2013}.\\
\indent
In addition to the differences in the slope of the mass loading vs.\ halo mass relation, \citet{jacob_dependence_2018} also predict different normalization of this relation compared to other models. For example, while \citet{salem_cosmic_2014} and \cite{booth_simulations_2013} predict strong winds for halo mass $\sim\!10^{12}$~M$_{\odot}$, \citet{jacob_dependence_2018} find negligible mass loading at this halo mass. This difference is unlikely to be due to the neglect of CR cooling in \citet{salem_cosmic_2014} because both \citet{jacob_dependence_2018} and \cite{booth_simulations_2013} include CR losses and yet \citet{booth_simulations_2013} and \citet{salem_cosmic_2014} essentially find the same mass loading factor value at this halo mass. For smaller mass halos of $10^{10}$ to $10^{11}$~M$_{\odot}$, MHD simulations suggest that while the mass outflow rates increase significantly in the presence of diffusing and streaming CRs, the mass loading factor falls short of that found in observations \citep{dashyan_cosmic_2020}. Despite similar physics, mass loading values reported by \citet{dashyan_cosmic_2020} are an order of magnitude smaller than those reported by \citet{jacob_dependence_2018} for this same mass range. However, the exact comparisons of the mass loading factors inferred from the above simulations is difficult due to (i) different mass loading factor definitions adopted by various authors and, perhaps more importantly, (ii) because of different ways of initializing the simulations either via a pre-existing ``disk models'' \citep[e.g.,][]{booth_simulations_2013} or a ``collapsing halo'' models \citep[e.g.,][]{jacob_dependence_2018}). \\
\indent
The first global simulations to simultaneously incorporate both CRs and radiation feedback were presented by \citet{farcy_radiation-magnetohydrodynamics_2022}. In agreement with earlier studies, CRs were found to substantially boost wind mass loading factors and reduce the star formation rates with the efficiency decreasing with halo mass. However, the combined CR and radiation feedback was still less efficient than artificially boosted feedback designed to reproduce the observed galaxy luminosity function in cosmological simulations \citep{rosdahl_sphinx_2018}.\\
\indent
Nevertheless, the true potential of CR feedback has not yet been fully elucidated as current modelling of CR physics in simulations is incomplete. Specifically, in addition to radiation pressure, several missing CR physics factors that may boost the efficiency of feedback are required to increase the realism of feedback modeling and need to be \textit{simultaneously} incorporated in future simulations. These factors include (i) self-consistent and non-equilibrium computation of the CR diffusion coefficient and CR transport speeds \citep{thomas_cosmic-ray-driven_2023}, (ii) spectrally-resolved CR transport that not only results in a qualitatively different wind driving \citep{girichidis_spectrally_2022} but also enables detailed model comparisons to radio and $\gamma$-ray data \citep{werhahn_gamma-ray_2023}, (iii) CR transport in the multi-phase ISM with a self-generated diffusion coefficient that responds to the prevalent wave damping processes of the different phases of the ISM, (iv) in situ CR acceleration boost at the location of SNRs (\citealt{diesing_effect_2018}; cf.\ \citealt{jana_role_2020}, who do include in situ acceleration and find negligible effect of CR feedback on wind driving in general, but consider only early stages of feedback), and (v) cosmological effects (e.g., \citealt{chan_cosmic_2019,hopkins_but_2020,buck_effects_2020}; see also Section \ref{Cosmological effects}). 

\paragraph{Magnetic dynamo and CR transport.}
Galactic winds and fountain flows discussed above play a critical role in magnetic field amplification via the galactic dynamo process \citep{parker_fast_1992, hanasz_cosmicray_2000, brandenburg_astrophysical_2005,shukurov_galactic_2006}. Since CRs can launch the winds, they are an important ingredient in models aiming to explain the magnetic field amplification. The galactic CR dynamo model includes a number of interdependent elements such as differential rotation, Coriolis force, outflow driven by CRs, and fast magnetic reconnection. In the original Parker galactic dynamo model \citep{parker_fast_1992}, CRs are an essential ingredient of the model as they provide buoyancy to the gas. Unlike in the case of buoyancy due to magnetic fields, CRs do not introduce tension forces that tend to suppress the convective rise of the gas. Thus, the total buoyant forces can exceed magnetic tension forces when the CR energy density is sufficiently large compared to the equipartition energy density of the magnetic field. These buoyant forces play an important role in the dynamo process as they facilitate the production of the radial magnetic field component from the azimuthal one. As the magnetic loops rise and twist, they reconnect forming amplified magnetic structures that posses a radial magnetic field component. For this process to operate efficiently, reconnection needs to be fast in order to form large scale poloidal magnetic fields. Similarly, the azimuthal component can be generated from the radial one via differential rotation.\\
\indent 
Several aspects of this process can be captured via local zoom-in simulations of galactic disks. In particular, early work demonstrated that magnetic fields can grow exponentially fast, reaching equipartition with the turbulent energy on timescales comparable to 100 Myr even in the limit of vanishing explicit magnetic resistivity \citep{hanasz_amplification_2004,hanasz_building_2004}. Interestingly, these shearing box simulations showed that the mean radial component of the magnetic field can be generated from the mean azimuthal one and vice versa, consistent with the incoherent $\alpha-\Omega$ dynamo theory \citep{vishniac_incoherent_1997}. However, other aspects of the problem are better investigated using global simulations. For example, the fastest growing modes of the Parker instability have characteristic sizes close to $\sim 1$ kpc \citep{giz_parker_1993,rodrigues_parker_2016}, which is comparable to the sizes of zoom-in boxes \citep[e.g.,][]{simpson_role_2016}. This limitation may thus lead to reduced instability growth rates. Limiting the geometry of the computational domain to zoom-in boxes may also affect the distribution of CRs that play key role in the dynamo process. For example, in the stratified box cases with periodic boundary conditions in the horizontal directions, CRs will find it harder to escape the disk, which may result in CR energy density vastly exceeding that in the magnetic fields, which is contrary to observations \citep[e.g.,][]{hanasz_cosmic-ray-driven_2009}. This can be avoided in global models in which CRs can escape in all directions rather than only along the vertical axis, and where disk spiral structure can develop enabling more realistic CR distributions. 
As discussed below, general consensus that emerges from global simulations is that the field amplification is fast, magnetic fields reach saturation with (a fraction of) the kinetic energy of the gas, and that the level of anisotropy of CR transport has an important positive impact on the magnetic field amplification. While these conclusions are in line with the findings from those based on zoom-in simulations, there are important differences between the results based on these two approaches.\\
\indent
The dynamo process can generate and amplify large scale fields, but it requires weak initial seed fields to operate. These fields can be supplied by SNRs that can simultaneously inject CRs via diffuse shock acceleration. 
This process could result in small yet astrophysically relevant fields before the dynamo process could amplify them further. For example, explosions of $\sim 10^{6}$ pleirionic SNe could generate seed fields at the level of $10^{-3}~\mu$G \citep{rees_origin_1987}. 
Global simulations that emulate these processes by injecting weak small-scale dipole magnetic fields associated with individual SNe, rather than starting from a predefined initial field, demonstrate that the volume-filling magnetic field can indeed be amplified to equipartition with the gas energy and that small scale fields can be efficiently reordered on timescales comparable to the galactic orbital timescale forming large scale fields in the disk and galactic halo \citep{hanasz_global_2009}. In these simulations, the field reconfiguration is aided by explicit Ohmic resistivity.  \\
\indent
The tendency for the field amplification to reach saturation values in global simulations is a general one and is robust with respect to changes in the model initialization and assumptions. A number of global simulations demonstrated that a fluctuating small-scale dynamo \citep{wang_magnetohydrodynamic_2009,rieder_small-scale_2016} and differential rotation \citep{dubois_magnetised_2010}, can amplify the magnetic fields up to typical values of several to several tens of $\mu$G in dense ISM regions, a process that can be aided by CRs (e.g., \citealt{pakmor_galactic_2016,pfrommer_simulating_tmp_2017,pfrommer_simulating_2022,dashyan_cosmic_2020,nunez-castineyra_cosmic-ray_2022}). This is in agreement with the observed values \citep{beck_magnetic_2000,crutcher_magnetic_2012}.\\
\indent
Moreover, the conclusions regarding the formation of large scale fields extending to the halo also appear to be robust to changes in the magnitude and topology of the initial field. In simulations starting with large scale toroidal fields of $\mu$G strength, such fields develop despite continuous small scale tangling of the field by exploding SNe, and the winds expel magnetic energy into the galactic halo forming large-scale open magnetic field lines \citep{hanasz_cosmic_2013}. Compared to the CR advection-only case, the formation of the vertical component of the magnetic field can be enhanced as much as two orders of magnitude when active CR transport is also taken into consideration \citep{dashyan_cosmic_2020}.\\
\indent
Important quantitative differences in the field amplification emerge when CR transport anisotropy is taken into account. When diffusion is anisotropic (or suppressed), CRs are initially confined to the disk because the dominant component of the magnetic field is aligned with the plane of the disk due to galactic rotation. This confinement drives strong gradients in the vertical and radial magnetic fields close to the disk. These strong gradients are responsible for the rapid rate of magnetic field amplification in the anisotropic diffusion and purely advective cases. In the case of isotropic CR diffusion, the outflow rate is larger and hence it also carries more magnetic flux to the halo that has to be replenished by the dynamo in the disk, which results in a effectively suppressed dynamo \citep{pakmor_galactic_2016, nunez-castineyra_cosmic-ray_2022}. Depending on the outflow rate in the isotropic diffusion case, field amplification may not reach equipartition with turbulence or may only saturate at equipartition after a Hubble time, while the fields can be very quickly amplified (with a rate that depends on the magnetic Reynolds number) when diffusion is anisotropic or absent. This picture is consistent with the analytic galactic dynamo theory \citep{shukurov_galactic_2006} and the results from zoom-in simulations that suggest the existence of a critical level of anisotropy necessary for the CR dynamo to efficiently amplify magnetic fields \citep{hanasz_cosmic-ray-driven_2009}. However, contrary to observations, for Milky Way conditions, the magnetic energy density in these simulations is always significantly subdominant compared to the CR energy density which may be either due to the omission of CR losses or the geometrical effects mentioned above. An interesting by-product of the suppression of CR diffusion that leads to the increase in both CR and magnetic pressures is that it slows down star formation, and thus has consequences for the net energy losses of CR due to hadronic and Coulomb effects as described below.
\begin{wrapfigure}{r}{0.5\textwidth}
  \begin{center}
  \vspace{-2em}
    \includegraphics[width=0.4\textwidth]{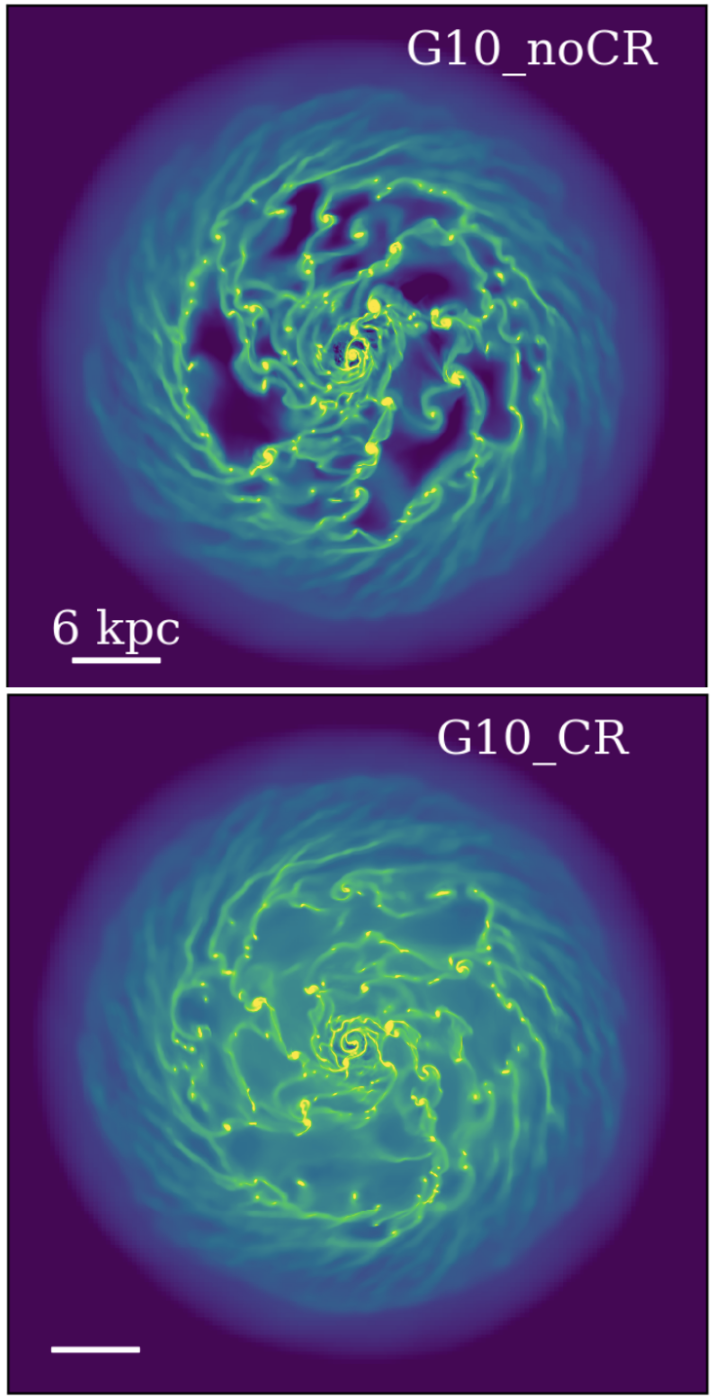}
  \end{center}
  \caption{Face-on distribution of the hydrogen column density in a Milky Way-like galaxy. Inclusion of CRs significantly smoothens the gas distribution (cf.\ top/bottom panel for the case without/with CRs; \citealt{farcy_radiation-magnetohydrodynamics_2022}); reproduced with permission from MNRAS.}
  \label{fig:farcy}
  \vspace{-1.5em}
\end{wrapfigure}
\indent
Further complexity to the problem of the field amplification can be introduced by considering cooling and collapsing galactic halos \citep{pfrommer_simulating_2022}. In this scenario, the field amplification occurs as a result of a fluctuating small-scale dynamo driven by (i) turbulence injected at the outward-propagating corrugated accretion shock front, and (ii) Kelvin-Helmholtz body modes that are excited at the interface of the supersonically (with respect to the warm ISM phase) rotating galactic disk and the hot CGM. In tandem with the CR energy injection at SNe, these modes source ISM turbulence throughout the entire star-forming disk. While this mechanism exponentially amplifies the magnetic field to reach saturation with the thermal and CR energy densities in Milky Way galaxies, in smaller galaxies the magnetic field energy density reaches sub-equipartition values with respect to these energy densities (though it saturates at the kinetic energy density of the gas). Interestingly, this model matches the global FIR-radio correlation across four orders of magnitude in luminosity (see Section~\ref{sec:radio emission} for more discussion). 

\begin{figure}
  \begin{center}
    \includegraphics[width=0.9\textwidth]{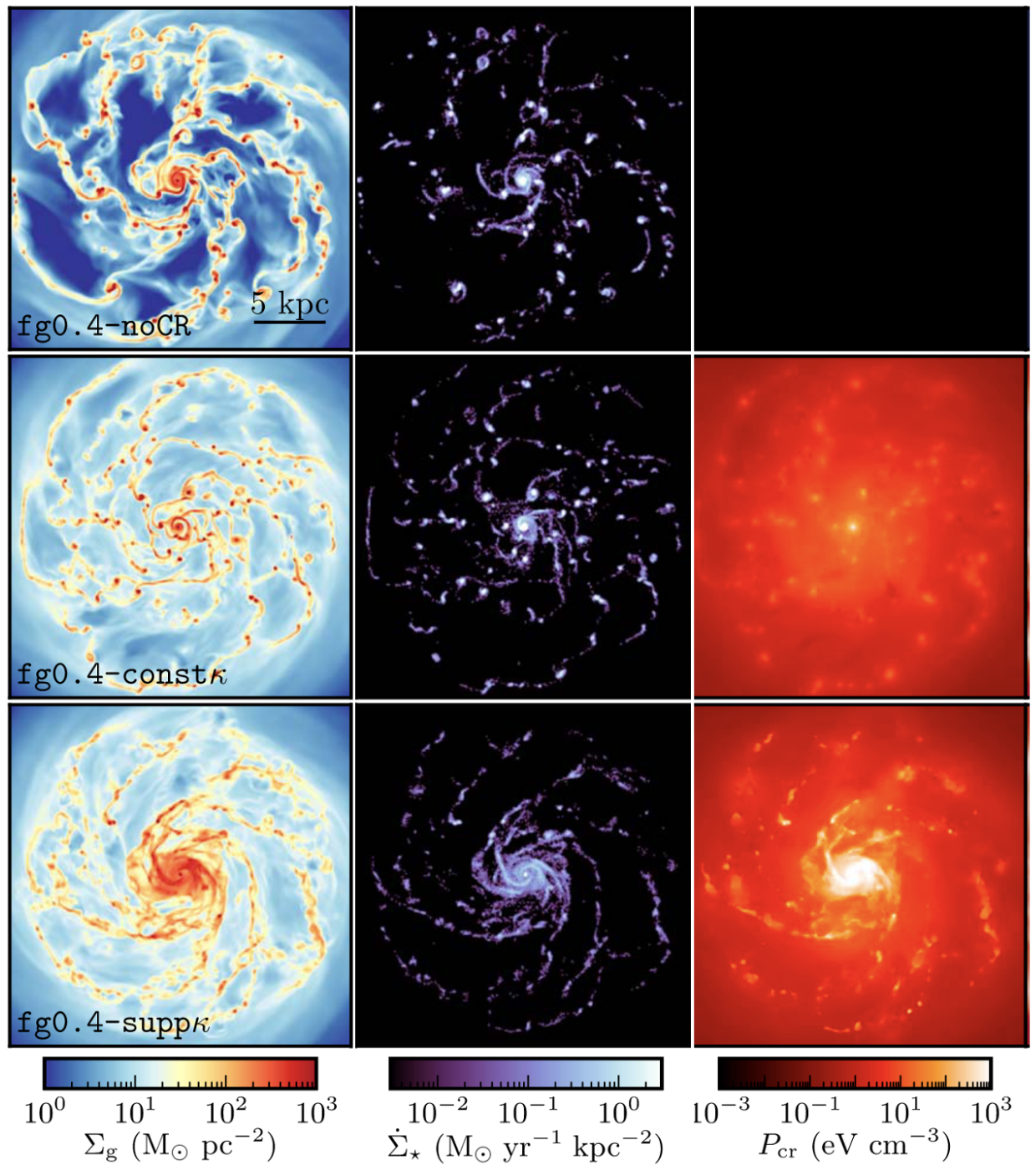}
  \end{center}
  \caption{Impact of CRs on global galactic structure. From left to right, columns show surface density of the gas, star formation rate, and CR pressure. From top to bottom, different rows correspond to cases without CRs, isotropic CR diffusion, and isotropic diffusion suppressed near the CR injection sites. Inclusion of CRs prevents dense clump formation and allows the disk to maintain a ``grand design'' spiral structure and the effect is particularly strong when diffusion is locally suppressed. Image based on \citet{semenov_cosmic-ray_2021}; reproduced with permission from ApJ.}
  \label{fig:semenov}
\end{figure}

\paragraph{Disk structure and CR transport.}
CR feedback can shape both the local and global structure of galactic disks through driving of galactic winds and suppressing star formation. As pointed out in several studies (e.g., \citealt{salem_cosmic_2014,ruszkowski_global_2017,pfrommer_simulating_tmp_2017}), the latter process occurs due to significant increase in the midplane pressure provided by CRs. This additional pressure stabilizes the disk, prevents the gas from collapsing to form very dense clouds, and thus reduces the star formation rate. Figure~\ref{fig:farcy} illustrates this effect and shows face-on distributions of the hydrogen column density in a Milky Way-like galaxy. Inclusion of CRs significantly smoothens the gas distribution (see bottom panel) compared to the case without CRs \citep[top panel,][]{farcy_radiation-magnetohydrodynamics_2022}. \\
\indent
As discussed in Section~\ref{dynamics_of_CR_near_sources}, observations of the enhanced $\gamma$-ray emission in the vicinity of SNRs \citep{casanova_molecular_2010, hanabata_detailed_2014} suggests that CR transport speeds can be significantly suppressed near these sources compared to the Galactic average. From the theoretical standpoint, this suppression could be due to the excitation of non-resonant modes that can locally amplify the magnetic field and drastically reduce the diffusion coefficient by orders of magnitude
\citep{bell_turbulent_2004, reville_universal_2013, caprioli_simulations_2014-1,caprioli_simulations_2014,schroer_dynamical_2021,schroer_cosmic-ray_2022}.
One consequence of this suppression of transport is that the accumulation of CRs near SNe and superbubbles further increases local CR pressure gradients that prevent the formation of massive gas clumps and significantly suppress star formation \citep{semenov_cosmic-ray_2021}. This is illustrated in Fig.~\ref{fig:semenov}, which shows the impact of CRs on global galactic structure. Most notably, this figure also demonstrates that when CR diffusion is locally suppressed in the vicinity of SNe, the disk can maintain a ``grand design'' spiral structure. This is consistent with recent discoveries of high-redshift galaxies with massive and dynamically cold disks \citep{neeleman_cold_2020,rizzo_dynamically_2020}.
From the point of view of feedback modeling, some aspects of the impact of CR diffusion suppression are qualitatively similar to those seen in the ``delayed cooling'' approach that is sometimes employed in galaxy formation simulations (e.g., \citealt{thacker_implementing_2000,agertz_formation_2011}; see also Section \ref{CR_and_star_formation}). In the latter approach, radiative cooling is switched off for some time following SN explosions. Rather than getting quickly radiated away, the feedback energy can locally accumulate and strong pressure gradients can develop that are capable of dispersing dense gas clouds, which leads to the suppression of star formation. This effect is also similar to that due to increased star formation efficiency or including an ad hoc SN energy and momentum boost. However, in the case of suppressed CR diffusion, this effect is modeled more from first-principles, is better physically motivated and, in contrast to the effective feedback models, CR feedback also affects the state of the ISM and wind launching in other fundamental ways. 

\paragraph{CR energy losses and $\gamma$-ray emission.}
The issues of magnetic field amplification, multiphase structure of the ISM, and CR transport all play important and interrelated roles in determining the net CR energy losses and, thus, the $\gamma$-ray emission due to pion production. Below we discuss the latest results from global simulations that make predictions for the $\gamma$-ray luminosities of galaxies. Since there is currently no consensus on the single most important factor responsible for the significant differences between these predictions, we follow this discussion by examining the various factors that may play important roles in determining the level of $\gamma$-ray emission.\\
\indent
Global non-cosmological simulations of dwarfs and $L_{*}$ galaxies employing the FIRE-2 model demonstrated that $\gamma$-ray luminosities are significantly overproduced compared to observations when CR transport occurs only via advection and streaming \citep{chan_cosmic_2019}. However, in these simulations, agreement with the observations can be achieved when very fast CR diffusion is included with diffusion coefficients in the range $10^{29-31}$~cm$^{2}$~s$^{-1}$. In these simulations the streaming transport is subdominant compared to diffusion. One fundamental reason for the CR energy losses to increase for slower CR diffusion, or when only streaming rather than diffusion is included, is that in this case CRs spend more time in the densest regions, where the hadronic and Coulomb CR cooling rates are the highest (e.g., \citealt{dashyan_cosmic_2020}, who confirm that streaming losses are generally subdominant in dwarfs). Even if CRs do not spend a significant amount of time and lose a lot of energy in the densest gas due to efficient wave damping \citep{farber_impact_2018}, large values of the diffusion coefficient may be needed in the tenuous warm ionized medium and CGM surrounding these clouds in order to reduce the losses and excessive $\gamma$-ray emission as demonstrated in cosmological FIRE-2 simulations (\citealt{hopkins_testing_2021}, see also Section \ref{Cosmological effects}).\\
\indent
However, these results are in tension with other reports suggesting that $\gamma$-ray luminosities are fully consistent with the observational \textit{Fermi} constraints even for typical diffusion coefficient values of $3\times 10^{28}$ to $10^{29}$~cm$^{2}$s$^{-1}$, equivalent to isotropic values comparable to or smaller than $3\times 10^{28}$~cm$^{2}$s$^{-1}$ \citep{pfrommer_simulating_2017,buck_effects_2020,nunez-castineyra_cosmic-ray_2022}. Specifically, these simulations suggest that typical luminosities are an order of magnitude smaller than those based on the above FIRE-2 simulations for the same level of diffusion, and about five times smaller than those based on AREPO simulations when measured per star formation rate \citep{werhahn_cosmic_ii_2021}. Moreover, global simulations that include suppression of CR diffusion near the sites of SN explosions also find agreement with the observed star formation vs.\ $\gamma$-ray luminosity relation \citep{semenov_cosmic-ray_2021}.\\
\indent
There are several factors that may be responsible for the discrepancies between the above results. The key factors that may have the most important impact on $\gamma$-ray emission are: (i) the level of magnetic field amplification that affects the speed of CR transport and energy losses, (ii) ISM structure and its relationship to the spatial distribution of CR feedback injection sites, and (iii) modeling of CR spectra.
The speed of CR streaming and the associated streaming losses increase with the strength of the magnetic field. As mentioned above, it has been suggested that streaming may be subdominant compared to diffusion in terms of CR transport speed and that CR streaming losses may be much smaller than Coulomb and hadronic CR cooling rates. While the exact comparisons of the magnetic field strengths in the simulations considering CR losses and the associated $\gamma$-ray emission are difficult due to the non-uniformity of the data sets (i.e., due to the different ways in which the fields are measured with groups reporting averages vs.\ profiles, volume- vs.\ mass-weighted values, averages in volumes of different sizes and shapes, or because the values are reported for halos of different mass), the strengths of magnetic fields present in the simulations likely vary substantially between simulations performed by different groups. Consequently, the streaming speed may be underestimated when the magnetic fields are relatively weak even if the streaming boost due to the efficient wave damping processes, or the increase in the Alfv{\'e}n speed due to gas ionization, is very large (in this case, the streaming losses would also be underestimated). Similarly, even if the magnetic fields are close to equipartition but the streaming boost is only moderate, the effects of streaming could be too weak. In addition to these factors, magnetic field amplification, in conjunction with CR pressure support, affects the level of star formation \citep[e.g.,][]{nunez-castineyra_cosmic-ray_2022} and thus indirectly reduces the $\gamma$-ray emission. We return to these issues in Section \ref{Cosmological effects} below.\\
\indent
Setting aside the issues of the magnetic field amplification, $\gamma$-ray luminosities can also be suppressed if the distribution of CRs correlates weakly with the gas density and vice versa. This can be most easily illustrated by comparing results from simulations performed for identical initial conditions setups, resolutions, and numerical tools, but with different prescriptions for the physics of the ISM. In this context, a comparison of global simulations that follow the development of thermal instability in the cold gas \citep{nunez-castineyra_cosmic-ray_2022} and those that do not lead to the multiphase structure of the ISM \citep{dashyan_cosmic_2020} reveals that the latter result in moderately reduced $\gamma$-ray emission per star formation rate. For similar reasons, in the simulations where CR diffusion is suppressed near feedback sites, $\gamma$-ray luminosites and CR energy losses are significantly suppressed 
\citep{semenov_cosmic-ray_2021} because CRs primarily occupy underdense superbubbles, where collisions between CRs and thermal ISM ions are less frequent. On a qualitative level, this is consistent with the conclusions from controlled zoom-in numerical experiments that allow for relative offsets between the locations of gas density peaks and the SNe injecting CRs \citep{simpson_how_2023}. Such offsets could be due to SN explosions in regions that have been evacuated by prior explosions, radiation pressure, or stellar winds. These models are broadly consistent with $\gamma$-ray observations.\\
\indent
Finally, incorporation of energy-dependent CR fluid can result in the reduction of the level of $\gamma$-ray emission if steeper spectra are considered \citep{nunez-castineyra_cosmic-ray_2022} or when the spectra are allowed to age due to CR losses \citep{werhahn_cosmic_iii_2021, girichidis_spectrally_2022}. 

\subsection{Cosmological effects of cosmic ray-driven winds}\label{Cosmological effects}
Complete understanding of CR feedback processes requires insights from cosmological simulations capable of simultaneously resolving the structure of the CGM and ISM. Such an extreme dynamical range is necessary because certain aspects of CR-driven outflows cannot be fully captured in idealized global models of isolated galaxies. Rather than relying on approximate prescriptions for the distribution of mass and angular momentum in the CGM in global models of isolated galaxies, cosmological simulations allow for self-consistent and most realistic modeling of the accretion of the gas onto galaxies from the CGM and the interaction of the winds with the accreting gas. The cosmological approach also offers the most straightforward way to model the interaction of a CR-dominated CGM with the accretion shocks, and to incorporate mergers and \textit{continuous} accretion of gas onto galaxies, which gives rise to extended (and perhaps even bursty) star formation histories rather than a single episode of accretion and one burst of CR injection as expected in global models of isolated galaxies. Such a long-term cosmological evolution of galaxies provides more robust predictions for the role of CR feedback in shaping structural properties of galaxies and the surrounding CGM, which is discussed next. \\
\indent
We caution that while the cosmological approach comes with clear advantages, such as the possibility to realistically model boundary conditions (that is not possible with zoom-in simulations of galactic patches or isolated global models) to simulate the interaction of outflows with the CGM or the formation of ``galactic fountains,'' it also comes with certain trade offs. These trade offs are related to limited dynamical range of the simulations and involve the need to resort to the effective modeling of the ISM physics as well as small scale physics of mixing using sub-grid models \citep{fielding_multiphase_2020,weinberger_modelling_2023}.

\subsubsection{Analytic considerations} 
\label{sec:anaytics}
Considerable insight into the role of CRs in shaping the properties of galaxies, the CGM, and the dependence of the solutions on key model parameters can be gleaned from a simple analytic model that incorporates CR injection and diffusion out of gaseous halos in dark matter potentials \citep{hopkins_but_2020,ji_properties_2020}. In this approximate model, it is assumed that CRs are generated by SNe at the energy injection rate of $\dot{E}_{\rm cr}$ at the center of an isothermal sphere characterized by a circular velocity $V_{\rm c}$, scale radius $r_{\rm s}$, and gas fraction $f_\rmn{gas}$. In steady state, and for a constant diffusion coefficient $\kappa$, this leads to a simple distribution of CR energy density $\eps_\rmn{cr}=\dot{E}_{\rm cr}/(4\pi\kappa r)$ and the corresponding CR pressure gradient $\bs\nabla P_{\rm cr}=-\eps_\rmn{cr}\bs{\hat{r}}/(3r)$, which can be used to compare the magnitude of the CR pressure force opposing the gravitational force $|\bs\nabla P_{\rm cr}|/|\rho\bs\nabla\Phi |\sim \dot{E}_{\rm cr} G r_{\rm s}/(3 \kappa f_{\rm gas}V_\rmn{c}^{4})$. Assuming that (i) the CR energy injection rate is proportional to the star formation rate $\dot{M}_\star$, i.e., $\dot{E}_{\rm cr} = \zeta_{\rm cr}\epsilon_{\rm sn}\dot{M}_\star$ (where $\zeta_{\rm cr}\sim 0.05$ is the CR acceleration efficiency and $\epsilon_{\rm sn}\sim 10^{51}{\rm erg}/(100 ~\rmn{M}_{\odot})$ is the SN yield per solar mass of star formation), and (ii) for sub-$L_{*}$ galaxies $\dot{M}_\star\sim \alpha M_\star/\tau_\rmn{H}$ (where $M_\star$ is the mass in stars \citep{mitra_equilibrium_2017}, $\tau_\rmn{H}\approx 14\,(1+z)^{-3/2}$Gyr is the Hubble time, $\alpha\sim 1$ which depends very weakly on halo mass), the ratio of the CR to gravitational force becomes
\begin{equation}
\frac{|\bs\nabla P_{\rm cr}|}{|\rho\bs\nabla\Phi |}\sim \frac{\alpha\zeta_{\rm cr,0.05}}{f_{\rm gas,0.1}\kappa_{28}(1+z)^{3/2}}\left( \frac{M_\star}{M(<r_\rmn{s})} \right),
\label{forcebalance2}
\end{equation}
where $\kappa_{28}$ is the diffusion coefficient in units of $10^{28}$~cm$^{2}$~s$^{-1}$, $\zeta_{\rm cr,\,0.05}$ is the CR acceleration efficiency in units of 0.05, and $f_{\rm gas,\,0.1}$ is the gas mass fraction in units of 0.1. Given that $M_\star\propto M(<r_\rmn{s})^{2}$ for sub-$L_{*}$ galaxies in the mass range $\sim 10^{11-12}~\rmn{M}_{\odot}$, equation~\eqref{forcebalance2} reveals the following interesting trends: (i) the relative importance of CR pressure forces can be high in Milky Way-mass galaxies. For a constant $\kappa$, the force ratio declines for dwarfs where $M_\star/M(<r_\rmn{s})$ is smaller. (However, CR transport could self-regulate in dwarfs so that the effective $\kappa$ decreases and compensates for the impact of smaller $M_\star/M(<r_\rmn{s})$ thus making CR feedback also important in smaller systems.) (ii) Slower CR transport speed allows for more CR pressure buildup (see equation~\ref{forcebalance2}), which helps to drive the wind. (However, slower transport also increases the average residence time of CRs in the ISM and CGM. This could either increase CR energy losses compared to the CR energy injection rate, which would in turn weaken the impact of CRs, or cause the region to adiabatically expand due to strong CR pressure buildup, which would then lower CR collisional losses.) (iii) The impact of CRs on the dynamics is maximized at lower redshifts provided the above simple relationship between $\dot{M}_\star$ and $\dot{E}_{\rm cr}$ applies. If instead the star formation history is burstier in dwarfs \citep[for which there is observational evidence;][]{weisz_modeling_2012,kauffmann_quantitative_2014}, this would decrease the timescale of a star formation event $\tau_\rmn{H}$ and thus temporarily increase $\eps_\rmn{cr}$. \\
\indent
We stress that while this simple model offers first insights into the impact of CRs, it does have limitations that stem from making simplifying assumptions related to the neglect of more realistic CR transport, CR energy losses due to streaming heating, and neglect of the impact CR inertia on CR forces imparted to the gas, all of which can affect the distribution of CRs or effective CR force acting on the ISM and CGM and CR energy losses. This in turn can shift the regime of parameters where the effect of CRs is maximized. Nevertheless, the model provides a useful reference framework that reveals some of the trends seen in the simulations. In particular, it suggests that CR feedback should be very important for halo masses ranging from LMC to Milky Way mass ($\sim10^{11-12}~\rmn{M}_{\odot}$). For smaller halo masses, more work is needed to study the self-regulation of CR transport and whether the small star formation rates are sufficient to build up sufficiently large CR pressure. Similarly, for halo masses much larger than that of the Milky Way, star formation rates are too low for CRs from SNe to produce dynamically important pressure gradients especially given generally larger CGM pressures. We will come back to this halo mass regime in Section~\ref{agntheorysection} and discuss CR feedback associated with AGN jets.

\subsubsection{Structural properties of galaxies}  While the first cosmological simulations to include the effects of CRs focused on CR production in galaxy clusters, their contribution to the extragalactic isotropic $\gamma$-ray background, and the physics of radio halos and relics \citep[e.g.,][]{miniati_cosmic-ray_p_2001,miniati_cosmic-ray_e_2001,miniati_numerical_2003,pfrommer_simulating_2007,pfrommer_simulating_2008}, an important role of CR feedback in shaping the structural properties of galaxies was recognized early on with the advent of first cosmological galaxy formation simulations that included CR physics \citep{jubelgas_cosmic_2008,wadepuhl_satellite_2010,salem_cosmological_2014,chen_cosmological_2016}. These early hydrodynamic simulations demonstrated that CR feedback can have a dramatic impact on the baryon distribution and kinematics in galaxies. Specifically, they showed that CRs may significantly reduce star formation rates \citep{jubelgas_cosmic_2008} and thus help to address the ``missing satellites'’ problem \citep{klypin_where_1999,moore_dark_1999} -- by substantially suppressing the star formation efficiency in the smallest galaxies, the relatively small number of observed satellites could be reconciled with the theoretically predicted numbers of dark matter satellites \citep{wadepuhl_satellite_2010}. Furthermore, CRs could help to solve the ``angular momentum problem,’’ in which the efficient loss of the baryonic angular momentum leads to galactic disk sizes that are too small compared to observations \citep{jubelgas_cosmic_2008}. This positive impact of CR feedback on producing thin and extended galactic disks persisting to low redshifts was seen both in dwarf systems \citep{chen_cosmological_2016} and Milky Way analogs \citep{salem_cosmological_2014}. CR feedback effects also led to flatter and more realistic rotation curves when CR transport was included in the simulations. Interestingly, in dwarf galaxies, the inclusion of CR physics, and in particular CR diffusion, resulted not only in flatter and more extended disks but also in a good match of the models to the observed baryonic Tully-Fisher relation \citep{chen_cosmological_2016}.\\
\indent
Adding magnetic fields could in principle already modify the structural properties of galaxies without CRs through angular momentum transport by magnetic tension forces. While this scenario does not seem to be realized in practice in cosmological simulations of the growth of isolated galaxies \citep[][using the MHD approximation]{pakmor_magnetic_2017,hopkins_but_2020}, spiral galaxies reforming after a major merger event show more extended disks and better developed spiral structure in comparison to their pure hydrodynamical analogues \citep{whittingham_impact_2021}. This different galactic morphology is caused by efficient angular momentum redistribution by magnetic fields, which increases the central baryonic concentration after the galaxy merger. This prevents the formation of a bar, which forms in the simulation without magnetic fields, and thus implies a different star formation surface density and feedback strength when taking into account magnetic fields \citep{whittingham_impact_2023}.\\ 
\indent
Early hydrodynamic models with CR feedback \citep{chen_cosmological_2016} suffered from too low gas fractions compared to the observed systems and from cuspy dark matter profiles that are also in tension with observations \citep[e.g.,][]{moore_evidence_1994,walter_things_2008}. Moreover, more recent CR MHD models with subgrid-scale feedback tailored to solve the above mentioned problems found either a weaker or even opposite trend regarding disk sizes \citep{buck_effects_2020,hopkins_but_2020}. In agreement with analytic trends presented above, these models adopt an (effectively) constant CR diffusion coefficient and suggest that the effect of CRs in dwarf galaxies should be minor while the impact of SN CR feedback could be very important in the halo mass range $\sim 10^{11}$ to $\sim 10^{12}~M_{\odot}$. Depending on the details of the implementation of stellar feedback and CR physics, these models predict that global or integrated properties of galaxies either vary little when CRs are included \citep{buck_effects_2020} or have a significant effect on global properties, such as star formation rates that can be reduced by as much as a factor of $\sim 5$, or total stellar mass that can be suppressed by a factor of a few due to CR pressure gradients that support the CGM and slow down gas accretion \citep{hopkins_but_2020}. These effects can consequently substantially reduce central peaks in the galactic rotation curves.\\
\begin{figure}[tbp]
\begin{center}
\includegraphics[width=0.9\textwidth]{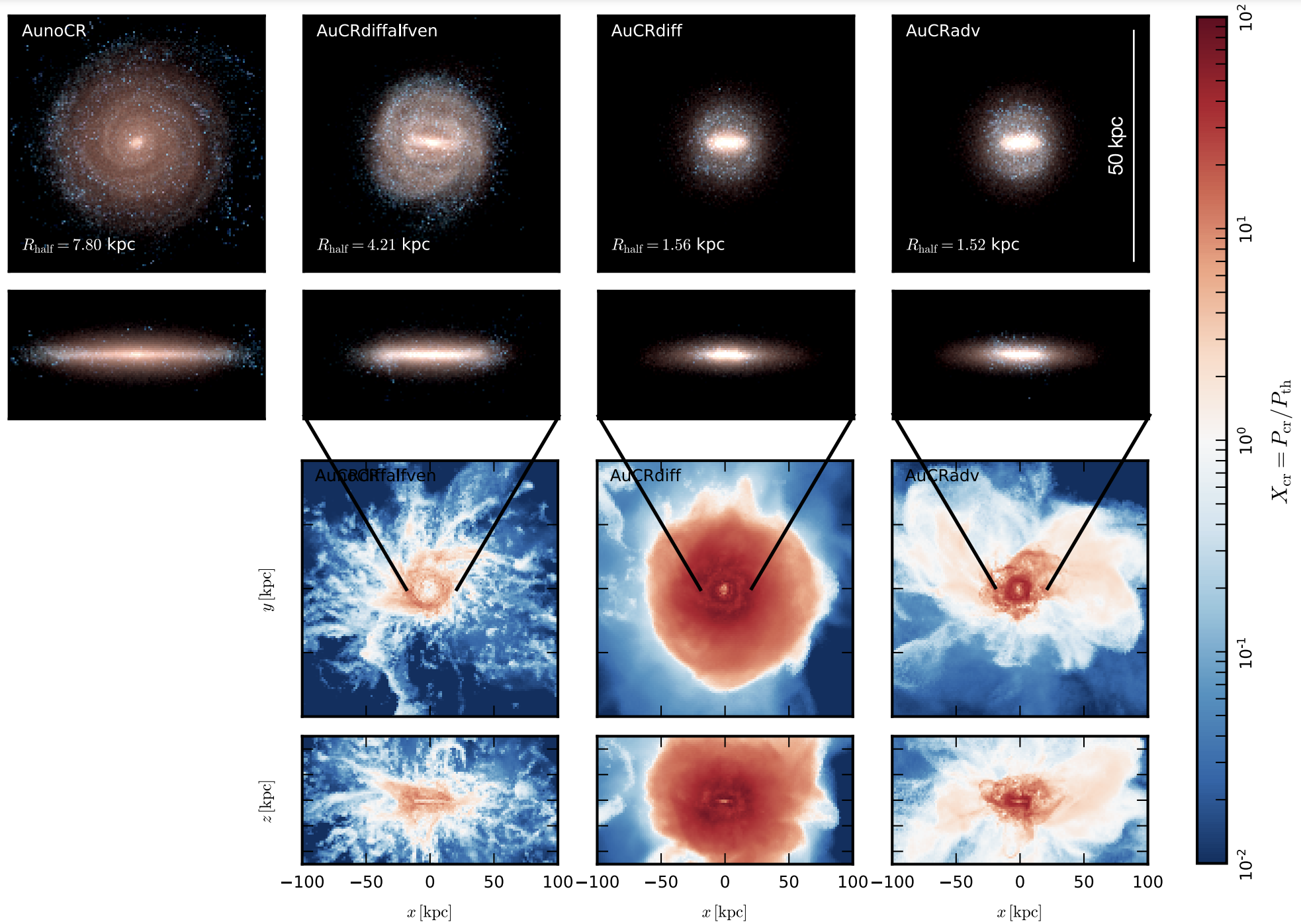}
\end{center}
\caption{Face-on and edge-on projections of stellar density (top two rows; K-, B- and U-band stellar luminosities, are shown in red, green, and blue, respectively) and the corresponding projections of the ratio of CR-to-gas pressure (bottom). From left to right, columns correspond to the following cases: (i) no CRs, (ii) anisotropic CR diffusion including Alfv{\'e}n wave heating of the gas, (iii) anisotropic CR diffusion, and (iv) CR advection only. Notice the dramatic impact of the CR physics on the properties of the galaxy. Image from \citet{buck_effects_2020}; reproduced with permission from MNRAS.}
\label{fig:buck}
\end{figure}
\indent
Irrespective of the magnitude of the impact of CRs on the global properties of galaxies, CRs substantially modify the structural properties of the CGM such as its density, temperature, and kinematics and, consequently, alter the angular momentum acquisition by the galactic disks during their formation in cosmological simulations. This can influence galactic properties such as disk size or the ratio of the disk mass to total stellar mass with outcomes strongly dependent on the physics of CR transport. For example, in simulations of \citet{buck_effects_2020}, when CR advection or anisotropic diffusion are included, the resulting CR-driven outflows tend to be more spherically symmetric compared to those that either include additional CR streaming losses or neglect CR physics altogether in their baseline Auriga model \citep{grand_auriga_2017,grand_gas_2019}. This outflow geometry interferes with coherent accretion near the disk midplane and thus suppresses accretion of high angular momentum gas leading to more compact disks. This is evident from the two top rows in Fig.~\ref{fig:buck} that show the projections of stellar density for the Auriga model: (i) baseline model without CRs; (ii) model that includes CR advection, anisotropic CR diffusion, and CR streaming heating; (iii) model with CR advection and anisotropic CR diffusion; (iv) model with CR advection only. The trend for the disks to decrease in size when CR physics is included is also present in FIRE-2 simulations \citep{hopkins_but_2020} though the outflow geometry is different. This illustrates that exact conclusions are sensitive to the physics of transport and how it shapes the properties of the CGM, which we discuss next.

\subsubsection{CGM thermodynamics, kinematics, and cosmic ray transport}\label{CGMtheory}
The CGM lies at the intersection of the inflowing pristine intergalactic medium  \citep[e.g.,][]{keres_galaxies_2009,ocvirk_bimodal_2008,van_de_voort_rates_2011} and the outflowing metal-enriched gas from galactic disks  \citep[e.g.,][]{oppenheimer_mass_2008,behroozi_comprehensive_2010,dave_analytic_2012} thus making it a useful probe of the processes governing galaxy evolution. First cosmological simulations of galaxy formation focusing on the impact of CRs from SNe demonstrated that the CGM pressure support can be dominated by CRs \citep{jubelgas_cosmic_2008,salem_role_2016,chen_cosmological_2016} just as in CR simulations of global models of isolated galaxies  \citep[e.g.,][]{booth_simulations_2013,salem_cosmic_2014,butsky_role_2018} and zoom-in simulations of stratified boxes and disk patches \citep[e.g.,][]{girichidis_launching_2016, simpson_role_2016}. In agreement with these conclusions, in the cosmological simulations, the CGM was also more diffuse and characterized by substantially lower temperatures (smaller than $\sim 10^{5}$~K) when CRs were included. While these cosmological simulations neglected Coulomb, hadronic, and streaming CR losses, they were nevertheless in broad agreement with the observations suggesting that a significant fraction of the CGM may be relatively cool
\citep{tumlinson_large_2011,werk_cos-halos_2013,werk_cos-halos_2014}.\\
\begin{figure}
  \begin{center}
    \includegraphics[width=0.9\textwidth]{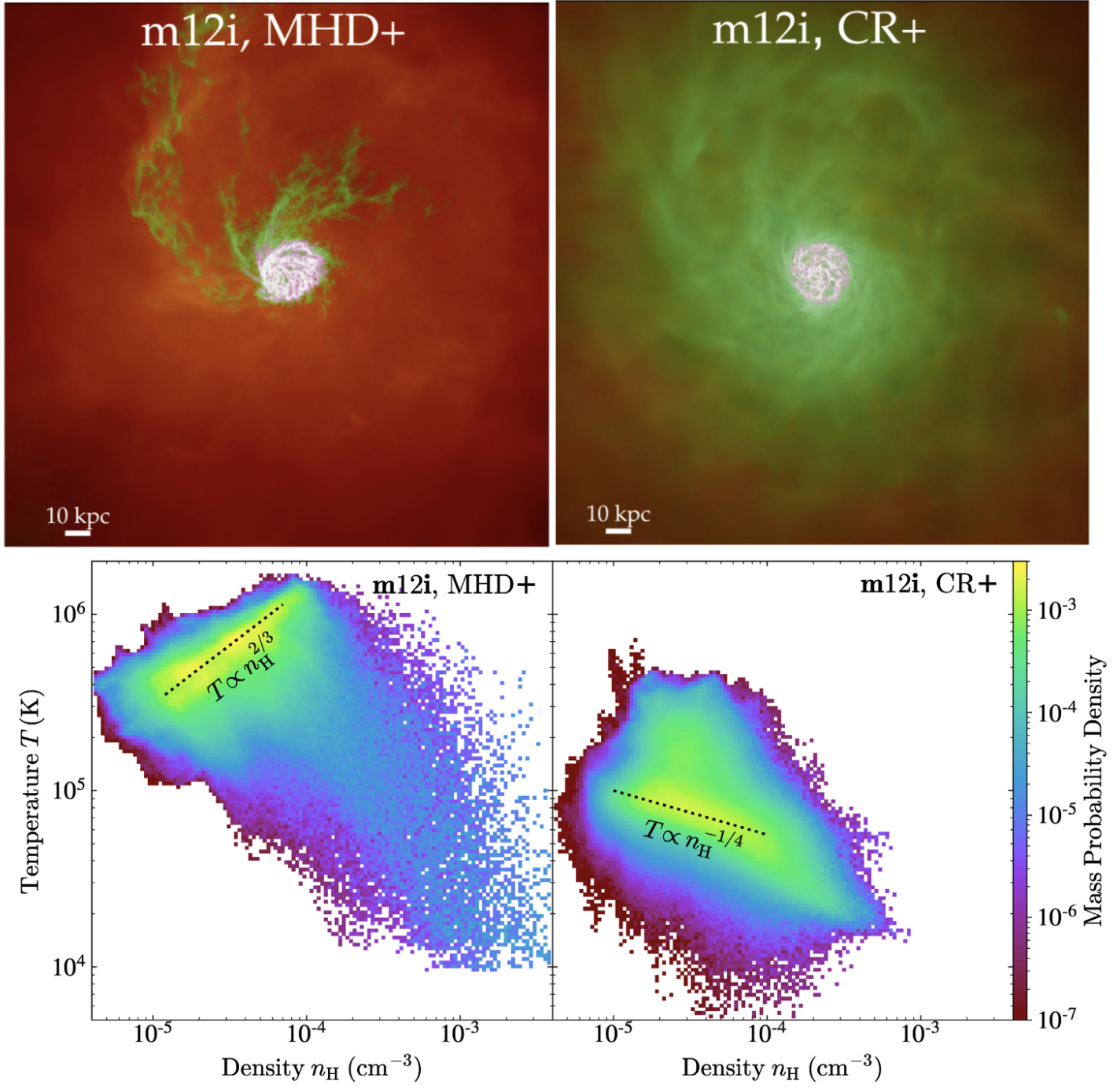}
  \end{center}
  \caption{
  \textit{Top row}: temperature distribution in the CGM of a simulated Milky Way-size galaxy without (left panel) and with CR physics (right panel). Warm/hot ($T \gg 10^{5}$~K; red), cool ($T\sim 10^{4}$--$10^{5}$~K; green), and cold neutral gas ($T \ll 10^{4}$~K; magenta/white). \textit{Bottom row}: phase-space diagrams of density and temperature in the CGM (only gas between 50 kpc and 200 kpc from the galactic center is included) from cosmological simulations of galaxy formation without CRs (left panel) and with CRs (right panel) at $z=0$. Notice that the gas in the CR run is generally cooler and follows a different $\rho$ vs.\ $T$ trend than that in the non-CR case. Panels based on figures in  \citet{ji_properties_2020}; reproduced with permission from MNRAS.}
  \label{fig:ji_cgm}
\end{figure}
\indent 
These findings are in broad agreement with cosmological MHD simulations that consider more realistic CR transport physics including CR streaming effects such as the FIRE-2 simulations \cite[e.g.,][]{ji_properties_2020,chan_impact_2022} and AREPO simulations \citep{buck_effects_2020} that emulate the effects of CR streaming through anisotropic diffusion and CR Alfv\'en wave losses. In these cases, the comparisons of otherwise identical MHD simulations with and without CRs confirm that, when CRs are present, the halo gas is cool with characteristic temperatures around several times $10^{4}$~K and typical thermal pressures smaller than those required for local thermal and hydrostatic pressure balance. In these CR-dominated halos, the overall gas density profile is shaped by the balance of gravity and CR pressure forces, and the CR and thermal pressures are spatially anticorrelated. Consequently, patches of gas characterized by very different temperatures but similar densities can occupy the same regions of the halo. This means that the cool phase can be out of local thermal equilibrium or ``underpressured'' with respect to the pressure needed for thermal balance. By comparison, in pure MHD counterparts of these simulations, the halo gas temperatures are much higher and typically exceed $10^{5}$~K; see top panels in Fig.~\ref{fig:ji_cgm} that show the distribution of the hot warm/hot ($T \gg 10^{5}$~K; red) and cool gas ($T\sim 10^{4}$--$10^{5}$~K; green).
These substantial differences between halo properties are also evident in the phase-space diagrams (see bottom row in Fig.~\ref{fig:ji_cgm}). From left to right, the panels show the temperature-density diagrams for the pure MHD and CR cases, respectively. While in the MHD case most of the gas closely follows the usual $T\propto\rho^{2/3}$ adiabat, in the CR case, the slope of the $T(\rho)$ relation is much shallower and the range of gas temperatures at a given density is much wider. In addition, the overall thermal gas pressures in the CR case are much reduced compared to the MHD case.\\
\indent
Interestingly, the above results are in stark contrast to the AREPO models when CR streaming losses are included \citep{buck_effects_2020}. Compared to AREPO models with advection and diffusion where large CR-dominated halos can form, inclusion of CR streaming in AREPO runs confines these CR-dominated regions to just the disk-halo interface at the very center of the galactic potential wells. In this case, CR streaming losses, and thus heating of the CGM gas, result in relatively warmer halo gas in excess of $10^{5}$~K rather than $10^{4} - 10^{5}$~K seen in the advection/diffusion cases. This difference remains despite the fact that FIRE-2 models also include CR streaming effects, possibly because of the larger magnetic field strengths, which are increased by more than an order of magnitude in the AREPO simulations and which imply a similarly stronger Alfv\'en cooling rate of CRs (and stronger heating of the thermal gas). The significant impact of CR streaming losses on reducing CR pressure support was also reported in the context of isolated models of galaxies \citep{wiener_cosmic_2017}.\\
\indent
In addition to the overall shift in the distribution of the gas temperature, the morphology of the cold gas phase is also significantly altered by CRs. When CRs are present, the cold gas can be diffuse and volume-filling. That is in stark contrast to the pure MHD case, where the cold gas is very filamentary, dense, and supported by thermal pressure (see top row in Fig.~\ref{fig:ji_cgm}). However, the extent to which the gas is filamentary also strongly depends on the physics of CR transport. While the CGM may be cooler and more smoothly distributed when advection and diffusion are dominant, streaming losses quickly sap CR energy at the locations where the magnetic field and CR pressure gradients are large thus leading to highly structured density and temperature distributions (\citet{buck_effects_2020}; see lower left panel in Fig.~\ref{fig:buck}).\\
\indent
Kinematics of the CGM on large scales can also be dramatically altered by the presence of CRs. Generally speaking, the impact of CRs on the kinematics is expected to be strong when significant CR gradients can build up. As argued above, whether this can happen depends on the halo mass and redshift \citep{hopkins_but_2020,ji_properties_2020} and it may also depend, even strongly, on the physics of CR transport \citep{buck_effects_2020,hopkins_effects_2021}. While this dramatic impact of CRs is seen in both the FIRE and AREPO-based cosmological CR feedback simulations, there are important differences between the morphology of inflows and outflows in these simulations. In FIRE-2 models, in the absence of CR feedback, galactic outflows are ``trapped'' and recycled near the disk-halo interface due to the combined effects of the strong gravitational field, efficient cooling, and the presence of hot CGM thus forming galactic ``fountains.'' In this case, most of the gas near the virial radius is infalling. On the other hand, when CR feedback is included and when halo pressure is dominated by CRs, fountain flows are suppressed as the CR-driven outflows can prevent the cooling gas from returning to the disk \citep{chan_impact_2022}, and slow large-scale bi-conical outflows develop, reaching $\sim$Mpc scales and becoming volume-filling at large distances, while inflows are filamentary and confined to a narrow solid angle near the disk midplane \citep{hopkins_cosmic_2021}. In general, with CR feedback present, the inflows are slower as they are decelerated by CR pressure gradients, and so in this sense, the feedback is ``preventative'' rather than ``ejective.'' This effect of CRs to prevent the gas from efficiently cooling onto the galactic disk is also seen in the simulations of \citet{buck_effects_2020} when CR transport is dominated by diffusion. However, in these simulations, the CGM kinematics depends critically on CR physics. Specifically, when CRs are neglected, the outflows are bi-polar and aligned with the spin axis of the galaxy as in the FIRE-2 model with CRs included. When CR diffusion is present, CR pressure gradients are most effective in reducing inflow velocities, but when CR streaming heating is introduced, the outflows are less ordered than in the non-CR case and not always perpendicular to galactic disks, so the streaming case lies ``in between'' the pure MHD and CR advective cases.\\
\indent
One interesting consequence of significant CR pressure support in the halos is that it may affect the formation of virial shocks. Absent CRs, virial shocks develop when the cooling time of the accreting gas is longer than the compression time of that gas. When this condition is met, the shocked gas provides pressure support for the accreting gas, which leads to the formation of a stable shock. In the opposite regime, the post-shock thermal energy is quickly radiated away and no stable shock can form. The halo mass separating these regimes is $M_{\rm crit}\sim 10^{11.5}$M$_{\odot}$ \citep{dekel_galaxy_2006} with stable shocks forming above this critical mass \citep[but see][for a different definition of these accretion modes]{nelson_moving_2013}. However, as argued above, this falls within the halo mass range where stellar CR feedback can significantly affect the state of the CGM, which leads to a natural question of the impact of CR pressure support on virial shocks. In FIRE-2 models of Milky Way-mass galaxies with constant CR diffusion and Alfv{\'e}nic streaming, CR pressure dominates over thermal and ram pressure near the virial radius and the CR pressure gradients are so effective that the inflowing gas becomes subsonic and the virial shock disappears \citep{ji_virial_2021}. However, as mentioned above, there is currently no consensus on the expected level of CR pressure support as a function of halo mass, and the halo mass regime where it is important likely depends on the physics of CR transport and on the saturation value of the magnetic field in the ISM/CGM as this quantity linearly enters the CR Alfv\'en cooling rate. Consequently, the criterion for the formation of virial shocks is also likely to depend on these factors.

\subsection{Thermal instability and cosmic rays in the CGM and ICM}\label{TI}\label{titheorysection}

There is significant observational evidence for the existence of large amounts of cold ($10^{4}$~K) gas embedded in the hot CGM \citep[e.g.,][]{werk_cos-halos_2014} with the cold gas contributing up to 50\% to the overall galactic baryon mass budget. While the morphology of this gas is uncertain, it may exist either in the form of filaments and small clouds (``mist”) or it could be volume-filling. Similarly, cold gas is also readily detected in the hot and volume-filling ICM \citep[e.g.,][]{salome_cold_2006}, where it assumes filamentary morphology. We now take a closer look at the physical origin of the cold gas in the hot halos of galaxies and galaxy clusters, focusing on the role of thermal instability and the impact of CRs on this process. We argue below that CRs may play a fundamental role in shaping the properties of the cold phase of the CGM and ICM and, specifically, that CRs may (i) hold the key to solving the puzzle of star formation quenching even in the presence of large amounts of cold gas in the CGM and ICM, (ii) regulate the amount of cold gas, (iii) shape the morphology and characteristic sizes of the cold gas clouds, and (iv) provide a source of excitation for the line emission from the cold gas.\\
\indent 
There are two main classes of solutions that have been proposed to explain the existence of the cold gas in the CGM and ICM. In the first category of solutions, the cold gas is dredged out from the centers of potential wells by hot and fast galactic outflows. Historically, the main challenge for this class of models was how to ensure the survival of the accelerated clouds. In a simple model, where the cold clouds are accelerated by a hot wind, the characteristic acceleration timescale is longer than the cloud 
crushing time (see also Section~\ref{Interaction of CR with dense clouds}). However, the clouds exposed to the galactic wind may survive the destruction in the CGM by hydrodynamical instabilities thanks to the stabilizing effect of the magnetic fields \citep{dursi_draping_2008,mccourt_magnetized_2015}. Similarly, the cold gas may be uplifted by buoyant AGN bubbles in the case of the ICM, where the dynamics of the uplifted gas may be strongly affected by magnetic fields \citep[e.g.,][see Section~\ref{agntheorysection} for the discussion of AGN feedback]{fabian_magnetic_2008}. 
Several mechanisms described below may be responsible for the effective transfer of momentum to the clouds. (i) The survival and acceleration of the clouds is more likely if sufficiently fast radiative cooling can regrow the clouds before they get completely disrupted and mixed with the hot wind \citep{gronke_growth_2018}. (ii) The cloud acceleration in the CGM may be also due to CRs streaming down their pressure gradients that encounter magnetic bottlenecks in a cold cloud. As discussed in Section~\ref{Interaction of CR with dense clouds}, this causes the emergence of CR pressure gradients across the cold gas cloud that accelerated it, thereby avoiding the destructive impact of CR streaming heating, which can also be counterbalanced by radiative cooling \citep[e.g.,][]{wiener_interaction_2017,thomas_finite_2021}. (iii) Draping of magnetic field around a cold cloud that is anchored in the hot wind is another means of transferring momentum from the hot to the cold phase, thereby accelerating it. Because the strong magnetic field in the draping layer pushes the field lines in the dimension perpendicular to the magnetic orientation over the draped cloud, the momentum coupling from the hot to the cold phase is incomplete, and only moderately increases the hydrodynamic drag by a factor of about 2.5 through the action of magnetic draping \citep{dursi_draping_2008}. (iv) Finally, the clouds could indeed be shattered into small cloudlets without completely mixing in with the hot wind gas. This shattering process increases the effective surface area of the cold clouds and thus strongly increases the drag force acting on them and facilitates rapid entrainment and acceleration of the cold gas \citep{mccourt_characteristic_2018}. This shattering process leads to very small cloudlets of characteristic sizes $\sim c_{\rm s}\tau_{\rm cool}\sim 0.1/n_0~\rmn{pc}$, where $c_{\rm s}$, $\tau_{\rm cool}$, and $n_0=n/(1~\rmn{cm}^{-3})$ are the sound speed, cooling time, and gas number density, respectively. The fragmentation process is analogous to the Jeans instability fragmentation with the main difference that the contraction is driven by radiative cooling and ambient pressure rather than self-gravity. However, because these cloudlets quickly mix and disolve in the hot wind via the Kelvin-Helmholz instability on several cloud crushing timescales, they need to be replenished on the same timescale for this cold phase to be in dynamical equilibrium. \\
\indent
The second category of solutions involves \textit{in situ} production of cold gas via thermal instability (TI). This process has been extensively studied in the context of the ISM starting with the seminal work of \citet{field_thermal_1965}. In the ISM, the gas is characterized by the thermally stable molecular ($\sim 10^{2}$~K) and atomic ($\sim 10^{4}$~K) phases. In addition, the thermally unstable hot phase ($\sim 10^{6}$~K) is sustained by SN heating. In the halos of galaxy clusters, the role of SN heating in maintaining the hot phase is played by the AGN feedback heating. Although TI in gravitationally stratified plasmas was originally suggested to address the issue of the origin of the cold gas in clusters of galaxies \citep[e.g.,][]{gaspari_cause_2012}, it may also play the same role in the CGM \citep{fielding_impact_2017,voit_circumgalactic_2019}. \\
\indent
In general, in gravitationally stratified media, such as the CGM and ICM, the development of TI is tightly linked to the details of the heating function. In particular, if the effective cooling function (i.e., the difference between the heating and cooling functions) is not an explicit function of position, and the sound waves are thermally stable, then the plasma can only be thermally unstable according to the Field criterion if it is also convectively unstable according to the Schwarzschild criterion \citep{balbus_theory_1989}. Thus, the very existence of multiphase medium in the ICM suggests that the heating of the gas is not solely describable in terms of local thermodynamical variables of the background plasma. In fact, in a simple globally thermally stable model, where the heating function is constructed such that the amount of heating at each radius equals the shell-averaged cooling rate (rather than being a function of local thermodynamical variables), the gas becomes locally thermally unstable \citep[e.g.,][]{mccourt_thermal_2012}. In this case, a linear stability analysis reveals that the heating function has no first order perturbation terms, and isobaric perturbations can cool unimpededly and grow exponentially on a characteristic timescale comparable to the radiative cooling timescale. According to this analysis, the instability is triggered when the locally computed ratio of cooling-to-free-fall timescales, $f_{\rm ff}=\tau_{\rm cool}/\tau_{\rm ff} \lesssim 1$. However, numerical simulations reveal that the plasma subject to isobaric perturbations can become unstable even for $f_{\rm ff}$ as high as $\sim 25$ locally \citep{sharma_thermal_2012,mccourt_thermal_2012,beckmann_dense_2019}. This discrepancy can be traced back to the fact that the suscibility of an atmosphere to thermal instability depends not only on the median $f_{\rm ff}$ but also on the dispersion of entropy fluctuations in the atmosphere. When the entropy fluctuations are small, buoyancy forces lead to oscillations of the perturbations around an equilibrium in the form of gravity waves. These buoyant oscillations are eventually damped. However, when the perturbations are large, radiative cooling timescale can be shorter than the buoyancy oscillation timescale in which case the perturbations can grow nonlinear and the plasma can become multiphase 
\citep{palchoudhury_multiphase_2019,voit_graphical_2021}. Thus, even if the median $f_{\rm ff}\gtrsim 1$, some perturbations may be large enough to trigger the formation of multiphase medium. From the observational standpoint, galaxy cluster observations reveal that the values of $f_{\rm ff}\lesssim 25$ do indeed coincide with the elevated H$\alpha$ emission from the cold gas \citep{mcdonald_origin_2010,hogan_onset_2017}.

\subsubsection{Onset of thermal instability in the presence of cosmic rays}
One fundamental reason why CRs can change the picture of TI is that they can provide additional heating and pressure support against cooling-induced collapse of the gas while not being as susceptible to cooling losses as the thermal gas. A linear stability analysis of atmospheres in hydrostatic and thermal equilibrium (where radiative cooling is offset either by background CR streaming heating or by an unspecified heating function designed to offset cooling by construction) reveals that the outcome depends on the properties of the cooling function, CR pressure support, relative transport speed via diffusion and streaming, and the impact of Alfv{\'e}nic heating compared to radiative cooling rate \citep{kempski_thermal_2020}. The importance of these factors can be quantified in terms of only three parameters: (i) the logarithmic slope $\Lambda_{\rm T}$ of the cooling function, (ii) the ratio of the CR-to-thermal pressure $X_{\rm cr}$, and (iii) $\xi\equiv \kappa/(\tau_{\rm cool}X_{\rm cr} {\varv}_\rmn{a}^{2})$, where ${\varv}_\rmn{a}$ is the Alfv{\'e}n speed, $\tau_{\rm cool}$ is the radiative cooling time of the thermal gas, and $\kappa$ is the CR diffusion coefficient. The plasma is stable when $\Lambda_{\rm T}>2$. In the opposite regime, when $\Lambda_{\rm T}<2$, the instability either grows at the rate $\propto(2-\Lambda_{\rm T})$ when $\xi\gtrsim 1$, or is suppressed on length scales below a critical ``CR Field length'' $\lambda_{\rm cr}\sim 2\pi|\bs{k\cdot \hat{b}}|[\kappa \tau_{\rm cool}{\rm min}(X_{\rm cr},X_{\rm cr}^{-1})]^{1/2}$, where $\bs{k}$ is the wave vector and $\bs{\hat{b}}$ is the unit magnetic field vector, when $\xi\lesssim 1$. Neglecting factors containing $X_{\rm cr}$, this critical length scale is analogous to the classic Field length for the case of thermal conduction \citep{field_thermal_1965} with the CR diffusion coefficient substituted for the thermal diffusion coefficient. The physical reason why CR diffusion is unable to suppress TI when $\xi\gtrsim 1$ is that diffusion is so important that it tends to suppress CR pressure perturbations and the rate at which CRs heat the gas on small length scales is always smaller than the radiative cooling rate. Consequently, there is no CR Field length in this case. In the regime where $\xi\lesssim1$, the dependence of the Field length on the diffusion coefficient and cooling timescale is intuitively correct: (i) when $\kappa$ is large, CRs can suppress TI over larger length scales, and (ii) shorter $\tau_{\rm cool}$ makes the plasma more unstable and reduces the scale below which the gas is stabilized. The character of the instability and its growth rates are strongly influenced by the level of CR pressure support. For $X_{\rm cr} \ll 1$ the instability is isobaric (i.e., cooling occurs at constant pressure), while for $X_{\rm cr} \gg 1$ it is isochoric (i.e, cooling proceeds at constant density). This is also intuitively correct as large CR pressures can oppose gas compression while the gas undergoes radiative cooling. It is also consistent with the results from nonlinear simulations where the gas densities are reduced when significant CR pressure support is present (see below).\\
\indent 
As stated above, the plasma is stable when $\Lambda_{\rm T}>2$. In particular, this condition is satisfied when the galactic halo plasma is in photoionization equilibrium with no significant background CR heating (even though CRs could still affect the evolution of entropy fluctuations). In this case, the gas is unstable irrespectively of the value of $X_{\rm cr}$. If instead the plasma is in collisional equilibrium and radiative cooling is offset by background CR heating, the TI is expected for realistic parameter choices. 
Specifically, in the gas temperature range $10^{5}\lesssim T \lesssim {\rm a\; few}\times 10^{7}$~K, that is relevant for galactic and cluster halos, $\Lambda_{\rm T}\lesssim 0.5$ and the gas can be stabilized against TI as long as $\xi\lesssim 1$, i.e., when CR pressure support is important and CR streaming dominates over diffusion. However, while the conditions in galaxy halos may be more favorable to suppressing the TI because of (possibly) significant CR pressure support (which nevertheless still does not preclude the onset of TI), in cool cores of galaxy clusters, where we generally expect $X_{\rm cr}\lesssim 0.1$ on observational grounds \citep[i.e., in order not to overproduce non-thermal $\gamma$-ray and radio emission generated by hadronic CR interactions with the ambient medium;][]{pfrommer_toward_2013,jacob_cosmic_2017,jacob_cosmic_2017-1}, $\xi$ is likely to significantly exceed unity and the TI is expected to develop.

\begin{figure}
  \begin{center}
    \includegraphics[width=1.0\textwidth]{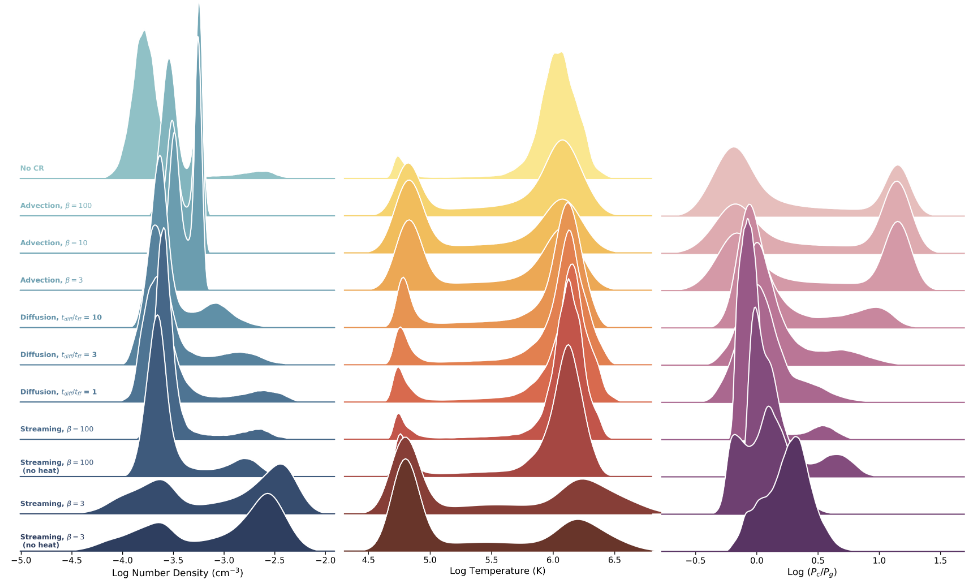}
  \end{center}
  \caption{The mass-weighted probability density functions of the gas density (left panel), temperature (middle panel), and CR-to-gas pressure ratio (right panel) for the case when the radiative cooling time equals the free-fall time. The top row correspond to the case without CRs. Rows 2 to 4 correspond to the cases with CR advection for different plasma $\beta$. Rows 5 to 7 illustrate the impact of CR transport via diffusion for different diffusion timescales. Rows 8 to 11 show the impact of streaming for different values of $\beta$, which corresponds to different streaming speeds. One general trend emerging from these simulations is that the density contrast between the hot and cold phase decreases significantly in the presence of adiabatic CRs. Including CR transport frees CRs and enables gas condensation, which tends to restore the density contrast. Figure from \citet{butsky_impact_2020}; reproduced with permission from ApJ.}
  \label{fig:butsky_ti}
\end{figure}

\subsubsection{Non-linear outcome of thermal instability with cosmic rays}
We now turn our attention to the non-linear evolution of the TI in the presence of CRs. Three-dimensional simulations of the TI in gravitationally stratified atmospheres including CR diffusion and streaming transport and heating demonstrate that CRs can have a significant impact on the properties of the cold CGM phase \citep{butsky_impact_2020,tsung_impact_2023}. Since CR pressure tends to oppose the collapse of the gas that looses thermal pressure due to radiative cooling, the gas can cool isochorically if CR pressure is sufficiently high. Thus, for high CR pressures, the density contrast between the cold and hot phases can be substantially lower, i.e., the density of the cold phase can be only slightly elevated beyond that of the hot phase and lower than it would have been absent CRs. 
The impact of CRs on the density contrast is maximized if CRs are only subject to advective transport. When CR diffusion or streaming transport and heating are included, CRs can propagate out of the cooling and collapsing gas clouds (or lose energy due to streaming losses) thus diminishing the impact of CRs and enabling further cloud collapse. Not surprisingly, this leads to a bimodal distribution of gas density as in the non-CR case (and closer to single phase when CRs with advection as the only transport process are included). The impact of CRs on the gas pressure distribution is opposite. That is, when CRs are only subject to advection and follow the gas, CR pressure can increase in the cold phase as the gas begins to compress and, consequently, the CR pressure distribution is bimodal. 
However, as active CR transport processes are included, CRs can avoid compression and the CR distribution is narrower and closer to the initial conditions. These trends are evident in Fig.~\ref{fig:butsky_ti}, which shows the mass-weighted probability density functions of the gas density (left panel) and CR-to-gas pressure ratio (right panel) for the case when radiative cooling time equals the free-fall time.\\
\indent
The above findings have fundamental consequences for the properties of the CGM. One fundamental implication of these results is that, in the presence of significant CR pressure support, the cold thermal gas can appear to be out of pressure equilibrium with the ambient hot CGM, i.e., the pressure of the cold gas may be present at sub-equipartition levels (see Section~\ref{sec:CGM_feedback}, where we discuss observational aspects of CR feedback in the CGM).\\
\indent
Furthermore, CR pressure may affect characteristic cold cloud sizes. As mentioned in the beginning of Section~\ref{titheorysection}, cloud sizes are expected to be comparable to $c_{\rm s}\tau_{\rm cool}$. If the pressure is dominated by CRs and the density of the gas is comparable to that of the ambient medium (due to isochoric cooling out of the hot phase when CR pressure support is high and transport is mostly advective), then both the sound speed and the cooling time are increased in the clouds and the clouds can be larger compared to the case without CRs. The expected sizes can range from 1 to 10$^{3}$~kpc. This effect may improve convergence properties of cosmological simulations that include CR physics. By contrast, cosmological simulations that do not include CRs reveal that increasing numerical resolution results in the increase of the amount of cool gas in the simulations \citep[e.g.,][]{van_de_voort_cosmological_2019,hummels_impact_2019}.\\
\indent
Significant CR pressure gradients can also suspend cold gas in the hot halos in the relatively slow cooling regime ($\tau_{\rm cool}/\tau_{\rm ff}\gtrsim 1$, where $\tau_{\rm ff}$ is the free-fall time). This is evident in the middle column in Fig.~\ref{fig:butsky_ti}, where the MHD case (top of the column) exhibits much smaller amount of the cold gas compared to the cases where CRs are included. This difference is due to precipitation that efficiently removes cold gas from the atmosphere in the MHD case. The suspension of the cold phase in the CR cases is also made more likely by the fact that the density contrast between the hot and cold phases can be lower in these cases, which makes it more difficult for the colder gas to dynamically decouple from the hot phase. These CR effects have the potential to explain the quiescence (lack of significant star formation) of galaxies in the presence of copious amounts of cold gas in the CGM \citep{berg_red_2019}. The suspension of the cold gas by CR pressure gradient forces can increase the amount of cold gas in the CGM and reduce mass accretion rate thereby suppressing star formation.\\
\indent
In agreement with the linear analysis of \citet{kempski_thermal_2020}, non-linear simulations of thermal instability development in the ICM in the presence of AGN feedback heating and CRs demonstrate that CRs are initially unable to stabilize most of the hot and dilute plasma against TI \citep[i.e., the gas is mostly characterized by $\Lambda_{\rm T}<2$ and $\xi>1$;][]{beckmann_cosmic_2022}. However, once the gas has cooled out of the hot phase, it either settles down to a thermally stable state with $\Lambda_{\rm T}>2$, or to a state where $\Lambda_{\rm T}<2$ but $\xi<1$ and further collapse can be prevented by significant CR pressure \citep[see Fig. 15 in][]{beckmann_cosmic_2022}. Such significant pressure support ($X_{\rm cr}\gg 1$) in the cold phase is expected even if the initial $X_{\rm cr}\ll 1$ in the hot phase because of adiabatic compression of CRs combined with dramatic loss of thermal pressure due to radiative cooling. This is seen in simulations inspired by the observations of cold filaments in the ICM \citep{sharma_thermal_2010}. This effect leads to the decrease in the density of the cold gas (compared to the non-CR case) as confirmed by the simulations of TI in the context of the CGM \citep{butsky_impact_2020}.

\subsubsection{Impact of magnetic fields on thermal instability in hot halos}
The impact of CRs on the TI is strongly intertwined with that of the magnetic fields. Even in the absence of CRs, MHD simulations of gravitationally stratified atmospheres in thermal and hydrostatic equilibrium show that magnetic fields can have a strongly destabilizing impact on the onset of the TI \citep{ji_impact_2018}. Specifically, magnetic tension reduces buoyant oscillations of the cooling gas, which prevents disruption of the condensing cold gas clouds and, thus, increases the amplitude of the density fluctuations, which makes the gas more prone to the TI. This effect destabilizes the gas below a characteristic length scale of $\sim\varv_\rmn{a}\tau_{\rm cool}$, where $\varv_\rmn{a}$ is the Alfv{\'e}n speed. Compared to the pure hydrodynamical case where the steady-state amplitude of density fluctuations is predicted to be $\delta\rho/\rho\sim 0.1 (\tau_{\rm ff}/\tau_{\rm cool})$, the amplitude in the MHD case is $\delta\rho/\rho\sim 0.1 (\beta/10^{3})^{-1/2}(\tau_{\rm ff}/\tau_{\rm cool})$, where $\tau_{\rm ff}$ is the free-fall time and $\beta$ is the ratio of the thermal-to-magnetic pressure in the plasma. This demonstrates that the enhancement in density fluctuations, and the TI, is expected even for very weak magnetic fields provided that $\beta\lesssim 10^{3}$, i.e., stronger fields help to thermally destabilize the gas. Given that the magnetic field strengths in the bulk of the CGM can be as high as $\sim\mu$G \citep[e.g.,][]{lan_constraining_2020,pakmor_magnetizing_2020}, and even higher in the ICM \citep[e.g.,][]{govoni_magnetic_2004}, it is plausible that plasma $\beta\ll 10^{3}$ for typical densities and temperatures in the CGM and ICM, which may help to explain the ubiquity of the cold gas in these environments.\\
\indent
Non-linear evolution of the TI leads to the amplification of magnetic fields due to flux freezing at the locations of cold gas clumps \citep{wang_chaotic_2020,nelson_resolving_2020}. This in turn leads to $\beta\ll1$ in the cold phase, where the pressure support can be dominated by the magnetic fields. Therefore, as in the case of CR pressure support discussed above, the non-thermal magnetic pressure support can also help to explain the sub-equipartition thermal pressure paradox in the CGM.
The strong magnetic field amplification can also affect the dynamics of the cold phase. This situation is somewhat reminiscent of the coronal rain in the Sun's corona, where the magnetic field can also affect the dynamics of the cooling and condensing gas sliding along strong magnetic fields \citep{mason_observations_2019}, though the physics of the impact of the magnetic field on the cold gas is different in this case as we discuss in more detail in Section~\ref{self-regulated}.\\
\indent
Magnetic fields can also shape the morphology of the thermally unstable plasma. The large aspect ratios of the filaments observed in galaxy clusters suggest that magnetic fields may indeed play a key role in the dynamics of these structures \citep{fabian_magnetic_2008}. Even in the absence of anisotropic CR diffusion, magnetic-field-aligned thermal conduction can stabilize the plasma against TI along the magnetic field on scales smaller than the classical Field length. Consequently, because thermal conduction is severely suppressed in the direction perpendicular to the local magnetic field, the morphology of the condensing gas is expected to be highly filamentary \citep{sharma_thermal_2010}. In the saturated state, the magnetic fields are aligned with the filaments and both the magnetic and CR pressures are expected to dominate over thermal pressure. Note that the alignment of the magnetic field with the filaments is possible because plasma $\beta\gg 1$ in the ambient CGM/ICM. In the opposite case, i.e., when $\beta$ in the ambient plasma is smaller or comparable to unity (as can be the case in the ISM), such alignment is not guaranteed.

\subsubsection{Cosmic ray heating of the cold thermally unstable gas}\label{filaments}
The characteristic cooling timescales of the cold phase in the CGM and ICM can be very short. In the ICM, where the cold gas filaments are known to be copious H$\alpha$ emitters, this timescale is much shorter than the dynamical timescale. Thus, in order to sustain this level of emission, the energy has to be continuously resupplied to the filaments. A number of powering mechanisms have been proposed to solve this long-standing problem, and subsequently ruled out: (i) conduction of heat from the hot ambient ICM to the filaments
\citep{voit_conduction_2008} may be ineffective as thermal conduction is likely to be severely suppressed by whistlers \citep{roberg-clark_suppression_2016,komarov_self-inhibiting_2018,drake_whistler-regulated_2021}; (ii) filament heating by X-rays from the ambient ICM is in conflict with the observed high H$_{2}$/H$\alpha$ ratios \citep{donahue_hubble_2000}; (iii) photoionization, where the gas is either photoionized by massive stars or the central AGN, is in disagreement with the line ratios in HII regions \citep{kent_ionization_1979} and relative lack of star formation \citep{canning_collisional_2016} in the former case, and does not predict the decline of H$\alpha$ emission with the distance from the AGN, which is expected in the latter case \citep{johnstone_extended_1988}; (iv) shock heating (e.g., due to AGN) overpredicts OIII emission \citep{voit_deep_1997}. Other viable heating mechanisms include heating by mixing in turbulent layers \citep{begelman_turbulent_1990} and reconnection of anti-parallel magnetic fields in the wakes of buoyant AGN bubbles \citep{churazov_powering_2013}.\\
\indent
The limitations of some of the above models motivate a search for alternative solutions. Since, as argued above, the filaments are likely to be the sites of significantly enhanced magnetic and CR pressures, this creates perfect conditions for CR heating to supply energy precisely where the resupply is needed due to significant H$\alpha$ losses. Heating by CR streaming along magnetic fields may supply adequate and sustained power to the filaments \citep{ruszkowski_powering_2018}. This mechanism powers the emission in situ rather than relying on the filaments to be magnetically connected to the ambient ICM or CGM and may operate even if the magnetic and CR pressures in the bulk of the volume are much smaller than the thermal pressure. Interestingly, the filament spectra in general show unusually strong molecular, atomic, low-ionization emission lines relative to molecular clouds in the vicinity of O-stars in the Milky Way. While heating by thermal particles fails to explain these observations, a model in which the plasma is energized by non-thermal particles can match the observations \citep{ferland_origin_2008,ferland_collisional_2009}. Furthermore, and as mentioned in Section~\ref{CR_ionization}, CRs can affect the chemistry of the cold gas. As suggested by \citet{beckmann_cosmic_2022} in the context of the ICM, secondary CR electrons may dissociate CO molecules, which may help to explain strong CII lines \citep{mittal_herschel_2012}
and high CII/CO abundance \citep{mashian_ratio_2013}.

\subsection{Impact of cosmic rays from AGN in massive hot halos}\label{agntheorysection}

Up to now we have focused predominantly on the impact of CRs associated with the stellar feedback in halos up to the Milky Way-mass scale ($\sim 10^{12}~\rmn{M}_{\odot}$). Stellar feedback, with or without a CR component, becomes ineffective in more massive halos primarily due to the relatively lower ratio of stellar-to-halo mass and higher density and pressure of the CGM that the galactic winds need to work against \citep{su_failure_2019,hopkins_cosmic_2021}. However, it is precisely in this high halo mass range that an additional CR heating source becomes dominant, namely the heating associated with CRs supplied by the AGN at the centers of massive elliptical galaxies and galaxy clusters. Specifically, while radio AGN are present in only 10 percent of galaxies with stellar masses below $3\times 10^{10}~\rmn{M}_{\odot}$, essentially all galaxies with masses above $10^{11}~\rmn{M}_{\odot}$ show evidence for radio emission \citep{sabater_lotss_2019}. Interestingly, the ratio of the stellar-to-halo mass peaks within this stellar mass range \citep{moster_constraints_2010} and the peak corresponds to the halo mass $\sim 10^{12}~\rmn{M}_{\odot}$ (see Fig.~\ref{fig:motivation}). In order to put this promising heating mode in context, we first turn our attention to the discussion of the mechanisms that historically have been invoked to offset radiative cooling and provide feedback at this halo mass range.

\subsubsection{Thermalization of the AGN energy and heating of the CGM and ICM}
In the absence of any heating mechanism, the effective cooling times of the $\sim 1$ keV gas can be much shorter than the ages of the halos. The gas can then quickly lose pressure support against gravity and accrete onto the center at rates of $\sim 10^{2-3}$~M$_{\odot}$~yr$^{-1}$, which would significantly exceed observational constraints \citep[e.g.,][]{peterson_x-ray_2006}. This ``cooling catastrophe'' can be prevented by invoking some heating mechanism to offset the fast cooling. The most promising mechanism involves feeding the central supermassive black hole and AGN outflows associated with it. The support for this picture comes from a strong correlation between the observed radiative cooling rates and the AGN power required to inflate the X-ray cavities seen in elliptical galaxies and galaxy clusters \citep[e.g.,][]{churazov_asymmetric_2000,birzan_systematic_2004,rafferty_feedbackregulated_2006,werner_hot_2019}. The AGN power determined this way is commensurate with the radiative cooling losses and is sufficient to prevent global thermal instability of cool cores and provide globally thermally stable heating \citep{ruszkowski_heating_2002} especially in the presence of conduction. However, how this energy is thermalized in the CGM/ICM is still an open question. Several possibilities have been suggested, including (i) dissipation of sound waves and weak shocks induced by the inflation of AGN lobes \citep[e.g.,][]{fabian_deep_2003,ruszkowski_cluster_2004,ruszkowski_three-dimensional_2004,forman_reflections_2005}, (ii) excitation and subsequent turbulent dissipation of internal gravity waves \citep[e.g.,][]{zhuravleva_turbulent_2014,li_direct_2020}, (iii) uplift and mixing of hot thermal plasma from AGN lobes with the ambient medium \citep[e.g.,][]{churazov_evolution_2001,hillel_gentle_2017,yang_how_2016}, (iv) CR escape from AGN cavities and subsequent heating of the CGM/ICM by CRs \citep[e.g.,][see detailed discussion below]{guo_feedback_2008,pfrommer_toward_2013,ruszkowski_cosmic-ray_2017,ehlert_simulations_2018}. While there is consensus in the literature that any successful AGN feedback model needs to be self-regulating and sufficiently gentle so as not to introduce large variations in the monotonically increasing gas temperature and entropy profiles in cluster cool cores \citep[e.g.,][]{voit_global_2017}, it is currently an open question which of the above processes represents the dominant mode of heating in massive halos.\\
\indent 
Other processes that do not posses the potential for self-regulation of heating and cooling include turbulent mixing \citep{kim_turbulent_2003}, where the high entropy gas at larger distances from the center is mixed with the lower entropy gas close to the center via turbulent motions, and thermal conduction that can supply heat to the fast cooling central regions by redistributing thermal energy from the hot outer regions to the centers of cool cores \citep[e.g.,][]{zakamska_models_2003}. Both of these processes are complementary to the above AGN-related heating processes, i.e., they are likely to only partially reduce the burden on the AGN to offset radiative cooling losses.

\subsubsection{Limitations of the heating mechanisms} \label{heating_limitations}
All of the above mechanisms of AGN energy thermalization and heating of cool cores come with certain limitations. For example, the efficiency with which the AGN energy is channeled into weak shocks and sound waves (that could subsequently dissipate into heat) is uncertain and depends on a number of factors such as the power of the AGN \citep[ranging from 12\%, in the limit of instantaneous injection of the AGN energy, to zero in the limit of slow inflation of the AGN lobes; when modeled via one dimensional spherically symmetric simulations;][]{tang_sound_2017} or the jet velocity and opening angle \citep[reaching values up to $\sim$30\% for high-velocity wide-angle jets;][]{bambic_efficient_2019}. Similarly, the efficiency of conversion of the AGN energy to turbulent motions, and their subsequent thermalization, may also be inefficient as demonstrated via controlled three-dimensional numerical simulations involving AGN outburts and assuming one-dimensional gravity and inviscid pure hydrodynamics \citep{reynolds_inefficient_2015} and in self-regulating three-dimensional simulations of the AGN feedback loop \citep{yang_how_2016}. These simulations demonstrate that the energy transfer from the AGN to ICM turbulence is extremely  inefficient with only $\sim$1\% of the injected energy getting transferred to internal gravity waves and, subsequently, to volume-filling turbulent motions. In such purely hydrodynamical simulations, the inefficiency of driving the gravity modes is likely caused by the disruption of the AGN-inflated bubbles by means of Kelvin-Helmholtz and Rayleigh-Taylor instabilities  before the bubbles can excite these modes. However, even in MHD simulations, where these instabilities are partially inhibited by the draping of the magnetic fields around buoyantly rising bubbles (see below), the driving of gravity modes is still not efficient enough and magnetic tension may further limit the transfer of energy to turbulent motions \citep{bambic_suppression_2018}. The same magnetic draping mechanism may also suppress mixing of the AGN bubble content with the ambient gas, and thus suppress heating of the CGM/ICM \citep{ehlert_simulations_2018}. However, as demonstrated in idealized pure hydrodynamical simulations,  gravity modes can be efficiently excited if the AGN bubbles that drive motions are flattened in the radial direction leading to volume-filling turbulence \citep{zhang_generation_2018}. Interestingly, this can be accomplished by invoking CR-filled AGN bubbles (see below).\\
\indent
Just as in the case of the feedback mechanisms discussed above, the efficiency of the supplementary processes that assist the AGN feedback in offsetting radiative losses may also be reduced. Specifically, even when energetically significant turbulence is present, the dissipation rate of turbulence may be reduced compared to the classic Kolmogorov dissipation rate when gravitational stratification is significant \citep[i.e., when the Froude number $\rmn{Fr}\sim \varv/(L\omega_\rmn{BV})\ll 1$, where $\varv$ and $L$ are the characteristic turbulent velocity and turbulence driving scale, and $\omega_\rmn{BV}$ is the angular frequency of buoyant oscillations;][]{wang_turbulent_2023}. For related reasons, the entropy mixing can also be suppressed in the regime of low Froude number. The heat supply due to thermal conduction can also be limited due to plasma instabilities briefly discussed below.\\
\indent 
In dilute and magnetized plasmas such as the CGM/ICM, the particle mean free paths vastly exceed the Larmor radii of the particles and can be large even compared to the sizes of macroscopic thermodynamical fluctuations in the plasma (e.g., AGN bubbles or the wavelengths of sound waves excited by the AGN). In such weakly collisional plasmas, heat and momentum transport occurs predominantly along the direction of the local magnetic field. Under such conditions, gravitationally stratified plasma can become buoyantly unstable due to the magnetothermal instability \citep[MTI;][]{balbus_convective_2001} and heat-flux-driven buoyancy instability \citep[HBI;][]{quataert_buoyancy_2008}. These instabilities operate the fastest when the following hierarchy of mode frequencies $\omega_\rmn{r}$ (or, alternatively, the characteristic timescales $\tau\sim\omega_\rmn{r}^{-1}$) is realized: $\omega_{\rm cond}\gg\omega_{\rm dyn}\gg\omega_\rmn{a}$, where $\omega_{\rm cond}= 0.4\kappa_{\rm cond} (\bs{k\cdot \hat{b}})^{2}$ is the conduction frequency,  $\omega_{\rm dyn}$ is the dynamical frequency comparable to the Brunt-V{\"a}is{\"a}l{\"a} frequency, and $\omega_\rmn{a}=\bs{k\cdot {\varv}}_\rmn{a}$, where $\bs{k}$ is the wave vector, $\bs{\hat{b}}$ is the unit magnetic field vector, and $\bs{\varv}_\rmn{a}$ is the Alfv{\'e}n speed. In the case of the MTI, the instability occurs when the initially horizontal magnetic fields are perturbed in the positive radial direction in the background plasma, where the temperature increases in the direction of gravity. Fast conduction along the field lines heats the radially displaced gas and makes it more buoyant, which in turn increases the magnitude of the radial perturbation. In the case of the HBI, the instability occurs when the heat from the outer hot regions of a cool core flows in the direction of radial gravitational acceleration toward the cooler gas at the center. 
Obliquely propagating fluctuations in the initially radial magnetic field lead to local focusing of the heat flux that increases the buoyancy of the gas and consequently amplifies the horizontal magnetic field component. This reorientation of the magnetic field effectively shuts down the global conductive heat flux and makes conductive heating of cool cores ineffective \citep{parrish_nonlinear_2008}. However, the fields can be randomized by turbulent motions with kinetic energies at the level of 1\% of thermal energy as demonstrated in controlled simulations of driven turbulence \citep{parrish_turbulence_2010,ruszkowski_shaken_2010,ruszkowski_galaxy_2011} and self-regulated simulations of AGN feedback \citep{yang_interplay_2016} with anisotropic thermal conduction. Magnetic tension can also inhibit the HBI and MTI \citep[e.g.,][]{yang_interplay_2016}. \\
\indent
Furthermore, electron scattering can be controlled not only by classical collisions but also by self-generated whistler waves \citep{levinson_inhibition_1992,komarov_self-inhibiting_2018,drake_whistler-regulated_2021}. The latter dominate when the classical electron mean free path $\lambda_\rmn{c} > L/\beta_\rmn{e}$, where $L$ is the temperature length scale and $\beta_\rmn{e}\gg 1$ is the ratio of the electron thermal pressure to the magnetic pressure. When whistlers dominate scattering, the energy flux is primarily driven by advection/diffusion with the effective speed $\sim \varv_{\rm te}/\beta_\rmn{e}\ll \varv_{\rm te}$, where $\varv_{\rm te}$ is the electron thermal speed. Consequently, conduction along the magnetic field in the CGM/ICM can be severely limited by whistlers down to values well below the Spitzer level, thus limiting the efficiency of the HBI and MTI. However, self-regulated simulations of AGN feedback that do include whistler suppression of conduction suggest that the development of the HBI is not completely prevented \citep{beckmann_agn_2022}. Interestingly, the mathematical form of the heat transfer in the presence of whisters bears similarity to the equations governing CR energy flux. In addition to the usual diffusion term, the former also include a term representing advection of heat by the whistler waves, which is analogous to the CR streaming term \citep{drake_whistler-regulated_2021}.

\begin{figure}
  \begin{center}
    \includegraphics[width=\textwidth]{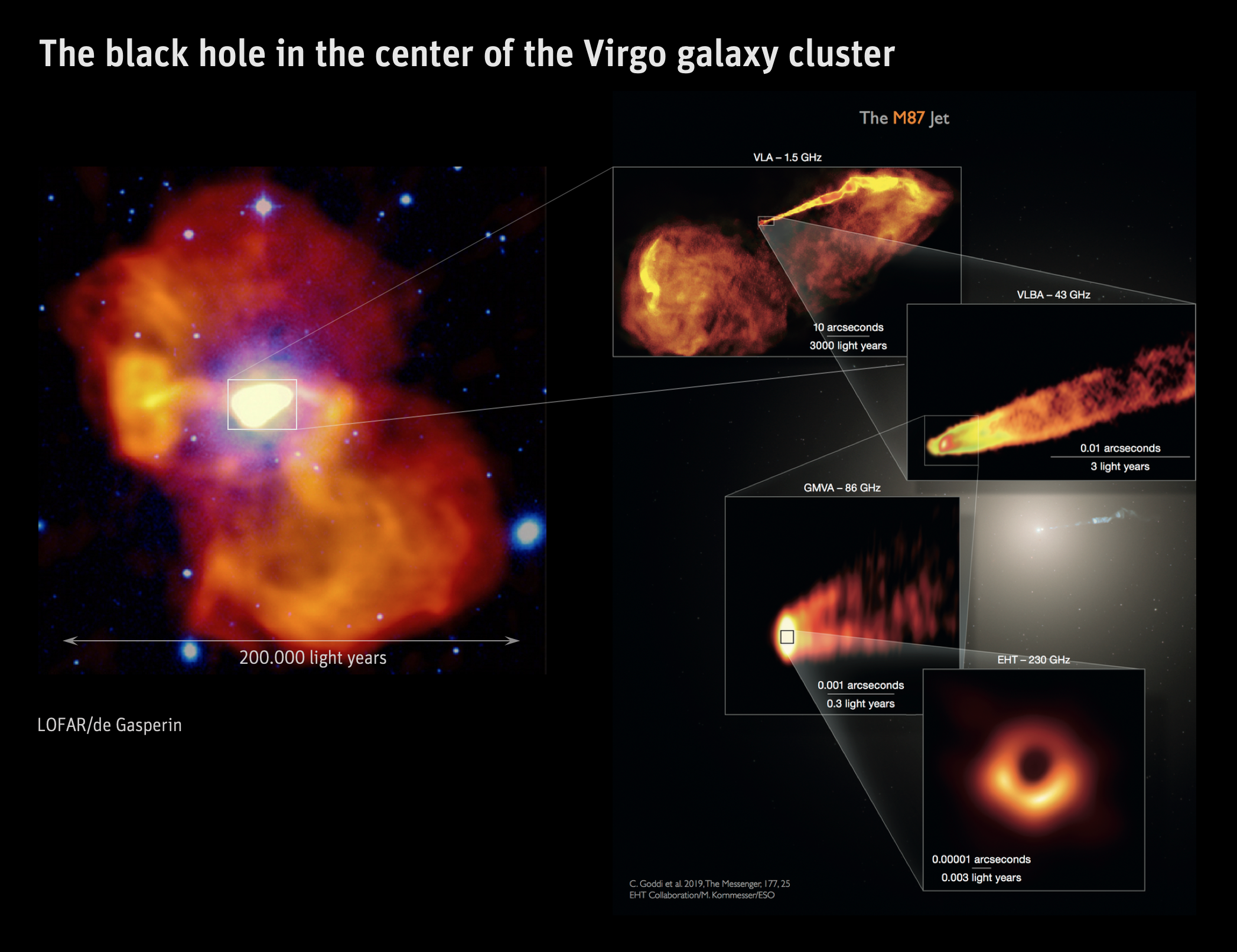}
  \end{center}
      \caption{\textit{Left}: Radio image from LOFAR of M87 in the Virgo cluster. The image shows AGN-inflated bubbles and diffuse emission distributed over the cool core of the Virgo cluster. \textit{Right}: VLA, VLBA, and EHT images of the central jet and the supermassive black hole. Image credits: LOFAR \citep{de_gasperin_m_2012}, VLA (Biretta, J. A. \& Owen, F. N., 1993, in prep.), VLBA \citep{walker_structure_2018}, GMVA \citep{kim_limb-brightened_2018}, ETH \citep[][ETH Collaboration]{akiyama_first_2019}; collage of images on the right: 
      \citep[][and ETH Collaboration, M. Kommesser, ESO]{goddi_first_2019}}.
  \label{fig:aLOFAR}
\end{figure}

\subsubsection{The role of AGN-injected cosmic rays in the CGM and ICM in massive halos}

The above discussion demonstrates that our understanding of the energy transfer from the AGN outflows/bubbles to the ambient CGM/ICM and the spatial redistribution of heat is incomplete. More specifically, it suggests that processes going beyond ideal MHD may be necessary to correctly model CGM/ICM heating and AGN feedback. At the same time, there is clear observational evidence for a significant population of non-thermal particles (see Fig.~\ref{fig:aLOFAR}) in the form of diffuse radio synchrotron emission \citep[e.g.,][]{de_gasperin_m_2012}, emission from Fanaroff--Riley~I jet lobes \citep[e.g.,][]{mingo_revisiting_2019}, and Sunyaev--Zel'dovich (SZ) signal deficit spatially coincident with the X-ray bubbles
\citep{abdulla_constraints_2019} in galaxy clusters. We refer the reader to Section~\ref{sec:AGN_feedback_clusters}, where we discuss the observational aspects of AGN CR feedback in more detail.\\
\indent
As argued above, the heating processes discussed at the beginning of this section (see Section~\ref{heating_limitations}) may be inefficient. Interactions of CRs with the CGM/ICM may offer a viable alternative to these processes. Some of the fundamental reasons why CR heating may be promising (apart from the very evidence for the presence of CRs in the ICM) include the fact that, for CR-filled AGN bubbles, the energy injected by the AGN can be better utilized. In the CR case, the bubble enthalpy is $\gamma_\rmn{ad}/(\gamma_\rmn{ad} -1)PV = 4PV$, where $\gamma_\rmn{ad}$, $P$, and $V$ are the adiabatic index of the bubble plasma ($\gamma_\rmn{ad} =4/3$ for relativistic CRs), ambient pressure, and bubble volume, respectively. This enthalpy is larger compared to that in the case of the bubble interior filled with thermal gas (in which case $\gamma_\rmn{ad} =5/3$ and the entalphy is $2.5PV$). Part of the enthalpy is used on $P\rmn{d}V$ work needed to inflate the bubbles. The reminder of the enthalpy could be used on $P\rmn{d}V$ work if the buoyantly rising bubbles never shred due to Rayleigh-Taylor and Kelvin-Helmholz instabilities (which is unlikely) or it could be efficiently utilized via mixing aided by diffusion and streaming of CRs into the ambient plasma. Furthermore, radially-flattened bubble morphologies of light and CR-filled bubbles are in better agreement with observations \citep[e.g.,][]{ehlert_simulations_2018} and, importantly, only a very small CR pressure support in the diffuse CGM/ICM is needed to offset radiative losses (e.g., \citealt{guo_feedback_2008,jacob_cosmic_2017,jacob_cosmic_2017-1}; see below). Additionally, CRs can affect the state and stability of the CGM/ICM plasma in a number of surprising ways. We discuss these topics in more detail below.

\paragraph{AGN CR heating of the CGM and ICM.}
It is instructive to begin the discussion of the physics of the CGM/ICM heating by CRs escaping from AGN lobes by considering the following simple estimate. The ratio of the CR streaming heating to radiative cooling rates is
\begin{equation}
\frac{\mathcal{H}_{\rm heat}}{\mathcal{C}_{\rm cool}} = 
\frac{|\bs{\varv}_\rmn{a}\bs{\cdot}\bs{\nabla} P_{\rm cr}|}{n^{2}\Lambda(T)}\sim \frac{1}{\beta^{1/2}}\frac{P_{\rm cr}}{P_\rmn{th}}\frac{H_\rmn{th}}{H_{\rm cr}}\frac{\tau_{\rm cool}}{\tau_{\rm ff}}\sim\mathcal{O}(1),    
\end{equation}
where $\beta = P_\rmn{th}/P_{B}$ is the plasma $\beta$ parameter, i.e., the ratio of thermal to magnetic pressure; $P_{\rm cr}$ is the CR pressure; $H_\rmn{th}$ and $H_{\rm cr}$ are the scale heights of thermal and CR pressures, respectively; $\tau_{\rm cool}$ is the radiative cooling timescale of the ICM/CGM; and $\tau_{\rm ff}$ is the free-fall timescale (assuming approximate hydrostatic equilibrium). Given that (i) typically $\beta\sim 10^{2}$, (ii) the CR pressure profile can be locally much steeper than the thermal gas profile (e.g., \citet{ehlert_simulations_2018}; we assume $H_\rmn{th}/H_{\rm cr}\sim10$ for the sake of the argument), (iii) $\tau_{\rm cool}/\tau_{\rm ff}\sim 10$ when the gas is locally thermally unstable (as is often the case in cool cores; see also discussion of TI in Section~\ref{TI}), and (iv) the ratio of CR-to-gas pressure is small (here assumed to be 10\% for the sake of the argument; see more discussion in Section~\ref{sec:AGN_feedback_clusters}), this estimate shows that CR heating can be competitive with radiative cooling for reasonable choices of model parameters.\\
\indent
Early one-dimensional models of ICM heating via CR streaming \citep{loewenstein_cosmic-ray_1991}
reached the conclusion that, in combination with thermal conduction, this mode of heating can substantially reduce the cooling flow. However, these models were inconsistent with the observations because they either overpredicted CR pressure or required too strong magnetic fields in order to reduce CR energy density via CR streaming heating. These problems stemmed from the assumption that CRs propagate only via diffusive CR transport, and can be circumvented by noting that the dispersal of CRs can be to large extent driven by buoyant advection of CRs before they leave the AGN lobes some distance from the potential center and interact with the ambient plasma via streaming. Relaxing the assumption of the diffusive-only transport leads to successful models that match the ICM temperature and density profiles and requires only a small CR pressure support in the bulk of the ICM \citep{guo_feedback_2008,enslin_cosmic_2011,fujita_non-thermal_2012,wiener_cosmic_2013, pfrommer_toward_2013,jacob_cosmic_2017,jacob_cosmic_2017-1}. Interestingly, the functional form of the heating function in some of these one-dimensional simulations is similar to the one that relies on the $P\rmn{d}V$ work to heat the ICM (both in terms of the proportionality of the heating function to the pressure gradient and the suppression of heating close to the center of the potential well; \citealt{ruszkowski_heating_2002}).

\paragraph{Three-dimensional models of CR AGN heating.} The general premise of the above models, that the streaming of CRs and the associated with it heating of the CGM/ICM is sufficient offset cooling, can be tested by performing controlled three-dimensional simulations of the AGN feedback. Specifically, one of the key assumptions of the above one-dimensional models is that the contents of the AGN-inflated cavities can come into contact with the ambient plasma and heat it efficiently. This requirement for sufficiently fast mixing of CRs with the ICM is not guaranteed a priori because of the formation of magnetic draping layers on the leading edge of the buoyantly rising AGN bubbles \citep{ruszkowski_impact_2007,dursi_draping_2008}. These layers are formed when enough ambient magnetic field is swept up in front of the bubbles to balance the ram pressure associated with the upward motion. Thus, the magnetic pressure in these thin layers is dynamically important even if typical magnetic pressures in the ambient plasma are small compared to thermal pressures. The significance of these layers lies in that they can suppress the diffusion and streaming of CRs from the bubble interiors to the ambient plasma ahead of the bubbles. The mathematical condition for the formation of such layers is $\lambda_{B}\gtrsim R/(12\mathcal{M}_\rmn{a}^2)$ \citep{pfrommer_detecting_2010,sparre_interaction_2020}, where $\lambda_{B}$ is the coherence length of the ambient magnetic field, $R$ is the curvature radius of the bubble at the stagnation point (where the flow lines bifurcate to pass left- and right-ward of the upward rising bubble), and $\mathcal{M}_\rmn{a}$ is the Alfv{\'e}n Mach number of the bubbles \citep[$\mathcal{M}_\rmn{a}\sim\beta^{1/2}\mathcal{M}\sim$ a few, where $\mathcal{M}\sim 0.5$ is the Mach number of the bubbles; see, e.g.,][]{churazov_evolution_2001}. 
Given that characteristic sizes of the bubbles are a few tens of kpc and the coherence length of the magnetic field is a factor of a few smaller
\citep{vogt_bayesian_2005,kuchar_magnetic_2011}, the condition for the formation of the draping layers can be easily satisfied. Some observational evidence for such layers is seen in the radio polarization data \citep{adebahr_polarised_2019}.
\begin{figure}[tbp]
\begin{center}
\includegraphics[width=1.0\textwidth]{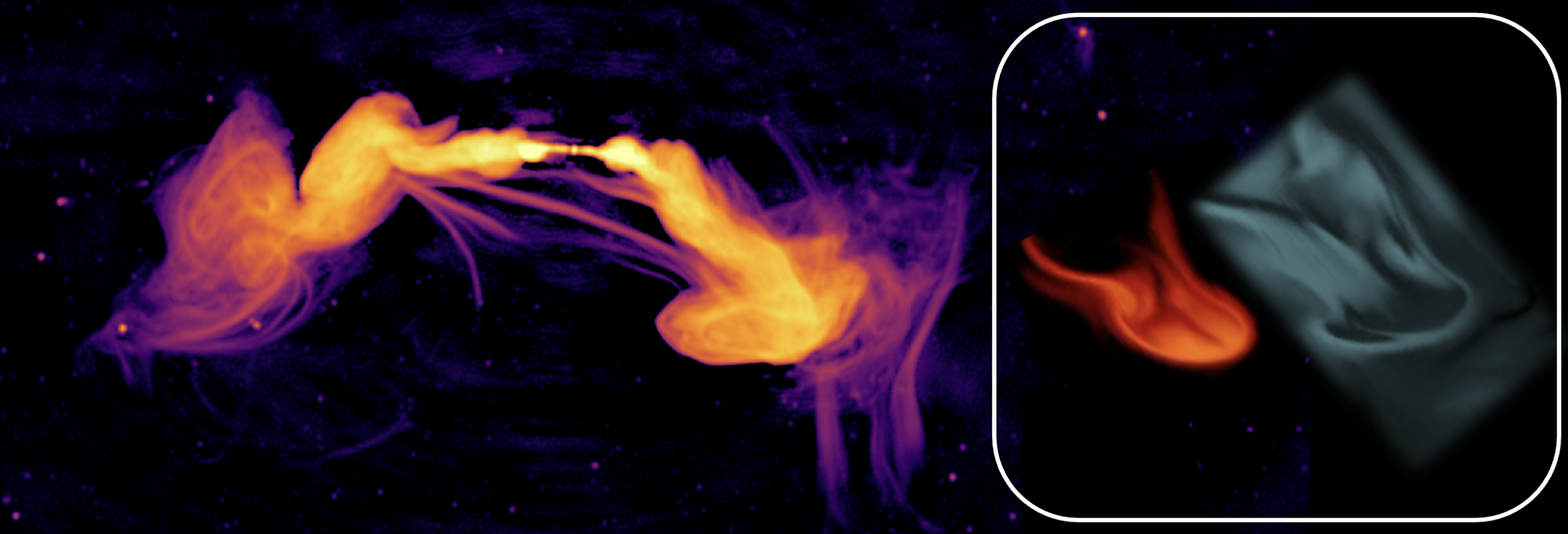}
\end{center}
\caption{\textit{Left}: Image of radio lobes in ESO 137-006 obtained with the MeerKAT telescope. Image credit: Rhodes University, INAF, SARAO; 
Image from \citet{ramatsoku_collimated_2020}; reproduced with permission from A\&A.
\textit{White box}: CR energy density distribution in a buoyantly rising AGN bubble (left) and magnetic pressure (right). The bubble is moving towards bottom right. Notice the features reminiscent of those seen in the data -- formation of a magnetic draping layer on the leading edge of the bubble prevents CR escape in the forward direction while CR are seen escaping in the wake of the bubble. Images based on data from  \citet{ruszkowski_cosmic_2008}; reproduced with permission from MNRAS.}
\label{fig:ramatsoku}
\end{figure}
The picture of the suppression of CR transport out across draping layers is supported by radio observations (see left panel in Fig.~\ref{fig:ramatsoku}) that reveal sharp edges in the radio emission on the sides of the radio lobes most distant from the galactic center. The draping phenomenon is also seen in numerical simulations of AGN feedback \citep[see right panel in Fig.~\ref{fig:ramatsoku};][]{ruszkowski_cosmic_2008}. Interestingly, the distribution of the diffusing CRs in the wake of the bubbles is reminiscent of the observed collimated synchrotron threads that connect the AGN lobes to the galaxy \citep{ehlert_simulations_2018}.\\  
\begin{figure}[tbp]
\begin{center}
\includegraphics[width=1.0\textwidth]{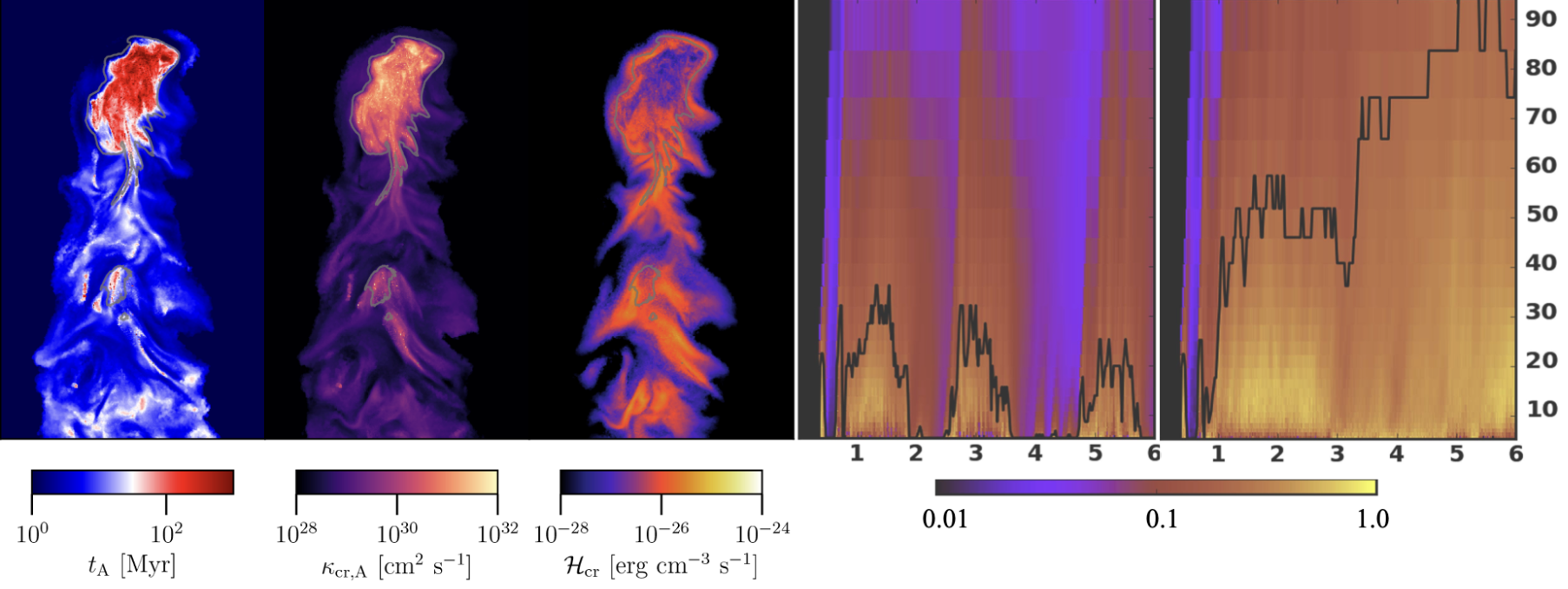}
\end{center}
\caption{\textit{Three left panels}: Slices of various quantities from feedback simulations of an individual AGN outburst with CR physics included. The slice plane is parallel to the jet. From left to right, the panels show CR loss timescale due to the streaming instability (same as gas heating timescale), instantaneous CR diffusion coefficient, and CR heating rate per unit volume. Notice that characteristic CR heating times are comparable to typical radiative cooling times in cool cores and that significant amount of heating occurs in the wake of the buoyantly rising bubbles. \textit{Two rightmost panels}: Evolution of the CR-to-gas pressure ratio $X_{\rm cr}$ from simulations of self-regulated AGN feedback with CR physics included. 
The right panel corresponds to a simulation with Coulomb and hadronic losses and the left one to the case that also includes CR streaming heating and streaming transport. 
The units on the horizontal and vertical axes are Gyr and kpc, respectively. The color bar shows the CR-to-gas pressure ratio; the black line corresponds to $X_{\rm cr}=0.25$. Notice that including CR streaming heating dramatically reduces CR pressure support in the ICM as CR energy is expended on preventing radiative losses of the gas. Images from \citet{ehlert_simulations_2018} and \citet{ruszkowski_cosmic-ray_2017}; reproduced with permission from MNRAS and ApJ.}
\label{fig:ehlert_ruszkowski}
\end{figure}
\indent
Based on the above evidence, mixing is suppressed when the magnetic field effects are included. Simulations of long-term evolution of AGN outbursts show that this is indeed the case (\citealt{ehlert_simulations_2018}; see also \citealt{weinberger_simulating_2017}). These simulations demonstrate that the mixing efficiency decreases with AGN jet power. More powerful jets result in high-density-contrast and high-Mach-number outflows with bipolar and lobe-brightened morphologies resembling Fanaroff--Riley~II sources, while weaker jets result in low-density-contrast AGN cavities that are more easily deflected close to the cluster center and more flattened in the radial direction, resembling Fanaroff--Riley~I sources. The mixing that does occur for these low power AGN may facilitate CR heating of the CGM/ICM. Most of the CR escape occurs in the wake of the buoyantly rising bubbles as seen in the three left panels in Fig.~\ref{fig:ehlert_ruszkowski}. Leftmost of these panels shows the CR streaming instability loss timescales that are comparable to typical radiative cooling times in cool cores, which demonstrates that CR streaming heating is competitive with radiative cooling. Importantly, the volumetric heating in these simulations approximately matches that assumed in one-dimensional steady-state solutions discussed above \citep{jacob_cosmic_2017,jacob_cosmic_2017-1}. We note that while mixing is an important element of the model \citep[see also][]{soker_requirement_2019}, CR transport to larger distances from the center by buoyancy as well as CR streaming itself play very important roles in facilitating high heating rates. The energy exchange between CRs and ambient plasma via mixing alone followed by Coulomb and hadronic interactions is much slower than via the streaming instability \citep[e.g.,][]{guo_feedback_2008,ruszkowski_cosmic-ray_2017}.

\paragraph{Self-regulated CR AGN feedback models of the CGM/ICM heating.}\label{self-regulated} A successful model of the CGM/ICM heating by AGN has to account for the long-term global thermal stability of cool cores of massive ellipticals and clusters. This can be achieved thanks to the positive AGN feedback cycle, where the central black holes are fed by the cooling-induced accretion of the gas that triggers AGN outbursts that in turn prevent excessive cooling. AGN feedback models that demonstrate that such self-regulation can be achieved have been considered by many authors in the framework of purely hydrodynamical simulations  \citep[e.g.,][]{sijacki_unified_2007,sharma_thermal_2012,gaspari_cause_2012,dubois_self-regulated_2012,prasad_cool_2015,yang_how_2016,meece_triggering_2017,li_agn_2017,martizzi_simulations_2019}. In these models, heating occurs predominantly via a combination of weak shocks and very efficient mixing of hot thermal AGN bubble gas with the ambient medium. From the observational perspective, these models, while successful in establishing the self-regulating feedback, reducing cooling rates to observable values, and explaining average thermodynamical profiles, tend to overproduce spatial fluctuations in the temperature profiles \citep[e.g.,][]{gaspari_cause_2012} compared to observations \citep[e.g.,][]{zhuravleva_turbulent_2014}. From the theoretical point of view, these models overestimate mixing due to the neglect of the magnetic field and ignore CR heating due to CRs diffusing out of AGN-inflated cavities into the ambient medium.\\
\indent
Self-regulated MHD-CR models of AGN jet feedback demonstrate that CR heating can indeed offset radiative losses but the establishment of steady-state feedback loop may depend on how CR heating and transport are modeled. Two rightmost panels in Fig.~\ref{fig:ehlert_ruszkowski} show the ratio of CR-to-gas pressure $X_{\rm cr}$ from simulations of AGN jet feedback in a galaxy cluster. This quantity is shown as a function of time (horizontal axis; in Gyr) and radius (vertical axis; in kpc) for the case (i) with CR streaming and streaming heating (second to last panel) and (ii) neglecting these effects (rightmost panel). In both cases, Coulomb and hadronic losses are included. While in the former case, $X_{\rm cr}$ has generally low values $\lesssim 0.1$ and the AGN is intermittent, in the latter case, $X_{\rm cr}$ systematically grows over time reaching dynamically-important values beyond 25\% in the entire the cool core. The latter case is in disagreement with the observational constraints on non-thermal cool core content based on $\gamma$-ray emission \citep{pfrommer_toward_2013}, radio emission \citep{jacob_cosmic_2017,jacob_cosmic_2017-1} and comparisons of the gravitational potential profiles obtained from optical and X-ray data \citep{churazov_measuring_2008}. Moreover, in the latter case, a substantial cooling flow develops with mass accretion rates exceeding observational estimates.
The dramatic differences between these results can be attributed to CRs from AGN lobes more easily coming into contact with the ambient medium via streaming transport and to the increased CR energy losses associated with the streaming instability \citep{ruszkowski_cosmic-ray_2017}. We reiterate that the exchange of energy via CR Coulomb and hadronic losses is relatively slow compared to mixing of the hot thermal gas with the ambient CGM/ICM in pure hydrodynamical simulations, and this inefficiency, combined with the presence of magnetic fields that reduce mixing, may explain why CR pressure is building up when streaming losses are neglected.
\begin{figure}[tbp]
\begin{center}
\includegraphics[width=1.0\textwidth]{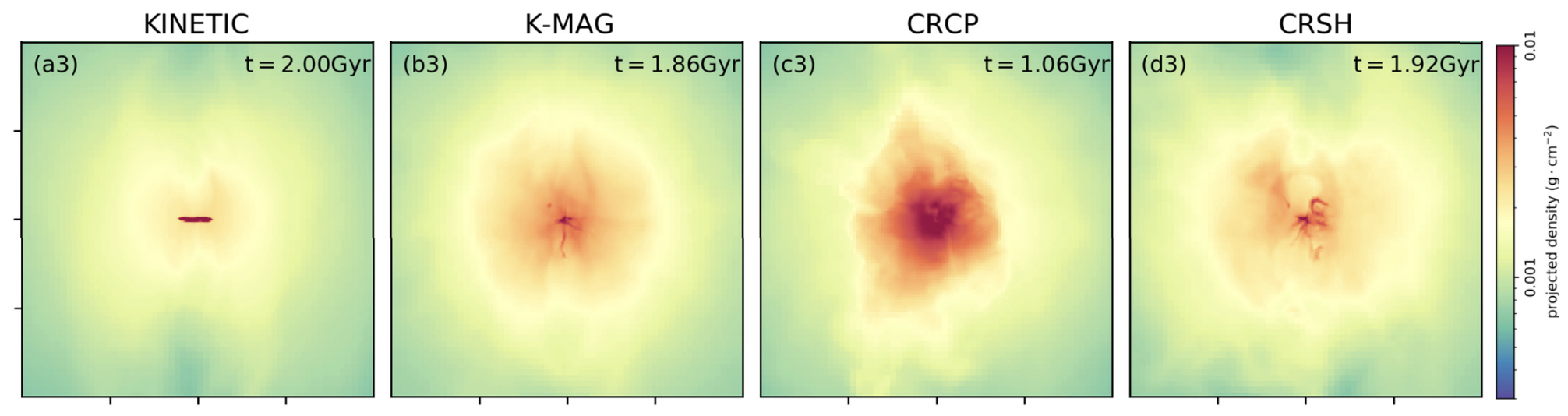}
\end{center}
\caption{Projected gas density from MHD-CR AGN jet feedback simulations. From left to right the panels show cases corresponding to (i) kinetic jet feedback, (ii) same as (i) but with magnetic fields added, (iii) same as (ii) but with CR advection and Coulomb and hadronic losses, and (iv) same as (iii) but adding CR streaming transport and the associated CR streaming losses.  AGN jet is approximately along the $z$ direction. Panels are 60 kpc on each side. Image from \citet{wang_chaotic_2020}; reproduced with permission from MNRAS.}
\label{fig:wang}
\end{figure}
Similar conclusions regarding CR heating and transport were reached in the case of AGN feedback in massive ellipticals \citep{wang_chaotic_2020}.  Figure~\ref{fig:wang} shows projected gas density for various AGN feedback models with and without CR physics. The pure hydrodynamical ``kinetic'' jet case (left panel) does not result in a cooling catastrophe (though it leads to an over-massive cold disk in the center and overheating of the gas). In the MHD counterpart of this case (second panel from the left), large cooling flows do not develop as mixing of the thermal AGN lobe content, albeit reduced by the magnetic fields, is able to offset cooling in the time-averaged sense, and the feeding of the AGN is more efficient due to magnetic breaking acting on the cold gas (see next paragraph). When the AGN jet content is replaced by CRs (third panel from the left), the slow Coulomb and hadronic losses lead to inefficient heat transfer from the lobes and the CGM cools and accretes forming a large central overdensity in disagreement with observations. However, as in the case of clusters, including streaming transport and heating (right panel) elevates the efficiency of heat transfer from the AGN outflow to the ambient gas and restores global thermodynamical balance.\\
\indent
The conclusions regarding the average steady state of the CGM/ICM (such as globally stable atmosphere vs.\ uncontrolled cooling flow, amount of cold gas in the CGM/ICM, morphology of the cold gas, etc) are likely dependent on the specific choices concerning AGN jet power (or jet efficiency), jet reorientation, jet/bubble composition (i.e., CR fraction or jet density contrast/lightness). For example, \citet{beckmann_cosmic_2022} show that the cold gas continues to build up over the course of the simulations in the MHD case, while in their CR case, with or without CR transport, the amount of gas reaches a steady state consistent with observations. Compared to the results discussed above, the differences in the conclusions are likely to be due to a combination of factors related to magnetic tension and jet physics. As mentioned above, magnetic fields decrease the efficiency of mixing and, thus, the transfer of the energy from the AGN lobes to the ambient medium. This process can increase net cooling and the amount of cold gas in the CGM/ICM. Furthermore, as discussed in Section~\ref{TI}, even in high plasma $\beta$ environments, weak magnetic tension can enhance local thermal instability \citep{ji_impact_2018}, which may further increase the amount of cold gas. On the other hand, magnetic tension acting on the clouds can also remove the angular momentum of the clouds and facilitate faster accretion/removal of the cold gas (\citealt{wang_chaotic_2020,ehlert_self-regulated_2023}; cf.\ first and second panel in Fig.~\ref{fig:wang} that show that overmassive dense disks present in the pure hydro case are removed when the magnetic fields are present; purely hydrodynamical simulations of AGN feedback often exhibit the formation of such disks \citep{li_modeling_2014,eisenreich_active_2017,qiu_using_2019} that are inconsistent with observations \citealt{russell_driving_2019}). Competition between these processes may influence the actual amount of cold gas present in the CGM/ICM.\\
\indent 
The simulations of \citet{beckmann_cosmic_2022} also include a sub-grid model for black hole spin magnitude and orientation \citep[which is missing in][]{ruszkowski_cosmic-ray_2017,wang_chaotic_2020}, that allows for the jet reorientation and power to be determined on-the-fly depending on the properties of the accreting gas. When the amount of cold gas is large, and the motions of this gas phase are predominantly chaotic rather than ordered, accretion of the clouds may help to reorient the jet. This, in turn, may help to distribute and mix the CRs injected by the AGN jet more evenly reducing the need to rely on CR transport that is otherwise needed to ensure that CRs come into contact with the ambient CGM/ICM to heat it. Furthermore, accretion of the clouds onto the central black hole may result in partial cancellation of the angular momentum added to the black hole, which could limit the maximum spin that the central black hole could acquire (and thus limit the jet efficiency). Both the fast reorientation of the jet and the reduction in AGN efficiency, as well as allowing for black hole wandering, may affect how far the jet can propagate and thus how it heats the CGM/ICM.\\ 
\indent
In the case of relatively low-density jets, the re-distribution of CR heating throughout the cool core can also be facilitated by the deflection of the bubbles by cold gas filaments \citep{ehlert_self-regulated_2023}. Unlike in the higher-density jet case, the deflection of the bubbles, acting together with the magnetic fields, can prevent the formation of the cold disks mentioned above. Interestingly, gentle heating by such low-density jets leads to temperature profiles that show remarkably little small-scale spatial variability, which is in good agreement with X-ray observations. 
A detailed investigation of the different modes of AGN feedback (kinetic, thermal, CR) in $10^{14}$~M$_{\odot}$ halos that explored the impact of varying model assumptions regarding the magnetic fields, jet opening angle and precession, and AGN duty cycle, was also performed in the framework of non-cosmological FIRE-2 simulation suite that included stellar feedback, viscosity, and conduction \citep{su_which_2021}. This work demonstrated that CR-dominated jets are most efficient in quenching galaxies. One of the key quenching criteria identified in that work is that the jet must be able to expand to subtend a large solid angle, which can be accomplished if the mode of feedback is (i) non-kinetic, i.e., dominated by a combination of CRs, magnetic and thermal energy; or (ii) kinetic jet is extremely ``light''  \citep[see also][]{weinberger_active_2023}; or (iii) kinetic jet is characterized by a wide opening angle or exhibits significant precession.\\
\indent
The ultimate goal of the AGN CR feedback models is to establish if this mode of heating plays an important role in shaping the properties of the most massive galaxies in fully cosmological setting. First cosmological hydrodynamical simulations of AGN feedback that included adiabatic CRs were performed by \citet{sijacki_simulations_2008}, who showed that CRs can provide significant pressure support in galaxy clusters. However, these early simulations did not include magnetic fields and active CR transport processes. Recent work based on a large suite of cosmological FIRE-2 zoom-in simulations included much richer physics and explored different AGN feedback models (CR, mechanical, and radiative feedback) for a wide range of halo masses \citep[$10^{10-13}$~M$_{\odot}$;][]{wellons_exploring_2023}. This work demonstrated that, while several models are consistent with the observational constraints, CR physics indeed plays an important role in many of the models.

\subsubsection{Impact of cosmic rays on CGM/ICM instabilities}
CRs diffusing and streaming out of AGN lobes and dispersing throughout the CGM and ICM can affect the dynamical and thermal stability of the plasma. In Section~\ref{TI}, we discussed the impact of CRs on thermal instability, which has consequences for the formation of the cold gas phase that plays an important role in chaotic cold accretion and feeding of central supermassive black holes. Here we extend this discussion to include other CR instabilities that may shape the properties of the CGM and ICM -- the CR buoyancy instability (CRBI) and CR acoustic Braginskii instability. The importance of these instabilities lies in that they can affect the transfer of energy from the AGN to the ambient medium by determining how the energy injected by the AGN is partitioned between sound waves, turbulence, and direct heating of the gas by CRs.\\
\begin{figure}[tbp]
\begin{center}
\includegraphics[width=0.8\textwidth]{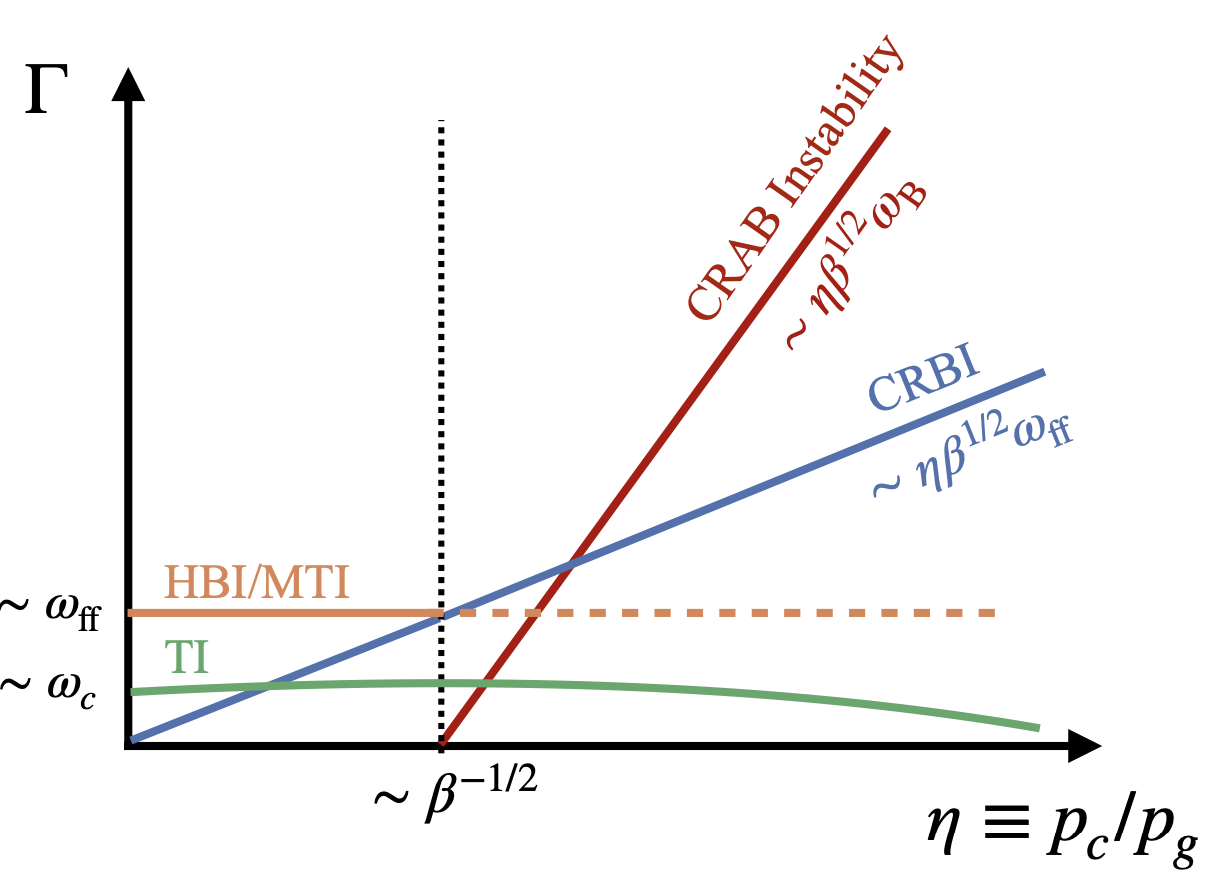}
\end{center}
\caption{Schematic representation of the growth rates $\Gamma$ of the various instabilities in dilute plasmas as a function of the CR-to-thermal gas pressure ratio $P_\rmn{cr}/P_\rmn{th}$, defining the abbreviations for thermal instability (TI); heat flux buoyancy instability (HBI); magnetothermal instability (MTI); CR acoustic Braginskii (CRAB) instability; CR buoyancy instability (CRBI). The characteristic frequencies $\omega_{\rm ff}$ and $\omega_\rmn{c}$ represent free-fall and cooling frequency, respectively (and are inversely proportional to the corresponding timescales). The viscous frequency $\omega_{B}\approx (P_\rmn{th}/(\rho\nu_{\rm ii})(\bs{k\cdot \hat{b}})^{2}$, where $\nu_{\rm ii}$ is ion-ion collision rate, and $\bs{k}$ and $\bs{\hat{b}}$ are the wave and unit magnetic field vectors. TI is not significantly suppressed for $P_\rmn{cr}/P_\rmn{th}\lesssim 1$. A typical value of the ratio of the thermal pressure-to-magnetic pressure is $\beta\sim 10^{2}$. Since CRs can suppress the HBI/MTI, and HBI/MTI and CRBI can correspond to the same mode, the extension of the HBI/MTI to larger values of $P_\rmn{cr}/P_\rmn{th}$ is marked with a dashed line. Figure from \citet{kempski_new_2023}; reproduced with permission from MNRAS.}
\label{fig:kempski}
\end{figure}
\indent
In the presence of CR pressure support, the fast magnetosonic mode may become unstable \citep{kempski_sound-wave_2020} leading to CR acoustic Braginskii instability. The instability arises due to an additional dependence of the CR streaming speed on pressure anisotropy of weakly-collisional thermal plasmas. This pressure anisotropy does positive work on CRs where CR fluctuations are positive, thus enhancing the fluctuations. The instability is very fast with growth rates on the order of the sound frequency. When CRs are absent, the acoustic waves are damped by anisotropic viscosity. The critical minimum ratio of the CR-to-thermal gas pressure necessary to trigger the instability is $\sim\beta^{-1/2}$, and thus the instability is likely to develop even for relatively low CR pressures in the high-$\beta$ CGM/ICM. An important consequence of this instability is the possibility of exciting and amplifying sound waves in the CGM/ICM that could resemble the X-ray ripples seen in the Virgo and Perseus clusters.\\
\indent
In the presence of gravity, streaming CRs can also make the plasma unstable. The generalization of the MTI to include the impact of anisotropic CR diffusion, when magnetic fields are initially perpendicular to gravity, demonstrated that the plasma is more buoyantly unstable in the presence of CR pressure gradients anti-parallel to the gravitational acceleration \citep{chandran_convective_2006,dennis_parkerbuoyancy_2009}.
Further extensions to include arbitrary speed of CR diffusion and orientation of the initial magnetic field with respect to gravity demonstrated that in the limits of (i) fast diffusion, negative CR pressure gradients enhance the growth rates of the MTI and HBI, and (ii) slow diffusion (adiabatic CRs), the criterion for driving buoyant instability by CRs becomes analogous to the classic Schwarzschild criterion with entropy replaced by CR ``entropy'' $(\equiv P_\rmn{cr}/\rho^{4/3})$ \citep{sharma_buoyancy_2009}.\\
\indent
For both limits, the above instabilities require CR pressures to exceed a significant fraction of the gas pressure to operate. Moreover, they do not include CR transport via streaming. However, the inclusion of CR streaming is crucial as it can render the gas buoyantly unstable by increasing the compressibility of pressure-balanced waves \citep{kempski_new_2023}. This leads to an instability (the CR buoyancy instability; CRBI) as the unbalanced gravitational forces can accelerate the enhanced density fluctuations. The CRBI is triggered when $\beta (\omega_\rmn{c}/\omega_{\rm ff})\gtrsim 1$, where $\omega_\rmn{c}$ and $\omega_{\rm ff}$ are the cooling and free-fall frequencies, respectively. For typical conditions in cool cores of galaxy clusters, where the ratio of the cooling-to-dynamical timescales is around 10 and plasma $\beta\sim 10^{2}$, this condition can be easily met. The maximum growth rate of the CRBI ($\sim\beta^{1/2}\omega_{\rm ff}P_\rmn{cr}/P_\rmn{th}$) can be very fast and comparable to the free-fall frequency $\omega_{\rm ff}$ in high-$\beta$ plasma even in the case of low CR pressure support. As the CRBI is particularly fast at small scales, where the effects of compressibility due CR streaming are the strongest, it operates efficiently exactly where the HBI and MTI are suppressed by the magnetic tension. Consequently, CRBI/MTI/HBI can all act together across a wide range of scales in the CGM/ICM. Figure~\ref{fig:kempski} summarizes many aspects of the above discussion and shows a schematic representation of the relative growth rates of the thermal instability in comparison to the various 
instabilities 
mediated
by anisotropic thermal conduction and CRs
as a function of the CR-to-thermal gas pressure ratio.

\clearpage

\vspace{0.25in}

\section{Observational signatures}\label{Observational signatures}\label{observational_signatures}

This chapter provides an overview over several observational signatures of CR feedback. In Section~\ref{sec:CRprop}, we start by reviewing how studies of CR propagation in the Milky Way provide insight into CR transport properties and the preferred value of the CR diffusion coefficient. In Section~\ref{sec:CR_driven_outflows} we summarize the evidence for CR-driven/aided outflows in the Milky Way. In moving to extragalactic sources, we explain in Section~\ref{Non-thermal emission from galaxies} how observations of the non-thermal emission from galaxies ranging from the radio to $\gamma$ rays provide a measure for the calorimetric energy fraction lost to radiation products, which is then missing from feedback. This is followed by a review of the observational signatures of CR feedback in the CGM in Section~\ref{sec:CGM_feedback} and AGN feedback in galaxy clusters in Section~\ref{sec:AGN_feedback_clusters}. The chapter ends with Section~\ref{sec:multi-messenger_observatories} that presents an outlook for current and future multi-messenger observatories that can be used for studying CR feedback.

\subsection{Cosmic ray propagation in the Milky Way}
\label{sec:CRprop}

Before we review more elaborate (two- and three-dimensional) models of CR propagation in our Galaxy, we introduce two important concepts, namely (i) how the CR residence time is inferred from observations of momentum spectra of different CR nuclei and (ii) how we can derive the CR diffusion coefficient in the simple leaky box model of CR transport. To this end, we define the rigidity of a particle via equation~\eqref{eq:rg}:
\begin{align}
   R \equiv r_\rmn{g} B = \frac{p_\perp c}{Ze},
\end{align}
which has the units of Volts, e.g., a 10 GeV proton has a rigidity of 10 GV.

\subsubsection{Cosmic ray residence time and the boron-to-carbon ratio}
\label{sec:B-to-C_ratio}

\begin{figure}
  \begin{center}
    \raisebox{0.5em}{
    \includegraphics[width=0.45\textwidth]{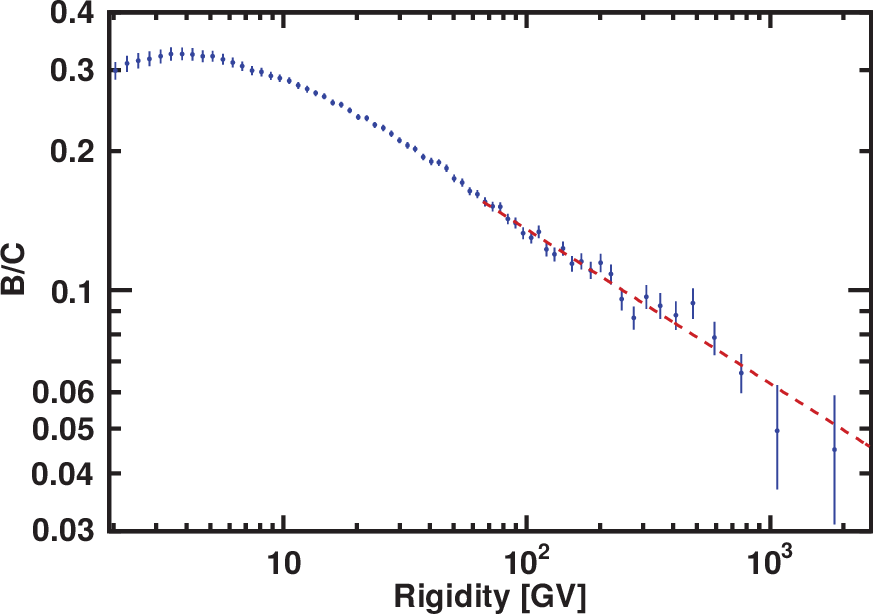}}\hfill
    \raisebox{0.35em}{        
    \includegraphics[width=0.53\textwidth]{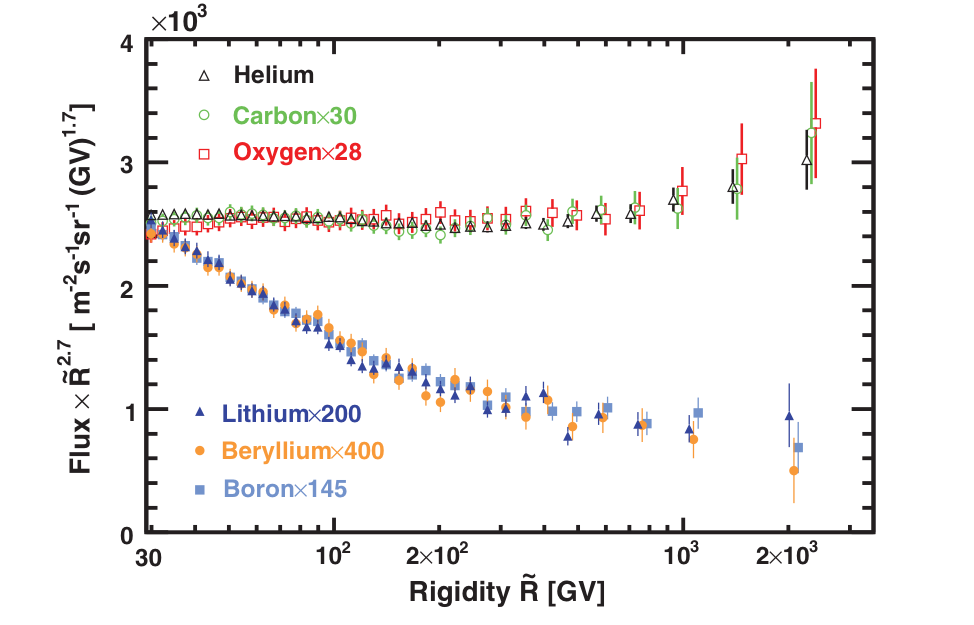}}
  \end{center}
  \caption{\textit{Left:} The AMS boron-to-carbon ratio (B/C) as a function of observed rigidity $\tilde{R}$, supplemented with a power-law fit of spectral index $-1/3$ (dashed line, starting from 65 GV). \textit{Right:} Comparison of the primary CR fluxes He, C, O with the secondary CR fluxes Li, Be, B, scaled by $\tilde{R}^{2.7}$ as a function of rigidity above 30 GV. Note that the various spectral compositions have been rescaled as specified, and for visual clarity, the data points for He, O, Li, and B above 400 GV have been horizontally shifted. Furthermore, it is important to highlight that both the primary and secondary fluxes exhibit an identical dependence on rigidity, but clearly differ from each other due to CR propagation effects, see Section~\ref{sec:leaky_box}. Figures from \citet{aguilar_precision_2016,aguilar_observation_2018}; reproduced with permission from PRL.}
  \label{fig:cosmic_ray_composition}
\end{figure}

The chemical composition of galactic CRs is similar to the element abundances in the Sun. This clearly indicates that CRs have been accelerated from stellar material, which defines the so-called primary CR population. However,  there are also important differences between CR and solar abundances: Li, Be, B are so-called secondary CR nuclei produced in the spallation reaction of heavier CR elements (C and O) with interstellar hydrogen. Similarly, Mn, V, and Sc nuclei of the CR population are the result of fragmentation of Fe. All these over-abundant CR elements (compared to solar abundances) are usually referred to as secondary CRs. By measuring the primary-to-secondary ratio, we can infer the propagation properties of CRs and their residence time in the Galaxy, as we will show in the following. Boron is mainly produced in spallation reactions by carbon and oxygen with an approximately conserved kinetic energy per nucleon. The boron source production rate is
\begin{align}
  Q_\rmn{B}(E) \approx \bar{n}_\rmn{H}\beta c \sigma_{\rmn{C\to B}}\,n_\rmn{C},
\end{align}
where $\bar{n}_\rmn{H}$ is the mean ISM number density, $n_\rmn{C}$ is the carbon density, $\beta c$ is the carbon velocity and $\sigma_{\rmn{C\to B}}$ is the  $\rmn{C\to B}$ spallation cross-section. In steady state, the boron production rate balances its number density divided by the lifetime in the Galaxy, $\tau$, before it escapes the scattering halo around the Galaxy or experiences spallation itself:
\begin{align}
  Q_\rmn{B}(E) = \dot{n}_\rmn{B} = \frac{n_\rmn{B}}{\tau}
  \quad\Rightarrow\quad
  \frac{n_\rmn{B}}{n_\rmn{C}} 
  \approx\bar{n}_\rmn{H}\beta c \sigma_{\rmn{C\to B}}\,\tau
  = 0.65 \beta \left(\frac{R}{\rmn{GV}}\right)^{-0.33}.
\end{align}
In the last step we used the results of the AMS (Alpha Magnetic Spectrometer) and CALET (CALorimetric Electron Telescope) collaborations \citep{aguilar_precision_2016,adriani_cosmic-ray_2022}, as shown for AMS data in the left-hand panel of Fig.~\ref{fig:cosmic_ray_composition}. Using the values of the cross-section, this leads to a lifetime gas density of
\begin{align}
  \bar{n}_\rmn{H}\tau \approx 10^{14} \left(\frac{R}{\rmn{GV}}\right)^{-0.33}
  \rmn{s~cm}^{-3}.
\end{align}
To obtain the diffuse escape time of CRs from the total life time, we need to account for the spallation or fragmentation process of the respective ions. We find a correction factor of order unity (at CR rigidities of around GV) and obtain the diffusive escape time of GV CRs from the disk of $\tau_\rmn{esc}\approx3\times10^7$~yr for a mean hydrogen density along the CR column in the CR scattering halo of $\bar{n}_\rmn{H}\approx0.1~\rmn{cm}^{-3}$. The boron-to-carbon ratio decreases with increasing rigidity (see Fig.~\ref{fig:cosmic_ray_composition}), which suggests that high energy CRs spend less time than low energy CRs in the Galaxy before escaping or that the height of the scattering halo depends on the CR energy so that higher-energy CRs probe lower average densities at larger heights.\footnote{Note that the dependence of the observationally-inferred diffusion coefficient on halo height is discussed in Section~\ref{sec:CGM_feedback}.} Studying CR diffusion in regular and turbulent magnetic fields, \citet{giacinti_reconciling_2018} find that CRs would stay for too long in the Galactic disk and overproduce secondary nuclei like boron. In consequence, CRs are anisotropically transported in galaxies either in regular magnetic fields or in anisotropic turbulent fields. As a corollary, the number of sources supplying the local CR flux is thereby reduced by two orders of magnitude in comparison to isotropic CR diffusion.

\subsubsection{Leaky box model: derivation and limitations in light of new cosmic ray data}
\label{sec:leaky_box}

The leaky box model is a simplified phenomenological picture of CR transport in the Galaxy \citep{davis_diffusion_1960,ginzburg_origin_1964,berezinskii_astrophysics_1990,ptuskin_using_1996,jones_modified_2001}. In this model, primary CRs of (solar) composition (including carbon) are exclusively accelerated in the (thin) galactic disk at SNR shocks. After escaping their source regions, those CRs random walk through the magnetized plasma in a scattering halo\footnote{This scattering halo includes the ISM and the inner CGM where the CR scattering rate is sufficiently high so that CRs are coupled to the gas and stream/diffuse along the mean magnetic field with a sufficiently low effective CR diffusion coefficient, which ensures a finite return probability from the halo to the disk.} surrounding the disk with a step size that is equal to their effective mean free path. Hence, they are diffusively transported within a scattering halo that is assumed to be a cylinder of height $2H$ and radius $R$, and are reflected at its boundaries. CRs are assumed to have a non-zero escape probability as they encounter the boundary (modelled by a Poisson process). Spallation occurs when C, N, O, Fe nuclei of the CR population impact on interstellar hydrogen. As a result, the large nuclei are broken up into smaller nuclei, including boron. These secondary CRs continue to diffuse through the ISM. The leaky box model describes this process by a diffusion equation in one dimension (the vertical height of the disk): 
\begin{align}
  \frac{\partial f_0(p,z)}{\partial t} = \frac{\partial}{\partial z}\left[\kappa(p)\frac{\partial f_0(p,z)}{\partial z}\right] + Q(p)\delta_\rmn{D}(z),
\end{align}
where $f_0(p,z)$ is the isotropic part of the CR distribution, $\kappa(p)$ is the diffusion coefficient and the last term is a source term where $\delta_\rmn{D}$ is Dirac's delta distribution. This equation can be readily derived from the quasi-linear CR transport equation \eqref{eq:f0} by constraining the transport to the vertical direction and neglecting advective and streaming transport as well as first- and second-order Fermi acceleration. Assuming a steady state ($\partial f_0/\partial t=0$) and that $\kappa(p)$ is independent of height, we find $\partial f_0/\partial z=\rmn{const}.$ for $z>0$, which implies the solution
\begin{align}
  f_0(p,z) = n_0(p)\left(1-\frac{z}{H}\right),\quad\mbox{for}\quad0<z<H,
\end{align}
and $f_0(p,0)=n_0(p)$ and $f_0(p,H)=0$. Integrating the steady state diffusion equation, we find
\begin{align}
  Q(p)=\int_{z_-}^{z_+}Q(p)\delta_\rmn{D}(z)\dd z = -\kappa \int_{z_-}^{z_+}\frac{\partial^2f_0(p,z)}{\partial z^2}\dd z%
  =\left.-\kappa\frac{\partial f_0(p,z)}{\partial z}\right|_{z_-}^{z_+}=\frac{2n_0(p)\kappa(p)}{H}.
\end{align}
Assuming that $H\neq H(p)$, a momentum-dependent diffusion coefficient $\kappa\propto p^\delta$, and a power-law momentum spectrum for carbon CRs, $f_\rmn{0,C}(p,z)\propto p^\alpha$, we obtain the carbon spectrum in the midplane of the disk:
\begin{align}
  n_\rmn{C}(p)=\frac{Q_\rmn{C}(p)H}{2\kappa(p)}\propto p^{\alpha-\delta}.
\end{align}
Spallation of carbon and oxygen yields boron with $Q_\rmn{B}(p)=\bar{n}_\rmn{H}c\sigma_{\rmn{C\to B}}n_\rmn{C}(p)$ and, after the diffusive transport of boron, we obtain:
\begin{align}
  n_\rmn{B}(p)=\frac{Q_\rmn{B}(p)H}{2\kappa(p)}\propto\frac{n_\rmn{C}(p)}{\kappa(p)}
  \propto\frac{Q_\rmn{C}(p)}{\kappa(p)^2}\propto p^{\alpha-2\delta}.
\end{align}
Hence, the leaky box model explains the momentum dependence of the diffusion coefficient, $\delta$ via
\begin{align}
  \label{eq:B-to-C}
  \frac{n_\rmn{B}(p)}{n_\rmn{C}(p)}\propto\frac{p^{\alpha-2\delta}}{p^{\alpha-\delta}}
  \propto p^{-\delta}\propto \kappa(p)^{-1}.
\end{align}
The measured carbon spectrum $n_\rmn{C}(p)\propto p^{\alpha-\delta}$ in combination with the inferred $\delta$ from B/C enables determining the source-intrinsic spectral index, $\alpha$. Comparing to precision data measured by AMS and its successor AMS-02 shows that this general picture applies (right-hand panel of Fig.~\ref{fig:cosmic_ray_composition}) and that $\delta\approx1/3$ (left-hand panel of Fig.~\ref{fig:cosmic_ray_composition}). 

Adopting a modified leaky box approach to the latest AMS-02 data and studying various ratios of secondary-to-primary CR nuclei, \citet{evoli_ams-02_2020} find that CRs diffuse in a halo of half-height $H\gtrsim5$~kpc. Like all secondary-to-primary ratios (including B/C), the positron fraction $n_{\rmn{e^+}}/(n_{\rmn{e^-}}+n_{\rmn{e^+}})$ decreases at energies $<8$ GeV, which can be understood in the leaky box model according to equation~\eqref{eq:B-to-C}. By contrast, the positron fraction increases for energies $>8$ GeV \citep{adriani_cosmic-ray_2011,ackermann_fermi-lat_2012,aguilar_first_2013}, which might hint towards contributions from individual nearby sources (SNRs or pulsars) emerging above a background. Note that lepton spectra should be suppressed at high lepton energies by synchrotron and IC losses (see Section~\ref{sec:CR_lepton_interactions}), implying a finite horizon\footnote{Here, the horizon is defined to be the largest path a CR of a given energy can travel in a cooling time.} of order 1~kpc at TeV energies. In fact, AMS-02 discovered an excess of electrons over the lower-energy power-law spectrum for energies $\gtrsim42$~GeV \citep{aguilar_towards_2019}, which can be attributed to IC losses of the electrons interacting with the ultraviolet, infrared, and cosmic-microwave-background photon fields, in addition to synchrotron losses in the Galactic magnetic field. The excess over the power-law is caused by the transitioning to the Klein-Nishina regime that renders electron-photon interactions less efficient, which in turn lowers the electron cooling rate and thus causes an excess of electrons over a power-law \citep{evoli_signature_2020}.

Recent CR data show further spectral breaks \citep[e.g.,][]{ahn_discrepant_2010,aguilar_precision_2015,aguilar_observation_2018,adriani_cosmic-ray_2022}, implying that secondary CRs harden more than the primaries, and suggest (i) a change in the energy dependence of the CR diffusion coefficient at energies $\sim 300$~GV and (ii) a similar injection spectrum for all CR nuclei heavier than helium and a significantly different injection spectrum of CR proton and alpha particles \citep{evoli_galactic_2019}. Moreover, \textit{Fermi} $\gamma$-ray data show rich spatial and spectral variation along latitude and longitude (see e.g., Fig.~\ref{fig:Fermi_sky}) which cannot be satisfactorily described in the leaky box model. Moreover, our discussions in Chapter~\ref{sec:physics} and Section~\ref{sec:B-to-C_ratio} make a strong case for anisotropic CR transport along the dominant toroidal magnetic field in the disk, implying that CR streaming and diffusion along the local mean magnetic field as well as CR advection with the background plasma (as a result of SN explosions or in a galactic outflow) cannot be neglected. Hence, those observational and theoretical considerations provide a strong incentive for moving beyond the leaky box model and suggest solving the CR transport equation in a multi-dimensional galactic density and magnetic field either inferred from various Galactic observations or derived from self-consistent MHD simulations.

\subsubsection{(Semi-)analytical cosmic ray propagation models}

In order to study the dominant effects of CR propagation, and the emerging energy spectra and CR composition, various methodologies of increasing levels of complexity have been developed. Assuming simplified one- or two-dimensional geometries for the Galaxy, the CR transport equation~\eqref{eq:f0} can be solved with Green's function or Fourier/Bessel techniques \citep{ginzburg_origin_1964,schlickeiser_cosmic-ray_1989,maurin_cosmic_2001}. This allows one to study the impact of stochastic fluctuations resulting from the spatial and temporal distribution of SNe in combination with spallation processes of nuclei on the energy dependence of the CR diffusion coefficient and show that the leaky box result of equation~\eqref{eq:B-to-C} is mildly violated \citep{blasi_diffusive_2012}. This efficient approach allows one to (analytically) understand the dependence of individual parameters on the solution and enables a quick scan of parameter space via Monte-Carlo methods \citep[e.g., with the USINE code,][]{putze_markov_2009,putze_markov_2010}. The best-fit CR diffusion coefficient at 1~GV is found to be $2\times10^{28}~\rmn{cm}^2~\rmn{s}^{-1}$ and the energy dependence of the diffusion coefficient $\delta$ is robust to changes in the model description of Galactic CR transport that is varied from the leaky box model to (two-dimensional) models of cylindrical symmetry with homogeneous or inhomogeneous source distributions \citep{genolini_theoretical_2015}. 

Coupling the one-dimensional CR transport equation with the equation for waves responsible for resonant CR-wave interactions enables one to infer the energy dependent diffusion coefficient and, hence, the Galactic CR energy spectrum. The model predicts a transition from CR streaming with Alfv\'en waves to a regime of prevalent CR diffusion in self-generated turbulence, which is associated with a  CR spectral hardening at a rigidity $R \simeq 10$~GV \citep{blasi_spectral_2012}. As the growth rate of the CR streaming instability becomes too weak, CRs diffuse in externally generated turbulence (e.g., by expanding SNRs) that cascades from large to small spatial scales, implying a second CR spectral hardening at a rigidity $R \simeq 300$~GV \citep{evoli_origin_2018}. 

The work of \citet{evoli_origin_2018} adopts a more general treatment of the wave spectra, which also provides evidence for a CR scattering halo with a height of a few kpc that naturally arises from CR scattering at resonant Alfv\'en waves. The level of Alfv\'en waves necessary for setting the required CR diffusion coefficient arises by balancing wave generation due to the CR streaming instability and external turbulence with a loss of resonant wave energy due to cascading to smaller scales \citep{evoli_origin_2018}. Alternatively, the height of the CR scattering halo could be set by the following two processes: (i) an increasing Alfv\'en velocity with distance to the disk due to the decreasing gas density, which implies increased adiabatic losses of the turbulence and a decrease of CR scattering at external turbulence in comparison to CR scattering by the self-generated MHD waves \citep{dogiel_formation_2020} or (ii) a dominating wave losses due to non-linear Landau damping \citep{chernyshov_formation_2022}. Independent of the exact mechanism responsible for setting the CR halo size, the resulting size is compatible with the best-fit values of simple parametric approaches to CR diffusion. While those (semi-)analytical approaches are powerful tools to interpret the increasingly more accurate CR data, CR transport is still represented as an effective model and the resulting solutions are difficult to interpret physically. 

\subsubsection{Numerical integration of cosmic ray transport in the static Galaxy} 

To improve the realism of the CR propagation models, three-dimensional models of the Galaxy are used to find a steady-state solution of the CR transport equation with numerical finite difference schemes that employ isotropic CR diffusion using GALPROP \citep{moskalenko_production_1998,strong_propagation_1998} or PICARD \citep{kissmann_picard_2014} codes. However, isotropic CR diffusion models are in conflict with B/C ratios \citep{giacinti_reconciling_2018} and cannot reproduce the observed spectral hardening of the Fermi-LAT $\gamma$-ray data towards the inner Galaxy. Modeling anisotropic diffusion with the DRAGON code \citep{evoli_cosmic_2008,evoli_cosmic-ray_2017} in a cylindrically symmetric model of the Galaxy reproduces the harder $\gamma$-ray slope towards the Galactic center because of parallel diffusive escape of CRs along the assumed poloidal component of the large-scale magnetic field \citep{cerri_signature_2017}. Assuming that local CR transport properties are representative for the entire Galaxy, the predicted $\gamma$-ray flux from GALPROP modeling towards the inner Galaxy falls short of the $\gamma$-ray flux above a few GeV measured by the \textit{Fermi} Large Area Telescope \citep{ackermann_fermi-lat_2012}. Using a phenomenological parameterization with a position dependent rigidity scaling of the CR diffusion coefficient and a radius dependent CR advection velocity, the observed $\gamma$-ray spectrum and longitude profile at 10~GeV can be reproduced with DRAGON modeling \citep{gaggero_gamma-ray_2015}, while still maintaining a good agreement with local CR data.

The various spectral breaks in CR spectra discussed in Section~\ref{sec:leaky_box} pose challenges to standard modeling of CR propagation. Adopting phenomenological scenarios for CR transport with a density dependent rigidity scaling and various breaks in the CR diffusion coefficient enables one to systematically explore their effect on local CR observables \citep{vladimirov_testing_2012}. Using a different phenomenological model with a spatially dependent diffusion coefficient, \citet{guo_spatial-dependent_2016} reproduce the spectral hardening of CR protons, the secondary-to-primary ratios (includinging $\bar{\rmn{p}}/\rmn{p}$) as well as the CR anisotropy from $\sim100$~GeV to $\sim100$~TeV. In fact, a systematic study adopting a Bayesian scan of the parameter space of the main CR propagation parameters used in GALPROP revealed that those are significantly different for low-mass isotopes (p, $\bar{\rmn{p}}$, and He) in comparison to the other light elements Be, B, C, N, and O \citep{johannesson_bayesian_2016}. This may suggest that each set of species probes different ISM conditions and argues that the standard approach of calibrating CR propagation parameters with B/C can lead to biased results. Moreover, the employed modeling shows degeneracies between CR advection and reacceleration, which suggests that successful modeling of CR data could give rise to a biased physics interpretation \citep{korsmeier_galactic_2016}. This makes the case for improving the modeling by adding a more accurate data over a larger range in (CR and $\gamma$-ray) energies as well as chemical compositions. Modeling the Galactic diffuse PeV $\gamma$-ray emission as recently observed by the Tibet AS$\gamma$ and LHAASO (Large High Altitude Air Shower Observatory) collaborations favors CR transport models with spatially dependent diffusion \citep{de_la_torre_luque_galactic_2023}.

Alternatively, stochastic differential equations are employed to solve the CR transport equation in three-dimensional galaxy models by evolving a large number of individual macro particles representing CR sub-populations \citep{webber_monte_1997,farahat_cosmic_2008,kopp_stochastic_2012,merten_crpropa_2017}. The underlying Markov process enables sampling of the statistical properties along the CR path through the Galaxy while it requires to simulate a large number of particles to control the inherent Poisson noise, in particular for modeling spectra at high CR energies, which start to starve of missing statistics. The approach of numerically integrating the CR diffusion equation with a Crank-Nicholson scheme or by means of stochastic differential equations can be adopted to any density and magnetic field distributions (including inferences from observational data) and allows for varying the ab initio unknown source distributions. However, the increasing complexity of CR and $\gamma$-ray data requires the use of phenomenological models of the CR diffusion coefficient, which limits the predictability of the models and may mask other physical effects such as spectral effects from local ISM structures. Moreover, these models do not include any back-reaction of the CR pressure on the structure of the ISM nor do they account for the time evolution of the dynamically evolving ISM and galactic fountains/winds. 

There has been recent evidence that the standard approach of modeling CR propagation described thus far is too simplistic. The isotopic composition of beryllium nuclei and its energy dependence contains important information on the residence time of CRs inside the Galaxy: while beryllium--10 (Be10) has a relatively long half-life of 1.4~Myr that is comparable with the CR residence time, Be9 and Be7 are stable isotopes. Modeling preliminary AMS-02 data on beryllium isotope ratios as a function of energy could imply an increasing average CR age in the Galaxy with energy \citep{lipari_beryllium_2022} or a more structured CR diffusion coefficient with regions of suppressed diffusion in the Galactic disk (possibly associated with $\gamma$-ray sources) embedded within a more extended halo \citep{jacobs_unstable_2023}. Such CR isotope studies provide an independent measurement on the age of CRs that is independent of other secondary-to-primary CR ratios \citep{evoli_ams-02_2020,weinrich_galactic_2020}. This reinforces the need to progress towards more realistic models of CR transport that could be realized by fully dynamical and self-consistent MHD-CR models.

\subsubsection{MHD simulations of active cosmic ray transport in evolving galaxies} 

To move on and cure these short-comings, the most recent developments employ fully dynamical three-dimensional simulations of forming galaxies that account for CR transport and its active feedback on the MHD, while self-consistently growing the galactic magnetic field through a small-scale dynamo \citep{pfrommer_simulating_2022}. Solving for steady state CR spectra of protons as well as primary and secondary electrons \citep[using the CRAYON+ code in combination with the moving mesh MHD code AREPO,][]{werhahn_cosmic_i_2021} produces evolving CR distributions in time and space. Averaging the CR spectra around the solar circle of a galaxy that matches the Milky Way's star formation rate reproduces (i) the observed declining positron fraction for energies $<8$~GeV and (ii) matches the decline of ion spectra below GeV energies due to Coulomb interactions so that electrons become the dominant component of the total particle spectra at these energies, in agreement with Voyager-1 and AMS-02 data \citep{werhahn_cosmic_i_2021}. In addition to Coulomb losses, the exact location of the turn-over in the ion spectra can be caused by the discrete nature of sources, which imply a stochastic realization of the CR flux close to the Sun's position (as inferred by \citealt{phan_stochastic_2021} through numerical integration of the CR transport equation).

Using the FIRE-2 model and performing cosmological simulations of galaxy formation with spectral CR physics also reproduce observations without the need to assume ad-hoc spectral breaks in the diffusion coefficient but instead use CR injection spectra represented by single power-laws in momentum and rigidity-dependent scattering coefficients \citep[][using the meshless finite mass code GIZMO]{hopkins_first_2022}. In these simulations, the height of the scattering halo is of order 10 kpc and adiabatic processes associated with the dominant CR advection in turbulent and fountain flows dominate over diffusion and turbulent reacceleration. Notably, CR spectra vary substantially across the galaxy and spectral features can often be attributed to local ISM structures rather than CR transport physics. While these MHD-CR simulations have demonstrated the enormous potential encoded within them to separate local source from CR transport effects and to study CR propagation and feedback self-consistently, they cannot be used to directly probe the non-thermal emission maps and local CR observables because they do not exactly resemble the Milky Way structure. More work is needed to unify the MHD-CR simulation approach with a concrete realization that resembles the Milky Way. 

\subsection{Cosmic ray driven/aided outflows in the Milky Way}\label{sec:CR_driven_outflows}
\subsubsection{Observational evidence for Galactic center outflow}
We now review observational evidence for the outflow from the Milky Way and argue that CRs may play an important role in driving the wind. Various non-thermal processes associated with the interactions of CRs with the ISM and CGM of the Galaxy provide important constraints on the properties and origin of the outflow. We discuss the evidence for the  outflow by starting with the largest structures and then progressively zooming in towards the smallest scales.\\
\begin{figure}
  \begin{center}
    \includegraphics[width=\textwidth]{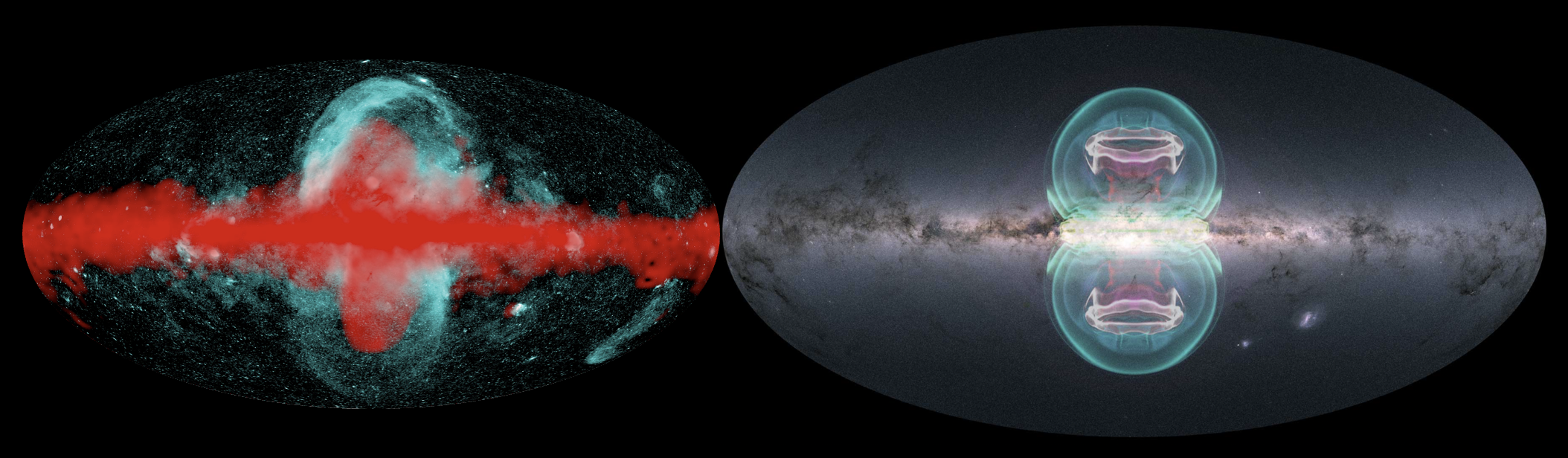}
  \end{center}
  \caption{\textit{Left:} An overlay of the diffuse $\gamma$-ray emission including the \textit{Fermi Bubbles} (red) and the diffuse X-ray emission including the eROSITA  bubbles (cyan). The image is highly suggestive of a physical connection between these two Galactic feedback structures. Figure from \citet{predehl_detection_2020}; reproduced with permission from Nature. 
  \textit{Right:} Three-dimensional rendering of gas density (cyan) and CR energy density (red) from three-dimensional simulations of an AGN outburst in the Milky Way based on the model of \citet{yang_fermi_2022}. The image is superimposed on Gaia image of Milky Way; credit: GAIA/DPAC/NASA/ESA.}
   \label{fig:predehl}
\end{figure}
\indent
Extending up to $|b|\sim 80^{\rm o}$ above and below the Galactic midplane are very large bipolar X-ray bubbles detected with eROSITA \citep[][see cyan structures in the left panel of Fig.~\ref{fig:predehl}]{predehl_detection_2020}. Related X-ray structures were previously detected with ROSAT \citep{snowden_rosat_1997} and suggested gas compression due to shocks surrounding  underdense bubbles. 
Embedded in the eROSITA bubbles are large bipolar \textit{Fermi Bubbles} observed in the $\gamma$-ray band that extend up to $|b|\sim 60^{\rm o}$ away from the Galactic equator (see red structures near the center of the left panel in Fig.~\ref{fig:predehl}). Their morphology is highly suggestive of a physical relationship between the above large-scale X-ray and $\gamma$-ray features. The $\gamma$-ray bubbles are characterized by almost flat intensity distribution and sharp edges \citep{su_giant_2010}. The $\gamma$-ray spectral slope, normalization, and high energy cutoff (at $\sim 100$ GeV) are also approximately independent of galactic latitude \citep{ackermann_spectrum_2014, narayanan_latitude-dependent_2017}. 
The high-energy spectral cutoff is also consistent with HAWC (High-Altitude Water Cherenkov) upper limits in the 1 to 100 TeV energy range \citep{abeysekara_search_2017}. 
At lower frequencies, a prominent feature called Loop I (extending up to $b\sim 70^{\rm o}$) is present near the east (left) side of the \textit{Fermi Bubble}. At latitudes $b\lesssim 30^{\rm o}$, this feature is approximately spatially coincident with the \textit{North Polar Spur} visible in X-rays. The physical origin, or even the distance to these structures, is unknown \citep{lallement_north_2022} with hypotheses ranging from a local SNR \citep{das_constraining_2020} to an outflow from the Galactic center \citep{kataoka_x-ray_2018,larocca_analysis_2020}.
However, S-PASS (S-band Polarization All Sky Survey; 2.3 GHz) reveals large-scale and highly polarized lobes extending up to $|b|\sim 50^{\rm o}$ above the disk midplane that surround the bubbles and suggest a high degree of magnetic field ordering likely caused by an outflow from the Galactic center \citep{carretti_giant_2013}.\\
\indent
Gamma-ray emission from the bubbles is also spatially correlated at lower latitudes ($|b|\lesssim 35^{\rm o}$) with the microwave \textit{Haze} detected by the WMAP (Wilkinson Microwave Anisotropy Probe) and \textit{Planck} missions \citep{finkbeiner_microwave_2004, planck_collaboration_planck_ix_2013}. Unlike the $\gamma$-ray bubbles, the microwave emission decays with the distance from the Galactic center. 
\begin{figure}
  \begin{center}
     \includegraphics[width=\textwidth]{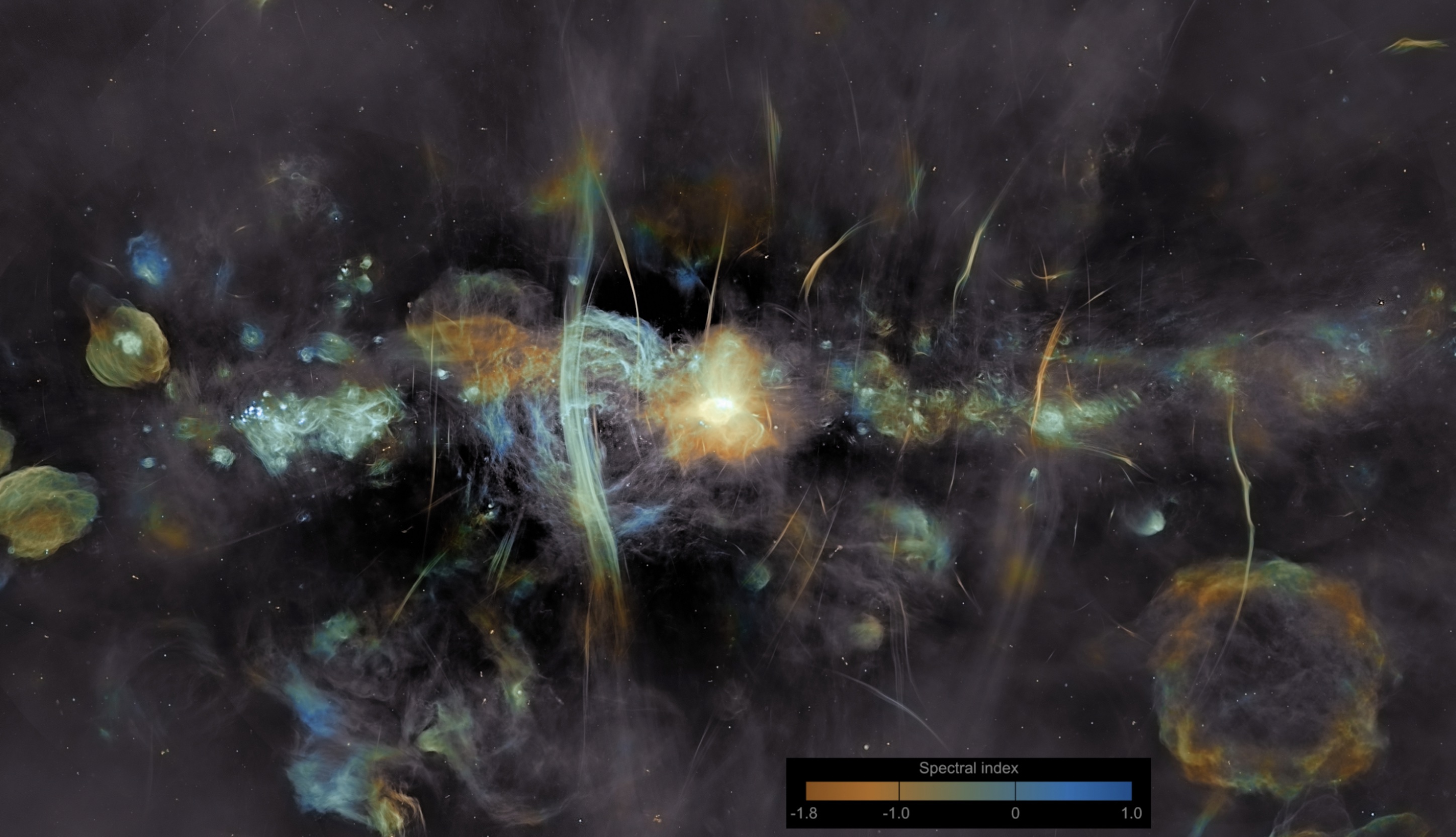}
  \end{center}
  \caption{MeerKAT radio spectral index map of the Galactic Center ($\sim3^{\circ}\times2^{\circ}$ or $\sim$~420~pc $\times$ 260~pc). The image reveals a number of features such as multiple SNRs, narrow radio filaments parpendicular to the disk, and strong emission from Sgr A$^{*}$ (at the center of the image). Figure from \citet{heywood_128_2022}; reproduced with permission from ApJ.}
  \label{fig:heywood}
\end{figure}
Closer to the disk midplane ($b\lesssim 20^{\rm o}$; for heights $\lesssim 3$ kpc above the disk midplane at the distance to the Galactic center), there is a clear spatial correlation between bipolar ROSAT 1.5 keV emission and the edges of the $\gamma$-ray bubbles \citep{bland-hawthorn_large-scale_2003,bland-hawthorn_large-scale_2019} that is indicative of an outflow originating at the center of the Galaxy. 
Similarly, on sub-kpc distances from the Galactic center there is evidence from \textit{XMM Newton} for ``chimneys'' below and above the disk that connect the Galactic center to the \textit{Fermi Bubbles} \citep{ponti_x-ray_2019}, and from MeerKAT for bipolar bubbles created by an energetic event a few million years ago \citep{heywood_inflation_2019} that is likely associated with a jet \citep{cecil_tracing_2021}.\\
\indent
The above outbursts may be responsible for boosting the CR energy density in the Galactic center by up to three orders of magnitude compared to the values typically measured in the rest of the Galactic disk \citep[][based in part on H$_{3}^{+}$ absorption line measuremens; see also Section~\ref{CR_ionization}]{oka_central_2019} and for creating very strong CR pressure gradients that could help drive an outflow \citep{yusef-zadeh_cosmic-ray-driven_2019} and power the narrow magnetized radio filaments that are approximately orthogonal to the Galactic disk \citep[see Fig.~\ref{fig:heywood}; this figure also reveals strong emission from the Galactic center,][]{heywood_128_2022}. 
The filaments could arise from the interactions of stellar winds with the galactic outflow as the magnetic field wraps around the stellar outflow and is compressed in the downstream region with respect to the star, and when the CRs from the galactic wind remain confined in the amplified magnetic field tail \citep{yusef-zadeh_cosmic-ray-driven_2019}. This picture is consistent with spectral steepening of the filament emission as a function of the distance from the disk as would be expected in the CR-driven outflow scenario \citep{yusef-zadeh_statistical_2022}. Alternatively, the non-thermal filaments could also be due sources of CRs (e.g., massive stars or pulsars) that cross intermittent magnetic field lines thus allowing CRs to stream along the magnetic fields, which could also explain why the filaments of so-called radio harps are ordered according to their length \citep[see Fig.~\ref{fig:harps},][]{thomas_probing_2020}.\\
\indent
Apart from the morphological evidence for the outflow, independent constraints come from gas kinematics. Shifts in UV absorption lines in the spectra of stars and quasars at high latitudes near the top of the \textit{Fermi Bubble} demonstrate that the gas is moving at speeds on the order of $10^{3}$~km s$^{-1}$ \citep{bordoloi_mapping_2017}. Closer to the base of the \textit{Fermi Bubbles} ($|b|\sim 10^{\rm o}$), H$\alpha$ emission lines reveals velocities on the order of $200$ km s$^{-1}$ \citep{krishnarao_discovery_2020}, and on somewhat smaller scales there is evidence for a bi-conical clumpy neutral hydrogen outflow that suggests that the clouds are accelerating as they move away from the center \citep{di_teodoro_blowing_2018}.\\
\indent
The models aiming to explain the above outflow phenomena fall into two broad categories of starburst and AGN driven outflow models, which themselves can belong to either hadronic or leptonic category depending on which process dominates the emission. Below we briefly discuss two extreme models that likely bracket the range of possibilities -- a hadronic star formation driven galactic wind model and a leptonic AGN jet model -- before providing examples of other models that combine aspects of both scenarios.

\subsubsection{Hadronic star formation--driven wind model}
Early models of the Galactic center outflow were motivated by the desire to explain the mismatch between the relatively centrally concentrated distribution of CR sources (SN, superbubbles) and a wider distribution of the diffuse $\gamma$-ray emission. Using a one-dimensional magnetic flux tube model (see Section~\ref{1D_models}), \citet{breitschwerdt_gradient_2002} demonstrated that this discrepancy can be explained by considering CR transport due to diffusion, and advection possibly aided by CRs.
This model was extended to include wind driving due to both thermal and CR pressures, which resulted in a better agreement between the model and the observed Galactic soft X-ray emission compared to earlier models that assumed a static gas distribution \citep{everett_milky_2008}, and further modified to jointly fit X-ray and synchrotron emission \citep{everett_synchrotron_2010}.\\
\indent
The discovery of the \textit{Fermi Bubbles} in $\gamma$ rays \citep{su_giant_2010} put significant new constraints on the outflow models. One of the key constraints comes from the following timing argument. 
Gamma-ray emission can arise from collisions of CR protons with thermal gas, which produce neutral pions that decay generating $\gamma$-ray photons (see equation~\ref{eq:hadronic} and Section~\ref{sec:CR_ion_interactions}). Winds driven by high-surface-density star formation activity in the central 200 pc of the Galaxy, and the SN explosions associated with it, can inflate the bubbles and explain their $\gamma$-ray emission via this process as long as the hadronic loss timescale is smaller than the CR injection and confinement timescales. Since the hadronic loss timescale is very long (on the order of $10^{10}$~yr for typical conditions inside the \textit{Fermi Bubbles}; see equation~\ref{eq:tau_pp}), this imposes rather low power requirement on the source of the wind which translates to about $\sim0.1$~M$_{\odot}$~yr$^{-1}$ sustained over this very long period \cite{crocker_non-thermal_2012}.\\
\indent
This \textit{hadronic wind model} for the origin of the \textit{Fermi Bubbles} can explain (i) the hard $\gamma$-ray spectrum \citep{crocker_fermi_2011}, (ii) the departure from the FIR-radio correlation and the suppression of $\gamma$-ray emission given the current SFR \citep[][see Section~\ref{Non-thermal emission from galaxies} for a general discussion of the relationship between radio and $\gamma$-ray emission and SFR]{crocker_-rays_2011}, (iii) spatially extended 511 keV electron-positron annihilation line signal due to advection of secondary positrons in the wind \citep{crocker_-rays_2011}, and (iv) flat $\gamma$-ray surface brightness \citep{giacinti_galactic_2018}.\\
\indent
Two main challenges facing the hadronic wind model are (i) the requirement for CR injection to occur over very long timescale of $\gtrsim8$~Gyr, which are significantly longer than typical stellar evolutionary timescales and which render the \textit{Fermi Bubbles'} survival over such a long timescale unlikely in the highly turbulent CGM and when exposed to past galaxy mergers of the progenitor system of the Milky Way and its satellite galaxies (the injection timescale can however be significantly lowered in the presence of dense clumps dominating the pion production within the bubbles, \citealt{crocker_unified_2015}) and (ii) underpredicting the microwave Haze emission, addressing which necessitates introducing an additional population of primary CR leptons as opposed to relying on the secondary ones generated via hadronic processes (but see \citet{crocker_unified_2015}, who suggest that re-acceleration of CR electrons a reverse shock could explain the spatial extent and level of the microwave Haze emission).

\subsubsection{Leptonic AGN jet model}
On the opposite spectrum of possibilities suggested to explain the \textit{Fermi Bubbles} is the leptonic AGN jet model. 
In this case, the $\gamma$-ray emission mechanism is the IC scattering of the CMB and optical starlight photons. Leptons that produce this emission are also subject to synchrotron cooling. An estimate for the shortest lepton cooling time can be obtained from the observed high-energy $\gamma$-ray cutoff near 100 GeV in the bubble spectrum \citep{ackermann_spectrum_2014} and assuming typical magnetic and radiation energy densities of $\sim 1$~eV cm$^{-3}$. This yields a leptonic cooling timescale $\sim$ a few Myr (see, e.g., \citealt{yang_spatially_2017}), which is much shorter than the hadronic cooling timescale discussed above. Given this very short cooling timescale, the bubble inflation timescale must be comparably short, which imposes high power demand on the source of the outflow. Therefore, the models that rely on the leptonic emission often postulate that the bubbles are rapidly inflated by a past episode of AGN jet activity in the Galactic center.\\
\indent
The required timescales are in line with some observational estimates of the Galactic center variability. Specifically, based on the modeling of OVII and OVIII emission line strengths and assuming continuous energy injection, \citet{miller_interaction_2016} argue that the outburst started $\sim4$ Myr ago. Similarly, observations of highly ionized gas clouds within bipolar ionization cones associated with Sgr A$^{*}$ suggest that the ionization must have been triggered by a powerful Seyfert-like flare with luminosity 10\% to 100\% of the Eddington rate some $\sim2$ to 8 Myr ago \citep{bland-hawthorn_large-scale_2019}.\\
\indent
The \textit{leptonic AGN jet model} is successful in explaining several aspects of the \textit{Fermi Bubbles} (listed below with a description of the modeling approach in parentheses) such as: (i) overall morphology including the sharpness of the bubble edges (two-dimensional hydrodynamical simulations with CRs; \citealt{guo_fermi_2012}), (ii) flat $\gamma$-ray surface brightness, sharp bubble edges, and X-ray emission matching ROSAT data (two-dimensional hydrodynamical simulations with viscosity and CRs (\citealt{guo_fermi_2012-1}), and three-dimensional MHD simulations with anisotropic CR diffusion;  \citealt{yang_fermi_2012}), (iii) broadband spectra (radio, microwave, $\gamma$ rays, including the high-energy $\gamma$-ray cutoff), radio polarization, latitude-independent distribution of the $\gamma$-ray spectrum (three-dimensional hydrodynamical simulations with on-the-fly modeling of CR energy spectra, all based on a single population of leptons; \citealt{yang_spatially_2017}), (iv) morphology of eROSITA bubbles (associated with a forward shock) and \textit{Fermi Bubbles} (associated with a contact discontinuity), Haze emission, and broadband spectra again simultaneously explained by a single population of CR electrons associated with a single AGN jet outburst (three-dimensional hydrodynamical simulations with on-the-fly modeling of CR spectra; \citealt{yang_fermi_2022}; see also right panel in Fig.~\ref{fig:predehl}).\\
\indent
One of the main challenges facing the leptonic AGN jet model is a possible inconsistency between the predicted high velocities (up to $\sim$~2000 km s$^{-1}$) of the hot gas, caused by the significant power of the AGN outburst, and the velocities inferred from X-ray \citep[$\sim$~200 to 300 km s$^{-1}$;][]{kataoka_suzaku_2013,fang_high_2014}
and UV spectroscopy \citep[$\sim$~900 to 1300 km s$^{-1}$;][]{fox_probing_2015,bordoloi_mapping_2017}. The discrepancy between the model and X-ray measurements could be at least partially attributed to the possibilities that the very hot gas with temperatures in excess of $10^{8}$~K is not easily detectable in the X-ray band of the instruments used so far, and that shock Mach numbers inferred from spectral fitting may underestimate the true Mach number because ions rather than electrons may be preferentially heated at strong shocks. The discrepancy between the model predictions and UV data may be accounted for by noting that cold UV-emitting clouds accelerated by the hot wind may be slower due to the inherent inefficiency of the cloud acceleration process. Furthermore, metallicity measurements of the clouds suggest that both super-solar and sub-solar metallicities are detected in the clouds. This suggests that while some clouds may be accelerated by the outflow, the more pristine clouds may be infalling from the CGM thus further complicating the interpretation of the data \citep{ashley_diverse_2022}.\\
\indent 
Several other model variants and combinations have been proposed in order to address some of the deficiencies of the hadronic starburst-driven wind and leptonic AGN jet models. These models often fall into the ``in-between'' categories in terms of the outflow speed and employ hybrid leptonic/hadronic emission mechanism. Some notable examples include:
(i) a starburst wind model of relatively short $\sim$~30 Myr duration including primary leptonic emission \citep{sarkar_multiwavelength_2015}; 
(ii) an AGN wind model (motivated by detailed MHD simulations of hot accretion flows) that includes hadronic processes and primary lepton injection \citep{mou_accretion_2015};
(iii) in situ second-order Fermi acceleration models (that can operate in either starburst or AGN cases), where CRs are accelerated throughout the bubble volume via plasma wave turbulence \citep{mertsch_fermi_2011};
(iv) models in which the \textit{Fermi Bubbles} are attributed to a forward shock and the eROSITA bubbles are caused by an earlier energetic outburst
(AGN jet and collimated outflow model by \citealt{zhang_simulating_2020,mondal_fermi_2022}; hybrid hadronic-leptonic model with in situ first-order CR acceleration at the forward shock, \citealt{fujita_hadronic-leptonic_2014}).
Overall, the problem of the origin of the eROSITA/\textit{Fermi Bubbles} remains unsolved and is an active field of study. New $\gamma$-ray data from the \textit{Fermi} telescope and potential constraints on the kinematics of the hot phase from upcoming X-ray missions such as XRISM and Athena may further help to distinguish between the various theoretical possibilities discussed above.

\subsection{Non-thermal emission from galaxies}\label{Non-thermal emission from galaxies}

\begin{figure}
  \begin{center}
    \includegraphics[width=\textwidth]{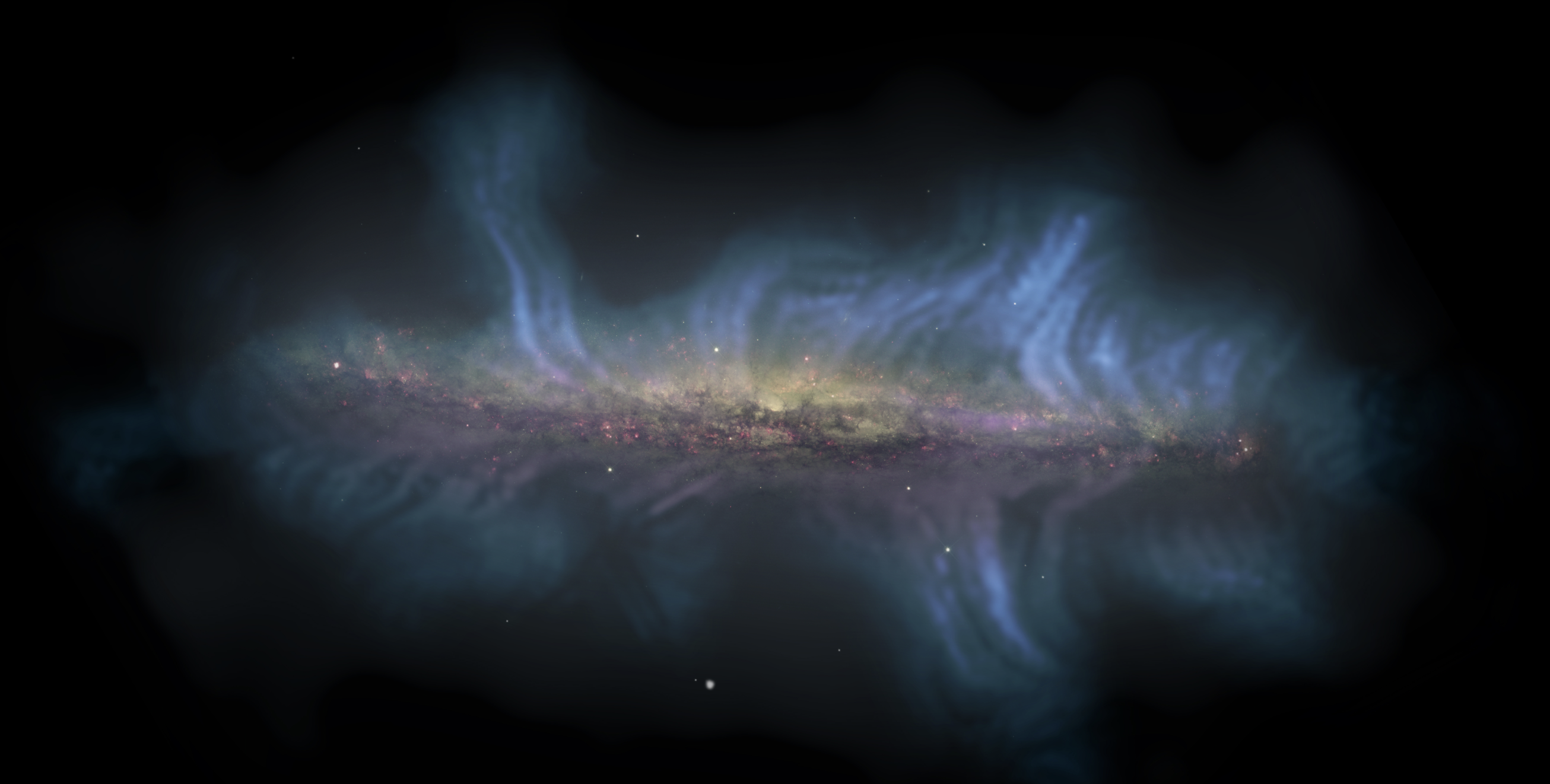}
  \end{center}
  \caption{A composite optical--radio image of the edge-on disk galaxy NGC 5775. Hot gas in the disk (magenta) approximately marks the location of CR injection sites. CR accelerated near the disk are transported away and form a radio continuum halo (blue). The linear structures in the halo follow the orientation of  magnetic field lines (obtained from radio polarization measurements). Composite image by English, Stein, Miskolczi (CHANG-ES collaboration), using NRAO VLA and HST WFC3 data.}
  \label{fig:irwin}
\end{figure}

The most direct way to constrain feedback in external galaxies is the observations of non-thermal radio synchrotron and $\gamma$-ray emission. In fact, global correlations of the \textit{non-thermal radio emission} with star formation indicators enable us to probe how much CR electron energy is lost to radiation and how much energy is still available for driving plasma waves, which suffer collisionless damping, thus transferring CR momentum and energy to the background plasma. The radio-emitting leptonic component contributes a negligible pressure support and as such cannot provide much energetic feedback. However, it can inform us about CR transport processes (which are identical to those of CR ions in the relativistic regime, barring helicity effects) and as such enable insights into the feedback provided by the pressure-carrying CR ions. The spatially resolved radio synchrotron maps (e.g., Fig.~\ref{fig:irwin}), spectra, and spatial correlations to star-formation estimators probe CR source regions, cooling times, and specifics of CR transport \citep{peretti_cosmic_2019,schmidt_chang-es_2019}.
Complementarily, synchrotron polarization in combination with Faraday rotation measurements enables one to decipher the three-dimensional topology of the magnetic field and potentially reveals the Parker instability or the presence of galactic outflows. Observations of polarized radio emission in edge-on galaxies show that poloidal magnetic fields connect the disk to the galactic halo \citep{tullmann_thermal_2000,heesen_cosmic_2009,damas-segovia_chang-es_2016,miskolczi_chang-es_2019,irwin_chang-es_2019,stein_chang-es_2019,stein_chang-es_2020,krause_chang-es_2020}, providing direct evidence for mechanisms that open up the dominant toroidal disk fields, such as galactic outflows, and enable CR electrons and ions to escape into the CGM. Another observational window into the magnetic field topology are the observations of the polarized thermal dust emission. Data from SOFIA/HAWC+ in combination with a potential field extrapolation show a dominant large-scale ordered poloidal field associated with the galactic superwind of two nearby starburst galaxies M82 and NGC~253 \citep{jones_sofia_2019,lopez-rodriguez_strength_2021}.\\
\indent
Unfortunately, synchrotron emission does not provide a clean probe of the CR electron population but is degenerate with the magnetic field strength. Hence, this complicates the task and requires us to also understand the amplification and saturation of magnetic dynamos that grow tiny magnetic seeds to observable strengths. On the other hand, \textit{galactic $\gamma$-ray emission} does not directly depend on the magnetic field strength and could shed light on the energetics of CRs. However, the $\gamma$-ray emission could be either produced by CR leptons via the IC process or by CR ions through hadronic interactions with the ISM. In the Milky Way, GeV $\gamma$ rays are mainly hadronic with a possible transition to leptonic IC emission above several tens of GeV \citep[with the exact transition energy depending on the region of interest;][]{ackermann_fermi-lat_2012}. As will be detailed below, the $\gamma$-ray emission in other nearby star-forming galaxies also seems to be dominated by (hadronic) pion decay. Hence, the FIR--$\gamma$-ray correlation (at GeV energies) probes CR ion calorimetry and could be used to calibrate CR feedback.

\subsubsection{Radio emission}
\label{sec:radio emission}

\paragraph{Correlation between radio emission and star formation.}  
There is a tight relation between the FIR emission and radio synchrotron luminosity in star-forming galaxies -- the FIR--radio correlation \citep{van_der_kruit_observations_1971,van_der_kruit_high-resolution_1973,de_jong_radio_1985,helou_thermal_1985,condon_radio_1992,yun_radio_2001,bell_estimating_2003,molnar_non-linear_2021,matthews_cosmic_2021}, which is slightly super-linear. It is believed to trace the physics of galactic star formation and extends over five decades in total luminosity. In addition to this global relation, there also exists a local FIR--radio correlation on scales down to a few 100~pc inside of galaxies \citep[as demonstrated in small galaxy samples, see][]{beck_far-infrared_1988,bicay_60_1990,murphy_connecting_2008,heesen_radio_2014,heesen_calibrating_2019}.\\
\indent
Young stellar populations emit copious amounts of ultra-violet photons that are absorbed in the dust-rich star-forming environment, where they are reprocessed and subsequently re-emitted in form of FIR photons. If the dust column seen by most ultra-violet photons is optically thick, the FIR emission scales with the star formation rate. Stars more massive than about eight solar masses explode as core-collapse SNe and drive powerful shocks into the ambient ISM, which can accelerate primary CR electrons and ions through the process of diffusive shock acceleration during the different phases of a SNR, starting with the remnant's free expansion, followed by the energy-conserving Sedov--Taylor phase, to the momentum-conserving snowplow phase (typically lasting for $\sim10^5$ years; see Section~\ref{sec:acceleration}). These electrons emit \textit{primary} radio synchrotron radiation. Once the newly generated CR ions escape from the source regions, they can hadronically interact with nuclei of the ISM and produce pions. Those decay into secondary electrons and positrons (hereafter referred to as secondary electrons; see equation~\ref{eq:hadronic}), which emit \textit{secondary} synchrotron radiation. On scales much larger than individual star-forming regions ($\gtrsim100$~pc) and averaged over a few lifetimes of massive stars (several million years), the primary and secondary radio emission is physically related to star formation and the associated FIR emission.

Clearly, if CR electrons radiate all their energy in the form of synchrotron and IC emission before they escape into the halo, galaxies can be considered ideal electron ``calorimeters,'' which yields a linear FIR--radio correlation \citep{voelk_correlation_1989,lisenfeld_quantitative_1996}. In the steady-state cooling regime, the spectral indices of CR electrons are steeper by unity in comparison to their injection spectral indices, $f_\rmn{e}(E)\propto{E}^{-\alpha_{\rmn{inj}}-1}$, where $\alpha_{\rmn{inj}}\approx2.3$ (see Section~\ref{sec:equilibirum} for a derivation). This implies a radio synchrotron spectrum of $I_\nu\propto\nu^{-\alpha_\nu}$, where $\alpha_\nu=\alpha_{\rmn{inj}}/2\approx1.15$. Surprisingly, this is in contradiction to the observed radio continuum spectra in starburst galaxies, which show hard spectral indices of $\alpha_\nu\approx0.5$--$0.8$. This inconsistency put the calorimeter theory of galactic radio emission under severe pressure. \\
\indent
Models that treat the galaxy as a single zone suggest that while the CR electron population is calorimetric, strong relativistic bremsstrahlung and ionization/Coulomb losses in dense starburst galaxies could flatten CR electron spectra because of their weaker energy dependence in comparison to synchrotron and IC losses \citep{thompson_magnetic_2006,lacki_physics_i_2010,lacki_physics_ii_2010,basu_synchrotron_2015}. However, in starburst galaxies, these emission processes would radiate a fraction of the CR electron energy into the X-ray and $\gamma$-ray regime rather than in the radio band, thus significantly flattening the FIR--radio correlation so that it would become a sub-linear relation\footnote{This relation is traditionally plotted such that the radio and FIR luminosities are shown on the vertical and horizontal axes, respectively. Consequently, any reduction in the radio emission leads to the flattening of the relation.}. The linear relationship can be restored by requiring that different CR electron populations dominantly contribute to the radio emission at different star-formation rates: while the primary synchrotron emission would dominate at low values of star-formation rates, in starburst systems, the secondary radio emission would significantly contribute to the total radio emission in starburst galaxies \citep{lacki_physics_i_2010}, thus requiring a ``conspiracy'' of the secondary emission component to take over at exactly that star-formation rate where bremsstrahlung and ionization losses reduce the primary synchrotron emission. These one-zone models are excellent tools for transparently linking the various multi-messenger signals in individual galaxies from the radio to $\gamma$ rays to neutrinos \citep{domingo-santamaria_high_2005,persic_very-high-energy_2008,de_cea_del_pozo_multimessenger_2009,lacki_physics_i_2010,lacki_physics_ii_2010,yoast-hull_winds_2013,yoast-hull_cosmic_2015,yoast-hull_equipartition_2016,eichmann_radio-gamma_2016} as well as for exploring the global and local FIR--radio correlation \citep{lacki_physics_i_2010,lacki_physics_ii_2010,vollmer_deciphering_2022}. Very dense starburst systems such as Arp~220 are optically thick to FIR photons so that there are more CRs produced given the emitted FIR flux. This increased CR injection rate is required to simultaneously fit the radio synchrotron and $\gamma$-ray spectra \citep{yoast-hull_breaking_2019}. However, all these models require fitting the density, magnetic and radiation energy densities, and the CR electron and proton abundances as a function of particle energy as free input parameters (see Table~1 of \citealt{lacki_physics_i_2010}, for an extensive list of required parameters), which limits the predictive power of these models. A first step toward a more complex modeling are two-dimensional (axisymmetric) models, which integrate the CR distributions using the CR propagation codes \textsc{CRPropa} \citep{dorner_cosmic-ray_2023} or GALPROP \citep{martin_interstellar_2014,buckman_cosmic_2020}. Adopting parameterized distributions of CR sources, gas density, and the magnetic field, they confirm the results of one-zone studies that (i) strong bremsstrahlung losses significantly flatten the synchrotron spectrum and (ii) predict that secondary electrons dominate the total synchrotron emission at high star-formation rates, as required by the linearity of the FIR--radio correlation. 

\begin{figure}
  \begin{center}
    \includegraphics[width=\textwidth]{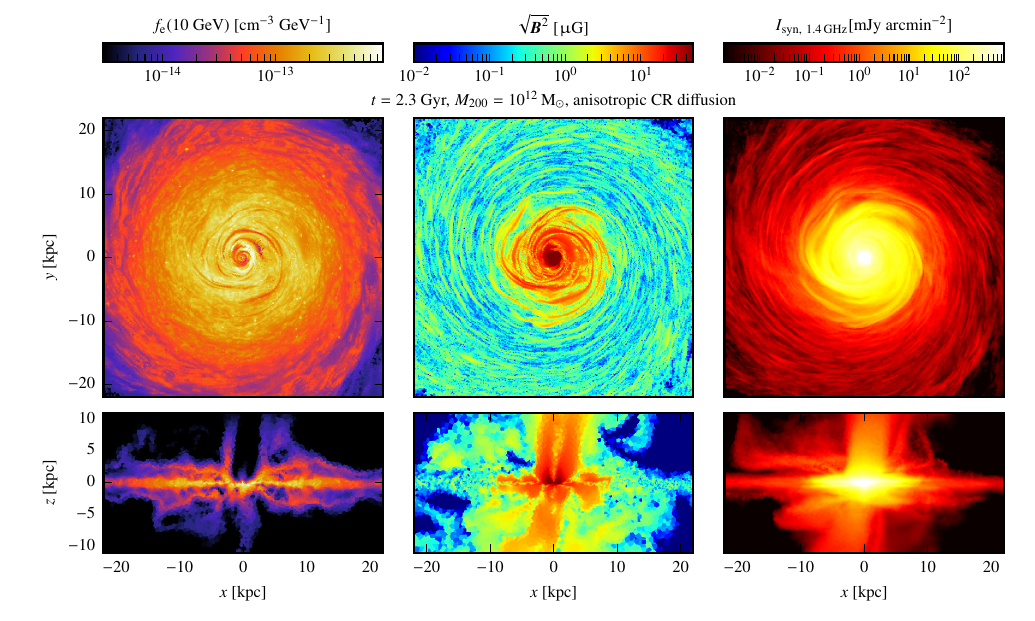}
  \end{center}
  \caption{Modeling the radio synchrotron emission in a three-dimensional MHD-CR simulation of a forming Milky Way-sized galaxy. The figure displays slices of various quantities in a face-on orientation (upper panels) and an edge-on perspective (lower panels). Shown are the CR electron distribution at 10~GeV (left), the magnetic field strength (middle), and total radio synchrotron intensity maps at 1.4~GHz (right). Figure from \citet{werhahn_cosmic_iii_2021}; reproduced with permission from MNRAS.}
  \label{fig:werhahn_maps}
\end{figure}
\begin{figure}
  \begin{center}
    \raisebox{0.5em}{
    \includegraphics[width=0.43\textwidth]{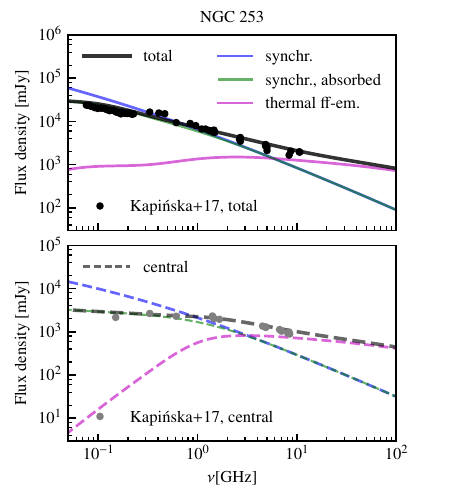}}\hfill
    \includegraphics[width=0.55\textwidth]{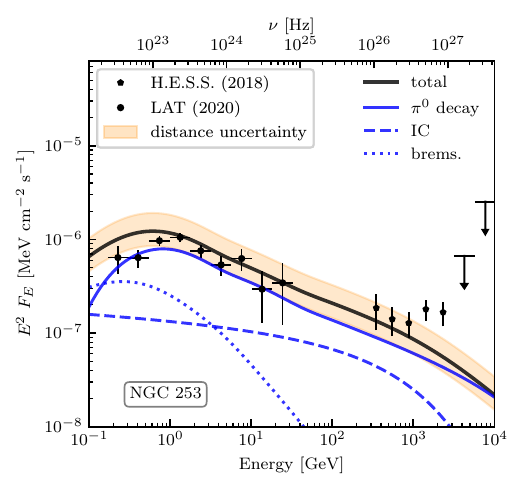}
  \end{center}
  \caption{\textit{Left:} Simulated radio spectrum of the entire galaxy (top) and the central region (bottom) based on CR+MHD simulations. Data for NGC~253 \citep{kapinska_spectral_2017} are well matched by a model that includes emission from primary and secondary electrons and synchrotron and free-free emission and absorption. \textit{Right:} A simulated $\gamma$-ray spectrum for the same model that includes pion decay, IC and non-thermal bremsstrahlung and matches \textit{Fermi} and H.E.S.S. data \citep{hess_collaboration_starburst_2018,ajello__2020}. Figures from \citet{werhahn_cosmic_iii_2021,werhahn_cosmic_ii_2021}; reproduced with permission from MNRAS.}
  \label{fig:werhahn_spectrum}
\end{figure}
\begin{figure}
  \begin{center}
    \includegraphics[width=0.5175\textwidth]{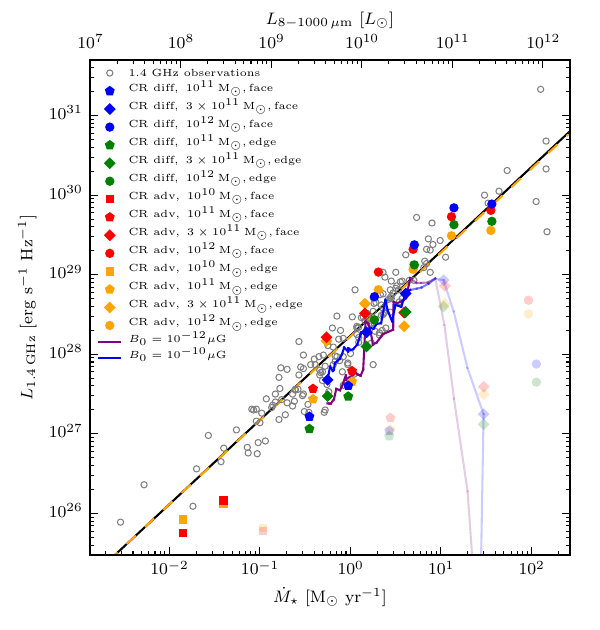}\hfill
    \raisebox{.5em}{    \includegraphics[width=0.455\textwidth]{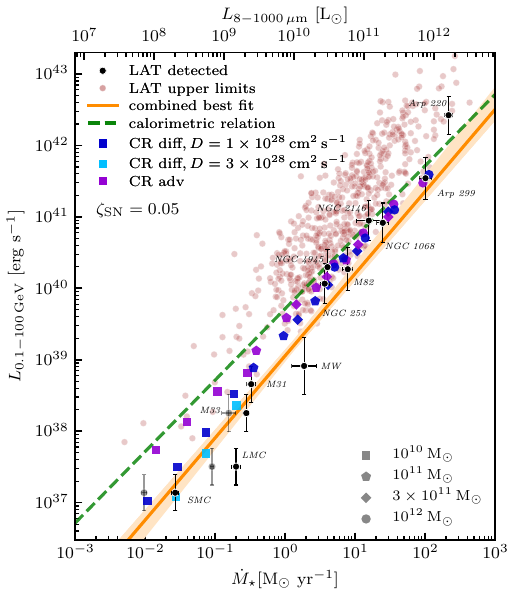}}
  \end{center}
  \caption{\textit{Left:} FIR--radio correlation of star-forming galaxies. The best-fit relation of the observed data \citep[open circles,][]{bell_estimating_2003} is slightly super-linear, $L_{1.4\,\rmn{GHz}}\propto L_{\rmn{FIR}}^{1.055}$ (dashed orange) and agrees with the analytical prediction \citep[solid black,][]{pfrommer_simulating_2022}. The different colored symbols show face-on and edge-on projections of the radio emission of MHD-CR simulations for different halos masses. Compared are models that only follow advective CR transport (`CR adv', red and orange) and those which additionally account for anisotropic CR diffusion (`CR diff', blue and green). Light symbols delineate the exponential growth phase of the galactic dynamo while fully-colored symbols show the emission in a saturated dynamo stage. Those agree with the observed relation of local galaxies (two evolutionary tracks are shown for different initial magnetic seed fields). \textit{Right:} FIR--$\gamma$-ray relation for the same simulations (colored symbols), employing two different values of the CR diffusion coefficient. The $\gamma$-ray emission is dominated by pion decay and only has a small contribution due to IC emission. While starburst galaxies are close to the calorimetric relation (dashed green line), galaxies with low star-formation rates fall short of it. The simulations match the mean relation (orange) of observed galaxies (black points) and obey the constraints imposed by $\gamma$-ray non-detections \citep[light red points,][]{ajello__2020}. Figures from \citet{pfrommer_simulating_2022,werhahn_cosmic_ii_2021}; reproduced with permission from MNRAS.}
  \label{fig:FRC}
\end{figure}

\paragraph{MHD models of galactic radio emission.} The resolved radio synchrotron maps (in total intensity and polarization, see Fig.~\ref{fig:irwin}) show a rich morphology while the FIR--radio correlation has outliers, such as colliding pairs of galaxies (e.g., the Taffy galaxies) that could signal the presence of an additional, freshly accelerated electron population at galaxy merger shocks \citep{lisenfeld_shock_2010}. These considerations make a strong case for moving towards fully three-dimensional MHD models of galaxies that also follow CR spectra in space and time. These simulations allow us to investigate the relation between galactic magnetic dynamo processes and radio emissions \citep{pfrommer_simulating_2022}, and directly examine widely-held assumptions, including the energy equipartition of magnetic fields, CRs, and turbulence. Performing MHD simulations of an isolated disk galaxy and assuming a uniform CR electron-to-thermal energy ratio throughout the galaxy (which neglects important CR electron cooling processes and their spectral evolution) reproduces general properties of edge-on radio halos \citep{vijayan_radio_2020}. Three-dimensional MHD simulations of forming galaxies in a dark matter halo that follow one-moment CR hydrodynamics and model (primary and secondary) CR electrons and ions in a steady-state approximation enables linking CR spectra in the Milky Way to radio and $\gamma$-ray correlations with the FIR emission. In addition, these simulations produce non-thermal radio and $\gamma$-ray maps and spectra of individual galaxies \citep{werhahn_cosmic_i_2021,werhahn_cosmic_ii_2021,werhahn_cosmic_iii_2021,pfrommer_simulating_2022}, thereby significantly reducing the available parameter space and improving the predictability of the models. The morphology of the total radio intensity maps reflects that seen in the data from the CHANG-ES collaboration (cf.\ Figs.~\ref{fig:irwin} and \ref{fig:werhahn_maps}). \\
\indent
Moreover, the MHD-CR model \citep{werhahn_cosmic_iii_2021,pfrommer_simulating_2022} does not require a change of the emitting CR electron component along the sequence of the FIR--radio correlation \citep[as proposed by][]{lacki_physics_i_2010} but finds a dominating primary synchrotron emission throughout \citep{werhahn_cosmic_iii_2021}. The increasing contribution of secondary radio emission in starburst galaxies (up to 30\%) is compensated for by the larger bremsstrahlung and Coulomb losses. While CR electrons are also in the calorimetric limit, the flat radio continuum spectra are a natural consequence of the additional contribution of free-free emission at high frequencies (which dominates at frequencies $\nu\gtrsim5$~GHz) and free-free absorption at low-frequencies, which is observed to be especially strong towards the central regions of starbursts \citep[see left-hand panel of Fig.~\ref{fig:werhahn_spectrum},][]{werhahn_cosmic_iii_2021}. \\
\indent
The slightly super-linear FIR--radio correlation can be explained by the following argument. Because the radio flux in the FIR-radio correlation is measured at a fixed radio frequency $\nu_\rmn{syn}\propto B\gamma_\rmn{e}^{2}$ (cf.\ equation~\ref{eq:nu_s}), and because the magnetic field saturates at higher values for higher star formation rates, the characteristic Lorentz factor of the radio-emitting electrons $\gamma_\rmn{e}$ must decrease towards larger star formation rates. Adopting a typical spectral indices for the injected electron population of $\alpha_\rmn{inj}\approx2.2$ (cf.\ Section~\ref{sec:equilibirum}), the reservoir of low-energy electrons is larger so that the radio luminosity is correspondingly increased, leading to a super-linear FIR--radio correlation \citep[see left-hand panel of Fig.~\ref{fig:FRC},][]{pfrommer_simulating_2022}. While very successful in explaining the bulk of the emission in the star-forming disk, the steady-state model fails in the galactic outflows and does not provide the correct spectra in the outer disks because of a dominant galactic CR gradient that implies a stronger CR diffusion at high energies, producing an inverted CR spectrum there \citep{werhahn_gamma-ray_2023,girichidis_spectrally_2023}. This shortcoming can be addressed with the next-generation models that evolve CR electron spectra in MHD-CR simulations \citep{winner_evolution_2019,ogrodnik_implementation_2021}. More accurate simulations of the effects of spatially and temporally varying CR transport on non-thermal emission maps require spectral models in two-moment formulations of CR hydrodynamics \citep{thomas_cosmic-ray-driven_2023} applied to cosmologically evolving galaxies \citep{hopkins_first_2022}.

\subsubsection{Energy equipartition between magnetic fields and cosmic rays}
\label{sec:magnetic equipartition}

The synchrotron emission $j_\nu\propto\eps_\rmn{cre}B^{\alpha_\nu+1}$ depends on the CR electron number density (at the electron energy relevant for emitting radio synchrotron photons) and on the magnetic field strength. There are two popular assumptions that are used estimate the magnetic field strength: (i) energy equipartition between magnetic field and CR ions, $\eps_B=\eps_\CR$, and a constant CR electron-to-ion density ratio, $\eps_{\rmn{cre}}=K_\rmn{ep}\eps_\CR\approx0.01\eps_\CR$ \citep[which is inferred for the solar circle in the Milky Way,][]{cummings_galactic_2016} and (ii) the minimum energy assumption. The minimum-energy magnetic field is obtained by minimizing the total non-thermal energy density with respect to the magnetic energy density with the constraint to reproduce the observed synchrotron emissivity (i.e., express the CR electron energy density in terms of $j_\nu$ and $B$):
\begin{align}
\label{eq:MEC}
\left.\frac{\partial \eps_\rmn{nt}}{\partial \eps_B}\right|_{j_\nu}=0
\quad
\mbox{where}
\quad
\eps_\rmn{nt}=\eps_B+\eps_\CR+\eps_\rmn{cre}
=\eps_B + (K_\rmn{ep}+1)\, \eps_\rmn{cre}.
\end{align}
Historically, the minimization was done on a fixed interval in radio frequency \citep{burbidge_synchrotron_1956,pacholczyk_radio_1970}. This biases the results if the sources exhibit different magnetic field strengths so that a radio frequency interval corresponds to different CR electron energy intervals, which contain different amounts of CR electrons for spectral indices larger than 2 (as is the case for steady-state electron cooling, cf.\ Section~\ref{sec:equilibirum}). Hence, the criterion has been modified to perform the minimization over a fixed interval in CR electron energy \citep{pohl_predictive_1993,beck_galactic_1996,pfrommer_estimating_2004,beck_revised_2005,beck_magnetic_2015}, which now depends on the unknown low-momentum cutoff of the CR electron distribution. Alternatively, if the electrons are secondaries injected as a result of hadronic CR reactions, this provides another constraint and eliminates the uncertainty of the electron distribution. In this case, the minimum-energy magnetic field only depends on the CR ions \citep{pfrommer_estimating_2004}, and the CR electron energy density is $\eps_\rmn{cre}\propto \eps_\CR / (\eps_B+\eps_\rmn{ph})$, where we adopted a steady-state CR electron distribution (equation~\ref{eq:fe_cooled}).\\
\indent
Assuming a dominant population of primary CR electrons, \citet{thompson_magnetic_2006} derive both magnetic field estimates along the star-forming sequence. While those estimates agree with each other for normal star-forming galaxies like our Milky Way, they differ significantly for increasing gas surface density (or star formation rate) and reach values for the ratio of the minimum energy-to-equipartition field strength of $10^{-4}$ in the most extreme local starburst Arp 220. Hence, if these equipartition strengths are realized as suggested by \citet{lacki_sturm_2013}, then the required non-thermal energy density is far from the energetically preferred minimum. Modeling the observed $\gamma$-ray spectra, which directly probe CR energy density, and the radio spectra, which offer insights into the magnetic field strengths within nearby starburst galaxies, allows one to validate the equipartition assumption in these extreme systems. \citet{yoast-hull_equipartition_2016} perform such an analysis and find that equipartition magnetic energy densities surpasses that of CRs by a significant margin, questioning the equipartition assumption for central molecular zones in such starbursting galaxies.
\\
\indent
In fact, there are a number of problems associated with these simple estimates for magnetic field strengths. (i) Clearly, there is no physical mechanism that ensures that either of the two conditions must be realized:  minimum energy (because we are talking about non-thermal processes) or equipartition between magnetic field and CR ion energy densities (where the CR electrons remain a universal constant fraction thereof). (ii) Particularly, in regions of strong magnetic field, the CR electrons cool quickly, which should lead to an anti-correlation between CR electrons and the magnetic field for times longer than the cooling time. Equating the cooling time to the timescales for CR advection, diffusion or streaming, depending on the dominant CR transport process, determines the CR electron cooling length above which we expect that the CR electron energy density neither traces the energy density of CR ions nor of the magnetic field. (iii) Because of particle trapping in compressible magnetosonic modes, energy equipartition is not valid on scales smaller than the turbulent driving scale, which is of order 100~pc in spiral galaxies \citep{seta_revisiting_2019}. (iv) Magnetic energy growing through a small-scale fluctuating dynamo is expected to saturate in (approximate) equipartition with the turbulent energy \citep{brandenburg_astrophysical_2005}. This seems to hold in spiral galaxies from dwarfs to normal star-forming and starburst galaxies \citep{pfrommer_simulating_2022}. However, the magnetic energy density only (approximately) equilibrates with that of CR ions on the scale of the Milky Way while it falls short in smaller galaxies to eventually differ by a factor of 50 in dwarfs \citep{pfrommer_simulating_2022}. This calls for a more efficient fluctuating small-scale dynamo resulting from SN driving (which is not explicitly modeled in those simulations) and/or questions the validity of equipartition magnetic fields at the dwarf scale. (v) Even for Milky Way-like disk galaxies, energy equipartition needs time to develop and may not be applicable for small times and spatial scales. Correlating radio continuum images of the Milky Way and M33, \citet{stepanov_observational_2014} show that magnetic equipartition is violated on scales $\lesssim1$~kpc, implying an underestimate of the equipartition field strength around CR electron sources and an overestimate in regions far away from the sources. Nevertheless, there is some observational support for equipartition on intermediate galactic scales from a joint analysis of $\gamma$-ray and radio continuum data, which provides a more independent  determination of magnetic field strengths in the Milky Way \citep{strong_diffuse_2000}, in the Large Magellanic Cloud \citep{mao_magnetic_2012} and in M82 \citep{yoast-hull_winds_2013}.

\subsubsection{Gamma-ray emission}

\paragraph{Gamma-ray emission from individual galaxies.} The launch of the ERGET $\gamma$-ray space telescope enabled us to carefully study the diffuse $\gamma$-ray emission of the Milky Way \citep{hunter_egret_1997}. Modelling CR propagation with the GALPROP code that includes nucleons, antiprotons, electrons, positrons, synchrotron and $\gamma$ rays demonstrates that nearly 90\% of the GeV $\gamma$-ray emission results from pion decay following hadronic CR-proton interactions \citep{strong_diffuse_2000}. While the associated $\gamma$-ray distribution is characterized by a thin-disk geometry, there is a large leptonic inverse-Compton halo with a typical size ranging from 4 to 10 kpc. The Large Magellanic Cloud is the only external galaxy detected by EGRET at $\gamma$-ray energies \citep{sreekumar_observations_1992}. Early on, it has been realized that the nearby starburst galaxies M82 and NGC~253 with their dense ISM ($n\sim300~\rmn{cm}^{-3}$) and high CR energy densities \citep[as inferred from the high radio luminosity,][]{voelk_correlation_1989} should also be bright $\gamma$-ray sources \citep{akyuz_m_1991,sreekumar_study_1994,volk_nonthermal_1996,paglione_diffuse_1996,romero_signatures_2003,torres_luminous_2004,domingo-santamaria_high_2005,thompson_starburst_2007,persic_very-high-energy_2008,de_cea_del_pozo_multimessenger_2009,rephaeli_high-energy_2010}, and that this argument should also carry over to starburst galaxies in general \citep{pohl_predictive_1994,torres_luminous_2004}. These theoretical predictions were confirmed by the detection of $\gamma$-ray emission from M82 and NGC~253 with \textit{Fermi} \citep{abdo_detection_2010} and imaging air Cerenkov telescopes \citep{acero_detection_2009,the_veritas_collaboration_connection_2009}. Over the course of its mission, \textit{Fermi} data enabled detections of $\gamma$ rays from nearby star-forming galaxies such as M31, M33, Arp 299 and the Sagittarius dwarf spheroidal galaxy \citep{ackermann_observations_2017,xi_gev_2020,crocker_gamma-ray_2022}.

\paragraph{Correlation between $\gamma$-ray emission and star formation.} Combining $\gamma$-ray flux measurements of individual galaxies and upper limits on the $\gamma$-ray emission of a large number of nearby star-forming galaxies enabled the discovery of a tight FIR--$\gamma$-ray relation \citep{ackermann_gev_2012,rojas-bravo_search_2016,ajello__2020}. Modeling the $\gamma$-ray spectra in one-zone models and comparing them to observational data of M82 and NGC~253 argues for the $\gamma$-ray emission that is mainly of hadronic origin \citep[see figure~1 of][for a compilation of several previous studies]{lacki_gev_2011}. Using cooling time arguments and the observed $\gamma$-ray fluxes, \citet{lacki_gev_2011} suggest that despite efficient pion losses of CR ions in both galaxies, a fraction of 0.6--0.8 of high-energy primary CR ions can escape before hadronically interacting with the ISM. Using three-dimensional MHD-CR simulations of isolated galaxies, \citet{pfrommer_simulating_2017} find that CR ions suffer substantial adiabatic losses in low star-forming galaxies, which are on par with the non-adiabatic CR losses. The combination of CR diffusion and adiabatic losses cause the FIR--$\gamma$-ray relation to deviate from the calorimetric relation at low star formation rates. While MHD-CR simulations with a parallel diffusion coefficient of $(1$--$3)\times10^{28}\rmn{cm}^2~\rmn{s}^{-1}$ match the observed relation for galaxies forming in isolation \citep{pfrommer_simulating_2017,werhahn_cosmic_ii_2021,nunez-castineyra_cosmic-ray_2022} and in a cosmological environment \citep{buck_effects_2020}, the FIRE simulations require a larger coefficient of $3\times10^{29}\rmn{cm}^2~\rmn{s}^{-1}$ \citep{chan_cosmic_2019,hopkins_but_2020}. Notably, simulations employing an isotropic CR diffusion coefficient of $10^{28}\rmn{cm}^2~\rmn{s}^{-1}$ and a local suppression in star-forming regions by a factor of 10--100 also match the observed relation \citep{semenov_cosmic-ray_2021}. While most of these studies calculate the $\gamma$-ray emission by assuming power-law momentum CR distribution, steady-state models \citep[that reproduce $\gamma$-ray spectra of individual galaxies; see right-hand panels of Fig.~\ref{fig:werhahn_spectrum}, and Fig.~\ref{fig:FRC} for the FIR--$\gamma$-ray correlation,][]{werhahn_cosmic_ii_2021} and simulations that follow the CR spectrum in the simulations and calculate the resulting pion-decay emission \citep{werhahn_gamma-ray_2023} come to the same conclusion. 

\paragraph{Gamma-ray background.} \textit{Fermi} observed the diffuse unresolved isotropic $\gamma$-ray background between 0.1 and 820 GeV \citep{ackermann_spectrum_2015}. Its photon spectrum is characterized by a decaying power-law with spectral index 2.3 below $\sim200$~GeV and an exponential cutoff above. For energies above $\sim50$~GeV, the majority of this flux (about $85\%$) is contributed by unresolved blazars \citep{ackermann_resolving_2016}. The exponential cutoff arises because the Universe is optically thick to very high-energy $\gamma$ rays over distances of several tens to hundreds of Mpc (depending on cosmic redshift), which pair-produce off of the extragalactic background light that is emitted by galaxies and quasars over cosmic history. Below 50 GeV, the isotropic $\gamma$-ray background could be made out of several extragalactic source classes, of which the main contribution is very likely also provided by blazars \citep{padovani_radio-loud_1993,stecker_high-energy_1993,stecker_predicted_1996,stecker_components_2011,broderick_cosmological_2012,broderick_implications_2014} and typically attains a small (but non-negligible) contribution from star-forming galaxies \citep{pavlidou_diffuse_2001,thompson_starburst_2007,stecker_are_2007,fields_cosmic_2010,makiya_contribution_2011,ackermann_gev_2012,tamborra_star-forming_2014,linden_star-forming_2017,ajello__2020}. This is usually modelled by combining the empirical scaling of the FIR and $\gamma$-ray emission in combination with a measured star-formation history (that is compatible with measurements of the extragalactic background light). Using a model of CR transport in star-forming galaxies \citep{krumholz_cosmic_2020}, which matches $\gamma$-ray spectra of nearby starbursts and quiescently star-forming galaxies, and applying it to the cosmologically evolving galaxy population, \citet{roth_diffuse_2021} calculate the diffuse isotropic $\gamma$-ray background and find it to match the amplitude and spectral slope of the observations. We note that there are still substantial uncertainties and whether this result still holds when confronted with fully three-dimensional MHD models of forming galaxies is an open question. In particular, this model leaves no room for the contribution of $\gamma$-ray emission of a blazar population that dominates the resolved population statistics of \textit{Fermi} (with several thousands of objects detected) while there are  only one dozen of individual star forming galaxies in the \textit{Fermi} data. In order for star-forming galaxies to completely dominate the background of unresolved objects, the model of \citet{roth_diffuse_2021} invokes a sharp cutoff in the blazar luminosity density and a leveling off of the cumulative flux distribution of blazars just below the current sensitivity limit of \textit{Fermi}, which however would be a chance coincidence.

\paragraph{CR feedback constrained by $\gamma$-rays.} By analyzing \textit{Fermi} $\gamma$-ray data of the Small Magellanic Cloud, \citet{lopez_evidence_2018} detect extended $\gamma$-ray emission below 13~GeV along the ``bar'' and ``wing'' structure that are associated with regions of active star formation. While the contribution of pulsars is not expected to exceed 10\%, most of the emission is likely due to hadronic CR ion interactions with the ISM \citep[in line with findings by][]{persic_diffuse_2022}. A comparison to the star formation rate shows that the Small Magellanic Cloud is well below the ``calorimetric limit,'' implying that CRs are escaping this dwarf irregular galaxy via a combination of advection, streaming and/or diffusion. The fact that the $\gamma$-ray luminosity is far below the calorimetric relation for galaxies with star formation rates $\lesssim10~\rmn{M}_\odot~\rmn{yr}^{-1}$ (see right-hand side of Fig.~\ref{fig:FRC}) makes these galaxies prime candidates for CR feedback \citep{crocker_cosmic_II_2021}, provided the CR inertia and pressure can be efficiently coupled to the background plasma via CR-wave scatterings. The calorimetric fraction increases towards greater star formation rates so that most of the CR energy between 1--100~GeV could be lost to hadronic interactions for starbursts such as Arp~220, powering the $\gamma$-ray emission \citep{wang_are_2018,krumholz_cosmic_2020}. Using different assumptions, \citet{lacki_gev_2011} find that a fraction of 0.6--0.8 of high-energy primary CR ions can escape the dense starburst region carrying substantial energy alongside that is available for feedback. This makes a strong case for performing three-dimensional MHD simulations of galaxies with self-consistent CR transport in multi-phase galactic media to eliminate some of the underlying assumptions and to quantify the efficiency of CR feedback in comparison to that of other processes such as SN and radiation feedback.

\subsection{Observational signatures of cosmic ray feedback in the CGM}
\label{sec:CGM_feedback}
\subsubsection{Gamma rays and constraints on cosmic ray transport in the inner CGM}
As argued in Section \ref{Cosmological effects} discussing the role of stellar CR feedback in the cosmological context, CR transport plays a decisive role in determining CR pressure support in the CGM and its thermodynamical state. When the effective transport speed is very low, CR pressure gradients may in principle be dynamically important (see equation~\ref{forcebalance2}), but because hadronic and Coulomb losses may become more important in this regime, this could limit the extent to which CRs drive winds and dominate the CGM pressure support. The $\gamma$-ray emission may violate observational constraints in this case. In the opposite regime of very high effective transport speed, these catastrophic CR energy losses and $\gamma$-ray emission are negligible but so are the CR pressure forces that could drive winds and slow down accretion in the halo. Consequently, it was argued that there exists an optimal range of the effective CR transport speeds that maximizes the impact of CRs on the galaxy and its CGM \citep{salem_role_2016,hopkins_but_2020,chan_impact_2022,hopkins_effects_2021}. \\
\indent 
However, typical CR transport speeds and their dependence on the thermodynamical state of the ISM and CGM are marred by significant uncertainties, both observational and theoretical. In particular, it has been suggested that in order to correctly infer the effective diffusion coefficient from observations, the CR propagation models need to take into consideration spatially extended CGM with sizes up to several tens of kpc rather than rely on simple ``leaky box'' approximation \citep{hopkins_testing_2021,heintz_galaxies_2022}. When considering such an extended CGM, the inferred diffusion coefficient increases with the characteristic size of the halo \citep[e.g.,][]{korsmeier_galactic_2016,sampson_turbulent_2023}
possibly reaching values as high as $\kappa\sim10^{29}$~cm$^{2}$~s$^{-1}$ \citep{johannesson_bayesian_2016}, i.e., about ten times higher than the ``canonical'' value of the diffusion coefficient.
Interestingly, this dependence of the effective diffusion coefficient on the scale of the system is broadly consistent with the results from idealized MHD simulations of streaming CRs \citep{sampson_turbulent_2023}. Coarse-graining of the results from these simulations reveals that CR transport can be better approximated as superdiffusion, where the characteristic time needed to transport CR by a distance $l$ is $\tau_\rmn{s}\propto l^{\alpha}/\kappa_\rmn{s}$ (where $\alpha\sim 3/2$) rather than $\tau_\rmn{c}\propto l^{2}/\kappa$ expected in the classical diffusion process. 
This has consequences for the observational determination of $\kappa$ because common methods to infer this quantity are closely linked to the time that CRs spend interacting with the gas. For example, the slope of the $\gamma$-ray and radio spectrum as a function of the distance from the galactic midplane depends on the time CRs spend losing energy as a result of hadronic, Coulomb, and additionally IC and synchrotron processes for CR electrons. Similarly, grammage, that can be inferred from the observed boron-to-carbon ratio, is a product of the density of the gas with which CRs interact, speed of light, and the CR propagation time. 
Since observables are set by these observationally constrained timescales, and $\tau_\rmn{c}$ must be approximately comparable to $\tau_\rmn{s}$, the characteristic timescales ratio $\tau_\rmn{c}/\tau_\rmn{s}\sim l^{2-\alpha}\kappa_\rmn{s}/\kappa\sim 1$. Consequently, if the coarse-grained CR transport is better approximated as superdiffusion rather than classical diffusion, then in order to recast the true CR transport speed in terms of $\kappa$, the classical diffusion coefficient would need to be larger than that inferred for smaller scales (for sufficiently large characteristic scales $l$) and $\kappa$ would increase with the size of the halo \citep[for fixed $\kappa_\rmn{s}$; see Fig.~13 in][]{sampson_turbulent_2023}. 
We note that these results correspond to the post-processing of MHD simulations that includes streaming of CRs. One limitation of this approach is that, by construction, the CRs neither exert forces on the gas nor heat it via the streaming instability.
Relatively large values of diffusivity have also been postulated in order to reconcile $\gamma$-ray detections and upper limits in dwarfs and $L_{*}$ galaxies (M33, SMC, LMC, Andromeda, Milky Way). Unlike in dense starbursts, where a substantial fraction of CRs experience catastrophic losses, in these objects most of CRs need to escape before significant losses can occur in order to avoid overproducing $\gamma$-ray emission \citep{wojaczynski_x--ray_2017,lopez_evidence_2018,wang_are_2018}. Nevertheless, the physics of CR propagation remains highly uncertain at large heights above disk midplanes. Further constraints on CR transport mechanisms on these scales may come from upcoming spatially extended $\gamma$-ray \citep{karwin_fermi-lat_2019} and synchrotron emission observations \citep{van_haarlem_lofar_2013}.\\
\indent
Given that spatially extended CGM may play an important role in determining the residence time of CRs, fully cosmological simulations may be needed to make reliable predictions for the $\gamma$-ray emission. Already based on the results from early cosmological non-MHD simulations of CR feedback, it was evident that, for constant ``canonincal'' values of the diffusion coefficient in the range $\kappa\sim 3\times 10^{27}$~cm$^{2}$~s$^{-1}$ to $10^{28}$~cm$^{2}$~s$^{-1}$, $\gamma$-ray emission from the CGM would exceed ``extragalactic'' background levels \citep{salem_role_2016}. Only for faster CR transport were these models consistent with the $\gamma$-ray observations from \textit{Fermi}.
In a qualitative sense, these findings were confirmed by cosmological MHD simulations with a more sophisticated CR transport approximated as a combination of constant diffusivity and simple prescriptions for the effective streaming speed including super-Alfv{\'e}nic streaming at a constant fraction of the Alfv{\'e}n speed \citep{hopkins_but_2020}. In agreement with earlier non-cosmological global models with the same physics \citep{chan_cosmic_2019}, these simulations suggested that very high values of $\kappa\sim 3\times 10^{29-30}$~cm$^{2}$~s$^{-1}$ are needed to reproduce the observed level of $\gamma$-ray emission.\\
\begin{figure}
  \begin{center}
    \includegraphics[width=1.0\textwidth]{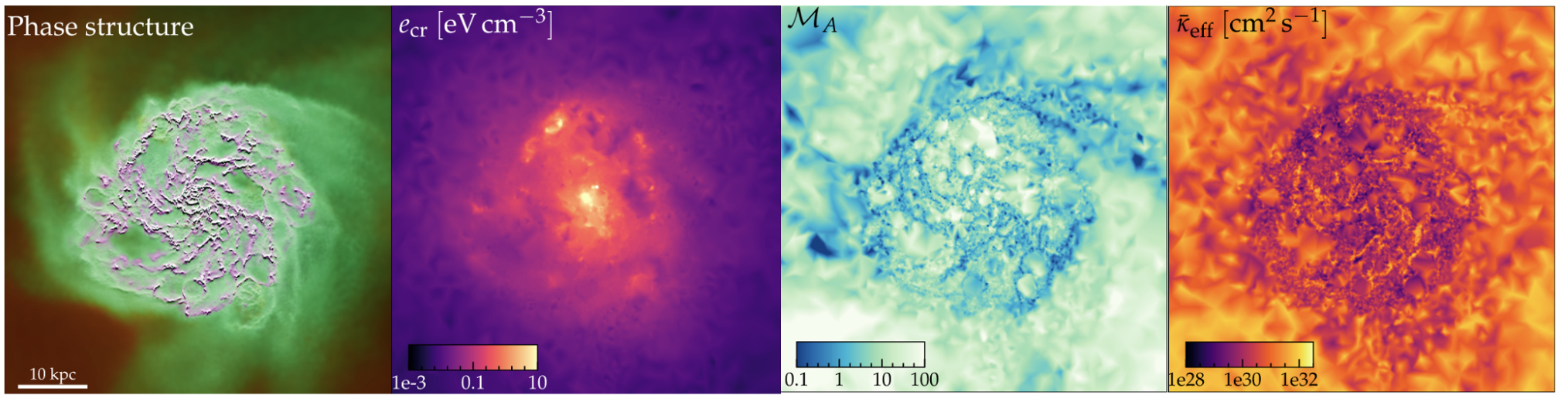}
  \end{center}
  \caption{Face-on images of a galaxy simulated in a cosmological context at $z=0$.  
  From left to right shown are phase structure (cold neutral in magenta, $T\lesssim 8000$ K; warm ionized in green, $10^{4} \lesssim T \lesssim 10^{5}$ K), CR energy density, Alfv{\'e}n Mach number ${\cal M}_\rmn{a}$, and the effective CR diffusion coefficient $\kappa_{\rm eff}$. The results correspond to a case where parallel diffusion is boosted by 100 beyond predictions from quasi-linear theory. Notice that CRs are distributed much more smoothly than the multiphase gas. Galactic outflows produce large ${\cal M}_\rmn{a}$ regions in the superbubbles in the disk and the surrounding CGM, which are associated with small-scale order-of-magnitude fluctuations in $\kappa_{\rm eff}$ in the disk and gradual increase in $\kappa_{\rm eff}$ in the CGM. Image from \citet{hopkins_testing_2021} reproduced with permission from MNRAS.}
  \label{fig:hopkins}
\end{figure}
\indent
Such high inferred values of $\kappa$ may be the result comparing the models to the data under the simplifying assumption that CR diffusion is spatially constant rather than resorting to more self-consistent modeling of the transport processes in the extrinsic and self-confinement scenarios. In particular, in a self-confinement scenario, a more complete model of CR transport should include CR streaming and relevant damping mechanisms such as ion--neutral, turbulent, linear Landau, and non-linear Landau damping as considered in cosmological simulations of galaxy formation by \citet{hopkins_testing_2021}. Since these processes are dependent on the local state of the plasma, including them leads to the effective CR transport speed varying by orders of magnitude on scales smaller than kpc in the ISM and CGM with patches of gas characterized by very high $\kappa$ (see Fig.~\ref{fig:hopkins}). Moreover, in this picture, the effective $\kappa$ increases significantly in the CGM as a function of the distance from the galactic center \citep{butsky_constraining_2023}, which reduces the ability of CR to accelerate the gas compared to the constant diffusivity case. The net result of this is that the physically-motivated transport leads to galaxy/CGM properties that are in-between those corresponding to non-CR and constant-$\kappa$ cases \citep{hopkins_effects_2021}.\\
\indent
Despite the fact that the wave damping processes increase the effective CR transport speed, CR tend to be too strongly confined in the ``inner'' CGM (within $\sim30$ kpc from the galactic center), thus violating $\gamma$-ray limits in the above models. Interestingly, compared to the constant diffusivity case, even very efficient ion--neutral damping and fast CR transport in the cold phase \citep[see, e.g.,][]{farber_impact_2018} results in only mild reduction of $\gamma$-ray emission (that depends on the cold-to-hot mass ratio) as CRs still experience excessive pionic losses in the hotter and more spatially extended surrounding CGM, where transport is slower. 
Since these results are obtained in the framework of quasi-linear theory, the true suppression of the scattering rate may be larger than expected based on this approximation. 
Interestingly, \citet{hopkins_testing_2021} demonstrate that galaxy models can be made consistent with $\gamma$-ray limits (and other observational constraints) if quasi-linear theory scattering rates in the self-confinement scenario are suppressed by $\sim 100$. This suggest that either quasi-linear theory is inadequate or that additional novel damping processes need to be incorporated into the models. \\
\indent
While the need for including the CGM in modeling CR feedback is universally accepted, there is no overwhelming consensus in the literature on the need for very large diffusion coefficients to explain the $\gamma$-ray constraints. As discussed in Section \ref{Role of CR in global models of isolated galaxies}, simulated emission consistent with $\gamma$-ray observations was predicted based on both zoom-in models of disk patches \citep{simpson_how_2023} and global models of galactic disks \citep{semenov_cosmic-ray_2021,nunez-castineyra_cosmic-ray_2022}. Admittedly, in these models the CGM was either absent or not modelled from first principles. However, even in some fully cosmological models of the formation of Milky Way-mass galaxies that employ typical values of CR diffusivity, the predicted level of emission can be reconciled with the $\gamma$-ray observations when Alfv{\'e}n CR losses are also included \citep{buck_effects_2020}. In this case, only the disk-halo interface is CR-dominated. When this effect is neglected, the models exhibit spatially extended CR-dominated halos and overpredict $\gamma$-rays and CR energy density at the solar circle. Interestingly, irrespective of whether the Alfv{\'e}n CR losses are included, the magnetic and CR pressures approximately trace each other within a factor of a few out to the virial radius even though they decline over orders of magnitude with the galacto-centric distance. An interesting implication of this is that non-resonant Bell instability \citep{bell_turbulent_2004} is not triggered in the CGM. The condition for the onset of this instability is that $\eps_{\rm cr}/\eps_{B,\rmn{sat}}\gtrsim 2c/\varv_{\rm drift}$ (see equation~\ref{eq:eps_B,sat}), where the left hand side is the ratio of the CR to magnetic energy densities, and $\varv_{\rm drift}$ is the CR drift speed, which is comparable to the Alfv{\'e}n speed. Since typically $2c/\varv_{\rm drift}\gg 1$, the instability is never triggered and, consequently, the diffusion coefficient is not expected to be reduced by the additional magnetic field fluctuations due to the Bell instability \citep{buck_effects_2020}.\\
\indent 
Possible resolution of the discrepancies between the levels of CR diffusivity required to match the observational constraints may involve considerations related to (i) the approach adopted to compare simulation predictions to the observations and (ii) detailed structure of the magnetic field in the ISM and CGM. 
On dwarf mass scales, some conclusions regarding $\kappa$ values appear to be the same for the Auriga \citep{buck_effects_2020} and FIRE-2 \citep{hopkins_testing_2021} simulations. That is, canonical values of the effective CR transport speed result in the overproduction of $\gamma$-ray luminosities $L_{\gamma}$ compared to observations and may be inconsistent with grammage estimates. This mismatch between the simulations and observations is exacerbated by extrapolating the relationship between the observed FIR luminosity and star formation rate to low FIR luminosities. This extrapolated relationship becomes less accurate in this regime because of lower metallicity and dust content, which results in less efficient reprocessing of the emission from massive stars to FIR radiation \citep{bell_estimating_2003}. Consequently, the observationally inferred star formation rates are underestimated, which ultimately increases the disparity between the simulated and observed $L_{\gamma}$ vs. star formation rate relation. In this low halo mass regime, faster transport may be needed to reduce the simulated $\gamma$-ray luminosities and bring them back into agreement with the observed relation.

On Milky Way-mass scales, there are significant discrepancies in the $\gamma$-ray emission predictions from different cosmological simulations. These stem at least partially from systematic differences in the magnetic field distributions predicted by these simulations. Specifically, FIRE-2 models \citep{hopkins_but_2020,ponnada_magnetic_2022} generally predict smaller average magnetic fields than the Auriga models \citep{pakmor_faraday_2018,buck_effects_2020}. In the latter case, 100 times larger average magnetic energy densities can be present compared to FIRE-2 simulations, which results in ten times larger Alfv{\'e}n cooling rates and faster CR streaming. In Auriga simulations, this leads to dramatic reduction in CR energy density and $\gamma$-ray emission compared to simulations neglecting Alfv{\'e}nic losses. In contrast, in FIRE-2 simulations, the impact of streaming is generally negligible in terms of its effect of reducing CR pressure and the effective bulk CR transport speeds required to match $\gamma$-ray limits translate to $\sim$10 to 1000 times the Alfv{\'e}n speed. Note that when the magnetic fields are generally weak, boosting the streaming speed by invoking damping processes would not simultaneously increase CR streaming losses.
\subsubsection{Interplay between cosmic ray feedback and magnetic fields in the CGM} The above considerations raise the question of which models lead to more reliable predictions for the magnetic field strength and distribution. Cosmological MHD simulations do approximately reproduce many aspects of the observations of magnetic fields \citep{pakmor_faraday_2018,ponnada_magnetic_2022}. These simulations predict magnetic field strengths that are generally consistent with the observations of pulsar dispersion and rotation measures
\citep{manchester_australia_2005,sobey_low-frequency_2019} and Zeeman splitting \citep{crutcher_magnetic_2010} in the Milky Way, and Faraday rotation measure maps of M51 \citep{fletcher_magnetic_2011}. In agreement with observations \citep{strong_diffuse_2000,beck_galactic_2001,stepanov_observational_2014}, the simulated magnetic pressures are in approximate equipartition with thermal and CR pressures for the ISM densities in the range $\sim1$--$10$~cm$^{-3}$ and on scales exceeding $\sim$1~kpc. However, there are very large variations in the observed relations between, e.g., rotation and dispersion measures or the Zeeman-inferred magnetic field and the density of the cold gas, which lessens their power to discriminate between the models. \\
\indent
As is the case with the observational data, differences between simulation results can be also be substantial, even for the MHD simulations that do not include CRs. The predicted volume-averaged values of plasma $\beta$ (ratio of the magnetic-to-thermal pressure) near the disk or at the disk-halo-interface vary between $\sim1$ and $\sim10$ (cf.~figure~13 in \citet{pakmor_magnetic_2017} and bottom panel of second column in figure~5 in \citet{hopkins_but_2020}; at the distance $\sim5$~kpc, the strength of the magnetic fields is $\sim 10^{2}$ larger in the former case). These differences become even more pronounced in the CGM (cf.\ middle panel in figure~9 in \citealt{pakmor_magnetizing_2020} and and bottom panel of second column in figure~5 in \citealt{hopkins_but_2020}; at the galacto-centric distance of 50 kpc, $\beta$ ranges from $\sim20$ (former case) to $\sim 1000$ (latter case); see also \citet{van_de_voort_effect_2021}, where $\beta$ can assume values below unity in the outflow region even beyond 50 kpc from the center). This is consistent with the fact that magnetic energy density is progressively more sub-equipartition as the density decreases below $\sim 1$~cm$^{-3}$ in the disk and CGM in the FIRE simulations \citep{su_feedback_2017,ponnada_magnetic_2022}, so the volume-averaged fields are weaker as the lower density gas is more volume filling. \\
\indent
Comparisons between models are made more difficult by the fact that complexities associated with CR transport can indirectly influence the strength of the magnetic field, which in turn can affect the expected $\gamma$-ray luminosity. For example, \citet{ponnada_magnetic_2022} demonstrated that in MHD simulations, magnetic field amplification in low density disk gas can be an order of magnitude larger compared to the predictions from CR simulations with anisotropic diffusion. In the MHD simulations, the outflows from the disk were ``trapped'' by the CGM pressure, resulting in the increased velocity dispersion of the gas in fountain flows \citep[e.g.,][]{muratov_gusty_2015,stern_virialization_2021,hafen_hot-mode_2022}, which in turn increased the magnetic field strength. In the CR counterparts of these runs, CR were able to drive outflows, which suppressed the ``churning'' motions of the gas near the disk and allowed the gas to adiabatically expand thus reducing the amplification of the magnetic field.\\
\indent
Overall, it remains an open question which models provide a more reliable description of the CGM and are fully consistent with the observational constraints on the magnetic field strength, $\gamma$-ray emission, grammage, etc. On the theoretical side, the answer to this question will likely involve a combination of considerations involving details of ISM models, non-equilibrium CR transport effects, wave damping physics,
as well as detailed comparisons of the resolution effects and the dependence of the results on numerical methods, and in particular on the cleaning method used to ensure $\bs{\nabla\cdot B}=0$. On the observational side, measurements of the magnetic field strength at the disk-halo interface and in the CGM are not sufficiently constraining to distinguish between different models despite the fact that model predictions for the field strengths can vary by over an order of magnitude. Future fast radio burst observations from the Canadian Hydrogen Intensity Mapping Experiment (CHIME; \citealt{chimefrb_collaboration_chime_2018,connor_observed_2022}) or the Deep Synoptic Array (DSA-10; \citealt{kocz_dsa-10_2019}) may significantly improve measurements of the magnetic field strengths in the CGM.

\subsubsection{Cosmic ray feedback and integrated X-ray emission from the CGM}
Independent constraints on the CGM models can be obtained from X-ray observations. The existence of X-ray emitting gas in halos more massive than $\sim10^{11.5}~$M$_{\odot}$ is predicted on theoretical grounds \citep[e.g.,][]{dekel_galaxy_2006} and has been detected not only in X-rays \citep{li_chandra_2008,anderson_detection_2011,fang_hot_2013,anderson_non-detection_2015} but also via the Sunyaev--Zel'dovich (SZ) effect \citep{planck_collaboration_planck_2013}. The amount of hot gas in the halos and, consequently, the level of X-ray emission, can be strongly affected by CR feedback effects. Generally speaking, the stronger the CR fraction in the CGM, the lower the expected temperature of the gas as the thermal pressure support becomes sub-dominant \citep[e.g.,][see Fig.~\ref{fig:ji_cgm}]{butsky_role_2018,ji_properties_2020}. Consequently, the CR fraction cannot be arbitrarily high in the CGM as this would underpredict the X-ray emission in comparison to the observed hot ($\sim 10^{6}$~K) Milky Way halo 
\citep{fang_hot_2013,faerman_massive_2017}.\\
\indent
FIRE-2 simulations generally exhibit significant CR pressure support in Milky Way-mass halos and consequently predict relatively low level of X-ray emission. For example, while the predictions from CR simulations \citep{chan_impact_2022} lie on the extrapolation of the observed relation between the soft (0.5--2 keV) X-ray luminosity and stellar mass \citep{anderson_unifying_2015}, they tend to somewhat underestimate soft X-ray luminosities at lower star formation rates compared to observations \citep{li_chandra_2013}.
MHD counterparts of these types of simulations tend to produce significantly stronger soft X-ray emission in the CGM. For example, in a Milky Way halo, the volume-integrated soft X-ray emission from the CGM within 200 kpc from the center, and in the same photon energy range as above, is about six times higher compared to what is seen in the CR counterparts of these simulations \citep{ji_virial_2021}. Furthermore, while at large galacto-centric radii ($\sim 200$~kpc) the \textit{average} gas temparature in the MHD case is $T_{\rm mhd}\sim 4\times 10^{5}$~K and exceeds the thermal gas temperature in the CR case, $T_{\rm cr}$, only by a factor of a few, this difference grows to $\sim 10^{2}$ in the inner CGM ($\sim 20$ kpc), where $T_{\rm mhd}\gtrsim10^{6}$~K and 
$T_{\rm cr}\sim {\rm a\; few}\times 10^{4}$~K \citep{ji_properties_2020}, which would render the X-ray emission from that region undetectable in the CR case. As mentioned earlier, CR streaming losses could sap CR energy, reduce CR pressure support in the CGM, and heat the gas (\citealt{buck_effects_2020}; but see \citealt{hopkins_but_2020} for an opposing view). If this mechanism is effective, it could elevate the X-ray emissivity and change its spatial distribution.
Significant observed X-ray emission from CGM on scales $\lesssim 20$ kpc \citep{bregman_extended_2018} could help to put constraints on these models.
Future simulations with a more complete and realistic treatment of CR physics are essential to resolve the significant differences in theoretical predictions and reconcile the results with the observations.

\subsubsection{CGM line diagnostics of cosmic ray feedback: thermodynamics}
\begin{figure}
  \begin{center}
    \includegraphics[width=0.99\textwidth]{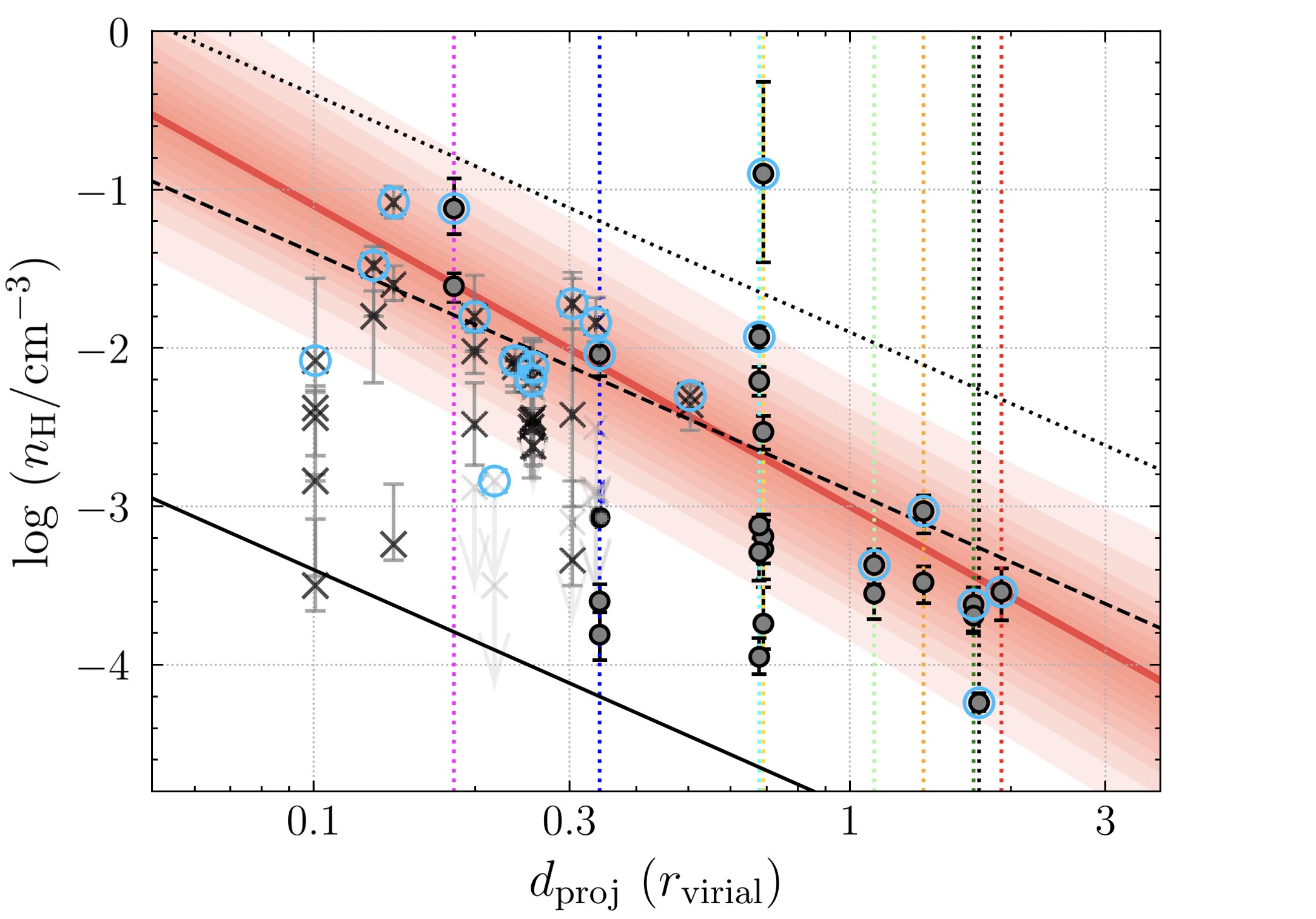}
  \end{center}
  \caption{Electron density of the cold CGM characterized by the temperature $\sim 10^{4}$~K as a function of the projected distances from halo centers from CUBSz1 and COS LRG samples (data points). Labelled with open blue circles are the highest densities in each absorption system, the values of which decline with projected distance as shown by the best-fit profile (red-shaded regions). The hot phase profile of the electron density is shown as a solid line. Density profiles of the $10^{4}$~K-gas in pressure equilibrium with the hot medium at $10^{6}$~K ($\sim$~virial temperature of $L_{*}$ galaxies) and $10^{7}$~K are shown as dashed and dotted lines, respectively. Clouds belonging to the same halo are connected by vertical lines. Notice the large scatter even in the highest electron densities (shown in blue), which can be attributed to a combination of line-of-sight projection of clouds in pressure equilibrium and non-thermal pressure support in the clouds. Figure adapted from \citet{qu_cosmic_2023}; reproduced with permission from MNRAS.}\label{fig:qu}
\end{figure}

In Milky Way-mass galaxies and smaller halos, the baryon fractions fall short of the universal baryon fraction predicted by structure formation models \citep[e.g.,][]{miller_structure_2013}. As discussed earlier, from the theoretical perspective, this inconsistency could be accounted for by invoking ``ejective'' feedback in form of stellar/CR driven galactic outflows that remove baryons from galaxies and/or ``preventative'' feedback, where the cosmological accretion of baryons is partially impeded, e.g., by substantial CR pressure forces in the CGM.
From the observational perspective, this tension is partially alleviated by the recent detections via Ly$\alpha$ emission at high-$z$ \citep{cantalupo_cosmic_2014,hennawi_quasar_2015,martin_ly_2015,cai_mapping_2017} and quasar absorption lines at low-$z$ \citep{stocke_characterizing_2013, werk_cos-halos_2014, stern_universal_2016, prochaska_cos-halos_2017} of a significant cold phase component of the halo mass budget (see \citealt{tumlinson_circumgalactic_2017}, for a review).\\
\indent
The multiphase state of the CGM is shaped by the processes operating during cosmological halo formation as well as the local thermal instability. To first order, gas accretion in the halos can proceed via ``hot'' or ``cold'' mode depending on whether halos are more or less massive than $\sim 10^{11.5}$~M$_{\odot}$, respectively (\citealt{dekel_galaxy_2006}, see also Section \ref{CGMtheory}). In the former case, the accreting gas is shock heated and does not have enough time to cool efficiently. However, accretion can also proceed via cold streams penetrating hot halos and supplying mass to galaxies  \citep[e.g.,][]{keres_how_2005,oppenheimer_feedback_2010,faucher-giguere_baryonic_2011,van_de_voort_rates_2011,nelson_moving_2013}.
The multiphase thermodynamical state of the CGM and the transition between the two modes of accretion can be further affected by stellar \citep[e.g.,][]{faucher-giguere_stellar_2016,fielding_impact_2017} and AGN feedback (e.g., \citealt{nelson_first_2019}, see also Section \ref{agntheorysection}, and our discussion on thermodynamical instabilities including the interplay between CRs and local thermal instability, \citealt{butsky_impact_2020}, see Section \ref{titheorysection}).\\
\indent
Observations of the abundances of cool photoionized low ions\footnote{Low, intermediate, and high ions are defined as having ionization energies of $\lesssim 40$ eV (corresponding to $T\sim 10^{4-4.5}$~K), between $\sim 40$ eV and $\sim 100$ eV ($T\sim 10^{4.5-5.5}$~K), and $\gtrsim 100$ eV, respectively \citep{tumlinson_circumgalactic_2017}.} suggest that the electron number densities of the cold phase may fall short of the values required to maintain thermal pressure equilibrium between the cold and hot CGM phases \citep{werk_cos-halos_2014, mcquinn_implications_2018}. 
The mismatch between the theoretical expectation and data suggest that either the hydrostatic assumption is invalid (i.e., the cold phase is underpressured), projection effects are at play (see below), or that additional non-thermal sources of pressure support are present in the CGM. 
However, these constraints may represent upper limits on the non-thermal pressure support. While the gas is likely clumpy with varying ionization conditions in individual absorption systems (see, e.g., Fig.~11 in \citealt{werk_cos-halos_2013}), the densities of the cool CGM 
were derived based on the integrated column densities across all density components \citep{werk_cos-halos_2014}. Because the intermediate-state ions (CIII, SiIII, etc.) tend to dominate the column density as a whole, the inferred mean density of the cold phase may therefore be skewed low. 
Interestingly, observations suggest that the upper envelope of the density distribution of individual absorption systems is consistent with the pressure equilibrium of the cold and hot halo gas \citep[see, Fig.~10 in][]{zahedy_characterizing_2019}. However, these measurements exhibit a very large scatter in the electron density at a given projected distance from the halo center.
This is also evident in Fig.~\ref{fig:qu}, which shows the distribution of the electron density in $10^{4}$~K clouds as a function of the projected distance in CUBSz1 and COS LRG samples \citep{qu_cosmic_2023}, with the best fit to the data points shown as a diagonal shaded region. For comparison, the density profiles of the diffuse hot gas, and the profiles with $10^{2}$ and $10^{3}$ higher normalization (corresponding to $10^{4}$~K cold gas in pressure equilibrium with $10^6$~K and $10^7$~K ambient gas) are shown as a solid, dashed, and dotted lines, respectively. We note that the integrated cold baryon budget corresponding to these solutions represents an upper limit due to the small filling factor of the cold gas.\\
\indent
While in the case of thermal pressure balance, the upper envelope of the distribution of data points would coincide with the dashed or dotted lines (depending on the temperature contrast between the hot and cold gas phases), there is a substantial scatter in the cold gas density at a given projected radius, and some of the clump density measurements are far from the expectation corresponding to the pressure balance with the ambient medium. This scatter could be attributed to a combination of (i) projection effects (where individual clouds at a given projected radius are indeed locally in pressure equilibrium with the ambient hot phase, but are physically located at very different distances from the halo center (for a fixed projected distance) and are therefore characterized by a range of densities, or (ii) a true non-thermal pressure support in the cold gas (due to a combination of turbulence, magnetic, and CR pressure).
Interestingly, the mismatch between the predicted and observed low and mid ion column densities is also seen in fully cosmological high-resolution hydrodynamical and MHD simulations 
\citep[e.g.,][]{liang_column_2016,oppenheimer_bimodality_2016,peeples_figuring_2019,hummels_impact_2019,vandevoort_cosmological_2019}. This is consistent with the hypothesis that CRs could provide the necessary non-thermal pressure support in the cold CGM phase and offer a possible solution to this potential discrepancy \citep{salem_role_2016,butsky_role_2018}.\\
\indent
The significant shift in the phase-space diagrams toward lower CGM temperatures seen in  cosmological simulations including CRs physics implies that there should be a marked difference in the relative proportions of low, intermediate, and high ions compared to the expectations based on purely hydrodynamical or MHD simulations. Indeed, depending on the physics of CR transport, dark matter halo mass, and redshift, the addition of CRs may result in significant CR pressure support in the CGM and, consequently, a decrease in the abundances of high ions from the collisionally ionized hot gas and a corresponding increase in the abundances of intermediate and low ions from the photoionized gas. Specifically, cosmological simulations predict that in CR-dominated halos, the presence of the cooler CGM component leads to larger ion column densities of (primarily photoionized) H I, Si IV, C III, O VI, and NeVIII and better fits to the \textit{Cosmic Origins Spectrograph} (COS) ion data \citep{salem_role_2016,ji_properties_2020,chan_impact_2022}. This effect can be attributed not only to the change in the CGM phase space but also to the increase in the CGM metallicity brought about by the CR-driven winds that expel the metal-enriched gas into the CGM.

\subsubsection{CGM line diagnostics of cosmic ray feedback: kinematics}\label{CGM line diagnostics of CR feedback: kinematics}
Many observational ultra-violet studies reveal that the velocity of the outflowing CGM is monotonically increasing with galacto-centric radius \citep[e.g.,][]{steidel_structure_2010,erb_galactic_2012,heckman_systematic_2015}. This gas acceleration in consistent with the expectations from CR-driven galactic outflow models, whereby CR pressure gradients can continuously accelerate the gas even at large distances. However, the general trend for the outflow velocity to increase with distance can be explained, for example, by radiation pressure forces \citep[e.g.,][]{murray_radiation_2011}. Alternatively, this could be explained by simply invoking outflows launched with a wide distribution of initial velocities, in which case faster gas parcels would travel the largest distance, thus mimicking a velocity field characteristic of accelerating gas \citep{hopkins_resolving_2013}.\\
\indent 
The level of CR pressure support in the CGM may also be probed by observing systematic kinematic offsets between low and high ions \citep{butsky_impact_2022}. While in CR-dominated halos the offset is relatively small, it can be substantial when CR pressure is negligible. Physically, this stems from the fact that the cold gas can be more easily supported against gravity by CR pressure forces when CR pressure support is significant, but in the absence of CRs the cold phase can be denser and thus more likely to dynamically decouple from the ambient hot gas and to develop a radial velocity component. In this respect, comparisons of models to the data for the offsets between OVI and SiIII \citep{werk_cos-halos_2016} favor CR-dominated halos.\\
\indent
Outflow morphologies could potentially also serve as an additional constraint on some CR wind models. Observations of quasar absorption lines reveal that the gas flows are bipolar
\citep{bouche_physical_2012,kacprzak_tracing_2012,kornei_properties_2012,rubin_evidence_2014,
kacprzak_azimuthal_2015,martin_kinematics_2019,schroetter_muse_2019}. Including CR physics in galaxy formation simulations can significantly affect outflow morphologies. For example, the outflows appear much more biconincal in CR simulations of \citet{hopkins_cosmic_2021} compared to their non-CR cases. The dependence of outflow morphology on CR physics is also evident in the simulations of \citet{buck_effects_2020}, which show that the outflow is irregular when only CR advection and diffusion are included, but becomes bipolar when CR Alfv{\'e}n streaming losses are considered, with the latter case resembling a non-CR morphology.\\
\indent 
Further progress in constraining galactic outflow models will likely be associated with the advent of new instruments, especially in the X-ray band. Various proposed missions, such as 
\textit{Arcus} \citep{smith_arcus_2016}, \textit{Athena} \citep{barcons_athena_2015}, \textit{Lynx} \citep{gaskin_x-ray_2016}, or the soft X-ray Line Emission Mapper (\textit{LEM}) mission \citep{kraft_line_2022}, may enhance line detection sensitivity and enable the observation of hot coronal gas at large galacto-centric radii.

\subsection{Observational evidence for cosmic rays in galaxy groups and clusters}\label{sec:AGN_feedback_clusters}
\begin{figure}
  \begin{center}
    \includegraphics[width=1.0\textwidth]{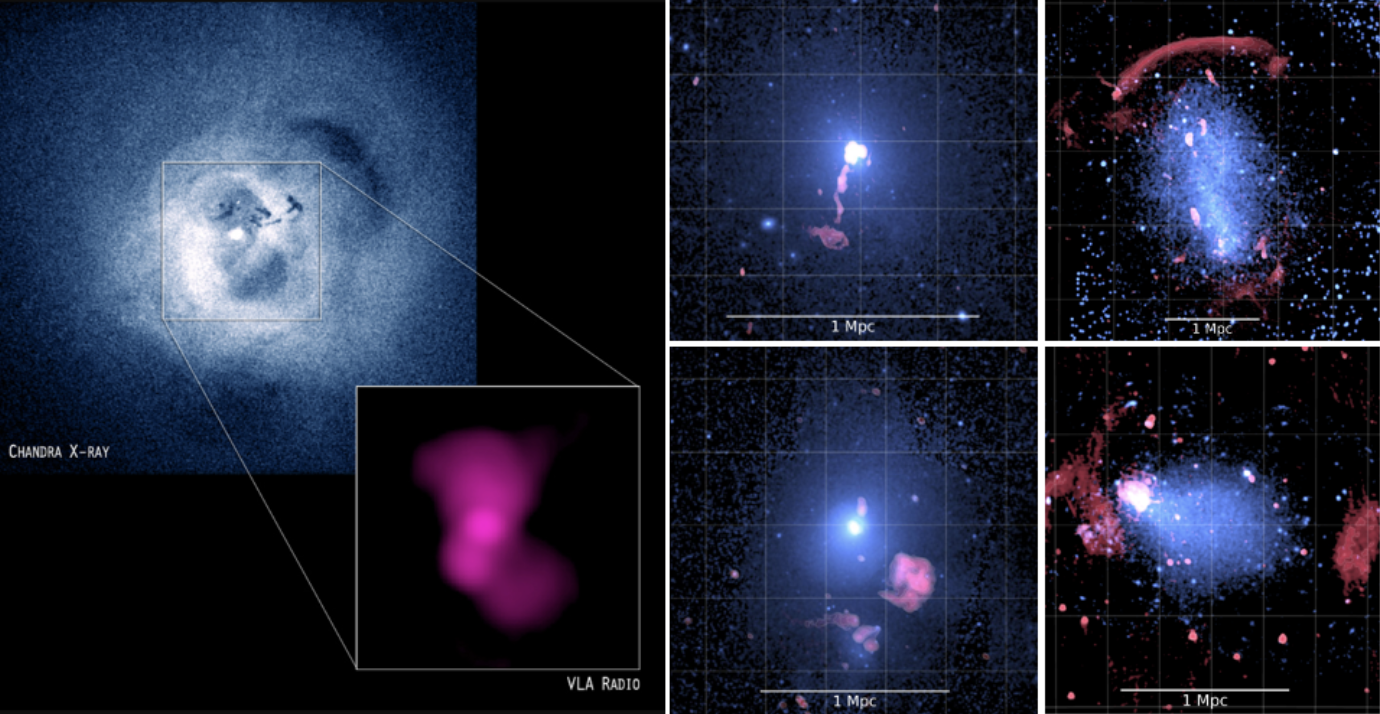}
  \end{center}
      \caption{Examples of radio relics. 
      \textit{Left}: \textit{Chandra} X-ray image of the Perseus cluster (upper left) and the zoom-in on the relic radio emission emission from inside the AGN-inflated bubbles (lower right). Credit: X-ray: NASA/CXC/IoA/A. Fabian et al.; Radio: NRAO/VLA/G. Taylor
      \textit{Middle}: Radio phoenix sources Abell 133 and Abell 85 (top and bottom, respectively). Radio (from GMRT) and X-ray (from \textit{Chandra}) emission is shown in red and blue, respectively. Credit: \citet{andrade-santos_fraction_2017,van_weeren_diffuse_2019}; reproduced with permission from ApJ and Space Sci Rev. \textit{Right}: Examples of radio gischt (or radio shocks): Sausage Cluster (top); radio (from GMRT) and X-ray (from \textit{Chandra}) emission is shown in red and blue, respectively \citep{van_weeren_particle_2010, ogrean_multiple_2014}. Reproduced with permission from Science and MNRAS; 
      Abell 3376 (bottom): radio (from GMRT) and X-ray (from XMM) emission is shown in red and blue, respectively \citep{kale_multi-frequency_2012,urdampilleta_x-ray_2018}. Reproduced with permission from ApJ and A\&A.
      }
  \label{fig:relics}
\end{figure}

\begin{figure}
  \begin{center}
    \includegraphics[width=1.0\textwidth]{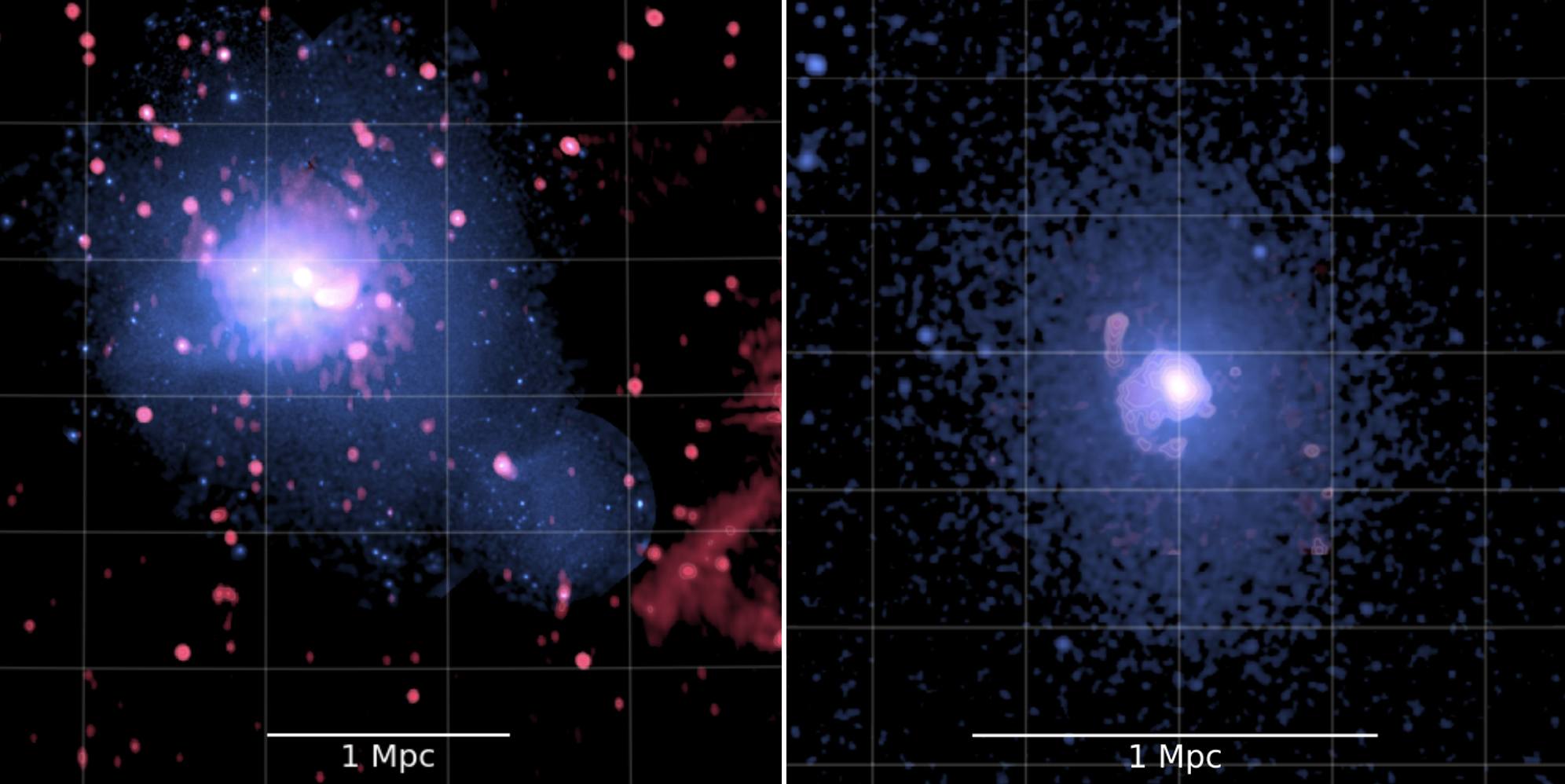}
  \end{center}
      \caption{Examples of radio halos. 
      \textit{Left}: Giant radio halo in the Coma cluster. Radio (from WSRT) and X-ray (from \textit{XMM-Newton}) emission is shown in red and blue, respectively. Credit \citet{brown_diffuse_2011,van_weeren_diffuse_2019}; reproduced with permission from Space Sci Rev. and MNRAS.
      \textit{Right}: Radio mini halo RX J1720.1+2638. Radio (from GMRT) and X-ray (from \textit{Chandra}) emission is shown in red and blue, respectively.
      Credit: \citet{giacintucci_mapping_2014,andrade-santos_fraction_2017}; reproduced with permission from ApJ.  
      }
  \label{fig:halos}
\end{figure}

Diffuse radio emission is commonly detected in galaxy clusters and signifies the presence of the magnetic fields and CR electrons (and possibly positrons) in the hot diffuse thermal plasma. Broadly speaking, the complicated taxonomy of radio emission phenomena can be divided into \textit{radio relics} and \textit{radio halos} \citep[][see Figs.~\ref{fig:relics} and ~\ref{fig:halos}]{kempner_taxonomy_2003, van_weeren_diffuse_2019}. Below we discuss these two categories of objects focusing on their physical origin and emphasizing the role of AGN feedback in this context. After briefly reviewing the physics underlying radio halos, we discuss the implication of cluster $\gamma$-ray emission limits for the CR pressure content.

\subsubsection{Radio relics} 
We start with general characteristics of radio relics before discussing specific types of relic sources. Radio relics are morphologically irregular objects characterized by significant polarization and steep spectra with spectral indexes $\alpha_{\nu}\sim 1-2.5$ (where $\alpha_{\nu}=-\partial\ln j_{\nu}/\partial\ln\nu$ and $j_{\nu}$ is the radio emissivity). Most of the radio band emission can be attributed to synchrotron radiation that is intrinsically highly polarized. However, the overall emission derives from the line-of-sight integration of the emission from different regions that in general possess magnetic fields of different magnitude and orientation. Consequently, the total emission may be depolarized. The fact that this depolarization is limited in radio relics implies that the magnetic fields may be partially ordered and/or that the emitting volume is relatively small compared to the characteristic length scale over which the fields vary. Radio relics fall into three distinct categories: relic AGN bubbles, radio phoenix, and radio gischt (or radio shocks) that we describe below.\\
\indent
The injection of energy by supermassive black holes at the centers of galaxy clusters leads to the formation of hot and underdense bubbles. These bubbles contain relativistic CR electrons and magnetic fields as evidenced by the detection of radio emission from within X-ray cavities in galaxy clusters (see left panel in Fig.~\ref{fig:relics}). Hence these radio structures are referred to as \textit{relic AGN bubbles}. Since the characteristic buoyant rise timescale of the bubbles is comparable to the sound crossing time \citep[see, e.g.,][]{churazov_evolution_2001}, and thus can be longer than the synchrotron plus IC cooling timescale for typical observing radio frequencies (see equation~\ref{eq:t_cool,max}), the radio emission from the bubbles may be quickly extinguished leading to the formation of ``ghost bubbles'' \citep{enslin_radio_1999,enslin_radio_2002}, particularly at larger distances from the cluster centers (see outer cavities devoid of radio emission in the same figure). This picture is consistent with low frequency LOFAR radio observations of nearby clusters \citep{birzan_lofar_2020} that reveal a spatial correlation between X-ray cavities and radio emission in about 50\% of massive clusters that show X-ray cavities associated with AGN feedback. A non-detection of radio emission in the remaining massive clusters could be the result of fast aging of CR electrons, in which case even lower frequency observations would be needed to detect older electron populations. This is because higher energy CR electrons radiating at relatively higher radio frequencies lose energy faster, which leads to spectral steepening and thus dimming of the source. However, the pressure support in the bubbles could instead be dominated by CR protons (see Section~\ref{content}) that do not produce appreciable radio emission. Alternatively, CR streaming in the high-plasma-$\beta$ ICM could be significantly enhanced \citep{wiener_high_2018} helping to remove the synchrotron emitting CRs from the bubbles and their vicinity thus suppressing radio emission.\\
\indent
The aging population of CRs in relic AGN bubbles and ``ghost bubbles'' could be revived by passing shock waves \citep{enslin_reviving_2001}, e.g., those associated with cosmological accretion. This leads to the structures termed \textit{radio phoenixes} (or phoenices). These objects are characterized by a steep radio spectrum, significant polarization, and extended low surface brightness distribution \citep[e.g.,][see also middle panels in Fig.~\ref{fig:relics} and references in the caption]{van_weeren_diffuse_2019}. The formation of these structures can be understood by considering gas flows in the frame of reference of the shock wave in the vicinity of the point of contact between the shock and the AGN bubble \citep{enslin_formation_2002}. In this frame, the ram pressure of the pre-shock plasma approximately balances the thermal pressure of the shocked gas when the shock Mach number is large. As the shock encounters the bubble, the ram pressure in the upstream region drops significantly because the bubbles are filled with light fluid (hot dilute gas, magnetic fields, CRs). The shock then accelerates inside the bubble and its speed is largest close to the axis connecting the bubble center and the first point of contact of the shock with the bubble. This implies the formation of a curved shock that injects axially symmetric vorticity, thereby transforming the quasi-spherical bubble into a torus structure with a substantially reduced volume. Consequently, CRs and the magnetic field in the bubble are adiabatically compressed and gain energy. Since the amplification due to adiabatic compression applies only to the magnetic field component perpendicular to the direction of compression, the resulting magnetic field is partially ordered so that it largely wraps around the outer radius of the torus. Adiabatic heating of CRs together with compression and ordering of the magnetic fields, leads to the enhancement in the radio emission that is characterized by significant polarization and a steep spectrum (note that adiabatic CR heating does not change the shape of the already aged CR population). This leads to the formation of toroidal radio-emitting objects on scales up to $\sim200$~kpc, which enables one to characterize the properties of the transforming shock wave such as Mach number and curvature radius  \citep{pfrommer_radio_2011}. Slow turbulent motions in the ICM can then easily distort the tori thus forming filamentary structures.\\
\indent
In addition to radio relic AGN bubbles and phoenixes, a related class of objects termed radio filaments has been identified. These narrow radio structures are seen in the radio galaxy lobes \citep{hines_filaments_1989}, peripheral radio relics \citep{rajpurohit_deep_2022}, and in the surroundings of radio galaxies \citep{ramatsoku_collimated_2020,botteon_beautiful_2020,van_weeren_lofar_2021,condon_threads_2021}. While the nature of these structure is not fully understood, it is likely that their emission is associated with the re-acceleration of seed electrons originally injected by current or past AGN activity \citep[][ see also discussion below]{rudnick_intracluster_2022}.\\
\indent
Structure formation shocks can accelerate CR electrons in situ. These merger and accretion shocks are typically found at cluster outskirts and CR acceleration at their location can lead to the formation of \textit{radio gischt} \citep[or radio shocks,][]{ensslin_cluster_1998}. Radio gischt is strongly polarized due to magnetic field predominantly aligned with the shock surface, and can often be observed in pairs on the opposite sides of clusters \citep[e.g.,][see also right panels in Fig.~\ref{fig:relics} and references in the caption]{van_weeren_diffuse_2019}. The origin of these structures is not completely understood. For example, CR acceleration is not expected to be efficient in the case of magnetic fields aligned with the shock surface \citep[e.g.,][]{winner_evolution_2020}. Furthermore, acceleration of CRs starting from a purely thermal pool of particles is unlikely to be sufficient to explain observations as it would require unphysically efficient CR acceleration \citep[e.g.,][]{macario_shock_2011}. An alternative solution may involve Fermi first-order re-acceleration of long-lived fossil CR electron population \citep{pinzke_giant_2013,kang_re-acceleration_2016}. At the outer cluster regions, these CR electrons have maximum lifetimes exceeding the age of the Universe near Lorentz factors $\gamma\sim 10^{2}$, and the lifetimes decrease both toward lower and higher electron energies due to (i) Coulomb and (ii) synchrotron and IC losses, respectively \citep[][see also Fig.~\ref{fig:cooling_times}]{sarazin_energy_1999}. The seed electrons for the shock re-acceleration process with $\gamma\sim10^{2}$ may have come from earlier structure formation shocks, galactic winds, or AGN feedback \citep{brunetti_cosmic_2014}. 

\subsubsection{Radio halos}
Radio emission in galaxy clusters can also come in the form that is more spatially extended than in the case of radio relics. This diffuse emission is unpolarized, is characterized by steep spectra with $\alpha\sim 1-1.5$, and tends to be centered on cluster centers. Objects belonging to this class are called radio halos \citep{brunetti_particle_2001,feretti_non-thermal_2005,de_gasperin_m_2012} and fall into the category of \textit{giant radio halos} (see left panel in Fig.~\ref{fig:halos}) or \textit{radio mini halos} (right panel in the same figure). The former occur in merging clusters and have sizes $\sim 1$ Mpc, i.e., are comparable to a large fraction of the sizes of clusters themselves. The latter are found in relaxed cool core clusters and their sizes are a few hundred kpc, which is comparable to the ICM cooling radii (i.e., the radii, where the cooling time is comparable to the Hubble time). Since significant spectral steepness makes it difficult to detect radio halos at high radio frequencies, low-frequency observations with telescopes such as Low Frequency Array (LOFAR) or Giant Metrewave Radio Telescope (GMRT) are best suited for studying these objects \citep{brunetti_low-frequency_2008}. Recently, LOFAR observations have revealed more extended radio emission surrounding four giant halos that reaches out to $R_{500}$ (i.e., a radius where the internal mean density equals 500 times the critical density of the universe). This radio emission exhibits an approximately flat emissivity profile at a level of about 20 times lower than the emissivity of classical giant radio halos \citep{cuciti_galaxy_2022}. 
Similarly, much deeper LOFAR observations of Abell 2255 revealed that non-thermal emission extends to even larger distances, up to $r_{200}$ \citep{botteon_magnetic_2022}.
High-resolution and signal-to-noise observations of radio mini halos reveal that the diffuse radio emission can also exhibit substructure in the form of filaments \citep{gendron-marsolais_deep_2017,botteon_magnetic_2022}, which indicates that the distributions of the magnetic fields and/or CR electrons may be localized. Furthermore, the radio mini halo emission can be confined to the center by sloshing fronts (see, e.g., right panel in Fig.~\ref{fig:halos} showing the radio mini halo RX J1720.1+2638).\\
\indent
Lack of significant observed polarization indicates that the intrinsically polarized synchrotron radiation undergoes Faraday depolarization and/or beam depolarization in the turbulent and magnetized ICM. This in turn suggests that the radio emission should be volume-filling. An additional argument in favor of a volume-filling emission mechanism in radio halos is as follows. We can estimate a characteristic CR electron cooling length defined as the distance travelled by a parcel of gas moving at typical turbulent ICM velocity during one synchrotron plus IC cooling time, so that CR electrons are advected with the flux-frozen magnetic field (assuming that the streaming/diffusion velocity is smaller or equal to this advection velocity). Adopting a typical level of turbulent pressure support in the ICM of $X_{\rm turb} = 0.1$, the corresponding turbulent velocity is $\varv_{\rm turb}\sim400\; {\rm km\; s}^{-1}(X_{\rm turb}/0.1)^{1/2}(c_{\rm s}/10^{3}~\rm km\; s^{-1}$). Since the cooling time $t_{\rm cool}$ at a typical observing frequency of 1.4 GHz is less than $\sim200$ Myr (see equation~\ref{eq:t_cool,max}), the cooling length $L_{\rm cool}=t_{\rm cool}\varv_{\rm turb}\lesssim 80$ kpc. As this is much smaller than the extent of giant radio halos, the observed emission must come from volume-filling and spatially uncorrelated emission regions rather than from localized sources embedded in the ICM. A similar argument applies to much smaller radio mini halos, where the cooling length is significantly smaller due to much lower level of turbulence. While the emission mechanism in radio halos is a hotly debated topic, any mechanism aiming to explain the emission must account for volume-filling in order to address the generic constraints imposed by the fast CR electron cooling and lack of net polarization.

\subsubsection{Hadronic and re-acceleration models of radio halos}
Two contenders for the emission mechanism in radio halos are the re-accleration \citep{jaffe_origin_1977,brunetti_particle_2001} and hadronic models \citep{dennison_formation_1980,blasi_cosmic_1999}. 
In the \textit{hadronic model}, CR protons collide with thermal ICM protons to produce pions, which subsequently decay to produce secondary electrons and positrons and $\gamma$ rays among other reaction products (see Section~\ref{hadronic}). The secondary electrons (and positrons) then interact with the magnetic fields to produce synchrotron radio emission. Since the proton cooling time due to such hadronic interactions is very long (see right panel in Fig.~\ref{fig:cooling_times}), CR protons can accumulate in the ICM throughout cluster lifetimes in the volume-filling fashion. In this model, CR protons originate in shocks due to structure formation \citep[or possibly also AGN and starburst outflows;][and references therein]{donnert_radio_2010}. This model naturally solves the problem of short CR electron cooling times mentioned above (as the secondary electrons can be essentially continuously produced throughout the ICM volume) in addition to correctly predicting radio halo luminosities and spectra \citep{wiener_cosmic_2013}. However, in its simplest incarnation, the model predictions contradict the observed bimodality of cluster radio distribution -- while the model predicts that all clusters should have giant radio halos (as all clusters experienced structure formation shocks that injected CR protons), only merging clusters possess giant radio halos \citep{feretti_clusters_2012}. A possible solution to this problem may require invoking very low CR shock acceleration efficiencies. Alternatively, fast streaming of CR protons out of cluster centers \citep{enslin_cosmic_2011}, facilitated by turbulent damping and strong ion Landau damping in high-$\beta$ plasma such as the ICM, could switch off radio emission in a population of clusters \citep{wiener_high_2018}.\\
\indent 
In the \textit{re-acceleration model}, electrons are accelerated via second-order Fermi processes involving electron interactions with MHD turbulence in the ICM, as a result of which the electrons gain sufficient energy to produce the radio emission. The seed electron population for the re-acceleration process may come from radio galaxies, past episodes of AGN activity, or CR protons that injected secondary electrons. This model also predicts spectra in agreement with observations \citep{brunetti_acceleration_2011}. Importantly, it naturally predicts the bimodality in cluster radio distribution -- giant radio halo emission is present when cluster mergers drive ICM turbulence which re-accelerates seed electrons, and subsequent decay of turbulence switches off radio emission. However, the model leads to radio emission profiles that are too centrally peaked compared to observations, thus requiring introducing changes to the standard model such as enhancements in the CR streaming speed and electron acceleration efficiency at shocks or turbulent-to-thermal energy density increasing with the cluster-centric radius \citep{pinzke_turbulence_2017}.\\
\indent
Some aspects of the above considerations have implications not only for giant radio halos but also for radio mini halos. Specifically, since minihalos are often found in relaxed cool core clusters not undergoing major mergers \citep{giacintucci_occurrence_2017}, the level of turbulence is lower and, consequently, the re-acceleration model, while not excluded, may be less promising in this case \citep[but see][]{gitti_modeling_2002}. 
While even relaxed cool core clusters can show signs of gas sloshing \citep[e.g., in the Perseus clusters;][]{ichinohe_substructures_2019} and exhibit signs of AGN feedback in the form of centrally located X-ray cavities \citep[e.g.,][]{donahue_baryon_2022}, the level of turbulence is typically low as evidenced by direct measurements of small gas velocities in the bulk of cool core ICM \citep{hitomi_collaboration_quiescent_2016}, especially given that only a fraction of the kinetic energy in the ICM is in the turbulent form. This inefficient driving of turbulence is consistent with theoretical expectations for turbulence excitation by AGN (see Section \ref{agntheorysection}).
However, the same AGN could serve as a source of CR protons. These CR protons could be distributed over the mini halo volume via CR advection by buoyant AGN bubbles and CR streaming. In the hadronic model, only a relatively small level of CR pressure buildup from intermittent AGN activity would be sufficient to power the mini halos \citep[e.g.,][and the discussion below]{pfrommer_constraining_2004,jacob_cosmic_2017-1}. Thus, unlike in the case of giant radio halos, AGN feedback may be directly responsible for powering the radio mini halos. This is consistent with the fact that mini halos are observed at the centers of cool core clusters most of which host AGNs.

\subsubsection{Gamma-ray emission limits}
Unlike in the case of the ISM, CRs with energies below $10^7$~GeV can be confined in the ICM. Given that CR proton cooling times due to hadronic interactions can be comparable to the Hubble time, CRs can accumulate in the ICM and have sufficient time to propagate away from their production sites to fill the entire ICM volume. Besides creating secondary electrons responsible for producing synchrotron radio emission discussed above, these same hadronic reactions also produce $\gamma$-ray photons \citep[e.g.,][]{volk_nonthermal_1996,colafrancesco_clusters_1998,pfrommer_simulating_2008}. This emission can then serve as a historical record of past CR shock acceleration due to cluster mergers and AGN feedback. In addition to the hadronic processes involving interactions of CR protons with the protons in the ambient gas, $\gamma$-ray emission can also be produced as a result of the interaction of CMB photons with very high-energy CR ions (with energies in excess of $10^{18}$~eV) accelerated in cluster accretions shocks. In this process, these interactions may efficiently create $e^{-}-e^{+}$ pairs with sufficient energies to produce $\gamma$ rays via IC scattering of the CMB photons. This mechanism also predicts that the interaction of electrons and positrons with the magnetic fields should generate hard X-ray emission \citep{inoue_hard_2005,vannoni_acceleration_2011}. More broadly, models of giant radio halos and radio mini halos predict that the same relativistic electrons that power the radio emission in these objects should also upscatter CMB photons to extreme UV and hard X-ray band \citep[e.g.][]{rephaeli_nonthermal_2008,nevalainen_xmm-newton_2009}. However, any conclusive detection of non-thermal X-ray emission is challenging due to its intrinsic weakness combined with significant thermal background emission from the hot ICM and instrumental effects.\\
\indent
Given that CR pressure support in the ICM is generally expected to be low, the above models predict $\gamma$-ray emission that is relatively weak. In fact, most attempts to detect $\gamma$-ray signal from clusters resulted in upper limits. This is the case for individual clusters \citep[e.g.,][]{ackermann_gev_2010} based on \textit{Fermi} data, the Perseus cluster \citep{aleksic_constraining_2012,ahnen_deep_2016} based on MAGIC observations, the Coma cluster based on \textit{Veritas} and \textit{Fermi} observations \citep{arlen_constraints_2012}, or stacked \textit{Fermi} data from multiple objects \citep[e.g.,][]{dutson_stacked_2013,ackermann_search_2014}. 
However, direct detections of $\gamma$-ray emission from cluster centers have also been reported. For example, there exist detections based on the \textit{Fermi} observations of M87, the central AGN of the Virgo cluster \citep{abdo_fermi_2009-1}, NGC~1275, the central cD galaxy of the Perseus cluster \citep{abdo_fermi_2009}, and from the Coma cluster \citep{xi_detection_2018}. However, the interpretation of these detections is challenging due to the difficulty in disentangling point-like and diffuse emission components \citep{ackermann_search_2015,colafrancesco_disentangling_2010,xi_detection_2018}.\\
\indent
The methods employed to put observational upper limits on the $\gamma$-ray emission rely on assumptions regarding the underlying distribution of CRs in the ICM. For example, in their stacking analysis, \citet{ackermann_search_2014} assume that CR distribution follows the prediction for the CR distribution from diffuse shock acceleration model at cosmological structure formation shocks and subsequent advection with the ICM \citep{pinzke_simulating_2010}. Similarly, \citet{huber_probing_2013}, consider a range of model distributions where the CR pressure distribution follows that of the thermal gas, or where the CR-to-thermal pressure is either described by a weakly increasing power law dependence with increasing distance from cluster centers (in order to emulate the expectations from models of CR acceleration in structure formation shocks) or a weakly decreasing power law function of radius (in order to attempt to model CR injection by AGN). While there are significant differences depending on the assumed profile, typical upper limits on CR pressure support reported in these studies are at the level of about a few percent of the thermal pressure when averaged over the entire cluster sample. Importantly, the CR-to-thermal pressure ratios represent averages over the entire cluster volumes up to cluster virial radii, potentially leaving room for more spatially-localized departures from these typical values. 
Nevertheless, with these caveats in mind, it is worth pointing out that the inferred limits on the level of CR pressure support in cool core clusters are significantly higher than in non-cool core ones, reaching values as high as a few tens of percent \citep{huber_probing_2013}. For example, in the closest cluster Virgo, where the radio morphology suggests that CRs are largely confined to central region within $\lesssim 30$~kpc, \textit{Fermi} and H.E.S.S. constraints allow CR-to-gas pressure ratio at the level of $\sim 30$ percent \citep{pfrommer_toward_2013}.\\
\indent
Upper limits on CR pressure support and $\gamma$-ray flux may help to constrain AGN feedback models. While simple one-dimensional steady-state models of CR feedback can offset radiative cooling and match the observed ICM density and temperature distributions \citep{jacob_cosmic_2017}, they systematically overpredict hadronically induced radio emission (and $\gamma$ rays in some cases) in a sub-sample of clusters hosting radio mini halos \citep{jacob_cosmic_2017-1}, while being fully consistent with the data for non-mini halo clusters. This suggests that CR heating may be insufficient to fully offset cooling in clusters with mini halos (i.e., that steady-state solutions are not possible in that sub-sample). The star formation rates and cooling radii in these objects are in fact larger, which is consistent with this suggestion and the idea of a feedback duty cycle where a fraction of cool cores is currently actively cooling without strong signatures of AGN radio bubbles \citep{ubertosi_waking_2023}, thus reconciling the overall model with the data.
Interestingly, the AGN radio fluxes in the sub-sample that does not host radio mini halos is exceptionally high, and the CR injection by these AGN may be adequate to offset cooling in this case. CR protons that have heated the cool core during the (earlier) non-mini halo phase may then provide secondary electrons to power the radio mini halos. 
However, the generally low observationally-inferred level of the volume-averaged CR pressure support may also suggest that CR injection is either relatively inefficient and/or that the redistribution of CRs via fast streaming \citep{wiener_high_2018} or other means, as well as changes in the ratio of hadrons to leptons in the model \citep{wiener_constraints_2019}, may be needed to reduce the radio emission and reconcile the model predictions with the observations.

\subsection{Cosmic rays from AGN jets in galaxy groups and clusters}\label{content}
While there is direct evidence only for the existence of CR electrons in the bulk of the ICM, we have better arguments for a hadronic CR population in and around AGN jets. As in the bulk of the ICM, radio synchrotron observations provide direct evidence for the existence of CR \textit{electrons} while there is 
tentative evidence for the CR \textit{proton} content in the AGN-inflated bubbles that comes from \textit{Chandra} X-ray observations that put lower limits on the temperature of the thermal gas component inside the AGN bubbles \citep{sanders_deeper_2007}, and from XMM-\textit{Newton} constraints on non-thermal X-ray spectral components \citep{sanders_non-thermal_2005,werner_possible_2007} as well as from Sunyaev--Zel'dovich observations \citep{abdulla_constraints_2019,orlowski-scherer_gbtmustang-2_2022}. In addition, there are hints for non-thermal CRs in ICM filaments from strong emission line measurements. Collectively, these observations suggest that the bubbles may be dominated by non-thermal pressure, which we discuss in more detail below. The content of the lobes and bubbles inflated by the AGN jets has implications for the heating of the ambient medium as the particles escaping from these structures begin to heat the gas via streaming heating and hadronic and leptonic processes (see Section~\ref{agntheorysection}). Below we discuss the phenomenology of AGN radio emission, CR content of AGN-inflated bubbles, and the consistency of the models with the observational constraints.

\subsubsection{Non-thermal content of lobes associated with FRI and FRII sources}
Before we discuss the role of CRs in AGN jet feedback, we review the different types of radio galaxies, which can be classified according the morphology of the radio emission associated with their outflows \citep{fanaroff_morphology_1974}. Fanaroff--Riley Type II (FR~II) sources are characterized by edge-brightened emission and the presence of ``hotspots'' at the ends of their lobes. These sources differ from Fanaroff--Riley Type I (FR~I) sources, which are brighter closer to the central galaxy and dimmer toward the lobes. The physical origin of the dichotomy is likely associated with differences in AGN jet power and environmental factors \citep[e.g.,][]{hardcastle_radio_2020}. FR~II sources tend to be more powerful than the ones in FR~Is \citep[][and references therein]{tchekhovskoy_three-dimensional_2016}. FR~II jets remain highly relativistic throughout their evolution up to the location of the hotspots. Consequently, these jets often appear one-sided due to relativistic Doppler beaming. While FRI sources also possess relativistic jets, their jets slow down significantly over much shorter distances from the central supermassive black hole. Furthermore, FR~II sources preferentially occur in isolated galaxies and FRI tend to occupy centers of groups and galaxy clusters \citep[e.g.,][]{rodman_radio_2019}. The interactions of FRI jets with relatively denser ambient medium leads to their deceleration on kpc scales. Consequently, the emission close to the centers of FRI sources is bipolar rather than one-sided.\\
\indent
Equipartition between the magnetic and particle energy densities is commonly invoked to put constraints on the non-thermal contributions to plasma pressure. More specifically, this assumption, combined with an estimate for the ratio of the total energy of CR ions to that of the emitting CR $e^{-}$ and $e^{+}$, allows one to infer the magnetic pressure (given the observed synchrotron flux) and the corresponding pressure of the non-thermal radiating particles \cite[e.g.,][]{pfrommer_estimating_2004,beck_revised_2005}. However, in FR~II sources the number density of leptons in lobes and hotspots can be measured directly from IC emission. In this case, the magnetic field can be directly obtained from synchrotron emission and the measured number density of electrons and positrons without relying on the above equipartition assumption. While in the radio lobes the sum of the magnetic and CR electron pressures obtained this way is often able to balance that of the ambient ICM pressure, in the hotspots it can exceed the ambient pressure. Therefore, CR protons do not have to be energetically dominant to explain the observations in FR~II sources \citep{croston_particle_2018}.\\
\indent
These kinds of measurements are very difficult in the case of FR~I sources, which tend to reside in denser environments. Most cool core clusters possess central AGNs \citep[e.g.,][]{donahue_baryon_2022} and many host FR~I sources. An example of radio emission from M87, the cD galaxy in the closest galaxy cluster Virgo, is shown in Fig.~\ref{fig:aLOFAR}. The image shows collimated jets, AGN-inflated bubbles, and diffuse emission distributed over the cool core of the cluster. Detecting the IC signal from FR~I sources is very challenging because the non-thermal signal can be easily swamped by thermal emission from the ambient ICM. This makes a direct measurement of the number density (and pressure) of the non-thermal electrons and positrons very difficult. However, future high-throughput X-ray missions may be able to constrain the contents of AGN bubbles by detecting the IC signal in hard X-rays \citep{ruszkowski_unravelling_2019,ruszkowski_supermassive_2019}. 
It is worth noting in passing that, despite the above challenges, direct detection of an \textit{extended} IC signal in hard X-rays is currently possible in galaxy groups. This is because thermal emission in groups is lower than in clusters and is expected to peak in soft X-rays. Consequently, non-thermal X-ray power-law emission can be directly detected as it dominates over the thermal component in hard X-rays. Combined with an independent detection of diffuse synchrotron emission, this allows one to directly measure the strength of the magnetic field \citep{mernier_discovery_2022}.\\
\indent
Estimates based on the equipartition assumption suggest that the pressure due to non-thermal electrons, positrons and magnetic fields in the AGN-inflated bubbles 
associated with FR I sources
is much smaller than the pressure of the ambient thermal ICM. Thus, in order to balance the pressure of the ambient ICM, these bubbles must be supported predominantly by non-thermal pressure component most likely in the form of CR protons \citep[e.g.,][]{morganti_low_1988,croston_xmm-newton_2008,croston_particle_2018}. While, in principle, the extrapolation of the power-law distribution of the radio-emitting electrons in the FR~I lobes to low energies could increase the pressure support due to CR electrons, the lack of LOFAR detection of significant very low frequency emission from such electrons rules out this possibility \citep[][in the case of an FR~I source in the Virgo cluster]{de_gasperin_m_2012,pfrommer_toward_2013}. Given that CR electron cooling times are much shorter than those of CR protons and that the acceleration efficiency of CR protons is expected to exceed that of CR electrons, it is very likely that indeed the dominant non-thermal pressure component in the AGN bubbles associated with FR~I sources comes from CR protons. Gas entrainment during the interaction of a Poynting-flux-dominated jet with the ambient ICM, and associated with it particle acceleration via internal shocks close to the jet base, could be responsible for the origin of these CR protons \citep{croston_particle_2014}.\\
\indent 
If the AGN bubbles are by construction assumed to be in pressure equilibrium with the surrounding gas, the inferred magnetic energy and particle energy densities are severely out equipartition \citep{dunn_particle_2004}. These arguments lead to constraints on $k/f$, in which $f$ is defined as the volume filling factor of the radio-emitting plasma, and $k$ which is the ratio of the total particle energy to that of the electrons radiating between 10 MHz and 10 GHz. The distribution of $k/f$ appears to be bimodal and characterized by peak values of 3 and 300. The origin of the bimodality could be due to systematic differences in the volume filling factor, particle acceleration efficiencies, or bubble composition.\\
\indent 
The above systematic differences in the particle composition of the FR~II and FR~I outflows (and more specifically the large differences between the masses of the constituent particles, i.e., electrons and protons) are consistent with the finding that jet power-to-radio luminosity ratios in FR~II radio galaxies are significantly smaller than in FR~I sources \citep{croston_particle_2018}.

\subsubsection{Gamma-ray constraints on the contents of AGN bubbles}
Further constraints on the composition of the AGN bubbles could be gleaned from comparing model predictions for the $\gamma$-ray emission to the observational constraints. If the bubbles are dominated by CR pressure, the $\gamma$-ray signal due to hadronic interactions of CRs with the residual thermal gas inside the bubbles is not expected to produce an appreciable $\gamma$-ray emission. However, once the CRs start diffusing out of the bubbles and interacting with the ambient ICM, they can produce $\gamma$ rays. It has been suggested that the existing upper limits on $\gamma$-ray emission could place severe lower limits on CR confinement times inside the bubbles \citep{prokhorov_confinement_2017}. 
These limits could be relaxed if CR streaming transport is included \citep[e.g.,][]{guo_feedback_2008,ruszkowski_cosmic-ray_2017,ehlert_sunyaevzeldovich_2019}. CR streaming can not only help to transport CR out of the cool core, but CR streaming heating losses are expected to dominate over hadronic losses, and are not associated with $\gamma$-ray production.
Using hydrodynamical simulations including CR diffusion and spanning relatively short evolutionary times $\lesssim 100$ Myr, \citet{yang_impact_2019} demonstrated that the predicted $\gamma$-ray flux falls below current observational limits in the hadronic CR model. On the other hand, in the leptonic model, the predicted fluxes exceeded observational upper limits by up to 50 times, suggesting that CR electron pressure contribution to the pressure balance should be below 2 percent, i.e., that the bubbles should be dominated by CR proton (or magnetic) pressure. However, these hydrodynamical simulations excluded synchrotron aging of electrons, which could relax these constraints. Further generalization of the hadronic model to consider self-regulated AGN feedback cycles demonstrated that CR-dominated jets violate upper $\gamma$-ray limits unless the CR fraction supplied by the jets is at the level of 10 percent \citep{beckmann_cosmic_2022}. Interestingly, even such low injected CR fractions are essential for preventing a strong cooling flow in these simulations and facilitate feedback self-regulation over billions of years due to CR streaming heating. The effective level of $\gamma$-ray emission may however depend on how the CRs escaping from the bubbles correlate spatially with the cold and dense ICM. Observational and theoretical arguments suggest that CRs interact with the cold ICM filaments \citep[][see also Section~\ref{filaments}]{ferland_origin_2008,ferland_collisional_2009,ruszkowski_powering_2018}, and that the $\gamma$-ray flux may be affected by the details of the physics of CR transport in the cold phase. 
\begin{figure}
  \begin{center}
    \includegraphics[width=1.0\textwidth]{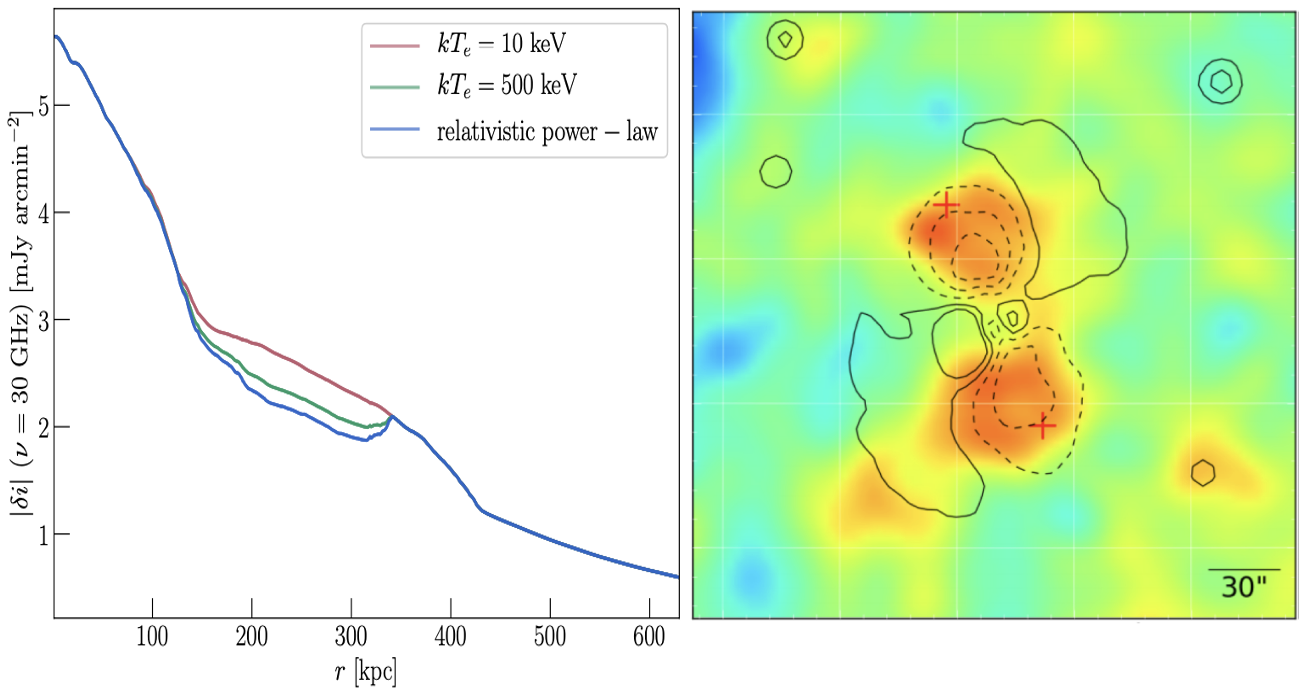}
  \end{center}
      \caption{\textit{Left}: Simulated profiles of SZ signal for different compositions of the AGN-inflated cavities. Notice that, unlike the bubbles filled with $\sim 10$ keV thermal gas, the ones filled with relativistic CR electrons show clear decrement in the SZ signal, with the strongest depression corresponding to the power-law distribution of electrons. Figure from \citet{ehlert_sunyaevzeldovich_2019}; reproduced with permission from ApJ. 
      \textit{Right:} Map of the residuals between the SZ data and model that only accounts for the cluster SZ effect. Positive values (shown in red) indicate a deficit of the SZ signal, which is consistent with CR-dominated AGN bubbles. For comparison, the solid and dotted black lines delineate the excess and dearth of X-ray surface brightness, respectively, after subtracting the best-fit double-$\beta$ model of the ICM density. Figure from \citet{abdulla_constraints_2019}; reproduced with permission from ApJ.
      }
  \label{fig:abdulla}
\end{figure}

\subsubsection{Probing non-thermal AGN bubble content using the Sunyaev-Zel'dovich effect} An independent approach to probing the contents of the AGN-inflated bubbles relies on the SZ effect \citep{sunyaev_observations_1972,pfrommer_unveiling_2005}. The magnitude of the intensity distortions $\delta i=g(x)y$ in the CMB due to the dominant thermal SZ effect is proportional to (i) the Compton-$y$ parameter 
\begin{align}
    y=\int^{L_\rmn{cmb}}_0 \sigma_\rmn{T} n_\rmn{e} \frac{k_{\rm B}T}{m_{\rm e}c^{2}}\rmn{d}l\sim \tau \frac{k_{\rm B}T}{m_{\rm e}c^{2}},
\end{align}
which to order of magnitude is a product of the optical depth of free electrons $\tau$ and the fractional energy gain per scattering $\sim k_{\rm B}T/(m_{\rm e}c^{2})$ experienced by the CMB photons due to the IC process, and (ii) the function $g(x)$ which describes the shape of spectral distortions of the CMB as a function of $x=h\nu/(m_{e}c^{2})$. As such, the thermal SZ effect measures the gas pressure integrated along the line of sight. Since $g(x)$ is of order unity for $x\lesssim 10$, and both $\tau$ and the fractional energy gain are very small, the magnitude of this effect is small albeit measurable. Depending on the observing frequency, the SZ effect can either produce a decrement or increment in the CMB signal.\\
\indent
The magnitude of the impact of CR electrons is easiest to understand in the ultra-relativistic limit (i.e., when electron energies significantly exceed $m_{\rm e}c^{2}$). Unlike in the case of the thermal SZ, the change in the intensity of the CMB due to photon upscattering is then always negative (in the vicinity of the CMB peak) and is simply proportional to the optical depth of the electrons that upscatter CMB photons to very large energies, i.e., $\delta i_{\rm rel}=-i(x)\tau_{\rm rel}$, where $i(x)$ describes the shape of the Planck spectrum and $\tau_{\rm rel}$ is the optical depth of the relativistic electrons \citep{enslin_comptonization_2000}. 
If AGN bubbles containing a non-thermal pressure component are present in the ICM, they may imprint significantly different distortions on the CMB than the thermal gas. If such bubbles are CR-dominated and in pressure equilibrium with the ambient ICM, the number density of CRs inside them will be vanishingly small compared to the number density of the thermal gas in the bulk of the ICM. In this case, and away from the characteristic crossover frequency 217 GHz where $g(x)$ changes sign, the ratio of the expected magnitude of the change in CMB intensity to that of the thermal SZ effect  $|\delta i_{\rm rel}/\delta i|\sim |i(x)/g(x)|(\tau_{\rm rel}/\tau)(k_{\rm B}T/m_{\rm e}c^{2})^{-1}\ll 1$. Consequently, non-thermal bubbles could become visible as regions characterized by SZ signal weaker than thermal SZ  \citep{pfrommer_unveiling_2005} or produce stronger-than-thermal SZ signal when observed near the crossover frequency, provided the magnitude of the kinematic SZ effect of the cluster and the relativistic corrections are perfectly known \citep{colafrancesco_sz_2005,prokhorov_comptonization_2010}.\\
\indent
Detailed modelling of the SZ effect using hydro and MHD simulations of AGN bubble inflation including dynamically important CR component inside the bubbles confirms that CRs leave a clear imprint in the overall cluster SZ signal. While in the case of kinetic jets dominated by hot thermal plasma, the SZ contrast produced by the jets is negligible, jets dominated by CRs show a detectable decrement in the SZ signal \citep[][see also the left panel in Fig.~\ref{fig:abdulla}]{yang_impact_2019,ehlert_sunyaevzeldovich_2019}. Constraints on the bubble content based on the SZ effect should carefully take into account degeneracies that stem from the jet orientation with respect to the line of sight. When the jet is in the plane of the sky, the SZ contrast introduced by the non-thermal component is maximized. For jets nearly aligned with the line of sight, some of the expected thermal decrement may be compensated for by the increase in the SZ signal associated with the shock-compressed gas outside the rapidly inflating bubbles. Not accounting for this effect would result in an underestimate of the non-thermal bubble content. However, this degeneracy could be broken if the kinetic SZ effect, characterized by different signs of the signal in opposite lobes, could be detected \citep{ehlert_sunyaevzeldovich_2019}.\\
\indent 
A clear SZ contrast associated with AGN bubbles has recently been detected. The right panel in Fig.~\ref{fig:abdulla} shows the residuals of the SZ signal at 30 GHz obtained with Combined Array for Research in Millimeter-wave Astronomy (CARMA) after removing a smooth ICM background model \citep{abdulla_constraints_2019}\footnote{Note that, at this observing frequency, the SZ effect manifests itself as a decrement in the CMB signal.}. The residuals shown in red coincide with the radio synchrotron lobes and X-ray cavities inflated by a powerful central AGN in this cluster (MS 0735.6+7421), and represent a first clear detection of non-thermal pressure component in AGN bubbles. These results indicate that the bubbles must be either completely filled with very diffuse thermal plasma that is hotter than several hundreds of keV or that the bubble pressure support is non-thermal, i.e., in the form of relativistic electrons/positrons/ions and/or magnetic fields (the SZ contribution due to ions is negligible also because of their high mass). 
Using MUSTANG-2, \citet{orlowski-scherer_gbtmustang-2_2022} find much smaller suppression of the SZ signal in this object  but nevertheless confirm that if the gas inside the bubbles is supported by thermal pressure, then its temperature must be at least $\gtrsim~200$ keV. It is worth pointing out that even if the fraction of the energy injected into the bubbles close to the jet basis is small, the relatively small effective CR adiabatic index (4/3) compared to that of the thermal gas (5/3) means that the CR pressure should be progressively more important as the bubble adiabatically expands and grows in size before significant mixing with the ambient ICM takes place. Future high-angular-resolution multi-frequency data from CCAT-prime, NIKA2 (The New IRAM Kids Arrays), MUSTANG2, TolTEC, AtLAST (The Atacama Large Aperture Submm/mm Telescope), CSST (Chajnantor Sub/millimeter Survey Telescope), or CMB-in-HD may provide further constraints on CR feedback in clusters \citep{mroczkowski_astrophysics_2019,mroczkowski_high-resolution_2019,sehgal_science_2019}.

\subsection{Current and future multi-messenger observatories}
\label{sec:multi-messenger_observatories}
We now focus on current and future observational capabilities associated with non-thermal physics that can help probe CR acceleration and feedback. There are a number of CR experiments on Earth that measure and characterize the secondary CR particles that result from air showers generated by a primary CR particle upon entering the Earth's atmosphere or directly interacting with the AMS experiment \citep[Alpha Magnetic Spectrometer,][]{aguilar_first_2013} or CALET \citep[CALorimetric Electron Telescope,][]{adriani_energy_2017} onboard the International Space Station. This allows studies of the CR electron and ion spectrum of different compositions at a rigidity larger than several GV \citep{aguilar_electron_2014,aguilar_precision_2015}, which are not affected by the CR flux modulation due to the solar wind (see Fig.~\ref{fig:fig1}). This is complemented by measurements of low-energy CR spectra with the \textit{Voyager} spacecraft \citep{cummings_galactic_2016}, which has by now transited the magnetopause of the Solar system and entered the ISM, where it directly characterizes the CR population from several MeV to hundreds of MeV in the ISM, more specifically in the Local Bubble surrounding the solar system. Together, these experiments are thus able to infer the CR spectrum and composition at the solar circle in our Milky Way from MeV energies to ultra-high energy CRs at $10^{20}$~eV. In addition to these in-situ measurements of the CR population in the Milky Way, non-thermal emission is able to probe regions beyond the solar vicinity in the Milky Way out to extra-galactic and cosmological scales.\\
\indent 
A new generation of \textit{radio interferometers} is particularly well suited for investigating CR acceleration and transport in SNRs, galaxies, clusters, and AGNs thanks to the unparalleled sensitivity and angular resolution. At low radio frequencies (30--300 MHz), a number of radio interferometers are in operation: LOFAR \citep[LOw Frequency ARray,][]{van_haarlem_lofar_2013},  MWA \citep[Murchison Widefield Array,][]{tingay_murchison_2013}, LWA  \citep[Long Wavelength Array,][]{ellingson_long_2009}, uGMRT \citep[upgraded Giant Metrewave Radio Telescope,][]{swarup_giant_1991}. At high radio frequencies (0.8--15 GHz), there is the JVLA \citep[Karl Jansky Very Large Array,][]{perley_expanded_2011}, MeerKAT \cite[originally Karoo Array Telescope situated in the Meerkat National Park,][]{jonas_meerkat_2009}, ASKAP  \citep[Australian Square Kilometre Array Pathfinder,][]{johnston_science_2008} and an even higher frequency range (84--950 GHz) is covered by the 
Atacama Large Millimeter/submillimeter Array (\href{https://almaobservatory.org/}{ALMA}). The radio community is already preparing for the next-generation radio telescopes: MWA and ASKAP (both located in Australia) are precursors to the SKA \citep[Square Kilometre Array Telescope,][]{grainge_square_2017} and will be integrated into the low-frequency component of SKA. In South Africa, MeerKAT and HERA \citep[Hydrogen Epoch of Reionization Array,][]{deboer_hydrogen_2017} are SKA precursors that will be integrated into the mid-frequency component of this gigantic radio observatory. The measurements of radio synchrotron emission, in particular total intensity, polarization, polarization fraction, the radio spectrum and Faraday rotation measures by these instruments enable one to address the following main CR-related science issues: (i) magnetic field growth via dynamo processes and their saturation mechanism(s) and (ii) CR electron acceleration, transport, and aging processes. Unfortunately, due to the observational complexity, this is an under-determined problem and it is impossible to fully break intrinsic degeneracies to solve for the magnetic field and the CR electron population separately because: (a) the observables are projected emissivities with different weighting functions, (b) primary and secondary CR electrons contribute to the radio emission, which shows the need to understand the leptonic and hadronic CR components, and (iii) CR electrons and magnetic fields are linked via cooling and transport. This underscores the need for modeling the emission in three-dimensional MHD simulations to improve our understanding of the underlying physics.\\
\indent
The \textit{X-ray band} (0.15--15 keV) probes CR acceleration in SNRs and AGNs as well as turbulence in hot, ionized media such as in galaxy clusters and SNRs. Today, the \textit{Chandra} \citep{weisskopf_overview_2002} and \textit{XMM-Newton} \citep{jansen_xmm-newton_2001} space telescopes provide (high-resolution) imaging and (comparably low-resolution) X-ray spectra. In particular, X-ray synchrotron emission enables probing CR electron acceleration in regions of high magnetic field strengths. XRISM \citep[X-Ray Imaging and Spectroscopy Mission,][]{xrism_science_team_science_2020} is a space-based X-ray calorimeter mission expected to be launched in 2023 and aimed at measuring high-resolution spectra to illuminate the physics of turbulence in X-ray emitting media. ATHENA \citep[Advanced Telescope for High ENergy Astrophysics,][]{nandra_hot_2013} is the future X-ray observatory mission selected by ESA, which combines and improves upon both observational capabilities.\\
\indent
The \textit{$\gamma$-ray regime} (from GeV to PeV $\gamma$ rays) is uniquely suited to probe CR acceleration and transport in SNRs, galaxies, and AGNs (mostly blazars). This is enabled by $\gamma$-ray emission mechanisms due to leptonic emission by IC and non-thermal bremsstrahlung, thus probing CR electrons, density and radiation fields, as well as hadronic $\gamma$-ray emission by pion decay, which probes CR proton transport and their energetics. While the $\gamma$-ray emissivities do not directly depend on the magnetic field, they indirectly depend on it, though, by guiding CR propagation, by enabling CR scattering and by shaping CR electron and thus IC spectra through the synchrotron cooling process. Space-based observations by the Large Area Telescope onboard the \textit{Fermi} Gamma-Ray Space Telescope \citep[at high-energy $\gamma$-ray energies from 0.1--300 GeV,][]{atwood_large_2009} are complemented by targeted observations with ground-based imaging air Cherenkov telescopes (at very-high energy $\gamma$-ray energies from 0.03--10 TeV). The major collaborations using this technique include H.E.S.S. \citep[High Energy Stereoscopic System,][]{bernlohr_optical_2003}, MAGIC \citep[Major Atmospheric Gamma-Ray Imaging Cherenkov Telescopes,][]{aleksic_major_2016}, and VERITAS \citep[Very Energetic Radiation Imaging Telescope Array System,][]{weekes_veritas_2002} which will join forces with the future CTA Observatory \citep[Cherenkov Telescope Array,][]{actis_design_2011}. \\
\indent
These imaging air Cherenkov telescopes perform targeted observations of typical fields of views of several degrees, which are complemented by observations with the HAWC Observatory \citep[High Altitude Water Cherenkov][]{hawc_collaboration_hawc_2013} at $\gamma$-ray energies 0.1--50 TeV. The latter technique uses photomultiplier tubes employed within water tanks instead of the Earth atmosphere to detect the Cherenkov radiation that is emitted by electrons/positrons moving at speeds faster than that of light in the corresponding medium. The leptons are generated in the electromagnetic cascade following the penetration of a very-high energy photon into the Earth's atmosphere. This water-Cherenkov technique allows one to continuously monitor the Northern sky, thus returning a large ensemble of TeV $\gamma$-ray sources, with their spectra and variability to characterize TeV scale acceleration processes. The LHAASO \citep[Large High Altitude Air Shower Observatory,][]{cao_large_2019} collaboration combines the imaging air and water Cherenkov techniques, complemented by electron and muon detectors on ground for exquisite background rejection and thereby started the ultra-high-energy $\gamma$-ray astronomy era in the energy range 10 TeV--1 PeV. The LHAASO experiment in the northern hemisphere will be augmented by the SWGO \citep[Southern Wide-field Gamma-ray Observatory,][]{Huentemeyer_southern_2019} in the southern hemisphere. Together, they will create a comprehensive map of the entire sky, including the plane of the Milky Way and the Galactic Center. This allows for initiating observations of transient and variable phenomena across the entire electromagnetic spectrum and other multi-messenger campaigns.\\
\indent
The detection of PeV neutrinos by the IceCube Neutrino Observatory in Antarctica has started neutrino astronomy \citep{icecube_collaboration_evidence_2013}. These neutrinos are generated by either proton-$\gamma$ or proton-proton collisions in astrophysical sources. As a result, charged pions are produced that decay into muon and electron (anti-)neutrinos as well as leptons that can produce radio synchrotron and $\gamma$-ray IC emission that we can detect. Hence, unambiguous cross-matching with an electromagnetic source counterpart enables one to study acceleration and transport of PeV CRs in those extreme particle accelerators. This can be achieved by successful timing coincidence searches of transient electromagnetic signals or spatial coincidences of angular neutrino overdensities with known astrophysical sources. A coincident $\gamma$-ray and neutrino flare event from a blazar has been interpreted as evidence for the acceleration of PeV protons in such sources and put the hadronic model into focus \citep{the_icecube_collaboration_multimessenger_2018,icecube_collaboration_neutrino_2018}. Statistical clustering of neutrino events points to high-energy sources; in particular there is a 4.2$\sigma$ coincidence signal with an active galaxy \citep{icecube_collaboration_evidence_2022}. Because this source does not emit detectable very-high energy $\gamma$ rays, this could either imply that the neutrino radiation is emitted from the very dense, opaque central region around a massive black hole or that the identification with the active galaxy is spurious and results from a different source class. More sensitive data is required settle this issue. This will be enabled with the upgrade to IceCube-Gen2, which enlarges the detector by a factor of ten and improves energy and spectral neutrino measurements as well as by the Cubic Kilometre Neutrino Telescope \citep[KM3NeT,][]{adrian-Martinez_letter_2016}, an upcoming European research infrastructure that is planned to be situated in the Mediterranean Sea below sea level.

\clearpage

\section{Open questions and future directions}\label{Open questions}
The last decade has witnessed a dramatic progress in our understanding of the role of CRs in stellar and supermassive black hole feedback processes and how these processes shape galaxy formation and evolution. From the theoretical perspective, major advances in the field include progress in modelling CR transport from first-principles and the application of these results in the context of galactic wind launching and AGN feedback. From the observational perspective, data from a number of observatories (e.g., LOFAR, MeerKAT, Jansky VLA, \textit{Fermi}, H.E.S.S., VERITAS, \textit{Voyager}) and new and upcoming missions (e.g., SKA, CTA, JWST) have played and/or will continue to play a pivotal role in answering some of the fundamental outstanding questions in the field.\\
\indent
To conclude this review, we have identified a number of such outstanding problems in the field of CR transport and feedback that need to be addressed before we can move to the next level and consolidate our picture of how CRs dynamically impact the formation and evolution of galaxies and galaxy clusters. To mirror the order of the review, we list these problems starting with the issues related to CR and plasma physics and then progress from small-scale (ISM) simulations to cosmological structure formation simulations.

\subsection{Plasma physics and cosmic ray transport challenges}\label{plasma_physics}

\subsubsection{Plasma physics challenges}
Simulating the plasma kinetics of CR driven instabilities and wave damping with first-principle PIC codes is an enormous numerical challenge because of the large separation of spatial and temporal scales, caused primarily by because of the low CR-to-thermal background number densities, $n_\rmn{cr}/n_\rmn{i}\sim10^{-9}$, and the low Alfv\'en speed, $\varv_\rmn{a}/c=\Omega_\rmn{i,0}/\omega_\rmn{i}\sim10^{-4}$ in the ISM. This requires the development of fluid-PIC codes that can reliably model Landau damping of plasma wave energy into thermal energy of the corresponding background fluid components \citep[e.g.,][]{lemmerz_coupling_2023}, which allows one (i) to study the impact of inhomogeneities in the background plasma density and magnetic field on the instability growth rates, microscopic CR-wave scattering rates and thus CR transport descriptions and (ii) to understand the non-linear saturated regime of the interplay of various CR driven instabilities and wave damping processes (non-linear Landau damping and ion--neutral damping). Owing to the numerical requirement of resolving light characteristics in these approaches, the simulation sizes and achievable dimensionality is naturally limited and argues for complementing those simulations with more efficient MHD-PIC codes \citep[e.g.,][]{bai_magnetohydrodynamic-particle--cell_2015} that however precludes modeling the displacement current and other electron dynamics as well as Landau damping into separate electron and proton background plasmas. Nevertheless, these approaches will enable us to pursue three-dimensional simulations of CR transport in driven turbulence and thus to study the impact of turbulent damping on CR transport. We envision that this numerical progress would have to be complemented by an analytical derivation of turbulent damping from first principles, which will enable a calculation of the streaming instability growth rate in a turbulent background -- something very important that has not yet been achieved.\\
\indent
Clearly, (MHD-)PIC simulations can at most study idealized cases and will not be able to probe global astrophysical settings. This calls for going beyond MHD at global scales, and to develop ``extended MHD'' models, which include Landau closures. To this end, several routes offer promising ways forward that include coarse graining fluid-PIC simulations to derive effective CR scattering rates and transport coefficients and/or to adopt algorithms developed for magnetic fusion plasmas. Those take into account the weakly collisional nature of most astrophysical diffuse plasmas, which leads to anisotropies of the distribution functions, which excites microscale instabilities that provide additional scattering for particle transport processes.

\subsubsection{Building a self-consistent cosmic ray transport theory}
The primary and secondary CR spectra observed by AMS-02 in Fig.~\ref{fig:cosmic_ray_composition} can be successfully described by the leaky box model of CR transport with an empirical power-law energy dependence of the CR diffusion coefficient (Section~\ref{sec:leaky_box}). However, it is unclear how to reconcile these phenomenological models with the physics underlying CR transport theory as explained in Section~\ref{sec:spatial_transport}. The current microphysical theories of CR transport, including self-confinement caused by the streaming instability and extrinsic turbulence resulting from a cascade of MHD fast modes, do not agree with observations \citep{kempski_reconciling_2022}. On the one hand, because CR scattering sensitively depends on the local plasma conditions in the multiphase ISM, a combination of CR self-confinement and CR scattering on extrinsic turbulent modes can potentially reproduce the main CR spectral trends as well as the boron-to-carbon ratio. On the other hand, this could perhaps point to an incomplete understanding of CR transport theory, which may either have to be extended to include new CR driven instabilities \citep{shalaby_new_2021,shalaby_deciphering_2023} and/or to include the impact of turbulent intermittency with small-scale field reversals on particle propagation \citep{kempski_cosmic_2023,lemoine_particle_2023}, which may alter CR transport to become non-Brownian up to the magnetic coherence scale.

\subsubsection{Incorporating complete theory of cosmic ray transport in MHD simulations} 
One of the key next challenges will be to develop a complete theory of CR transport with an improved modeling of the kinetic physics in CR hydrodynamic models. This theory must include all relevant CR driven instabilities and separately account for a spectral description of CR momenta and plasma waves that provide CR scattering. At the same time, the adopted numerical algorithm must be computationally efficient enough so that it may find application in three-dimensional simulations of the ISM and of entire galaxies. 

\subsubsection{Cosmic ray transport and non-thermal signatures in the Milky Way and other galaxies}
The implementation of such an improved CR hydrodynamic theory will enable us to perform self-consistent spectral CR hydrodynamics simulations with two-moment CR transport in galaxies not only to explain local CR data (H, He, B, C, O, electrons; see Section~\ref{sec:CRprop}), but also non-thermal emission signatures in the radio band and at $\gamma$-ray energies in the Milky Way and external galaxies. Comparison to observational data (FIR--radio and FIR--$\gamma$-ray correlations as well as individual non-thermal emission spectra) will reveal weaknesses in the theory and enable a first-principle understanding of CR transport in the ISM and CGM. This will enable us to solve a number of open problems such as the leptonic vs.\ hadronic origin of the \textit{Fermi} $\gamma$-ray bubbles in the Milky Way (Section~\ref{sec:CR_driven_outflows}) and to make reliable predictions for the future SKA radio telescope as well as the CTA $\gamma$-ray observatory. Moreover, this will enable quantifying the contribution of star-forming galaxies to the isotropic (extra-galactic) $\gamma$-ray background (Section~\ref{Non-thermal emission from galaxies}). Finally, these developments will be critical for calibrating the calorimetric fraction, i.e., the CR energy fraction lost to radiation (and neutrinos) and the remaining part that is in principle able to provide energetic feedback.

\subsection{Astrophysical challenges}

\subsubsection{Effective cosmic ray transport near sources}
CR transport in the ISM is expected to be highly spatially and temporally variable (see Section~\ref{sec:acceleration_escape} and Section~\ref{dynamics_of_CR_near_sources}). On the one hand, if CR transport is severely suppressed in the vicinity of SNe, as suggested theoretically, e.g., by \citet{schroer_cosmic-ray_2022}, or observationally by \citet{jacobs_unstable_2023}, this would (i) have profound consequences for the phase space of the ISM (such as a significant suppression of the high-density end of density probability distribution function), (ii) limit local fragmentation (because of the buildup of large local CR pressure gradients, which effectively increase the Jeans mass), and (iii) alter star formation rates and galactic morphology \citep{semenov_cosmic-ray_2021}. On the other hand, CR sources (SNe, stellar wind shocks, protostellar jets) can occur in or near denser and cooler ISM environments, where ion--neutral damping can increase CR transport speed \citep[e.g., Section~\ref{gmc_penetration},][see also below]{armillotta_cosmic-ray_2022}. Elucidating the picture of how CRs migrate from their injection sites, where they are well confined, to the larger ISM and galactic scales, where CR transport is faster, is one of key outstanding challenges.

\subsubsection{Interactions of low-energy cosmic rays with molecular clouds}
Realistic modeling of the CR spectrum in small-scale ISM simulations enables us to quantify the variability and time evolution of CR ionization by low-energy CRs (Section~\ref{CR_ionization}). One of the key challenges is to formulate a self-consistent theory capable of predicting energy-dependent penetration  of molecular clouds by low-energy CRs. Most current models tend to significantly underpredict CR ionization rates compared to observations even under the most optimistic scenario of free-streaming CRs that penetrate deep into the clouds without significant scattering. 
In other words, current models fail to self-consistently link the low-energy (2--10 MeV) proton spectrum to the CR proton spectrum at energies exceeding several GeV. Possible solutions to this challenge include refining models of CR transport and magnetic confinement of CRs to regions that dominate the observables of CR ionization, as well as  considering models with the additional presence of low-energy CR sources internal to the molecular clouds such as shocks associated with protostellar jets.
Addressing these problems will not only help to improve our understanding of the impact of CR ionization on the chemistry and dust formation but will also allow us to study and quantify the ambipolar diffusion rate in low-ionization regions and proto-planetary disks.

\subsubsection{Launching galactic winds}

\paragraph{Impact of wave damping on launching of CR-driven winds.} The friction forces between ions and neutrals in a partially ionized medium lead to strong damping of Alfv\'en waves \citep{kulsrud_effect_1969,blasi_non-linear_2018}. This ion--neutral damping is the dominant damping process for densities $\gtrsim 10^{-2}~\rmn{cm}^{-3}$, which applies to the cold and warm neutral phases of the ISM, and leads to quick CR diffusion and built up of homogeneous pressure distribution within most of the ISM mass \citep{armillotta_cosmic-ray_2021}. As the ionization fraction increases in the surrounding lower-density ISM (in the warm ionized and hot phases and at larger heights above the disk), ion--neutral damping becomes less important so that the CR scattering rates increase. This lowers the macroscopic transport speed and causes a larger momentum transfer from CRs to the ambient plasma (see Section~\ref{wind_launching}). Implications of this dynamical re-coupling of CRs to the gas need to be fully explored in models of wind launching that simultaneously include detailed modeling of a multi-phase ISM, a dynamically active CR pressure coupled to the MHD, time-dependent resonant wave energy, and wave damping mechanisms. Furthermore, it remains to be seen whether this is compatible with the observed low level of CR anisotropy or whether this calls for a revision of CR transport theory.

\paragraph{Elucidating the role of CR bottlenecks.} 
Streaming CRs can render sound waves unstable and steepen the waves to generate shocks and thus density discontinuities and peaks that create CR bottlenecks. This leads to the formation of staircase-like distribution of CRs with CRs exerting strong forces on and heating of the gas only at the location of stair jumps, while radiative cooling can facilitate further formation of additional shocks and bottlenecks. Even in the isothermal case, adiabatic sound waves in a stratified atmosphere are unstable in the CR streaming-dominated regime, which leads to shocks that trigger bottlenecks and staircase-like distribution of CRs.
This complex physics can change the effective CR equation of state and affect galactic mass loss rates. In this regard, extending the studies of CR bottlenecks in one-dimensional \citep[see Section~\ref{Spherically-symmetric wind models} and \ref{CR Eddington limit};][]{tsung_cosmic-ray_2022,quataert_physics_2022} and two-dimensional \citep[see Section~\ref{multiphase}
and \ref{CR Eddington limit};][]{huang_cosmic-ray-driven_2022} to fully three-dimensional stratified atmospheres that are subject to radiative cooling may be fundamental for elucidating the true role of CR bottlenecks in driving winds, heating the gas, and shaping the observational properties of galactic outflows.

\paragraph{Coupling CR and radiation hydrodynamics and their relative role in wind launching.} 
Because CRs propagate through a multi-phase ISM, the next generations of full galaxy models need to couple (two-moment) CR hydrodynamics to radiation hydrodynamics simultaneously. This effort will facilitate direct and self-consistent comparisons of the relative role of CR and radiation forces in driving galactic winds (see Section~\ref{Phase_space_structure}). These models will need to account for non-equilibrium cooling, photo-electric heating via radiation transfer and SNe energy injection while enabling sufficient resolution to allow for the magnetic field growth via an efficient small-scale dynamo. This will clarify the role of CRs in suppressing star formation and allow for distinguishing the contributions of the various physical processes in the ISM to setting the volume and mass fractions of the different phases \citep{rathjen_silcc_2021,simpson_how_2023}. Importantly, this approach will lead to cosmological simulations of dwarf galaxies with solar-mass resolution in which empirical feedback prescriptions will be superseded by a ``first-principle'' approach to modeling feedback.

\subsubsection{Impact of cosmic rays on the CGM}
CR pressure support in the CGM depends critically on CR transport physics (Section~\ref{Cosmological effects} and \ref{sec:CGM_feedback}). In the slow CR transport regime, significant CR pressure gradients could build up in the disk and drive outflows, but catastrophic inelastic CR energy losses may limit wind driving. In the opposite regime of very fast CR transport, CRs do not spend enough time in the disk to accelerate the wind efficiently. Consequently, there may be an optimal range of effective CR transport speeds that maximizes both the impact of CRs on galactic wind driving and the CR pressure support in the CGM \citep[e.g.,][]{ruszkowski_global_2017, hopkins_effects_2021}. These considerations also have a significant bearing on the predictions for γ-ray emission from the inner CGM, with models potentially violating observational γ-ray constraints in the slow CR transport regime. Furthermore, they may have a considerable impact on X-ray properties of galaxies. Specifically, while predictions from CR feedback models lie close to the extrapolated relation between stellar mass and soft X-ray luminosity, they tend underestimate X-ray emission in the low star formation regime. Finally, if CRs provide a significant pressure support of the CGM, the thermal gas could be substantially cooler and exhibit a smoother density distribution. As a result, the CGM should be characterized by a different set of metal emission and absorption lines.\\
\indent
Several avenues of investigation could be pursued to address these challenges including (i) exploration of the impact of the suppression of CR diffusion near the sites of SNe that shows promise in terms of reconciling the predicted and observed star formation vs.\ $\gamma$-ray luminosity relations \citep[because the suppression leads to partial anti-correlation between CR and gas distributions, which reduces $\gamma$-ray luminosity;][]{semenov_cosmic-ray_2021}, (ii) comprehensive investigation of the impact of numerical approaches on the level of magnetic field amplification, which may affect the magnitude of CR streaming losses, (iii) formulating a self-consistent theory that could predict significant suppression of CR scattering rates compared to those expected in quasi-linear theory for the conditions characteristic of the CGM (potentially needed to suppress γ-ray emission), and predict whether CRs predominantly stream or diffuse (see also Section~\ref{plasma_physics} above), and (iv) pursuing cosmological simulations with these improved models for CR hydrodynamics and studying the impact of changing CR transport prescriptions on the resulting metal line absorption and emission lines. Improved theoretical models will hopefully have the potential to enable critical exploration of the impact of CRs on the CGM as a function of cosmic redshift and halo masses in a cosmological context, and eliminate the tension between theory and observations. 

\subsubsection{Cosmic rays and AGN jet feedback}
The acceleration of CRs in relativistic jets launched by AGN likely produces a CR proton component in radio bubbles. As these CRs escape from the bubbles and stream into the ambient ICM, they can heat the surrounding plasma via excitation and damping of Alfv\'en waves to the point where the energy deposition potentially offsets strong radiative losses of the gas in the centers of cool core galaxy clusters (see Section~\ref{agntheorysection}). Because there are other plausible heating mechanisms such as turbulent mixing and heating as well as dissipation of shocks and sound waves, the challenge in the upcoming years will be to identify the relative importance of the different heating processes in the context of the ``cooling flow problem,'' as well as to identify the physical reason for the emergence of the cool core/non-cool core bimodality in galaxy clusters. Significant effort will need to be devoted to assessing if the CR jet feedback models are in agreement with multiphase observational constraints (radio, millimeter SZ emission, H$\alpha$ emission, X-ray, $\gamma$-ray).

\subsubsection{Impact of cosmic rays on the reionization of the Universe}
In addition to CR-driven galactic winds and Alfv\'en wave heating in cool core clusters, there could be other avenues to explore a potentially significant impact of CRs on cosmic structure formation. The ability of primordial gas to cool below $\sim 10^4~\rmn{K}$ and eventually to form stars depends on the availability of molecular hydrogen. Collapsing cosmic structures at redshift $z\sim20$--40 dissipate gravitational energy associated with the formation of the first generation of (population III) stars or proto-galaxies and thereby generate strong collisionless shocks. These should be able to generate a substantial CR flux that could ionize the ambient neutral gas. This process produces free electrons that catalyze the formation of molecular hydrogen, thus accelerating cooling and fostering the formation of the next generation of star formation \citep{jasche_cosmic_2007}. Recent improvements in the spectral and spatial descriptions of CR hydrodynamics enables scrutinizing these early results.

\subsection{Concluding remarks}
To conclude, we note that extraordinary progress has been achieved in this field in the last decade. New advancements have allowed us to gain a much improved understanding of the significance of CRs in shaping baryonic feedback in galaxies and galaxy clusters. Nevertheless, these accomplishments have simultaneously raised many more questions and paved the way for further research. We expect this field to become even more active in the upcoming decade to hopefully deliver answers to (some of) the challenges summarized here.

\begin{acknowledgement}
We thank the referee -- Ellen Zweibel -- for very a careful reading of our manuscript and for her excellent suggestions that helped to improve this review. We would also like to thank Timon Thomas and Mohamad Shalaby for comments on an earlier version of the manuscript and Lukas Platz for sending us the revised version of the all-sky \textit{Fermi} reconstructions. We thank Vladimir Lenok for help with preparing Figure 1 and David Maurin for help with interpreting and extracting data from the \href{https://lpsc.in2p3.fr/crdb/}{Cosmic-Ray Data Base}. 
We thank Tom Abel, Andrea Botteon, Hsiao-Wen Chen, Sean Johnson, and Martin Lemoine for helpful suggestions and clarifications.
MR thanks Volker Springel and the Max Planck Institute for Astrophysics (MPA) in Garching for their hospitality during his sabbatical stay at MPA. MR acknowledges Forschungstipendium from MPA and support from National Aeronautics and Space Administration grants 80NSSC20K1541 and 80NSSC20K1583 and National Science Foundation Collaborative Research Grants AST-1715140 and AST-2009227. CP acknowledges support by the European Research Council under ERC-AdG grant PICOGAL-101019746.
\end{acknowledgement}

\phantomsection
\addcontentsline{toc}{section}{References}
\bibliographystyle{spbasic-FS-etal}
\bibliography{extracted.bib}

\end{document}